\documentclass[printer,letterpaper]{tADP2e}
 \usepackage{graphicx}
\usepackage{algorithmic}
              
\newcommand{\beq}{\begin{equation}}
\newcommand{\eeq}{\end{equation}}
\newcommand{\bdis}{\begin{displaymath}}
\newcommand{\edis}{\end{displaymath}}
\newcommand{\bea}{\begin{eqnarray}}
\newcommand{\eea}{\end{eqnarray}}
\newcommand{\barr}{\begin{array}}
\newcommand{\earr}{\end{array}}
\begin{document}
\title{Statistical Models of Fracture}
\author{Mikko J. Alava$^1$, Phani K. V. V. Nukala$^2$, and Stefano Zapperi$^3$
\thanks{\newline\centerline{\tiny{$^1$ Laboratory of Physics,Helsinki University of Technology, FIN-02015 HUT,
Finland}}}
     \thanks{ \newline\centerline{\tiny{  $^2$ Computer Science and Mathematic Division,
Oak Ridge National Laboratory, Oak Ridge, TN 37831-6164, USA}}}
    \thanks{\newline\centerline{\tiny  $^3$ CNR-INFM, Dipartimento di Fisica,
    Universit\`a "La Sapienza", P.le A. Moro 2
        00185 Roma, Italy }}}

\maketitle

\begin{abstract}
Disorder and long-range interactions are two of the key
components that make material failure an interesting playfield
for the application of statistical mechanics. The cornerstone
in this respect has been lattice models of the fracture 
in which a network of elastic beams, bonds or electrical fuses 
with random failure thresholds are subject to an 
increasing external load. These models describe on a qualitative
level the failure processes of real, brittle or quasi-brittle
materials. This has been particularly important in solving 
the classical engineering problems of material strength:
the size dependence of maximum stress and its sample to sample
statistical fluctuations. At the same time, lattice models pose many 
new fundamental questions in statistical physics, such as the relation between
fracture and phase transitions. Experimental
results point out to the existence of an intriguing  crackling noise
in the acoustic emission and of self-affine fractals
in the crack surface morphology.  Recent advances in computer power have enabled considerable
progress in the understanding of such models. Among these partly
still controversial issues, are the scaling and size effects in material
strength and accumulated damage, the statistics of  
avalanches or bursts of microfailures, and the morphology of the crack 
surface. Here we present an overview of the results obtained with lattice models
for fracture, highlighting the relations with statistical physics theories and more 
conventional fracture mechanics approaches. 
\end{abstract}

\newpage
\centerline{\bfseries Table of contents}\medskip
\begin{itemize}
\item[1.]    Introduction 3
\item[2.]   Elements of fracture mechanics 6\\
\hspace*{7pt} {2.1.}  Theory of linear elasticity 6\\
\hspace*{7pt} {2.2.} Cracks in elastic media 8\\
\hspace*{7pt} {2.3.} The role of disorder on material strength 11\\
\hspace*{7pt} {2.4.} Extreme statistics for independent cracks 12\\
\hspace*{7pt} {2.5.} Interacting cracks and damage mechanics 14\\
\hspace*{7pt} {2.6.} Fracture mechanics of rough cracks 17
\item[{3.}]   Experimental background 23\\
\hspace*{7pt}{3.1.}  Strength distributions and size effects 24\\
\hspace*{7pt}{3.2.}  Rough cracks 27\\
\hspace*{7pt}{3.3.}  Acoustic emission and avalanches 35\\
\hspace*{7pt}{3.4.}  Time-dependent fracture and plasticity 39
\item[{4.}]   Statistical models of failure 42\\
\hspace*{7pt}{4.1.} Random fuse networks: brittle and plastic 43\\
\hspace*{7pt}{4.2.} Tensorial models 50\\
\hspace*{7pt}{4.3.} Discrete Lattice versus Finite Element Modeling of fracture 52\\
\hspace*{7pt}{4.4.} Dynamic effects 56\\
\hspace*{7pt}{4.5.}Atomistic simulations 61
\item[5.]Statistical theories for fracture models 63\\
\hspace*{7pt}{5.1.} Fiber bundle models 64\\
\hspace*{7pt}{5.2.} Statistical mechanics of cracks: fracture as a phase transition 71\\
\hspace*{7pt}{5.3.} Crack depinning 78\\
\hspace*{7pt}{5.4.} Percolation and fracture 81
\item[6.] Numerical simulations 89\\
\hspace*{7pt}{6.1.} The I-V characteristics and the damage variable 91\\
\hspace*{7pt}{6.2.} Damage Distribution 95\\
\hspace*{7pt}{6.3.} Fracture strength 107\\
\hspace*{7pt}{6.4.} Crack roughness 114\\
\hspace*{7pt}{6.5.} Avalanches 118
\item[7.] Discussion and outlook 121\\
\hspace*{7pt}{7.1.} Strength distribution and size effects 124\\
\hspace*{7pt}{7.2.} Morphology of the fracture surface: roughness exponents 126\\
\hspace*{7pt}{7.3.} Crack dynamics: avalanches and acoustic emission 127\\
\hspace*{7pt}{7.4.} From discrete models to damage mechanics 128\\
\hspace*{7pt}{7.5.} Concluding remarks and perspectives 128
\item[] Appendix: Algorithms 129
\item[] References 140
\end{itemize}
\newpage

\section{Introduction}

\begin{figure}[hbtp]
\centerline{\includegraphics[width=10cm]{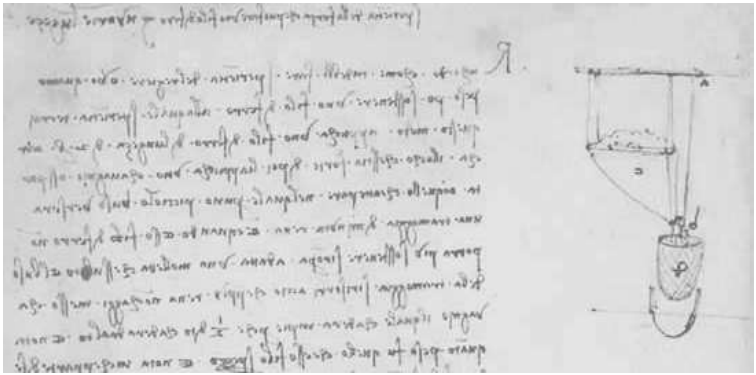}}
\caption{The setup devised by Leonardo da Vinci to test the strength
of metal wires. One end of the wire is suspended and the other end is attached 
to a basket that was slowly filled with sand. When the wire breaks, one can weigh the sand collected 
in the basket, determining the strength. From the original
notebook \protect\cite{leonardo40}.}
\label{fig:leo}
\end{figure}

Understanding how materials fracture is a fundamental
problem of science and engineering even today, although the effects of geometry,
loading conditions, and material characteristics on the strength of
materials have been empirically investigated from antique
times. Probably the first systematic mathematical analysis of the
problem dates back to the pioneering works of Leonardo da Vinci \cite{leonardo40} and
Galileo Galilei \cite{galilei58}. In Leonardo's notebooks one finds an interesting description
of a tension test on metal wires \cite{leonardo40}. Attaching a slowly filling 
sand bag to an end of the wire, Leonardo noticed that longer wires failed 
earlier (see Fig.~\ref{fig:leo}). A simple analysis based on continuum mechanics, however, would suggest
that wires of the same section should carry the same stress
and hence display equal strength. This lead some later commentators
to hypothesize that when Leonardo wrote {\it smaller length}, what he really meant 
was {\it larger diameter} \cite{parsons39}. As noted in Ref.~\cite{lund01}, there is
a simpler explanation: continuum mechanics treats the material as
homogeneous, but this is often very far from the truth, especially
for metal wires forged in the Renaissance.  Disorder has profound
effects on material strength, since fracture typically nucleates
from  weaker spots like preexisting microcracks or voids. The 
longer the wire, the easier it is to find a weak spot, hence the
strength will decrease with length.

What we have just described is probably the first experimental
verification of size effect in fracture. Size effects are rooted in
disorder and hence necessitates a statistical treatment.  In
particular, different samples of the same material could generally be
expected to display fluctuations in the strength. The importance of
defects and fluctuations in fracture was first realized in the early
twenties by the work of Griffith \cite{griffith20} and later
formalized from the statistical point of view by Weibull
\cite{weibull39}. An excellent discussion of the history of strength
up to the middle of the twentieth century can be found in the book by
Timoshenko \cite{timoshenko}.  At present the importance of a
statistical treatment in determining the strength of materials is
certainly widely accepted, but despite this fact standard textbooks on
fracture mechanics devote little attention to this problem. In
addition experimental evaluations of strength distribution and other
statistical properties are relatively scarce in the literature.

Besides the engineering aspects of the problem, in the last twenty years
the statistical properties of fracture have attracted a wide interest
in the statistical physics community \cite{herrmann90,chakrabarti,hansen00}. 
In this context, fracture can be
seen as the outcome of the irreversible dynamics of a long-range
interacting, disordered system. Several experimental observations have
revealed that fracture is a complex phenomenon, described by scale
invariant laws \cite{mishna97}. Examples notably include the acoustic emission
measured prior to fracture and the roughness of the fracture surface.
The observation of scaling makes it attractive to discuss breakdown 
processes in terms of statistical mechanics and in particular to
search for analogies with phase transitions and critical phenomena.
These give rise to probability distributions that bear resemblance
with the empirical observations on fracture, in particular ``power-law''
ones.

The statistical physics approach to fracture is often based on
numerical simulation of lattice models, which allow for a relatively
simple description of disorder and elasticity. The elastic medium is
represented by a network of springs or beams.  The disorder is modeled for
instance by imposing random failure thresholds on the springs or by removing a
fraction of the links. In this way the model retains the long range
nature and the tensorial structure of the interactions. The local
displacements can then be found by standard method for solving coupled
linear equations.  The lattice is loaded imposing appropriate
boundary conditions and the fracture process can be followed step by
step, in a series of quasi-equilibria.
The cornerstone in this respect has been for the last twenty
years the Random Fuse Network (or Model, RFM) \cite{deArcangelis85}, a lattice model of the fracture
of solid materials in which as a further key simplification vectorial
elasticity has been substituted with a scalar field.

As damage accumulates, one observes intriguing {\em history effects}:  
in brittle failure, microcracks grow via the failure of the weakest
elements \cite{weibull39}. In ductile fracture, dislocations or plastic deformation
accumulates, and in viscoelastic materials there is a variety of
complex relaxation mechanisms of stress in addition to irreversible
failure. Meanwhile, the stress field evolves and develops a
more and more inhomogeneous structure. This arises in the simplest
case from microcrack interactions that can shield each other or act
jointly to enhance the local stress at various points in the material.

In addition to numerical simulations, lattice models are useful to
obtain new theoretical ideas and approaches to the problem. 
The aim is to develop a
probabilistic description of fracture involving the interplay between
``frozen'' properties, such as microcracks, and ``thermal'' noise.  The 
former is independent of time, quenched into a sample and is only
affected by whatever external influences the sample is subjected to
and their internal consequences (consider growing microcracks). The
thermal noise comes from environmental effects and the coarse-grained
fluctuations at the atomistic level.
In this respect, so-called fiber bundle models represent a useful idealization and the
basis for more complex theoretical developments \cite{pierce26,daniels45}. The exactly solvable
global load sharing model can be seen as a mean-field version of the
RFM. As it is customary with phase transitions and critical phenomena,
mean-field theory provides a qualitative picture of the transition.
In general, however, for elastic, brittle failure the 
critical stress is ruled by the behavior of the
extremal values of the stress distribution, the highest stress(es)
present in any sample. This means that the
highest moments of the distribution are crucial, contrary to what
happens for other macroscopic properties such as the elastic moduli,
which depend only on average mean-field quantities. One should be aware
of this problem when trying to extend the results of fiber bundle
models to more realistic cases.

At a more application-oriented level, network models provide a valid 
alternative to standard finite element calculations, especially in cases where structural
disorder and inhomogeneities are particularly strong \cite{krajcinovic00}.  For instance,
models of this kind are extensively used to model concrete
\cite{vanmierbook,bazantbook,vanmier03,lilliu03,schlangen96}. This approach can be relevant for the
predictability of structural failure, the design of stronger materials
and optimization of structural composition.

The scope of this review is to present an introduction to statistical
models of fracture. At the same time, the goal is to provide an overview of the
current understanding gained with this approach. After at least twenty years of
intense research in the subject, we feel the need for a summarizing
work, complementing the early insight gained in the past with more
reliable results obtained by large scale simulations, which became
possible due to recent advances in computer power and novel computational 
algorithmic techniques. This allows to
clarify some controversial issues, such as the scaling in the 
damage dynamics, the avalanches or bursts of microfailures, and the scaling
of the fracture surface in lattice models. The access to better data
and experimental progress has also created new unsolved problems.
This work gives us thus an opportunity to outline both experimental and theoretical prospects.
A research path that will not be pursued here 
is that of dielectric breakdown \cite{niemayer84}, which has
many parallels with the RFM and the role of disorder is likewise important 
(see e.g. \cite{duxbury88,chakrabarti}). 

The review is organized as follows: a brief introduction to elasticity
and fracture mechanics is reported in Sec.~\ref{sec:frac-mec}. While
this does not represent by any means an exhaustive account of this
very extended field, it has the primary purpose to set the formalism
and the basic notation to be used in the following. In
Sec.~\ref{sec:exp} we discuss the experimental background that
motivates the introduction of statistical models of fracture. Thus
particular attention will be devoted to those aspects that call for a
statistical approach, such as the strength distributions and the
related size effects, the crack roughness, and the acoustic
emission. In Sec.~\ref{sec:models} we proceed with a detailed
description of network models for fracture, starting from the scalar
RFM up to more complex beam models. We will highlight the connections
with standard finite elements methods and atomistic calculations and
discuss dynamic effects. This complements the main emphasis here, of
general theories and scaling laws by pointing out what issues are
left out on purpose from the models at hand.

The main theoretical approaches are reviewed
in Sec.~\ref{sec:theory} starting from fiber bundle models as an
illustration.  We next discuss the general properties of phase
transitions and point out the connections for fracture. Finally, we
discuss the role of disorder in fracture model and introduce
percolation concepts.  Sect.~\ref{sec:simul} is devoted to a
discussion of recent simulation results.
We report results for the constitutive
behavior, fracture strength and damage distributions, size effects,
damage localization, crack roughness and avalanches.  The discussion
section (Sec.~\ref{sec:concl}) outlines current challenges and
possible future trends. An Appendix includes a description of the main
numerical algorithms. 

\section{Elements of fracture mechanics}
\label{sec:frac-mec}

Fracture mechanics describes the formation and propagation of cracks
in terms of macroscopic field equations, starting from the theory of
linear elasticity.  It is not our aim to review here the broad field
of linear elastic fracture mechanics (LEFM). We provide only a short
summary of the main concepts defining the basic terminology to be used
in the following sections. In addition, we present a general
discussion of the role of disorder in fracture.

\subsection{Theory of linear elasticity}
\label{sec:lin-elast}
The theory of linear elasticity describes the small deformations
of solid bodies under the action of external forces.
Consider a solid at rest and parameterize it by a set of coordinates
$\vec{r}$. Assuming that there is a direct relation
between the coordinates $\vec{r'}$ 
of the deformed solid with those of the undeformed solid, we 
define the displacement field as $\vec{u}\equiv \vec{r}\;'-\vec{r}$.
It is then convenient to introduce the the symmetric {\it strain tensor} as
\begin{equation}
\epsilon_{ik}\equiv \frac{1}{2}\left(\frac{\partial u_i}{\partial x_k}+
\frac{\partial u_k}{\partial x_i}\right).
\end{equation}

When a solid deforms under the action of external forces, 
interactions at the molecular scale provide restoring forces 
that tend to bring the solid back to its reference configuration. Each of the
material particles of the solid is in equilibrium under the action of both 
internal force interactions and external applied forces. 
Since all the internal forces should cancel each other
due the action-reaction principle, when the equilibrium of a volume
element $dV$ is considered, we are left only with surface tractions that act 
on the surface $dS$ that contains the volume element $dV$ and 
externally applied body forces (such as gravity) that act over the volume element $dV$.
In other words, the only non-vanishing contribution to the force 
should come from the tractions applied at the surface $dS$ and the external body forces that act 
over volume $dV$. Thus the external force vector ${\bf F}$ with components ${\bf F}_i$ is given by 
\begin{equation}
{\bf F}_i = \int_V {\bf b}_i dV + \int_S {\bf t}_i dS \label{forces}
\end{equation}
where ${\bf b}_i$ are the components of body forces per unit volume, and ${\bf t}_i$ are 
the components of the surface tractions (force per unit surface area). By notation, the bold symbols 
refer to vector or tensor quantities. Based on Cauchy's hypothesis, 
the surface tractions are given by ${\bf t}_i = {\bf n}_j {\bf \sigma}_{ij}$, where ${\bf n}$ is 
the unit outward normal to the surface $dS$ and ${\bf \sigma}$ is the Cauchy stress tensor with 
components ${\bf \sigma}_{ij}$. Using the divergence theorem, Eq. \ref{forces} can be 
written as ${\bf F} = \int_V \left[{\bf b} + \nabla \cdot {\bf \sigma}\right] dV$ (where $\nabla$ is the gradient 
operator), and in component form as
\begin{equation}
{\bf F}_i = \int_V {\bf b}_i dV + \int_V \sum_{j} \frac{\partial {\bf \sigma}_{ij}}{\partial {\bf x}_j} \label{eqbm}
\end{equation}
The equilibrium condition for the body implies ${\bf F}={\bf 0}$, which
in terms of the stress tensor can be written as 
\begin{equation}
\sum_j \frac{\partial {\bf \sigma}_{ij}}{\partial {\bf x}_j} + {\bf b}_i = 0.
\label{eq:stress_eq1}
\end{equation}
When the body forces ${\bf b}$ are absent, Eq. \ref{eq:stress_eq1} reduces to 
\begin{equation}
\sum_j \frac{\partial {\bf \sigma}_{ij}}{\partial {\bf x}_j} = 0.
\label{eq:stress_eq}
\end{equation}
 
In the limit of small deformations, there is a linear relation between
stress and strain
\begin{equation}
{\bf \sigma}_{ij}=\sum_{k,l} {\bf C}_{ijkl} {\bf \epsilon}_{kl},
\label{eq:hooke0}
\end{equation}
where ${\bf C}_{ijkl}$ is the elastic moduli tensor.
Eq.~(\ref{eq:hooke0}) is the most general linear expression
relating the strain and the stress tensor, but it
becomes much simpler in the case of isotropic media:
\begin{equation}
{\bf C}_{ijkl}=\lambda \delta_{ij} \delta_{kl}+ \mu
(\delta_{ik} \delta_{jl} + \delta_{il} \delta_{jk})
\end{equation}
where the parameters $\lambda$ and $\mu$ are known as Lam{\`e}
coefficients, and $\delta_{ij} = 1$ when $i = j$ and is $0$ otherwise. 
It is often convenient to write the strain tensor as
the sum of a compressive strain, involving a volume change, and 
a shear strain, which does not imply any volume changes. Then
the Hooke laws can be written as 
\begin{equation}
{\bf \sigma}_{ij}=K\delta_{ij}\sum_k {\bf \epsilon}_{kk} + 2\mu 
({\bf \epsilon}_{ij}-\frac{1}{3}\delta_{ij} \sum_k {\bf \epsilon}_{kk}).
\label{eq:hooke1}
\end{equation}
where $K$ is the bulk modulus and $\mu$ is the shear modulus.
Finally, another standard formulation of the Hooke law
involves the Young modulus $E\equiv \frac{9K\mu}{\mu+3K}$ and
the Poisson ratio $\nu\equiv 
\frac{1}{2}\left(\frac{3K-2\mu}{3K+\mu}\right)$
\begin{equation}
{\bf \sigma}_{ij}= \frac{E}{(1+\nu)}\left({\bf \epsilon}_{ij}+
\frac{\nu}{1-2\nu}\delta_{ij}\sum_k {\bf \epsilon}_{kk}\right).\label{eq:hooke3}
\label{eq:hooke2}
\end{equation}
The equation for stress equilibrium for an isotropic elastic medium
can be written, combining the Hooke law with Eq.~(\ref{eq:stress_eq}),
in terms of the displacement field $\vec{u}$ the Lam{\'e} coefficients
$\lambda$ and $\mu$ as
\begin{equation}
(\lambda+\mu) \vec{\nabla} (\vec{\nabla}\cdot \vec{u})+\mu \nabla^2 \vec{u}=0,
\label{lame}
\end{equation}
with appropriate boundary conditions. Note that we have used both $\vec(\dot)$ notation 
and the bold face notation interchangeably to denote a vector quantity.

\subsection{Cracks in elastic media}
\label{sec:elastic_cracks}

In mathematical terms, a crack can be seen as a boundary
condition for the elastic strain field, with the additional
complication that the precise shape of the crack is in principle
unknown, and the shape depends on the stress field acting on it. To overcome
this formidable problem, one can prescribe the crack geometry and
then compute the stress field. This approximation can provide a
useful guidance on the mechanism of crack propagation. 

\begin{figure}[t]
\begin{center}
\includegraphics[width=10cm]{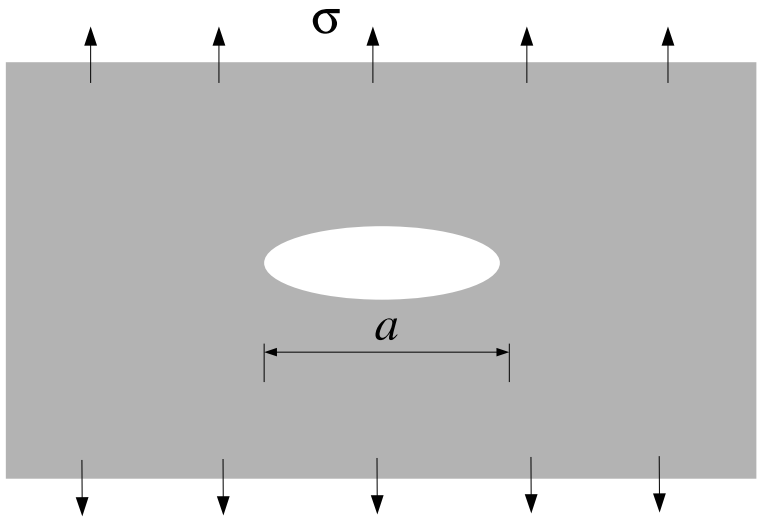}
\end{center}
\caption {An elliptic crack in a two dimensional medium.}
\label{fig:elliptic}
\end{figure}

A simple illustrative example of this procedure is provided by the solution
of the stress field around an elliptic crack placed under tensile stress
in an infinite two dimensional medium (see Fig.~\ref{fig:elliptic}).
The solution of the elastic equation of equilibrium yields for
instance  the stress concentration at the edge of the ellipse
\begin{equation}
K_T\equiv \frac{\sigma_{max}}{\sigma} = 1+\frac{2a}{b}.
\label{eq:KT}
\end{equation}
where $\sigma$ is the nominal applied stress and $\sigma_{max}$ is
the maximum stress, and $a$ and $b$ are major and minor radii of the ellipse. 
In reality cracks are not elliptical
but typically display a sharp tip. If we rewrite Eq.~(\ref{eq:KT})
in terms of the curvature radius $\rho\equiv b^2/a$, we see
that $K_T$ diverges as $1/\sqrt{\rho}$ as $\rho \to 0$, and hence 
the concept of stress concentration becomes useless.

\begin{figure}[t]
\begin{center}
\includegraphics[width=10cm]{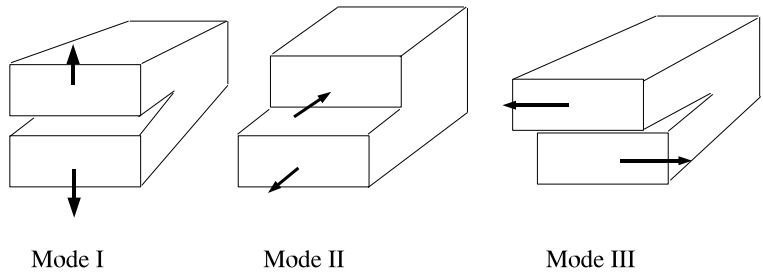}
\end{center}
 \caption {The three modes of fracture.}
\label{fig:modes}
\end{figure}

It was first shown by Irwin that the stress around a crack tip follows 
\begin{equation}
\sigma_{ij} = \frac{K_T}{\sqrt{2\pi r}} f_{ij}(\theta,\phi)
\label{sif}
\end{equation}
where $K_T$ is the {\em stress intensity factor} and $f_{ij}$ is a crack-tip 
function in terms of the angular variables $\theta$ and $\phi$.  The
particular values of $K_T$ and the functional forms of $f_{ij}$ depend on the particular loading
conditions (i.e. the way stress is applied to the body).  In this
respect, it is convenient to identify three fracture modes (see
Fig.~\ref{fig:modes}) corresponding to tensile (I), shear (II) and
tearing (III) conditions (i.e., $K_T$ equals $K_I$ in mode I, $K_{II}$ in mode II, and 
$K_{III}$ in mode III).  A generic loading rule can typically be
split into an appropriate combination of the three modes.  The
presence of singularity around the crack tip, however, is an artifact
of linear elasticity. Real materials can not sustain arbitrarily large
stresses and in general close to the crack tip deformation is ruled by
non-linear and inelastic effects. In practice it is customary to
define a fracture process zone (FPZ) as the region surrounding the
crack where these effects take place, while far enough from the crack
one recovers linear elastic behavior, that is the long-range decay of
perturbed displacement fields. 

Knowing the stress field around a crack does not tell us directly how
the fracture process would develop, but this information is necessary
to build a model. A simple and clear argument to understand the stability of
a crack was provided by Griffith almost a century ago \cite{griffith20}.
The idea is to compute the energy  needed to open a crack  
and see when this process becomes favorable.
Considering for simplicity a two dimensional geometry, the elastic
energy change associated with a crack of length $a$ is given by 
\begin{equation}
E_{el} = \frac{-\pi \sigma^2 a^2}{2E}, 
\end{equation}
where $\sigma$ is the applied stress. Forming a crack involves the
rupture of atomic bonds, a process that has an energy cost proportional
to the crack surface
\begin{equation}
E_{surf}=2 a \gamma,
\end{equation}
where $\gamma$ is the surface tension. Griffith noted that a crack
will grow when this process leads to a decrease of the total energy
$E=E_{el}+E_{surf}$. Thus the criterion for crack growth can 
be expressed as 
\begin{equation}
\frac{d E}{ d a} =  -\pi \sigma^2 a/E +2 \gamma <0,
\label{eq:griffith}
\end{equation}
which implies that under a stress $\sigma$ cracks of
size $a > a_c \equiv 2 \gamma E/(\pi \sigma^2)$ are unstable.
Equation~(\ref{eq:griffith}) can also be written as
$\sigma \sqrt{\pi a} > \sqrt{2E\gamma}$ and, noting that
$K=\sigma \sqrt{\pi a}$ is the stress intensity factor
of an infinite plate, we can reformulate the crack growth 
criterion in terms of a critical 
stress intensity factor $K_c \equiv\sqrt{2E\gamma}$, also
known as the material toughness (i.e. the crack grows when $K>K_c$).

\subsection{The role of disorder on material strength}
\label{sec:disorder_role}

The influence of randomness on fracture can have many disguises. It is
useful to leave aside inelastic effects for awhile, and consider only
brittle or ``quasi-brittle'' processes. In this case, we can
generalize the discussion of section~\ref{sec:elastic_cracks}, using
the concepts of long-range elastic fields, critical energy or
crack size. There are several independent subcases that present
different physics: one may consider a single crack and
analyze the effect of disorder on material strength or, alternatively, one
could study the dynamics of a whole crack population. 
In general, we can identify some key issues to address:

\begin{enumerate}
\item what is the effect of disorder on the stability of a single crack,
and how does the Griffith argument get modified?

\item in the presence of many cracks, how is the physics
changed from the case of a single crack?

\item what is the effect of disorder on the propagation of
a single crack?

\item what are the concepts, and statistical laws that operate in the
case of many cracks of which no single one dominates over
the others?

\end{enumerate}

These four fundamental questions also touch upon a number
of engineering mechanics and materials science issues, such
as fracture toughness, crack arrest, defect populations,
damage mechanics, and failure prediction. 

\begin{figure}[t]
\begin{center}
\includegraphics[width=10cm]{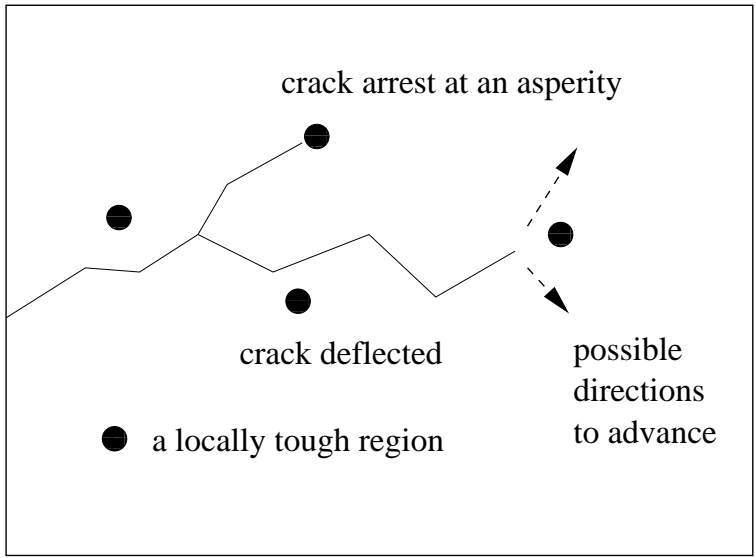}
\end{center}
 \caption {One-dimensional example of a crack propagating
in a disordered environment. Included are crack arrest
due to strong regions, deflection, branching and wandering.}
\label{fig:1ddemo}
\end{figure}

The first two issues from the above list deal with {\em size scaling}
and the statistics of strength. Figure ~\ref{fig:1ddemo} illustrates the
schematic advancement of a two-dimensional crack from a pre-existing flaw (or a
notch in a test). The presence of variation in the material structure
can be presented ideally - in brittle materials - by locally varying elastic
moduli, so that the stress-field becomes more complicated than the
homogeneous LEFM solution. It can also be seen as a crack surface
energy $\gamma_s(\vec{x})$ which has a fluctuating component.  The
issue of size effects amounts to computing the strength of a sample of
linear size $L$. The answer of course depends on whether the sample
has one dominating crack or more number of microcracks or flaws.

In the case of a single dominating defect, the Griffith argument can be easily
modified to account for a spatial variation of the elastic coefficients. The
technical problem is to consider the distributions of the surface energy $\gamma_s$
and elastic modulus $E$, or, even the presence of dislocations in the
sample. As a result, one gets instead of the deterministic critical
crack length (for constant $E$ and $\gamma_s$) a probability for the
crack to be stable under a given external stress. In other words, we
can characterize the problem by a {\em strength distribution} which in
the case of a single crack crucially depends on the microscopic
randomness. For instance, Arndt and Nattermann \cite{arndt01} showed that 
an uncorrelated random $\gamma_s$ results in an exponential strength distribution.
In general, however,  $\gamma_s$ could also display spatial correlations, 
as measured by the two point correlator $\langle \gamma_s (\vec{x}) \gamma_s
(\vec{y}) \rangle$, where the brackets denote a statistical average.
Correlations in $\gamma_s$ imply that an increased crack opening will
cost energy that not only depends on $\gamma_s$ at a single location 
but also depends on the presence of such correlations.

\subsection{Extreme statistics for independent cracks}
\label{sec:extr_stat}
\begin{figure}[t]
\begin{center}
\includegraphics[width=10cm]{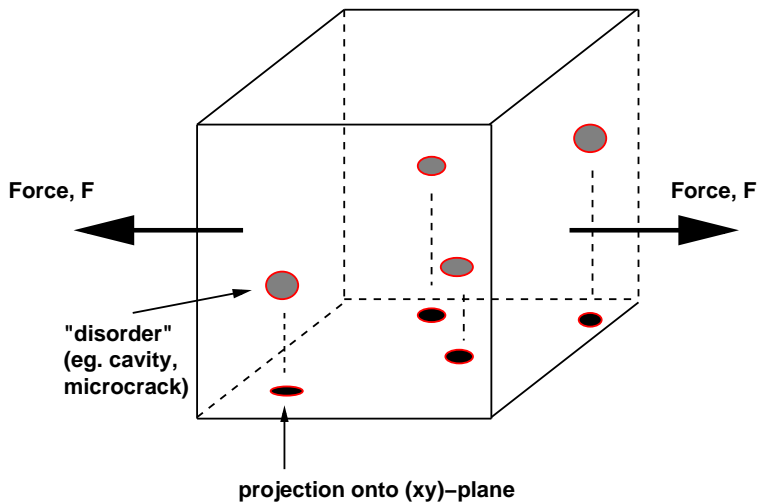}
\end{center}
 \caption {An example of a dilute defect population. These cracks
each of some size $l$ are assumed in the simplest case to have no influence on
each other, and they have a linear size distribution $P(l)$.}
\label{fig:cloud}
\end{figure}

In the presence of many non-interacting defects, it becomes apparent that the physics
is related to {\em extreme statistics} (see Fig.~\ref{fig:cloud}). The
simplest starting point is to find the largest defect - or the most
dangerous one - in the system. Later on in the context of random fuse
networks we will return to this argument in more detail. If the
critical crack fails instantaneously, there is no pre-critical crack
growth, and the cracks do not interact, then this is a sufficient
condition for determining the statistical strength properties. In
other words, we want to consider a dilute gas of non-interacting
defects, and find the worst one using a Lifshitz type argument 
\cite{weibull39}.

Consider a collection of $N_d$ defects, and assign to each of these
a failure strength $\sigma_i$, where $i = 1,\ldots,N_d$. 
The whole sample fails if one of these defects can 
not stand the applied stress. Thus for the cumulative strength
distribution $F(\sigma,N_d)$ one has that
\begin{equation}
1-F(\sigma) = \prod_{i=1}^{N_d} \left[1 -P_i(\sigma)\right]. 
\end{equation}
If the cracks are independent so that $\sigma_i$ are random
variables distributed according to $P_i=P(\sigma)$, then it follows that 
\begin{equation}
1-F(\sigma) \simeq \exp[-V \rho P(\sigma)]
\end{equation}
where $\rho$ is the crack density and $V$ the sample
volume. The crucial ingredients
here are the $P(\sigma)$ and $N_d$ since they rule both the asymptotic
form of the distribution $F$ as well as its form for finite $N_d$.
Of these two issues, more interesting is the behavior of $F(\sigma)$
in the large $N_d$ limit. Then, the question boils down to the
actual distribution of the $P(\sigma)$. For narrow distributions,
one has the {\em Gumbel} scaling form
\begin{equation}
F(\sigma) \sim 1 - \exp{[- V \exp{p(\sigma)}]}
\label{gumbel}
\end{equation}
where the function $p(\sigma)$ follows from $P(\sigma)$ and depends on
two things: the stress enhancement factor of a typical defect of
linear size $l$, and the size distribution $P(l)$ of the
defects. For example, an exponential $P(l)$ will thus result in this
{\em Gumbel} scaling form. 

The other possibility is to use a distribution $P(l)$ that makes
the local stresses at crack tips to have a wide distribution as
well. Using a length distribution $P(l) \sim l^{-g}$  
changes the limiting distribution of the sample strength to a 
Weibull distribution. The 
reason is simple: assuming that the stress $\sigma$ at the crack tips 
grows as a power of crack length $l$, then the $P(\sigma)$ also 
becomes a power law distribution since $P(l)$ is a power law distribution. 
As a result, the celebrated Weibull distribution \cite{weibull39}
(or its 3-parameter version) arises:
\begin{equation}
F(\sigma) \sim 1 - \exp{(-V \sigma^m / A^m)}
\label{weibull}
\end{equation}
where $A$ is a normalization constant, and $m$ is the Weibull
modulus, which can be related to the strength of a defect of size $l$. In particular,
we have that if $\sigma \sim l^{-1/2}$ then $m=2g$. As usual, a small $m$
implies a heterogeneous material/sample as $g$ is then small as well.
In any case, the distributions Eqs. (\ref{gumbel}), (\ref{weibull})
allow now to distinguish between microscopic details in experiments,
and predict the size-scaling of average strength. The Gumbel-form is
apt to give a logarithmic size-effect, but note that this is only true in
the absence of notches or dominating defects (see Sec. \ref{sec:notch}).

\subsection{Interacting cracks and damage mechanics}
\label{sec:damage}
The discourse on strength changes character if one allows for slow
{\em average} crack growth. Here, we would like to focus on the
case in which the fracture dynamics is continuous and slow, so that
in spite of possibly very fast local developments,
a sample can be considered in a quasi-equilibrium. This means that
the sample internal state is either independent of time or evolves slowly
with respect to the fast time-scale of the fracture process. 
Even in these conditions, treating explicitly the dynamics
of interacting microcracks represents a formidable task:
the growth of a microcrack can be either inhibited/screened or accelerated due to the presence
of a neighboring microcrack. The outcome in general depends on the loading 
conditions, the crack geometry and other details, as discussed 
in the engineering mechanics literature.

To avoid all these complications, one can disregard the dynamics of individual
microcracks and treat the damage at a coarse grained 
scale. This approach, known as "damage mechanics" \cite{Lemaitre,krajcinovic,bazantbook},
is valid when the interaction among cracks is moderate. In the simplest
case, one can
consider the evolution of a 
scalar valued damage field $D(\vec{x})$, describing the local variations of the Hooke law
\begin{equation}
\sigma(\vec{x}) = E_0 (1-D(\vec{x})) \epsilon.
\label{damage}
\end{equation}
The idea behind damage mechanics is that all the complications
of the fracture process can be encoded in the field $D$, that will
then evolve according to a prescribed law, reflecting the internal
microcracking in the material.

Simple models of statistical fracture can be used to check
some of the typical assumptions and possible outcomes in
damage mechanics.  This is in particular applicable in ``scalar fracture'', described by
the electrical analogue of the RFM. In order to apply the coarse-graining
procedure, implied by Eq. (\ref{damage}), one assumes
the presence of a length-scale over which the damage
self-averages, meaning that its fluctuations become much
smaller than the average. This scale is also known as  
the Representative Volume Element (RVE) above which  
it is possible to treat the damage as a smoothly varying 
variable.

Figure \ref{fig:cloud} depicts the typical scenario for damage
accumulation. In this example we imagine a set of microcracks that will grow
in a stable manner, and give rise to locally enhanced 
``damage'' $D(\vec{x})$. One may now naturally define
a {\em correlation length}, by measuring e.g. the two-point
correlation function of a scalar $D$ via
\begin{equation}
C_D(\vec{r}) = \langle ((D(\vec{x}+\vec{r})-\langle D \rangle)
-(D(\vec{x})-\langle D \rangle))^2\rangle
\end{equation}
which should decay, for a field $D$ which is uncorrelated
over longer distances, asymptotically exponentially and
thus defines a correlation length $\xi$ via $C_D \sim \exp{(-r/\xi)}$.
An interesting issue is the ``symmetry breaking''
in damage accumulation: the direction of crack growth is mostly 
dictated by geometry and loading conditions.
Thus, as illustrated by Fig.~\ref{fig:cloud}, it is necessary
to consider separately the directions perpendicular and parallel
to the external stress,  which would then imply {\em two
correlation lengths}, $\xi_\parallel$ and $\xi_\perp$.

Given a fracture process, in which damage grows in a controlled
fashion one can then - in principle, at least in simulations - 
follow the development of $\xi$. This quantity will increase, 
making a sample effectively ``smaller'': 
as $\xi$ (and the RVE size ) increases $L/\xi$ diminishes, where
$L$ is as usual the sample linear size. An additional complication comes
from the damage anisotropy, since the two correlation lengths 
$\xi_\perp$ and $\xi_\parallel$ will follow different dynamics. 

It is interesting to discuss what happens when one of the two
becomes much larger than the other. In this case the sample becomes effectively
much ``larger'' along the stress direction, becoming
almost one-dimensional, which in other words indicates the 
presence of {\em damage localization}.
The developments that lead to and beyond this point are of 
course an interesting issue, with also materials science twists. 
Consider for instance the mechanics of fiber-reinforced composites: eventually 
one crack will dominate, but still the specimen may have a finite
strength remaining thanks to fibers that bridge the crack and
induce cohesion.

\begin{figure}[t]
\begin{center}
\includegraphics[width=10cm]{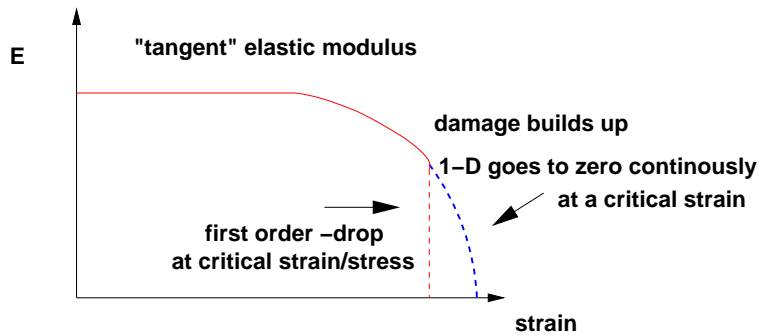}
\end{center} 
 \caption {An example of a ``first-order'' and a critical
scenario for damage build-up, via $E_{eff} = ED(\epsilon)$}.
\label{fig:damagelaw}
\end{figure}

In general, it is a complicated question as to what kind
of evolution would rule $\langle D \rangle$, even if the possibly
very important role of fluctuations is neglected. That is, as the
$\xi$ increases the damage inside the corresponding RVE should fluctuate
more and more.
Recall also the connection between $D$ and the sample effective
elastic modulus, which implies that the two follow  similar
dynamics. One can postulate several possible scenarios for the damage dynamics,
including so-called ``finite-time singularities''
(see e.g. \cite{gluzman01,shcherbakov03}). These imply
a divergence of the strain under a linearly growing applied stress $\sigma\propto t$
\begin{equation}
\epsilon \sim 1/(t_c-t)^\gamma,
\label{eq:ftimes}
\end{equation}
i.e. a power-law -like or pseudo-critical approach to the failure
time $t_c$ and to the corresponding critical stress $\sigma_c \propto t_c$. 
In the context of damage mechanics, Eq.~(\ref{eq:ftimes})
would result from  a stress dependent
damage law of the type $1-D \sim (\sigma_c - \sigma)^\gamma$. In 
cases with a dominant crack this law can be related to the crack growth
rate as a function of $\sigma$.
Another possibility is that the damage displays an abrupt jump at $\sigma_c$.
The two possibilities are depicted in
Fig. \ref{fig:damagelaw}. 

The statistical laws and geometry of damage accumulation
can be studied both via laboratory experiments and the models that are the main scope of
this review, and they often have also great importance for 
engineering applications. The typical engineering mechanics viewpoint of the previous discussion
is to look at the rate of crack growth given just a single crack. The assumption
is that the crack length $a$ evolves according to
\begin{equation}
\frac{\partial a}{\partial t} =  A K_{eff}^m
\label{eq:charles}
\end{equation}
where $K_{eff}$ is an effective stress intensity factor and 
the exponent $m$ typically varies between $30$ and $50$ \cite{atkinson87}.
Eq.~(\ref{eq:charles}), known as  the ``Charles law'', is 
a way of generalizing the standard LEFM expression, in Eq. (\ref{sif}),
for which $K_{eff} \sim a^{1/2}$. It implies that even a subcritical crack,
in the sense of the Griffith argument, can slowly grow in some conditions
such as creep or fatigue. Examples arise in non-brittle fracture, 
and in the presence
of a diffuse crack population \cite{straumsnes97}.  
For $K_{eff} \sim L$ one gets a power-law growth in time
by simple integration \cite{barenblatt96}.
The possible roles of disorder here is a question which has not been
studied systematically.

\subsection{Fracture mechanics of rough cracks} 

\subsubsection{Crack dynamics in a disordered environment:
self-affinity and anomalous scaling}
\label{sec:selfaff}

While damage mechanics treats fracture in disordered media at
a coarse grained scale, it is often necessary to treat the
dynamics of a single macroscopic crack. 
A simple example is to consider a medium containing regions with sufficiently
high $\gamma_s$, locally, so as to stop crack growth. In these
circumstances energetic considerations dictate that the crack path will
deviate from the straight direction (see Fig. \ref{fig:1ddemo}). 
The simple case is the two-dimensional one,
when the vertical coordinate of the crack tip, $h(t)$,
(where $t$ is the position along the horizontal projection of the crack path)
wanders with equal probability up and down. This process corresponds directly
to a random walk or Brownian motion. This fact implies that
the crack surface, corresponding to the trajectory of the crack tip,
displays a characteristic {\em roughness}. In the following we consider
the fracture surfaces $\vec{h}(\vec{x},t)$ in various cases.
The following sub-cases can be distinguished:
\begin{enumerate}
\item[(i)]
{\bf 2D crack}: the crack line grows as a {\em trail} of a zero-dimensional
point-like particle, the effective crack point. For example, consider 
the growth of a crack from a pre-made notch.
\item[(ii)]
{\bf 3D interfacial crack}: a crack line propagates in a plane ($xy$), so that 
the $z$-direction can be neglected. If the average direction of
propagation is in the $y$-direction, say, one is interested
in the statistics of the rough line $h(x,\bar{y})$ where 
$\bar{y}$ is the time-dependent average position, and $h$
varies with $x$, in the $y$-direction.
\item[(iii)]
{\bf 3D crack}: one has a crack line given by $\vec{r}(t)$. This leaves
a trail as the time increases, which is a two-dimensional 
surface  $h(x,y)$ where $h$ is along the $z$-axis. In general, 
$h$ is defined only for those ($x,y$) for which the projection of $\vec{r}$ on the $xy$-plane, ($(x',y')$), is such that that for $x=x'$, $y' > y$. One may now
study the projections of $h$ to the ($xy$)-
and ($(xz)$)-planes, defining thus the two components
$h_\parallel (x,\bar{y})$ and $h_\perp (x,\bar{y})$ which
lie in the respective planes.
\end{enumerate}

The geometry of rough surfaces becomes particularly interesting
when it turns out that they are self-affine fractals, or even
more complicated versions thereof as discussed below. There 
are three core questions related to this kind of ``roughening'',
in addition to the phenomenology of exponents and scaling relations,
and the origins of such scaling:
\begin{itemize}
\item
What are the the smallest length- (and time-) scales on which
one can see such behavior? 
\item
What is the largest scale, that of ``saturation'', where the
roughness becomes independent of scale? What leads to the 
saturation, in particular if one is concerned about the
microstructure?
\item
How does the roughness evolve? A crack is often
induced in the laboratory from a notch, which is ``flat''.
The roughness of the crack will develop as new crack
surface is created. Thus one must be careful in considering
a posteriori surfaces from experiment or simulations.
\end{itemize}

Let us now study the separate possibilities. Of these, case (ii)
from above is conceptually the easiest one. Assume 
that the crack front advances on average with a constant velocity $v$,
so that $\bar{y} = vt$. The simplest quantity used to characterize
the roughness is the global surface width 
\begin{equation}
W(L,t) = (\langle (h(x,t)-vt)^2 \rangle_x)^{1/2}
\label{eq:global_width}
\end{equation}
which simply measures the standard deviation around the mean position.
The global width is a quantity often considered in kinetic roughening
of surfaces, for instance in molecular-beam epitaxy or in the propagation
of slow combustion fronts in paper. In these cases, it happens that
the dynamics of $h(x,t)$ is ruled by a Langevin equation and
one can expect the so-called Family-Vicsek-scaling \cite{family91}.
This implies the presence of an unique spatial scale, the correlation
length $\xi$, and an associated time-scale. In practice one writes
\begin{equation}
W(L,t) \sim t^\beta f(L/t^{1/z}).
\label{FV}
\end{equation}
Eq.~(\ref{FV}) describes the roughening of $h$ from an initial condition,
which is usually a flat one,  although one can also consider the influence of
an arbitrary profile $h(x,t=0)$. The initial transient is governed
by the growth exponent $\beta$, until correlations build
up, as $\xi$ reaches $L$, and
the roughness saturates at the value $W \sim L^\zeta$. 
This implies that the scaling function $f(x)$ is a constant, 
except for $x$ small. The self-affine scaling relates now
$\beta$, $\zeta$, and the dynamic exponent $z$ through
\begin{equation}
\beta z = \zeta.
\label{eq:exprel}
\end{equation}
Note that one can have further constraints, such as the
so-called Galileian invariance for the Langevin equation,
coupling the exponents ($2-1/\beta = \zeta$ in 2D).

If we consider separately the profiles
$h_\parallel (x,\bar{y})$ and $h_\perp (x,\bar{y})$ (again,
one may substitute $\bar{y} = vt$), in the
simplest imaginable case, the Family-Vicsek picture 
applies to both, separately, with six separate
exponents $\beta_\parallel$, $z_\parallel$, $\zeta_\parallel$,
and $\beta_\perp$, $z_\perp$, $\zeta_\perp$, respectively, that are 
coupled by the relation Eq.~\ref{eq:exprel}.
In section \ref{sec:depinning}, we overview line models
of cracks, according to scenario (iii), such that they might
relate to this scenario.

The existence of a correlation length begs the
question of where it stems from in the crack growth dynamics,
or in other words if one measures the exponents experimentally
what is the equation underlying the roughening process. 
This is apparent if we consider the generalization of the
one-dimensional example of a Brownian ``crack'' trail 
to case (iii), which is the most relevant for experiments. 
The idea is to represent the
opening of a crack by a moving one-dimensional line
which leaves a two-dimensional surface $h(x,y)$ trailing after itself.
Thus in the intermediate fracture stage one has a line separating the 
untouched part and the crack - with two open faces. 
Considering the case slow fracture,
one could then write equations of motion for the fluctuating
components of the crack line, taking into account the
pertinent physics as the presence of randomness, and the coupling
due to elasticity of the fluctuations, both in-plane and out-of-plane.
Such an approach also contains the idea that the physics in the bulk,
as in the fracture process zone, can be neglected or incorporated again
as a long-range elasticity along the crackline. 
Similar models that include further complications such as dynamical
overshoots along the crack front have also been studied to a great
extent (see section \ref{sec:depinning}).

Scalings such as that implied by Eq.~(\ref{FV}) are to be
understood in the average sense. That is, one has to sample
the stochastic process sufficiently in order to ensure the convergence of 
quantities. This is naturally a problem in fracture,
where it may be difficult to collect enough samples. There are
other measures that can be used to extract the exponents, such
as the local width, the correlation functions, and the structure
factor \cite{barabasi95}, but notice
that the discussion assumes that the profile $h$ is translation-invariant,
which may be invalidated by the presence of boundary effects.

An example thereof is the case (i) from above, which in the
simplest imaginable case might be equivalent to Brownian motion.
Now, $h(x)$ resembles the case of a propagating one-dimensional
crack-front, but is actually different in several ways. First,
the already-formed profile is frozen in time (unless one has
microscopic processes that change the free crack surfaces).
Second, as in the Family-Vicsek picture of Eq.~(\ref{FV}),
one may assume that there are initial transients. Their
character and the appropriate scaling picture would depend
on the details. For instance, one can consider a Brownian
motion confined to a maximum excursion, so that eventually
the roughness would saturate. In reality, one should keep in mind
that due to the presence of long range elastic fields, the
dynamics of the crack tip is never a simple uncorrelated random walk
in 2D. The interplay between elasticity and disorder 
gives in general rise to {\em avalanches} or stick-slip motion,
as it is typical of non-linear driven systems. 
One may imagine that in 2D the crack is arrested at obstacles,
slowing it down till a deviating path is found and the crack tip
speeds up till the next obstacle. Such avalanches
can be detected through acoustic emission and we will
discuss this issue further in Sec. \ref{sec:exp}.

In real experiments, as in many that relate to
case (iii) from above, one can also observe two 
complicated scenarios than Family-Vicsek scaling:
{\it anomalous scaling} and {\it multiscaling}. 
To this end, we should consider more elaborate measures of 
rough surface profiles $h(\vec{x})$, measured over a window of size $l<L$
($\vec{x}$ implies, that we will not distinguish the various crack-inspired
scenarios). In full generality
we can define the q-moments and assume that they scale as power-laws
\begin{equation}
\label{eq:w_q_def}
w_q(l,t) \equiv \langle \langle \delta h(\vec{x})^q \rangle
\rangle^{1/q} \sim l^{\zeta_q} \mbox{ for
  } l < \xi(t).
\end{equation}
where $\delta h = h - \bar{h}$ is the deviation of $h$ {\em inside a window}
of size $l$, and $\xi$ is the correlation length at which $w_q$ becomes
a constant. The brackets imply again an ensemble average.  
The simplest case is encountered when all the $q-$order roughness exponents 
$\zeta_q$ are the same (i.e. $\zeta_q \equiv \zeta$  for all $q$),
corresponding to self-affine scaling. The opposite possibility is 
referred to as {\it multiscaling}.  In the context of kinetic 
interface roughening multiscaling implies that additional mechanisms
create large height differences which are measured more sensitively
by the higher $q$-moments (an example of this will be discussed in 
Sec.~\ref{sec:gradper}).

In terms of the local roughness, the Family-Vicsek scaling can be written
as 
\begin{equation}
\label{scaleform}
w(l,t) = \alpha(t) \; \xi(t)^\zeta \; \mathcal{W} \left( \frac{l}{\xi (t)}
\right).
\end{equation}
with the scaling function $\mathcal{W}(x) \sim x^\zeta$ for $x < 1$ and a constant for 
$x \gg 1$.  Notice that the exponent 
$\zeta$ can also be obtained by the structure factor,
or spatial power spectrum
\begin{equation}
\label{structfact}
S({\bf k},t) \equiv \langle | h({\bf k},t) |^2 \rangle \sim
\Biggl\{
\begin{array}{ll}
\! k^{-(2 \zeta + d)} & \mbox{for }k \gg 1/\xi(t) \\
\! \xi(t)^{2 \zeta + d} & \mbox{for }k \le 1/\xi(t)
\end{array}%\right
\label{eq:corr-funct1}
\end{equation}
where $h({\bf k},t)$ denotes the spatial ($d$-dimensional) Fourier transform
of $h(\vec{x},t)$. Alternatively, one can use the height difference correlation function 
\begin{equation}
\label{G_vs._S}
G_2(\vec{x},t) \equiv \langle | h(\vec{x},t) - h(0,t)|^2 \rangle^{1/2} =
\int_{\bf k} S({\bf k},t) \left( 1 - \cos({\bf k \cdot \vec{x}}) \right),
\label{eq:corr-funct}
\end{equation}
which for large separations scales as $|x|^{\zeta}$.

If in Eq. (\ref{scaleform}) the amplitude $\alpha(t)$ increases with time, 
{\it anomalous scaling}
ensues \cite{lopez97}, and if $\zeta > 1$ one talks about {\it superroughness}
\cite{lopez97,lopez99}. The quenched Edwards-Wilkinson equation in $d=1$, describing the
``depinning'' of a line with local elasticity is a classical  example
of this \cite{nattermann92,leschhorn97,narayan93,chauve01,chauve02}.
Anomalous scaling usually implies that the
roughness of a self-affine object (e.g. a crack) develops a power-law
dependence on the system size $L$ \cite{lopez97,lopez99}.  One should thus define 
two different
exponents $\zeta$ and $\zeta_{\rm loc}$: the first exponent describing the dependence
of the global roughness on the system size, while the second encodes the scaling
of the local roughness on the measuring window. In particular, the local width scales as 
$w(l) \sim l^{\zeta_{loc}}L^{\zeta-\zeta_{\rm loc}}$, so that the global 
roughness $w(L)$ scales as $L^\zeta$ with $\zeta>\zeta_{\rm loc}$. Consequently, the
power spectrum scales as $S(k) \sim k^{-(2\zeta_{\rm loc}+d)}L^{2(\zeta-\zeta_{\rm loc})}$.
Notice also that Eqs.~\ref{eq:corr-funct1} and \ref{eq:corr-funct}  
reflect only the local roughness exponent $\zeta_{\rm loc}$. 
This implies that for fracture surfaces with two (independent) projections, 
one can define no less than four independent roughness exponents, and
thus there could be eight exponents describing this case.

An interesting issue here is how ordinary LEFM changes in the 
presence of a crack with a ``rough'' or self-affine geometry. The
same problem can also be restated in the presence of general damage
or a population of self-affine microcracks. All in all, there are
at least three different possibilities. One can consider the stress-field
in the proximity of a ``fractal crack'', but then one has to remember that
Nature does not create a priori such objects - they grow because of some
stable crack growth process. This takes place in the presence of elastic 
fields possibly modified from LEFM predictions due to the fractal geometry
\cite{borodich97,vandembroucq97a,vandembroucq97b}.
A particular version is to consider the Lam{\'e} equation, 
Eq.~(\ref{lame}), in the presence of a given self-affine boundary.
This can be solved in some cases to illuminate possible changes from
LEFM, but is of course outside of fracture. Finally, a promising avenue
is given by the work of Procaccia and collaborators \cite{barra02,barra02b,leverman02}, in 
which conformal mapping techniques developed in the context of Laplacian growth problems
are combined with two-dimensional elasticity. This allows to grow relatively
large cracks iteratively, accounting for a stochastic ``annealed'' growth probability
on the crack perimeter. 

\subsubsection{Crack roughness and fracture energy}
\label{sec:rough-tough}

The reasons why the roughening of cracks is interesting are multiple.
First of all, as discussed below, there is the usual interest
in possible universality classes depending on dimension, fracture
mode, and material (disorder, average response). Second, it has
also an engineering implication. Consider a crack which ``starts
flat'' and then undergoes a process which indicates a scaling
behavior similar to Eq.~(\ref{FV}). An infinitesimal crack opening
will now cost {\em more surface energy} than in the usual case, and
thus the Griffith argument (Eq. (\ref{eq:griffith})) is modified in
a way that accounts for the roughness exponent $\zeta$, the maximal
extent of fluctuations $\xi_{max}$, and possibly also the modified
elastic energy due to the rough crack. In the case of anomalous
scaling (see Eq.~(\ref{scaleform}) again), the extra scale also
would enter the argument, further changing the amount of surface energy
needed for infinitesimal crack advancement. This implies a non-trivial
``R-curve'', or crack resistance, since the energy consumed via
forming new crack surface depends not only on the full set of exponents
but again on the distance propagated \cite{morel00,morel02}. 

An expression can be written
down using solely geometric arguments and the exponents as given below:
\begin{eqnarray*}
        \Delta h(l,y)\simeq A
        \left\{ \begin{array}{ll}
        l^{{\zeta}_{loc}} \: \xi (y)^{\zeta -{\zeta}_{loc}}  
        & \mbox{if \quad $l \ll \xi (y)$} \\
        \xi (y)^{\zeta} & \mbox{if \quad $l \gg \xi (y)$}
\label{bouchsca}
                \end{array}
\right. 
\end{eqnarray*}
where $\xi(y)=By^{1/z}$ is the distance to the initial notch $y$.
There are two different regimes: for $l \gg \xi (y)$, 
$\Delta h(l,y) \sim y^{\zeta/z}$, and for 
$l \ll \xi (y)$, which is characterized by the 
exponent $(\zeta-{\zeta}_{loc})/z$.  Important here is
presence of anomalous roughening  so that 
$\Delta h(l,y \gg y_{sat}) \simeq A \: l^{{\zeta}_{loc}} \: 
L^{\zeta -{\zeta}_{loc}}$.

Since the effective area from which surface energy follows is
\begin{equation} 
    G \ \delta A_p = 2\gamma\ \delta A_r
\label{Crit}
\end{equation}
where $\delta A_r$ is the real area increment and $\delta A_p$ its
projection on the fracture mean plane, the growth of 
roughness implies
\begin{eqnarray*}
        G_R(\Delta a \ll \Delta a_{sat}) & \simeq & 
2 \gamma 
        \sqrt{1+\left( \frac{A B^{\zeta-{\zeta}_{loc}}}
        {{l_o}^{1-{\zeta}_{loc}}} \right)^2  
        {\Delta a}^{2(\zeta-{\zeta}_{loc})/z}}
\label{Gr}   
\end{eqnarray*}
and the resistance to fracture growth becomes 
\begin{eqnarray*}
        G_{RC}(\Delta a \gg {\Delta a}_{sat}) &\simeq & 
2 \gamma 
        \sqrt{1+\left( \frac{A}
        {{l_o}^{1-{\zeta}_{loc}}} \right)^2  
        L^{2(\zeta-{\zeta}_{loc})}}.
\end{eqnarray*}
For rough cracks, this is the simplest analytical attempt
to consider size-effects (take $l/L$ constant) and the
effect of exponents describing self-affine roughness
on fracture energy.

Many authors have argued for the presence of ``fractal''
effects in the energy consumption (see e.g. \cite{borodich97}),
while it seems clear that the  
relevant framework for calculating the true surface energy ($\gamma_s$)
for the energy balance in the Griffith argument is a self-affine one
\cite{bouchaud94,weiss01} (here we neglect the complications of multiscaling
or anomalous scaling for simplicity). This can be argued for simply since the
experimental situation seems clear-cut: usually cracks are self-affine
(if not flat, but see also Ref. \cite{chiaia98}).  
The attraction of (self-affine) fractal scaling ideas
here is that they allow to understand the energy consumption mechanism(s)
simply via an interpretation of the roughness exponent $\zeta$ or
the fractal dimension $D_f$.
The models to be discussed in later sections are a toolbox for testing
the influence of disorder on strength, since these models lead to rough,
self-affine cracks. If one allows for mesoscopic plasticity, 
one would still expect the same kind of problems, but
further theoretical questions are expected.

\section{Experimental background}
\label{sec:exp}
To give a reference point for the theoretical and numerical
ideas discussed later, we next overview shortly some relevant
experimental quantities, focusing on
fracture properties for which disorder and stochasticity
are crucial. When necessary we refer to the concepts presented
in the Introduction, or point forward
to the theoretical state-of-the-art presented in later sections. 
We would like to emphasize rather generally that the statistical
aspects of fracture processes  - scaling, exponents, distributions -
are not understood even in the sense of a synthesis of experiments,
with the main exception of the roughness behavior. 
This is by force reflected in the discussion that follows.

The first of the issues is strength, both in terms of its
probability distributions and the scaling with sample size. As noted
in the previous section, from the theory viewpoint, the statistics of strength 
depend crucially on the presence of 
disorder. Next, we consider crack roughness, which has an
application to the question of the origins of the
fracture toughness. As pointed out already in Sec.~\ref{sec:rough-tough}, 
this is due to at least two reasons. The energy consumption that
takes place when a crack increases in size is due to the product
of surface energy and the new area. Both can be affected by disorder,
as in the case of Griffith-like energy arguments and in the idea of crack 
surface area increasing with scale for self-affine cracks.
In the case of fracture strength, disorder can be easily reinterpreted as a ``flaw
population'', whereas for cracks that roughen, disorder plays the role of ``noise''
during crack dynamics. This noise can either be
``quenched'', i.e., exists in the sample from the beginning
in many disguises, or be created dynamically via for instance
void-formation.

A very important topic from our point of view
is represented by acoustic emission (AE) 
in fracture experiments, a nice example of {\it crackling noise}. 
The released elastic energy in fracture is a revealing diagnostic 
for the temporal aspects of fracture processes, and is used in applications
as a monitoring tool for damage assessment.
In addition, it has been long since realized that the AE signal 
appears to be {\it critical}: the energy of AE events often exhibits a 
power-law -like probability distribution. In contrast to
thermal imaging or displacement field analysis, AE is not
limited to sample surfaces and is  sensitive to even very minute 
events.  The two fundamental questions that concern us are: 
i) how to connect the AE behavior to the other fracture properties, and 
ii) how to understand theoretically the AE statistics.

In the last part of this section, we briefly outline phenomena that exhibit 
time-dependent response, and often go  beyond the scope of brittle fracture. This concerns
the role of microscopic but collective effects on crack growth
resistance (``R-curve'') due to plasticity, 
and slow fracture processes with history effects such as creep
and fatigue failure. We mostly omit discussing the spatial aspects
of plastic deformation such as shear bands that bear some similarity to 
damage localization and fracture development in brittle and quasi-brittle fracture.

\subsection{Strength distributions and size effects}

Measuring the scale-dependent properties of strength is difficult 
mainly because of sample-to-sample fluctuations. In addition, 
samples have free surfaces, and though one often compensates 
by adjusting the test geometry for the resulting free boundary effects 
(Poisson contraction), materials often display surface defects, 
giving rise to additional disorder. 
We can thus expect that the strength distribution will be a 
superposition of cases where surface-induced defects lead to failure and cases 
where bulk defects rule the process. Obviously, separating the two 
mechanisms can be difficult. In the classical approach, the sample strength 
and the size effect is measured with a prepared notch. To obtain
reliable results, one should have access to a large enough range of 
sample sizes. 

\begin{figure}[hbtp]
\begin{center}
\includegraphics[width=10cm]{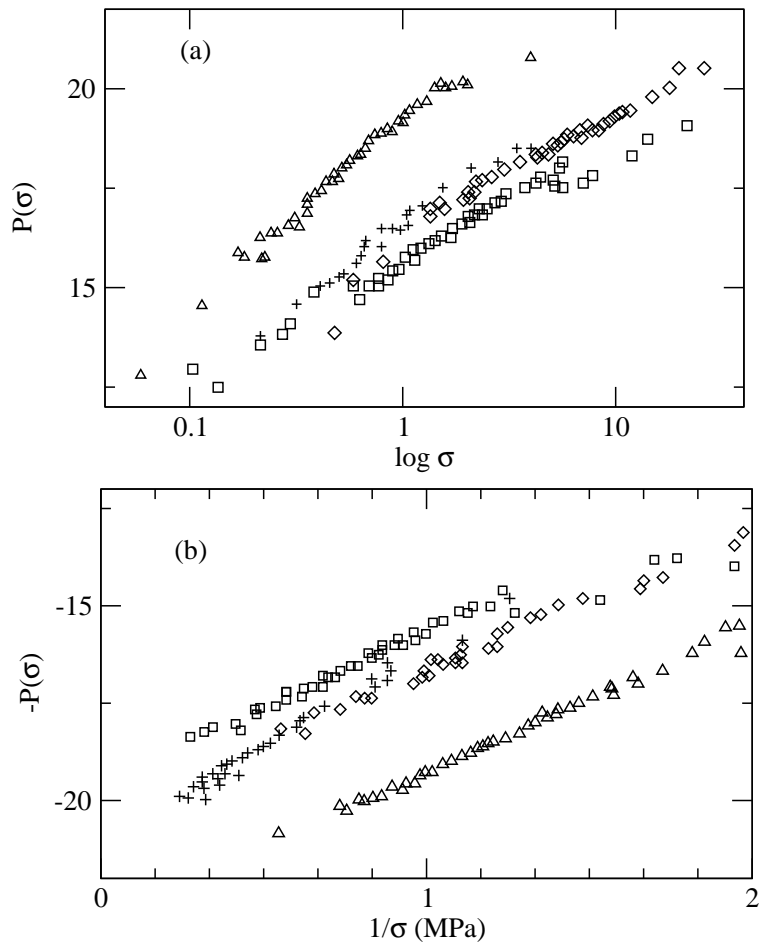}
\end{center}
 \caption {Examples of trial scaling plots for strength
distributions of porous, brittle ceramics (from Ref.~\cite{vandenborn91}).
By plotting the distributions in various scales one can try to
establish whether the shape resembles the Gumbel or Weibull form the best.}
\label{fig:strdistrs}
\end{figure}

Fig.~\ref{fig:strdistrs}
presents experimental data from brittle ceramics,
wherein the data are plotted 
according to different extremal statistics distributions \cite{vandenborn91}.
In this particular case, the Gumbel form fits the data
better than the Weibull distribution \cite{vandenborn91}.
The latter is traditionally used in the engineering
literature in the analysis of fracture statistics and
size effects in strength \cite{bazant99,sutherland99}. One should
pay attention to the fact that a Weibull scaling can be ascertained
more reliably by tests with varying sample sizes. One should note (see also 
Sec.~\ref{sec:extr_stat}) that there is a priori no reason why a material 
should fulfill the prerequisites of belonging to any of the classes 
following from extremal statistics, be it Weibull or Gumbel.

\begin{figure}[hbtp]
\begin{center}
\includegraphics[width=10cm]{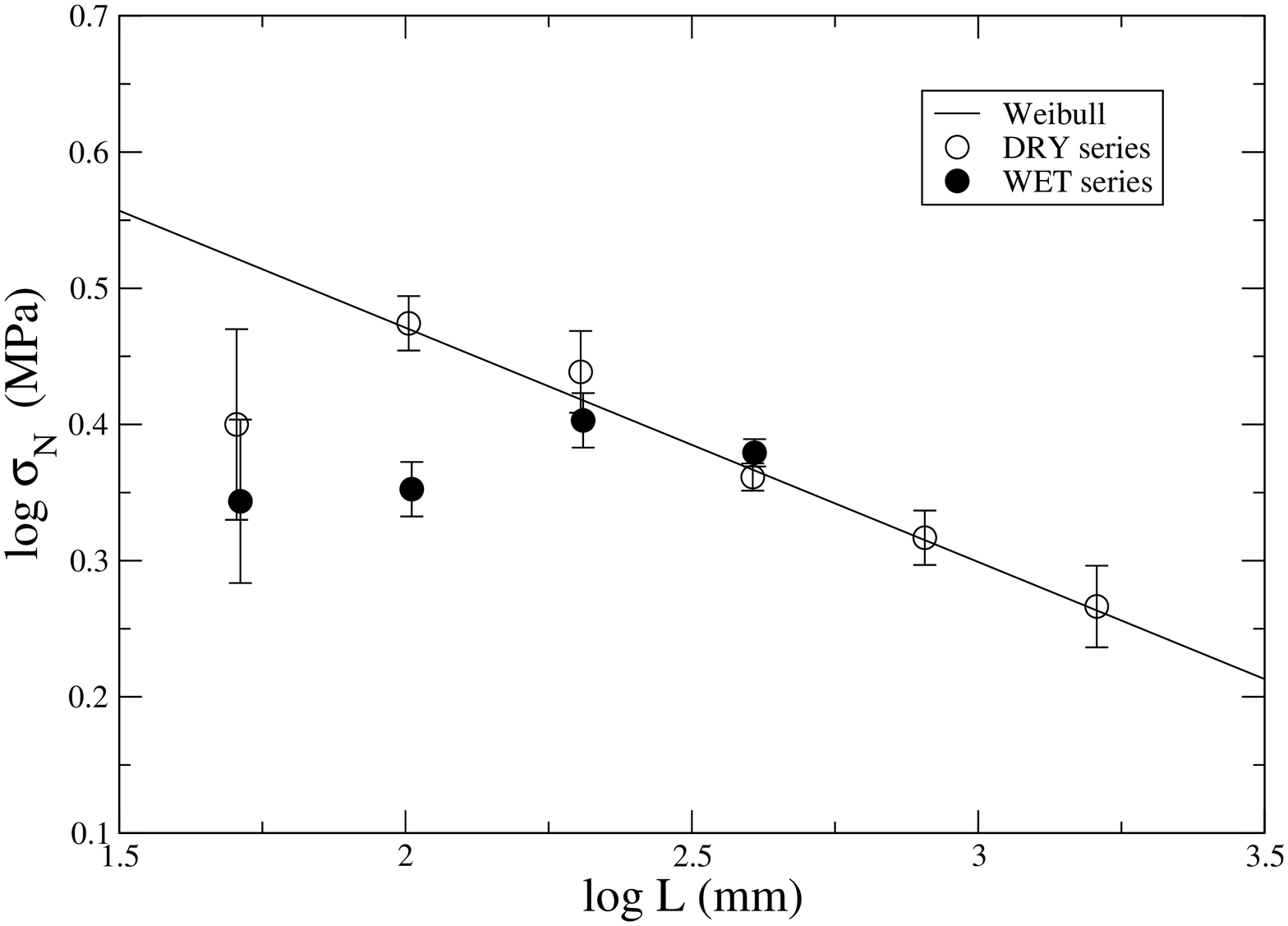}
\end{center}
 \caption {The size-dependence of concrete for two different
ambient conditions (adapted from \cite{vanvliet00}). As can be seen the strength
$\sigma_N$ follows a power-law scaling as a function of
sample size $D$ (for $D$ large). Similar data in other ambient
conditions however failed to conform to this picture.}
\label{fig:concreteweibull}
\end{figure}

Various recent attempts to characterize the strength distribution in materials
with intrinsic disorder such as paper \cite{korteoja98,korteoja99},
concrete \cite{vanvliet00} and others \cite{lu,doremus}
illustrate clearly the 
inherent difficulties in interpreting the data. Fig.~\ref{fig:concreteweibull} 
demonstrates this for concrete samples with a 1:32 size range. In this particular
case, in addition to the inherent disorder and randomness, one has to deal with systematic 
stress gradients present in these samples. This appears to be a generic 
problem that persists among various experimental studies
(see e.g.  \cite{lavoie00}).

\begin{figure}[hbtp]
\begin{center}
\includegraphics[width=10cm]{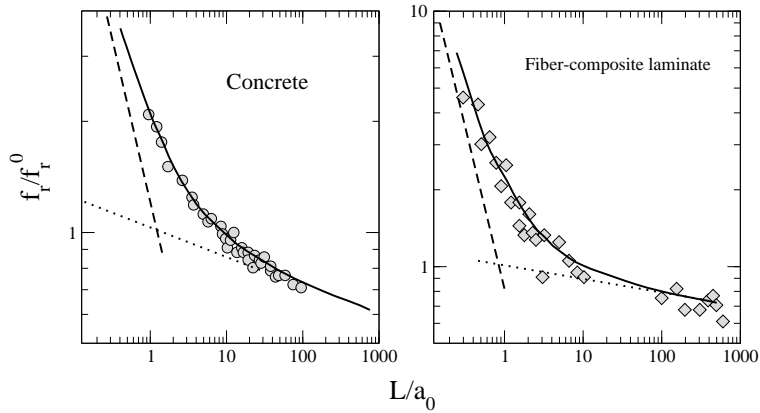}
\end{center}
 \caption {The typical size-effect from various experiments
on concrete. It is clear that the data for flexural strength, $f_r$,
exhibits a cross-over
to a scaling as a function of $D$, that may (perhaps) follow the
Weibull size effect for large $D$ values \cite{bazant04b}.}
\label{fig:sizescal}
\end{figure}

In the engineering mechanics literature, 
the scaling of strength with sample size is usually discussed with the aid
of three-point bending or tensile tests, done on samples 
in which the ratio $a_0/L$ of notch size to sample size is
constant. Clearly, in the absence of any other effects, 
the square-root -like enhancement $K \sim \sqrt{a_0}$ for the
stress intensity factor will eventually win the competition 
over the disorder, so that the
strength will decay as expected by an inverse square-root law: 
$\sigma_c \sim 1/\sqrt{a_0}$.
For small enough $a_0$ the situation is, however, not so simple
since the stress enhancement due to the notch competes with the
disorder. In this respect, Fig.~\ref{fig:sizescal} demonstrates 
an example of data from concrete, where one can observe 
clear deviation from the ideal behavior.

In the presence of disorder and/or plasticity the scaling
with a constant aspect ratio may be complicated by one issue.
Using engineering mechanics language, the fracture
process zone (FPZ) size $\xi_{FPZ}$ may not be constant as a function
of $a_0$ and $L$. Thus the energy balance (Griffith's) argument
develops a size-dependence, which can be understood considering that
the energy needed for crack growth changes with the volume of the FPZ.
Such complications should depend on the details of the processes that
make the FPZ change size. There is an extensive literature on the connection of microscopic
and mesoscopic mechanisms to the size-effect, analyzing the
possible behaviors of $\xi_{FPZ}$ for various materials.
Many of these exhibit ``strain-softening'', but other processes
as crack bridging by fibers or precipitates etc. 
are surprisingly general as well from composites to concrete 
\cite{foote86,cotterell87,mai87,wisnom91,hu92,li98,tabiei99,wisnom99,hu00,mai02,hu02,stevanovic03,duan03,bazant04,bazant04b,klock05}. This kind of
analysis often also takes into account the role of free surfaces,
which modify the FPZ geometry if the crack is close to a free surface.
From our viewpoint, one should note that 
the presence of disorder can simply make small enough
cracks irrelevant \cite{rosti01}, masking the presence of such defects.
Simply put, the randomness can correspond to an effective flaw
size $a_r$, so that for $a_0 < a_r$ a flaw is so small as to
not change the general behavior. 

In conclusion,  the issue of fracture strength from the statistical physics viewpoint
appears to still lack, to a large degree, physics-oriented experiments,
which would help to connect scaling theories to reality.
It would for instance be of immense interest to have empirical
characterizations of the ``disorder'' (as a microcrack population)
from sample to sample, and a concomitant measurement of the
sample strengths.

\subsection{Rough cracks}
\label{sec:rough-exp}
There are many different experimental signs of self-affine
scaling for crack surfaces in various scenarios reminiscent
of those overviewed in Sec. \ref{sec:selfaff}. An excellent
review by Elisabeth Bouchaud summarizes the major results up to
1998  \cite{bouchaud97}. In general, one has two main issues:
first, understanding the scaling, depending on the details such as
spatial dimension, material characteristics and loading state.
In the language of self-affine geometry, this implies determining
experimentally the scaling exponents, such as the $\beta$, $z$,
and $\zeta$ of the simplest Family-Vicsek scaling. Ultimately,
a determination of the behavior in terms of the exponents and
scaling functions would point out the right stochastic equation
that governs the process, including the noise that is present.
Second, whatever the truth be about universality, one can 
search for the mechanisms that lead to the formation of
rough interfaces, and try to understand the consequences
of self-affine crack surfaces. In other words, regardless
of whether one can write down an equation describing the
dynamics of a crack front, the process is intimately coupled
to other physical characteristics as acoustic emission, and 
damage mechanics.  We classify various experimental results
according to the dimensionality of the scenario described in 
Sec.~\ref{sec:selfaff}.

\begin{figure}[hbtp]
\begin{center}
\includegraphics[width=10cm]{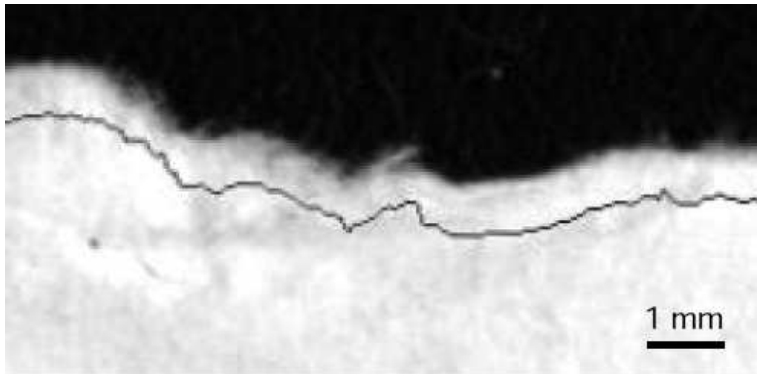}
\end{center}
 \caption {The diffuse crack surface in paper
(a two-dimensional material) and the thresholded
``interface'', shifted down for clarity (courtesy of L. Salminen, HUT). 
Here, the issue concerns how to define an unique $h(x)$ in the
case the damage field is continuous (one can see individual fibers
in the image on the other hand).}
\label{fig:fractline}
\end{figure}

\begin{figure}[hbtp]
\begin{center}
\includegraphics[width=8cm,angle=-90]{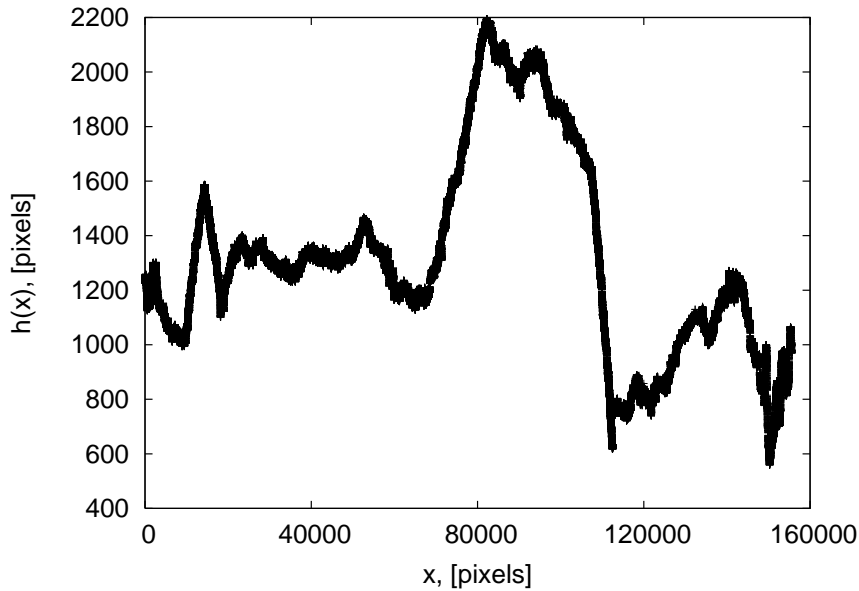}
\end{center}
 \caption {An example of a 2D crack-line from a paper web,
of width 6500 $mm$ \cite{salminen03}. The sample
has an {\em effective} exponent of about 0.6 as shown
by various measures, but only
above a scale of about 2-3 $mm$ ($\propto 30$ pixels in the lateral
direction).}
\label{fig:6500}
\end{figure}

The conceptually simpler case of crack roughening
is in general provided by case (i), where a the final crack path
is given by a one-dimensional curve $h(x)$ in two dimensions.
Two typical examples are shown in Figures \ref{fig:fractline}
and \ref{fig:6500} from paper fracture.
These illustrate the diffusive nature of fracture, such that
it is clear that the asymptotic {\it scaling} of $h(x)$
can only be attained beyond some microscopic lengthscale
$\xi_m$ in $x$ and $h$. That is, this might imply that $h(x)$ exhibits 
rms fluctuations (local width) with $w(l) \sim l^\zeta$
only for $l \gg \xi_m$. An example is provided of such behavior
in Fig. \ref{fig:6500}.

More quantitatively, the existing results point out towards a 2D (local) 
roughness exponent in the range $\zeta \simeq 0.6-0.7$ 
\cite{kertesz93,engoy94,salminen03}. A number of observations
can be made: the measured value is higher than a random walk
or a Brownian motion value $\zeta_{RW} = 1/2$, 
and $\zeta>1/2$ implies interesting 
positive correlations for the motion of the crack tip.
The value is also close to the numerical 
predictions of various models, both theoretical ones and the RFM
as discussed in Sec.~\ref{sec:rough}. However, even for ordinary, 
industrial paper itself there are numerous values available 
that are significantly higher than say $0.7$ (examples are found in Ref.~\cite{menezessobrinho05}).  
It is not known at this time whether
this reflects the difficulty of experimentally measuring $\zeta$,
or the fact that the result is not ``universal'' but depends on
material parameters as for instance ductility and anisotropy. 
In addition Bouchbinder et al. have indicated a scenario in which the
$h(x)$ has more complicated structure, exhibiting {\em
multifractal} behavior (implying a $q$th-order dependent
$\zeta_q$ for the various $q$-momenta of roughness measures,
see Sec. \ref{sec:selfaff}) \cite{bouchbinder05a}. An alternative
explanation is simply a crossover below $\xi_m$ as suggested
above \cite{salminen03}. Note that the values of $\zeta$  can 
not simply be interpreted as fractional
Brownian motion, since history effects in $h(x)$ are
more complicated.

Complicating our understanding of the problem is the fact that
the role of kinetics (velocity, dependence on the loading
scenario) has not been studied systematically.
When experiments are started from a notch, the useful conceptualization
is to consider the $h(x)$ to be a trail of a particle, the
crack tip. Then, there are two issues of immediate relevance: 
the role of the average crack tip velocity,
and the translational invariance of the profile. The former
is important since various kinetic effects might be crucial,
even for slow failure as in, for example, creep or fatigue fracture.
The crack tip velocity is even more relevant for fast fracture where the timescales of sound 
waves are not well separated from the velocity. In analogy to
the discussion in Sec.~\ref{sec:selfaff} the crack dynamics
is expected to follow a scaling that depends on the distance
to the crack tip,  so that translational invariance is broken
for $x<\xi$, before eventual saturation. This caveat is in particular relevant 
if one considers a case where there is a FPZ whose size, and thus 
$\xi_m$ changes with crack length $l$.

\begin{figure}[hbtp]
\begin{center}
\includegraphics[width=10cm]{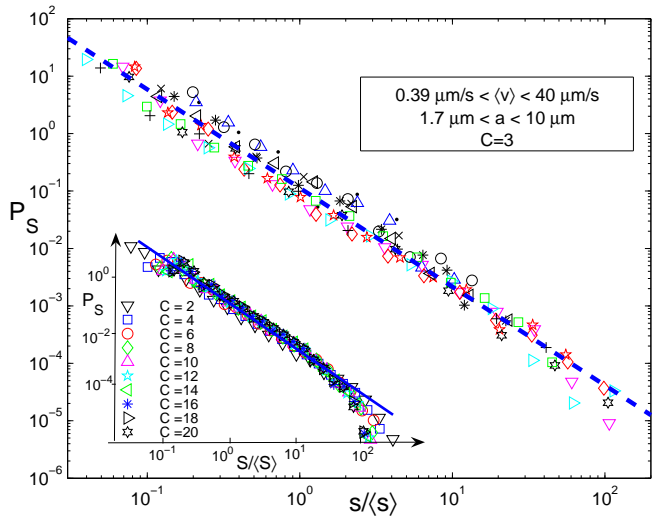}
\end{center}
 \caption {Examples of crack fronts in a quasi-two-dimensional
setup where a rough interface between two elastic plates is used
as the substrate for the crack line. Notice the avalanche-like
localized advancement of the line between subsequent frames. These
can be used to measure locally the roughness exponent $\zeta$
via the relation $\Delta h \sim l^\zeta$ where $\Delta h$ is the
average displacement, and $l$ is the lateral extent of the
region which moved  \cite{maloy06}.}
\label{fig:maloyline}
\end{figure}

\begin{figure}[hbtp]
\begin{center}
\includegraphics[width=10cm]{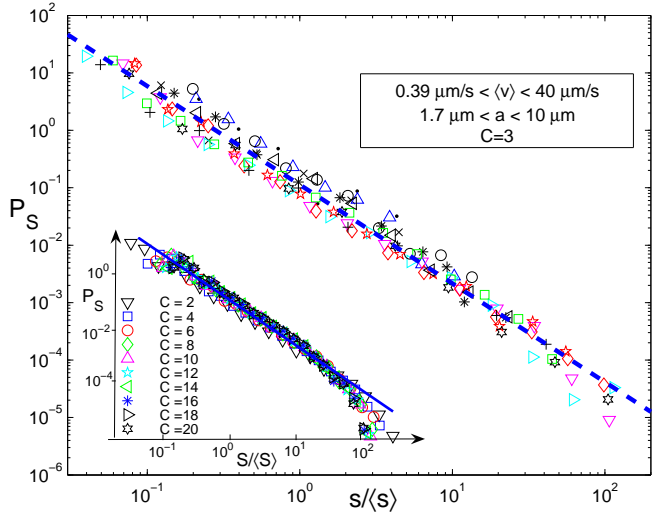}
\end{center}
 \caption {The avalanche size distribution $P(A)$ in the Plexiglas
experiment of Ref. \cite{maloy06}. Notice the independence of the
scaling exponent of the actual average crack velocity.}
\label{fig:maloystat}
\end{figure}

When a crack front propagates between two elastic blocks, 
the dynamics is enforced to be quasi-two-dimensional   
and one can again define an interface $h(x)$ separating the intact 
from the failed region. The theoretical models for this experimental setup, referred as case (ii) in Sec.~\ref{sec:selfaff}, are dealt with separately in Sec.~\ref{sec:depinning}.
The interest here lies with the roughness exhibited by $h$ and the associated dynamics
as $\bar{h}$ changes slowly with some average velocity.
Knut M\aa loy and collaborators have performed experiments on a couple of 
joined Plexiglas plates, previously sandblasted to induce a randomness. 
The dynamics of the crack is followed using a camera. 
As demonstrated in Fig. \ref{fig:maloyline},
the crack front moves by avalanches, whose areas are
distributed according to a power law distribution  $P(A) \sim A^{-1.8}$ 
(Fig. \ref{fig:maloystat}). The authors have also estimated 
the roughness exponent using different methods and the result is $\zeta \simeq 0.6$
\cite{schmittbuhl97,delaplace99,maloy01,maloy03,maloy06}. Notice that
this value is similar to the one measured in case (i) but the 
geometry and the elastic interactions
are completely different. Thus there is no apparent reason to
identify the two values.

\begin{figure}[hbtp]
\begin{center}
\includegraphics[width=10cm]{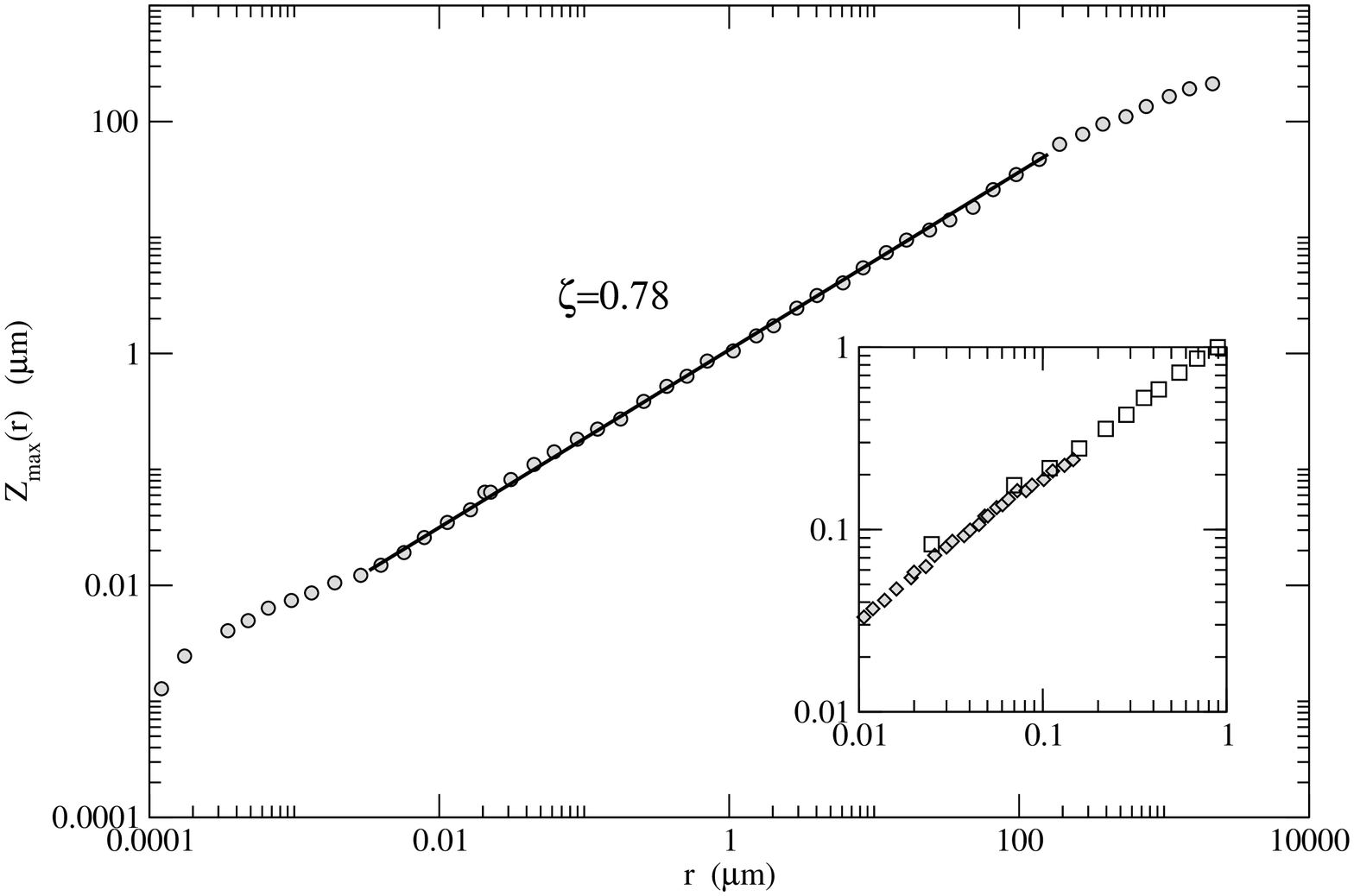}
\end{center}
 \caption{Roughness scaling measured in a metallic alloy broken in tension.
$Z_{max} (r)$, the maximum $h$-value difference
inside a window of size $r$ shows scaling over five decades with an
exponent of $\zeta \simeq 0.78$. The inset shows the overall between
two different measuring techniques (from Ref.~\cite{bouchaud97}).}
\label{fig:scaling08}
\end{figure}

\begin{figure}[hbtp]
\begin{center}
\includegraphics[width=10cm]{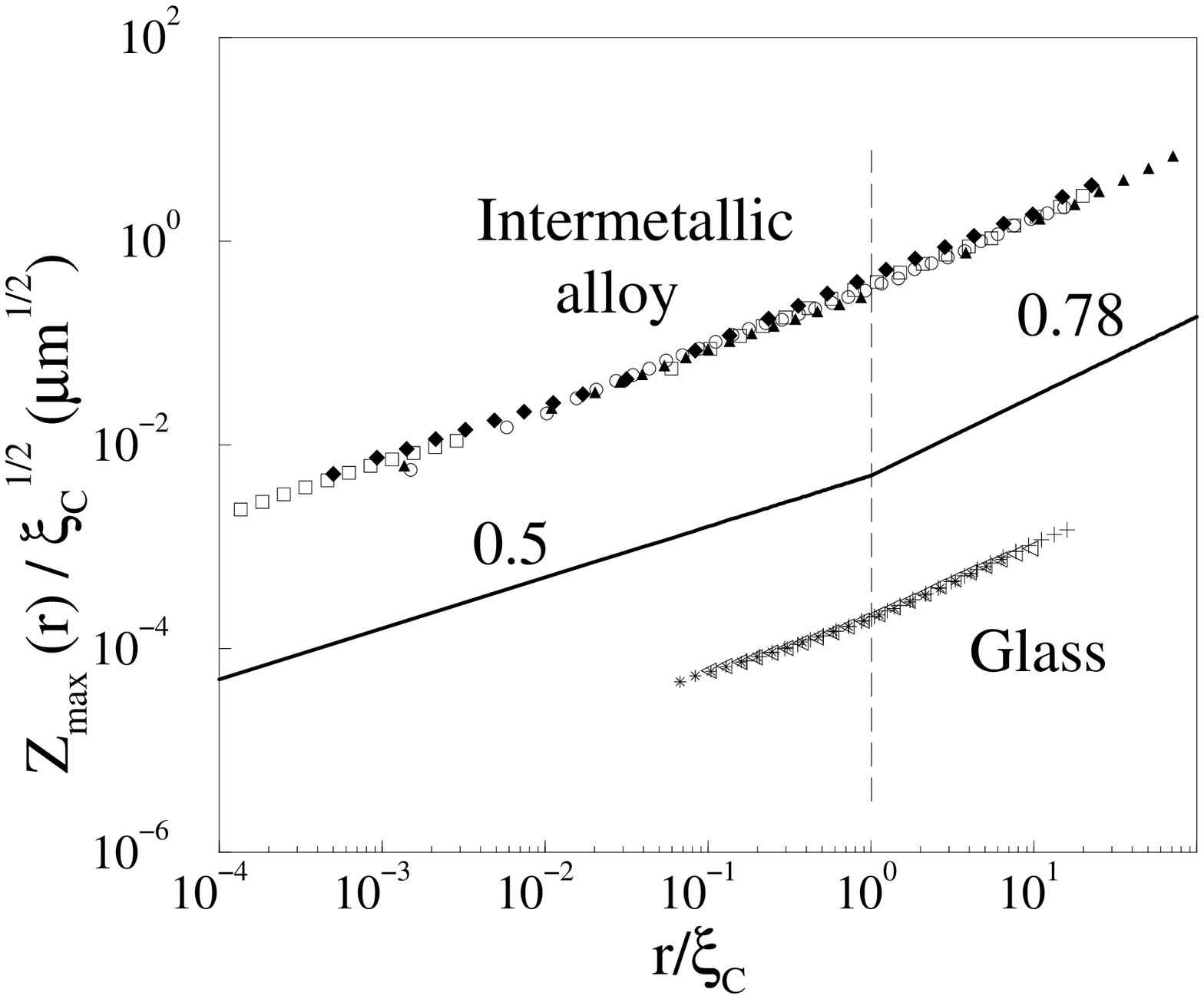}
\end{center}
 \caption{$Z_{max} (r)/\sqrt{\xi_c} $, the maximum $h$-value difference
inside a window of size $r/\xi_c$ for two 
materials. Here $\xi_c$ is a cross-over scale that can be determined 
separately and may be thought of as the FPZ scale. As can be seen 
for both cases and for various velocities of crack
growth, there seems to be two different regimes of roughness with
$\zeta\simeq 0.5$ at small scales and 
$\zeta \simeq 0.8$ at larger scales (\cite{daguier97}).}
\label{fig:twoexps}
\end{figure}

Most studies performed in the past, starting from the first pioneering
work of Mandelbrot et al. \cite{mandelbrot84}, refer to three dimensional fracture, 
case (iii) according to Sec.~\ref{sec:selfaff}. The evidence collected
during the first decade of experiments on several materials
suggested the presence of a universal roughness 
exponent in the range  $\zeta \simeq 0.75-0.85$ \cite{bouchaud97}.
The scaling regime over which this result is obtained is quite impressive,
spanning five decades as it is shown in Fig.~\ref{fig:scaling08}.
In some cases, namely metals, it has been shown that a different exponent 
(i.e. $\zeta \simeq 0.4-0.5$) describes the roughness at small scales,
with a crossover to the large-scale value. Fig. \ref{fig:twoexps} 
displays an example of both the short-scale and large-scale regimes in a 
typical (fatigue) experiment. This crossover has been initially explained 
as a dynamic effect: the small scale exponent would reflect
the quasistatic limit and the large scale exponent the effect of finite
velocities \cite{bouchaud97}. Notice, however, that the short-scale 
value is sometimes not observed, even at very low velocities 
as for silica glass \cite{ponson06}, while in other cases, 
namely in granite and sandstones, 
the large-scale exponent is not seen even at high velocities 
\cite{boffa98}.

\begin{figure}[hbtp]
\begin{center}
\includegraphics[width=10cm]{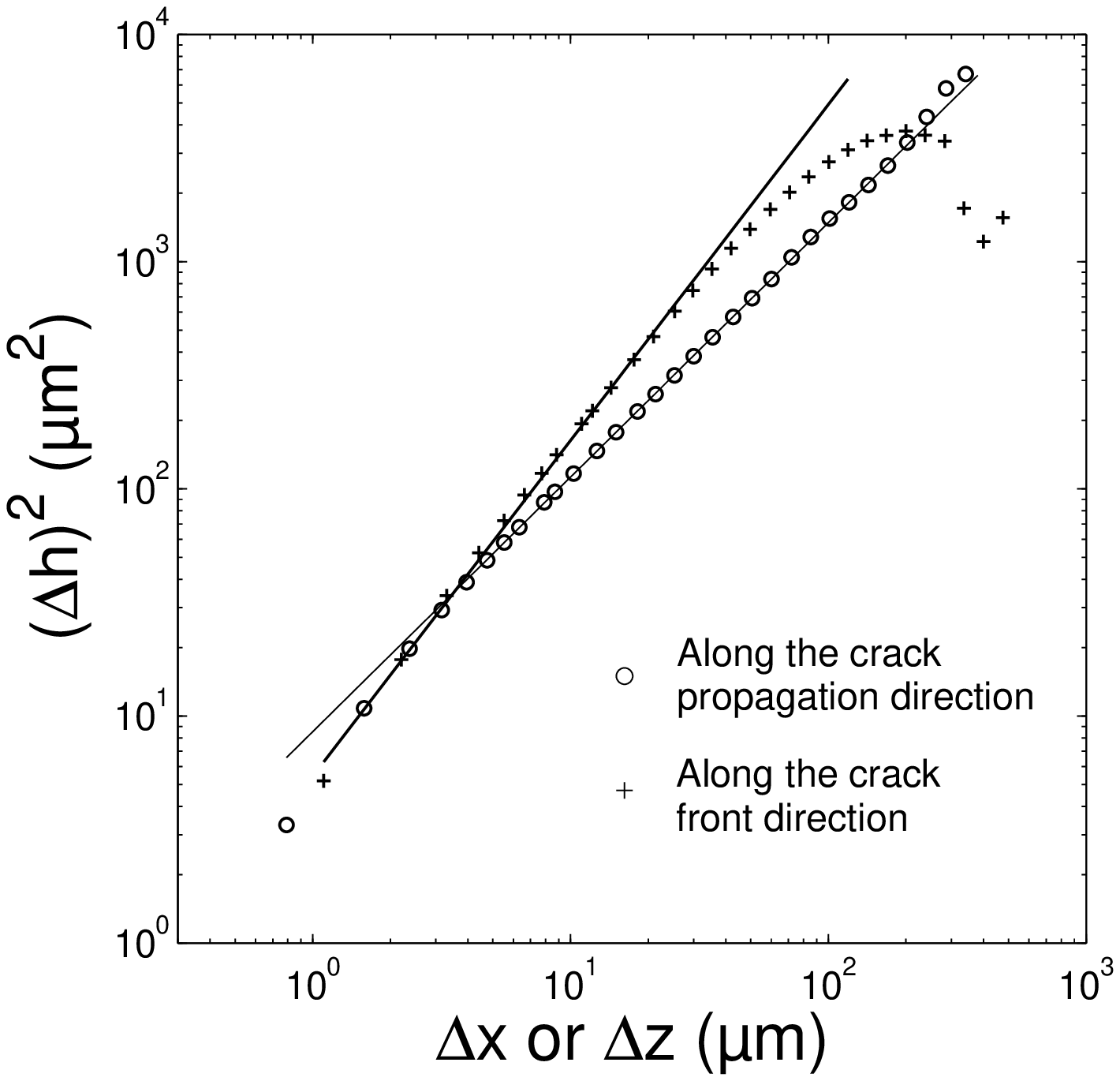}
\end{center}
\caption{1D height-height correlation functions for 
$h_\parallel$ and $h_\perp$, in the notation
of Sec. \ref{sec:selfaff} for an aluminum alloy
fracture surface \cite{ponson06}. The apparent
scaling exponents are $\zeta\equiv\zeta_\parallel\simeq 0.75$ and
$\beta\equiv\zeta_\perp \simeq 0.58$.}
\label{fig:ponson1}
\end{figure}

\begin{figure}[hbtp]
\begin{center}
\includegraphics[width=10cm]{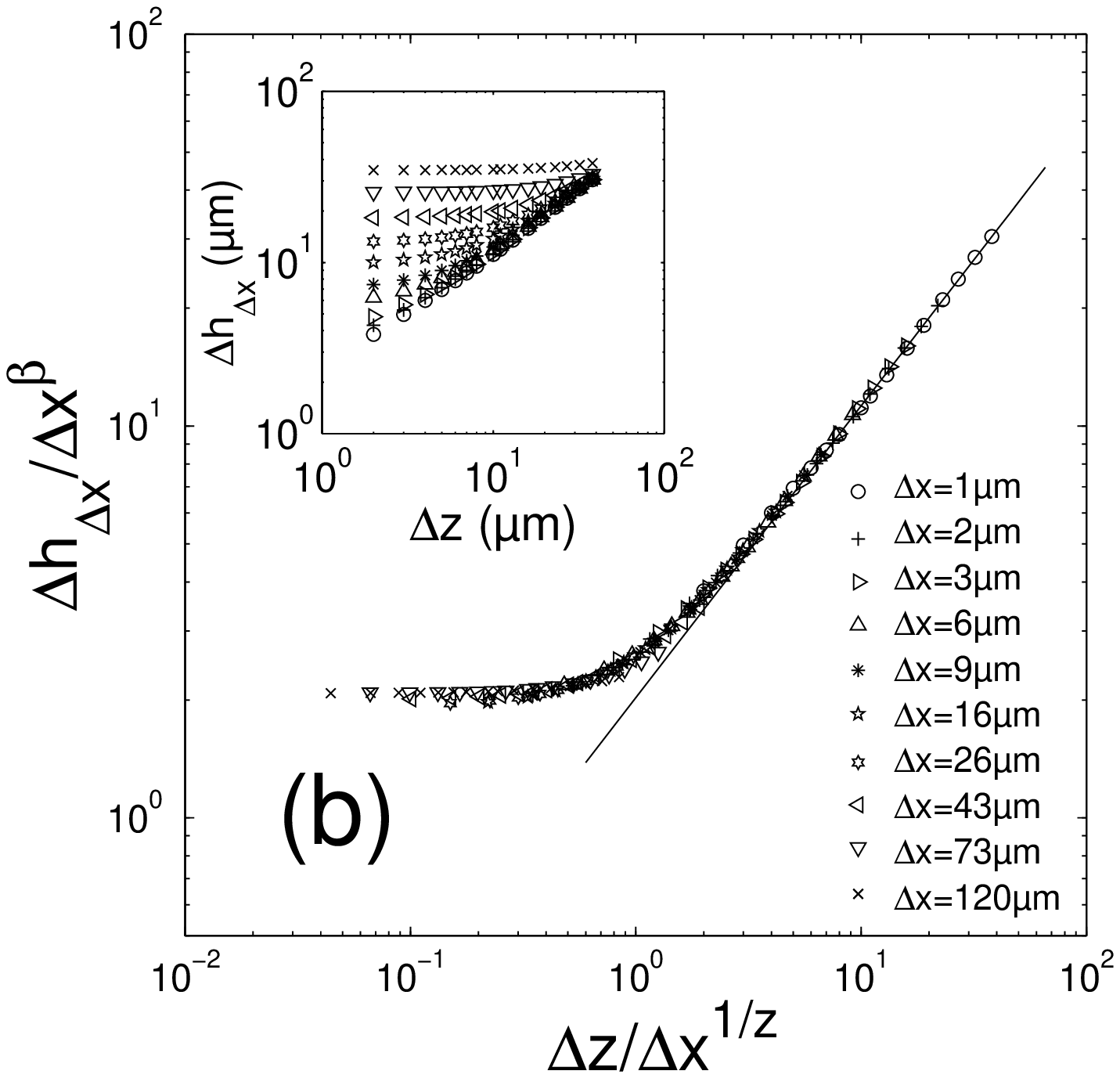}
\end{center}
\caption{2D height-height correlation functions $\Delta h_{\Delta z}(\Delta x)$ 
for various $\Delta x$, $\Delta z$ for a fracture surface of an
aluminum alloy (\cite{ponson06}). The collapse uses the form 
suggested by Ponson et al., Eq.~(\ref{eq:2dcor})}.
\label{fig:ponson2}
\end{figure}

Since the crack surface is separated from
the intact part of a sample by a line that can fluctuate
both in the out-of-plane and in-plane directions, one can in
principle measure two exponents $\zeta_\parallel$ and
$\zeta_\perp$ (see Sec.~\ref{sec:selfaff}). This was 
first attempted via elaborate experiments by stopping the experiment and cleaving 
the sample (\cite{bouchaud97}). 
Very recent research has brought new light into these issues. Ponson,
Bouchaud, and co-workers have tested the line propagation
scenario on four materials with very different microstructural
and brittleness/ductility properties \cite{ponson06,bonamy06}:
silica glass, an aluminum alloy, mortar, and wood.
Fig.~\ref{fig:ponson1} shows an example of the correlation
functions and the appropriate exponents, that seem to be valid
for all these cases, and independent of the detailed mode of fracture (tensile, two-point
bending, fatigue etc.). The parallel scaling is in
agreement with earlier $\zeta$-results. It is noteworthy 
that in both cases one has the same dynamical exponent $z$,
so that the two-dimensional correlation functions (to use
the notation of Ponson et al., where $x$ is the direction
of propagation) would follow the scaling Ansatz. Ponson et al.
find that $z \sim 1.25$ which together with the Family-Vicsek
relation then sets the three exponents (see Fig.~\ref{fig:ponson2})
\begin{equation}
\begin{array} {l}
   \Delta h(\Delta z,\Delta x)=\Delta x^{\beta}f(\Delta z/\Delta x^{1/z}) \\
\\
$where$\quad	f(u) \sim \left\{
\begin{array}{l l}
1 & $if u$ \ll 1  \\
u^{\zeta} & $if u$ \gg 1
\end{array}
\right.
\end{array}
\label{eq:2dcor}
\end{equation}
Thus, in this framework, the direction of crack propagation has
the role of the time axis and is thus described by a growth 
exponent $\beta\equiv\zeta_\perp$. This is because the fracture surface results indeed 
from the trail left by the crack front and the dynamic scaling properties
of the roughness are encoded in the perpendicular component.

The role of microstructural disorder 
for fracture roughness is still debated.  Recent experiments point out the
difference of intergranular-transgranular fracture
\cite{morel04}, and in many materials crack bridging should play a role in
roughening.  Some steps towards a more complete understanding
come from the seemingly irrelevant role of crack growth velocity \cite{plouraboue93}, 
and the fact that the same values for the roughness exponent can be measured
from a very brittle material such as amorphous glass
\cite{celarie03,marliere03} where the role of plasticity is negligible, and 
$\xi_{FPZ}$ being of the order of a hundred nanometers. 
Thus the roughness may originate from void formation, a process that
has not been studied much from the viewpoint of statistical fracture.
Bouchaud et al. \cite{bouchaud93} noted that in intergranular
fracture, one sees similar geometric phenomena as in tree-like 
directed percolation clusters of statistical physics. 
The ideas of branching critical cracks have not been studied
systematically since then. One should bear in mind that 
it is only the 2d projection of the 3d crack structure that is
directly analogous to such a "percolation cluster".

We should also mention that there are signs of anomalous
scaling from crack propagation experiments in three dimensions 
\cite{lopez97,lopez98,morel98,mourot05}. Thus
the commonly measured local roughness at a distance $l$ 
does not obey Family-Vicsek scaling, but we would 
need an additional independent exponent to describe
the coupling to sample size, $L$, as discussed in Sec.~\ref{sec:selfaff}. 
The origin of these scaling laws still remain to be understood,
possibly in the framework of line depinning -models. Note that Bouchbinder et al. have
recently suggested that  the actual scaling picture is not simply self-affine,
but one would need to resort to structure functions
as in turbulence \cite{bouchbinder05b}. 

\begin{figure}[hbtp]
\begin{center}
\includegraphics[width=10cm]{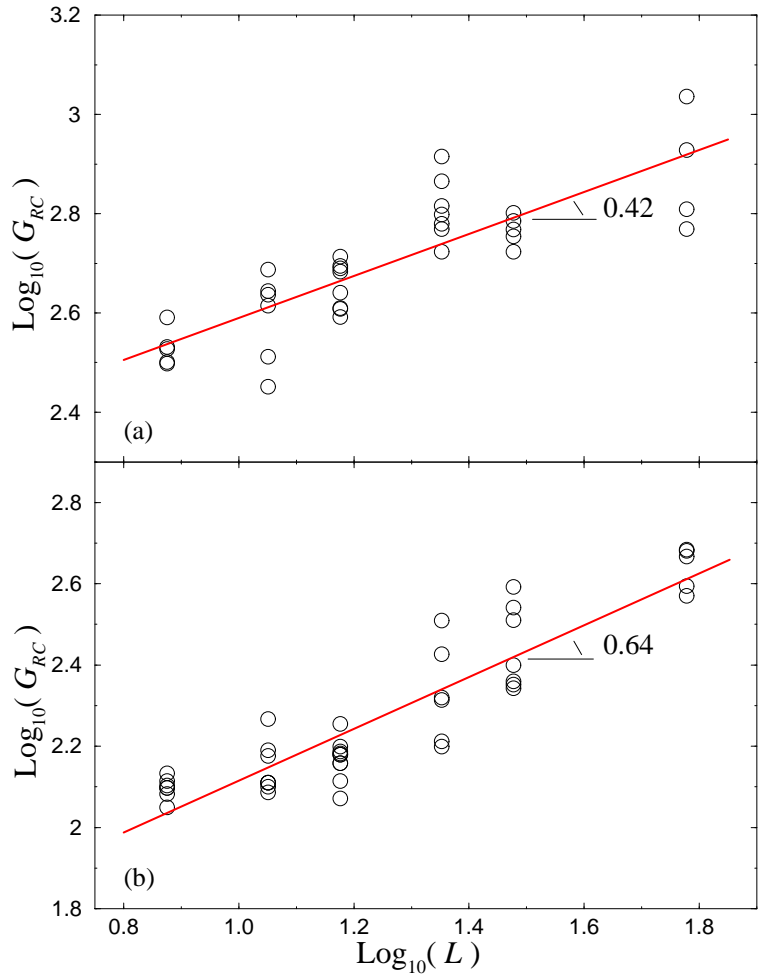}
\end{center}
\caption{Size effect on critical energy release rate $G_{RC}$ for 
two kinds of samples: pine (a), and spruce (b). 
The expected slopes are using the associated
roughness exponents $\zeta-\zeta_{loc}=0.47 \pm 0.17$ for pine (a), 
and $0.73 \pm 0.17$ for spruce (b) (from
\protect\cite{morel02}).}
\label{fig:GvsL}
\end{figure}

Finally, Fig. \ref{fig:GvsL} demonstrates 
how the crack growth resistance (as measured
by the energy release rate $G_{RC}$) depends on sample size $L$ 
in 3D experiments on two kinds of wood
\cite{morel00,morel02}. This can be compared to the theoretical
scaling given in Sec. \ref{sec:selfaff}.
This implies, that the self-affine picture of 
crack surface geometry would be directly useful in engineering
applications if a mathematical relationship can be established
between $G_{RC}$ (or $K$) and $\zeta$ and the material in question. 
One should keep in mind two issues: the complete
absence of a theory to explain the value of $\zeta$ and the 
anomalous scaling picture, and
the fact that quite possibly the microscopic mechanism that
makes $G_{RC}$ increase with $L$ also has a role to play in
setting the actual value of $\zeta$. An engineer's way to 
explain this is that it is simply the prefactor that matters
to first order, in $K$, not $\zeta$, in particular if
$\zeta$ would prove to be universal.

To conclude, we note that fractography is becoming a mature 
sub-branch of fracture mechanics, 
trying to establish links between the qualitative processes and the phenomenology
of surface geometry on one hand, and on the other hand between
crack topography and materials parameters such as fracture
toughness and orientational effects (a good reference is
the book by Hull \cite{hull99}). There has been little effort
to couple the statistical signatures of other quantities to 
those of crack surface roughening, however, regardless of whether
one starts from the statistical physics or materials science
viewpoints.

\subsection{Acoustic emission and avalanches}

The microscopic processes that take place in a fracturing
system with one or with many cracks are difficult to follow
experimentally. Some of the possibilities available are 
thermal imaging, electromagnetic diagnostics, surface strain measurements
including digital interferometry, elastic modulus diagnostics \cite{kouzeli01},
and three-dimensional tomography \cite{kawakata99}.
All of these have their disadvantages including lacking spatial
and temporal resolution, inability to access bulk dynamics,
and the presence of several noise sources that make interpreting data 
difficult. 

Here we concentrate on perhaps the best way to 
characterize microfracturing dynamics, acoustic emission (AE). 
This is an example of ``crackling noise'' which has gained
considerable attention in other contexts, from the viewpoint
of statistical physics. Acoustic emission arises in brittle
fracture due to the release of elastic energy, but can also
originate from internal friction or dislocation motion, to
name a few possible other causes. In the geophysics literature
\cite{lockner93} in particular it has been discussed in the context of
scaling and patterning of fault systems and fracture networks
\cite{berkowitz97,bonnet01,rundle99}, and the growth laws of the
same \cite{lyakhovsky01,goto04}. In analogy to the simple
statistical models discussed in this review the simplest
theoretical interpretations concern the relation of released
elastic energy to the detected AE. However, in most cases 
the models at hand do not take into account the discrete
statistical and noisy nature of the AE data which is, here as well as in
the context of earthquakes, really important.

\begin{figure}[hbtp]
\begin{center}
\includegraphics[width=10cm]{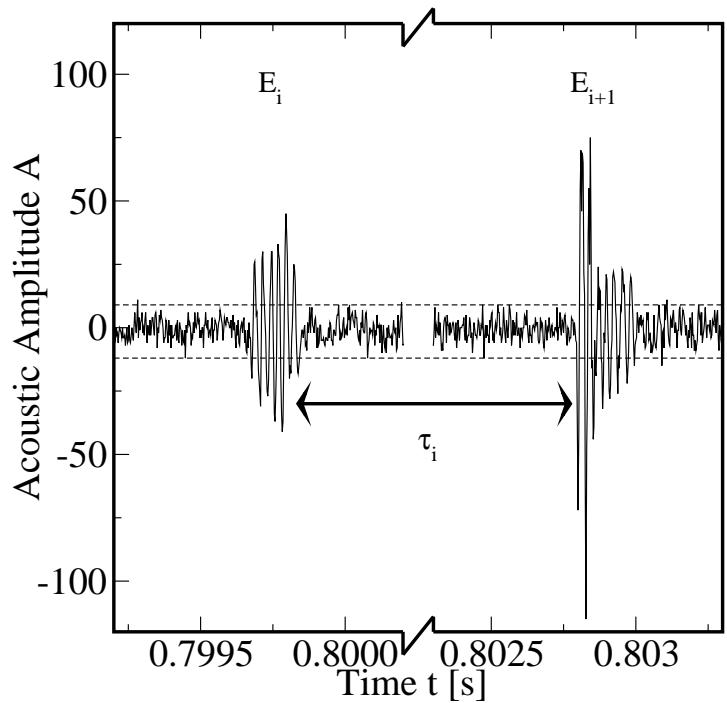}
\end{center}
\caption{A typical example of the crackling noise of acoustic
emission (courtesy of L. Salminen, HUT). The two events that are detected 
are often well-separated in time and are characterized by
duration intervals $\tau_i$ and energy $E_i$. Note the presence
of noise in the setup, which means one has to threshold the data
to define events.}
\label{fig:event}
\end{figure}

A typical AE experiment is based on detecting acoustic
activity via piezoelectric sensors that convert the sound
into an electrical signal. Various levels of sophistication
exist, that allow to distinguish between exact waveforms and
to triangulate the wavesource(s). After noise thresholding, 
the essential point is the existence of well-separated
``events'' such as depicted in Fig. \ref{fig:event}. It shows
two subsequent events from a tensile test on a paper sample,
and demonstrates the most important quantities that relate
to such a signal: quiet times, event durations and event energies.
In a slow fracture experiment one has a separation of timescales,
except perhaps close to the final catastrophic breakdown. The
events as detected by an experimental system are difficult to
interpret because of attenuation \cite{weiss97,lysak96}, 
measurement system response,
and dispersion including reflections from (internal) surfaces. Hence, 
only the simplest quantities such as energy, amplitude, and
silent intervals are relatively easy to analyze.
Nevertheless there is a considerable body of work
on the relation of AE to microscopic fracture mechanisms
and damage accumulation versus material composition,
including for instance concrete and metal-matrix and fiber composites 
where the contrast between the fibers and the matrix, 
allow such a distinction 
\cite{ouyang91,harris92,mummery93,barre94,degroot95,bohse00,dzenis02}.

\begin{figure}[hbtp]
\begin{center}
\includegraphics[width=10cm]{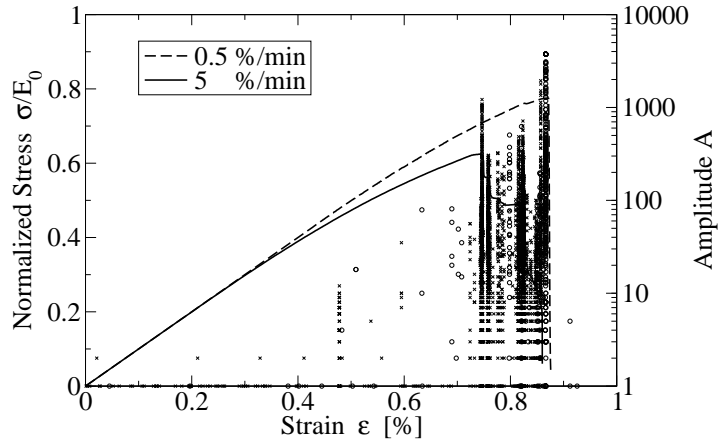}
\end{center}
 \caption{Two examples of acoustic data series (crosses and circles) 
for two strain rates of paper samples in tensile testing. 
The stress-strain curves change with the strain rate
(from Ref. \cite{salminen02}).}
\label{fig:aedata}
\end{figure}

\begin{figure}[hbtp]
\begin{center}
\includegraphics[width=10cm]{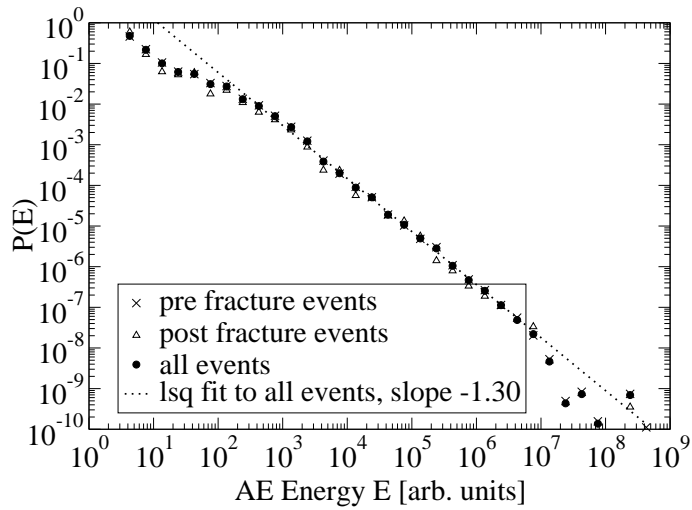}
\end{center}
 \caption{Three different
energy distributions from a tensile AE test for
paper ($\dot{\epsilon}$ = 1.0 [\%/min]). The data
is separated in addition to the total distribution for
the post- and pre-maximum stress parts (from Ref. \cite{salminen02}).}
\label{fig:aeenergystat}
\end{figure}

Figure \ref{fig:aedata}  depicts two tensile tests for
paper samples, illustrating the typical dynamics of AE
in a test. The interesting questions are, quite similar
to crack roughness, how does the statistics depend on
rate of loading, dimensionality, material, and fracture
mode? At the simplest level the integrated acoustic energy
should tell about the loss of elastic energy content,
and thus be proportional to $(1-D)$ (in terms of a scalar damage
variable) if one assumes that attenuation
and acoustic coupling do not change as failure is approached. 
From a  variety of
experiments that include a variety of loading 
conditions including tension, compression, and shear, 
and in addition under creep loading conditions, and different materials
it has been found that the silent times $\tau$ and the
acoustic event energies $E$ follow analogies of earthquake
statistical laws. For the energy, one has 
\begin{equation}
P(E) \sim E^{-\beta},
\label{eq:aeene}
\end{equation}
and for the intervals
\begin{equation}
P(\tau) \sim \tau^{-\alpha},
\label{eq:aetau}
\end{equation}
where the exponents $\alpha$ and $\beta$ are akin to those
of the Gutenberg-Richter and Omori laws of earthquakes. 
Typical values
for $\alpha \sim 1-1.5$ and $\beta \sim 1-1.3$
are then a target for theoretical models
\cite{petri94,garcimartin97,guarino98,maes98,guarino02,salminen02}. 
An example of
the energy statistics is given in Fig. \ref{fig:aeenergystat},
for paper as an example of a two-dimensional test material.
In an in-plane experiment, $\beta \sim 1.8$ has been found \cite{salminen05}
which is intriguingly close enough to the avalanche exponent of
crack lines in Plexiglas (see Fig.~\ref{fig:maloystat} \cite{maloy06}).

The example of Fig.~\ref{fig:aedata} hinted already about one
particular aspect of typical fracture AE: the lack of 
time-translational invariance since e.g. the activity 
gets stronger as the $\sigma_c$ is approached
(though in special cases, as in ``peeling'' \cite{salminen05} this
can be circumvented). This 
presents both a challenge in order to understand
the statistics - recall that often one does not have
access to large quantities of data - and implies that
the AE timeseries contains clues about the microscopic
dynamics. 

\begin{figure}[hbtp]
\begin{center}
\includegraphics[width=8cm]{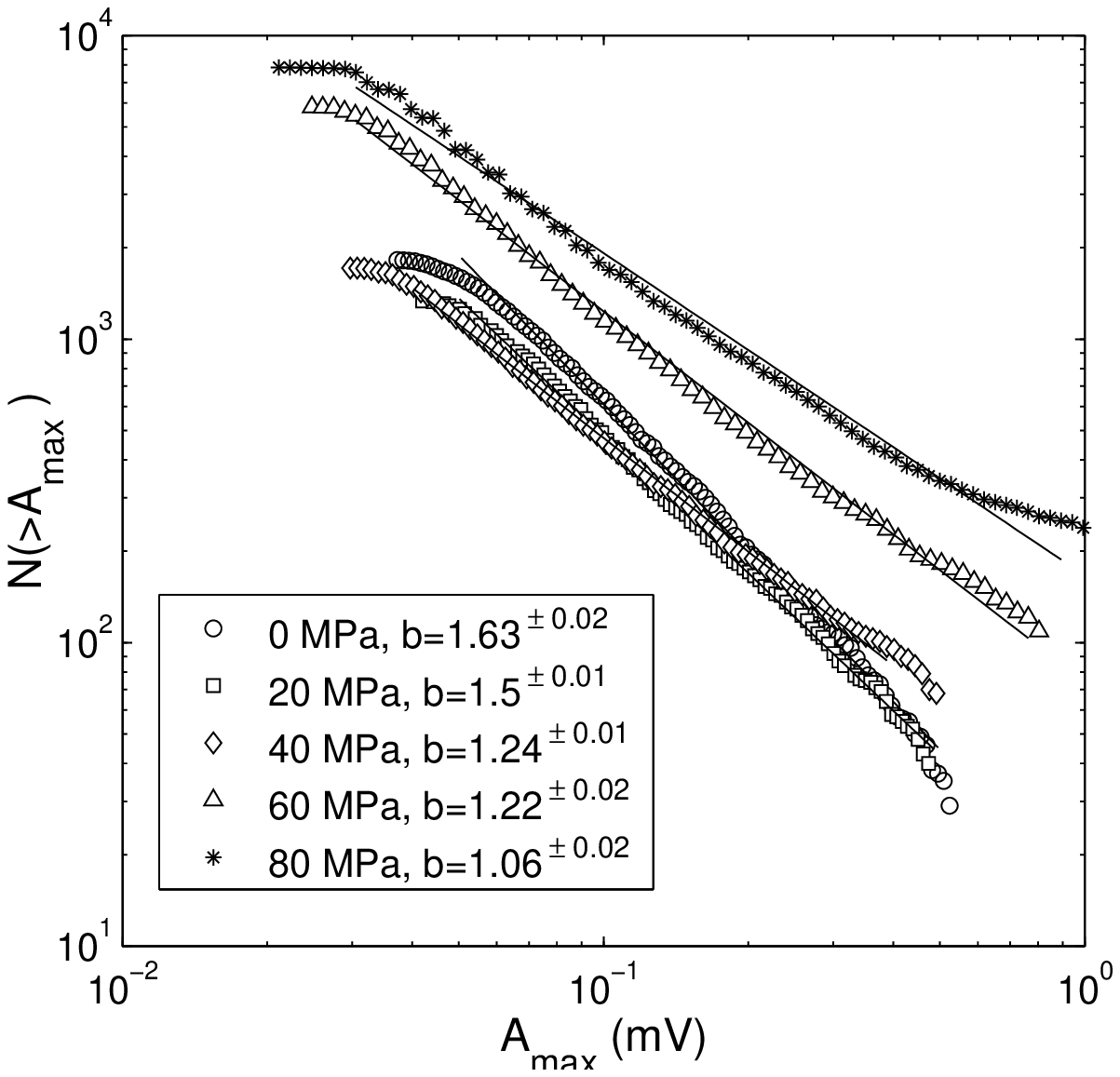}
\includegraphics[width=8cm]{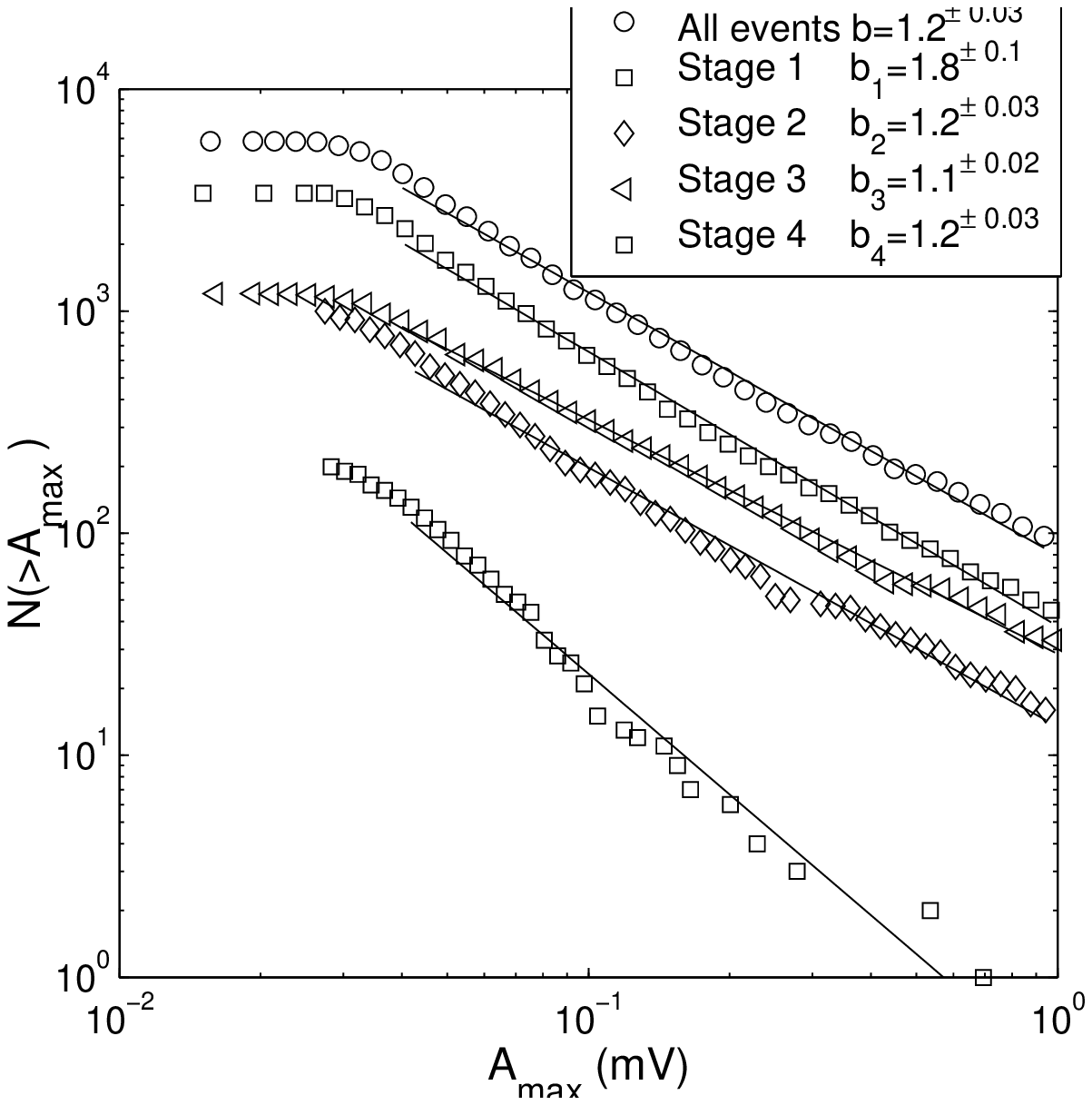}
\end{center}
 \caption {``b-values'' or the corresponding amplitude 
distributions for acoustic
emission from rock samples. It is noteworthy how the exponent
of the cumulative distribution becomes smaller (from ``stage 1'')
as the maximum
point along the stress-strain curve is approached \cite{amitrano03}.}
\label{fig:bvalues}
\end{figure}

Fig.~ \ref{fig:bvalues} (ref. \cite{amitrano03})
demonstrates how
the {\em amplitude} statistics changes as one follows
a fracture experiment along the stress-strain curve.
Here, the ``b-value'' or the analogous $\beta$-exponent
becomes smaller signaling that there are larger and larger
energy releases as $\sigma_c$ is approached. 
The b-value is one of the targets of theoretical analysis
based on averaged damage and crack-growth laws, and is based
on a ``$K_{eff}$'', which would predict energy release rates
as the catastrophic fracture gets closer. These models can
of course be tuned so as to match with the general trends
of detecting larger amplitudes as the final fracture is
approached.
Still, the
agreement is qualitative and the possible ``universality'' of
the signatures
remains unclear 
\cite{main91,main92,hatton93,main93,reches94,lyakhovsky97,main00}.
Similar b-value studies have outlined e.g. the role of pre-existing cracks
in statistics \cite{lei00}, followed damage \cite{colombo03} and discussed the crack velocity's
role \cite{aue98}.

The changes in behavior as sample failure is approached and
the concomitant changes in data can 
be discussed in a variety of ways. One way to
consider the matter is to try to do actual localization
experiments. These often suffer from the fact that
dead-times of the apparatus and uncertainties in
localization make it difficult to follow all the
developments (in contrast to plastic deformation \cite{weiss03}, 
or steady-state fracture 
where the possibility of a quasi-steady state exists
 \cite{salminen05}). Nevertheless, it can be convincingly
be demonstrated how damage localizes to the final
process zone, or to the proximity of an advancing
crack in the final stages of stable crack growth.

One can also check when
or up to which point 
the assumption of slow failure is valid, in that the
events remain uncorrelated on the fast timescale of
inverse sound velocity \cite{krysac98}. 
The spatial clustering of events has been extensively
discussed in rock mechanics \cite{lockner91,shah95,labuz98,zang98}
and correlation analysis is often performed on the
AE timeseries \cite{feng99,hayakawa04}. 
The latter presents a technical problem due to
the non-stationarity of the time-series, as the
AE event density often most varies during the
experiment, and thus one should apply caution.
In the spatial case, one can also do entropy analysis
to distinguish between random, Poissonian statistics
and real clustering \cite{guarino98}.  
This is demonstrated in
Fig. \ref{fig:guarino} where one can see the development
of correlations in the locations of AE events ``in the plane'',
when a planar sample is subjected to an effective tensile load.
The final failure surface is circular, as can be seen in
Fig. \ref{fig:guarino}a. The same samples (of wood)
also allow to look at the clustering of the event locations
via ``entropy analysis'' (Fig. \ref{fig:guarino}b) in analogy
to what we discuss about damage development in the numerical
models below.
An open question is, whether one can by spatial analysis
say something about the dynamics inside and
formation of the FPZ \cite{zietlow98} and determine the ``Representative
Volume Element'' size (see e.g. \cite{kanit03}).

\begin{figure}[hbtp]
\begin{center}
\includegraphics[width=8cm]{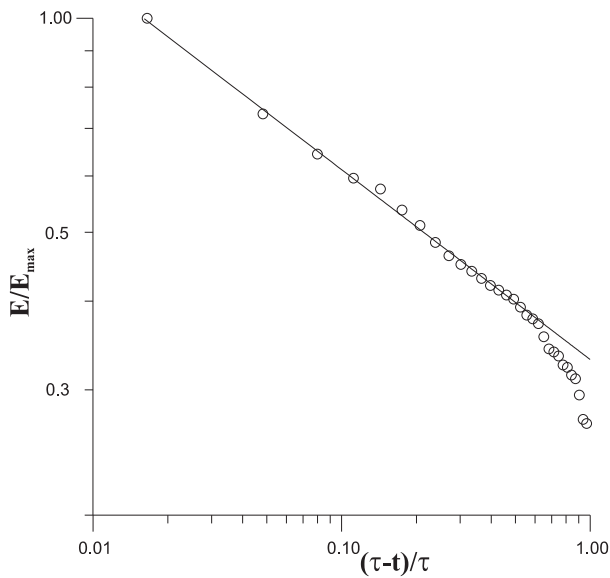}
\end{center}
 \caption{The cumulative energy E, normalized to E$_{\max }$, as a
function of the reduced control parameter $(\tau
-t)/\tau $, plotted in the proximity of the failure in the
case of a constant applied load (pressure). 
The circles are the averages, for 9 wood samples.
The solid line is a trial fit $%
E=E_0\left( \frac{\protect\tau -t}{\protect\tau} \right)
^{-\protect\gamma }$. According to 
Ref.~\cite{guarino02}, $\protect\gamma =0.26$,
is independent of the ``load''. Note the range of $E$ in which
the scaling seems to apply.}
\label{fig:enediv}
\end{figure}

\begin{figure}[hbtp]
\begin{center}
\includegraphics[width=8cm]{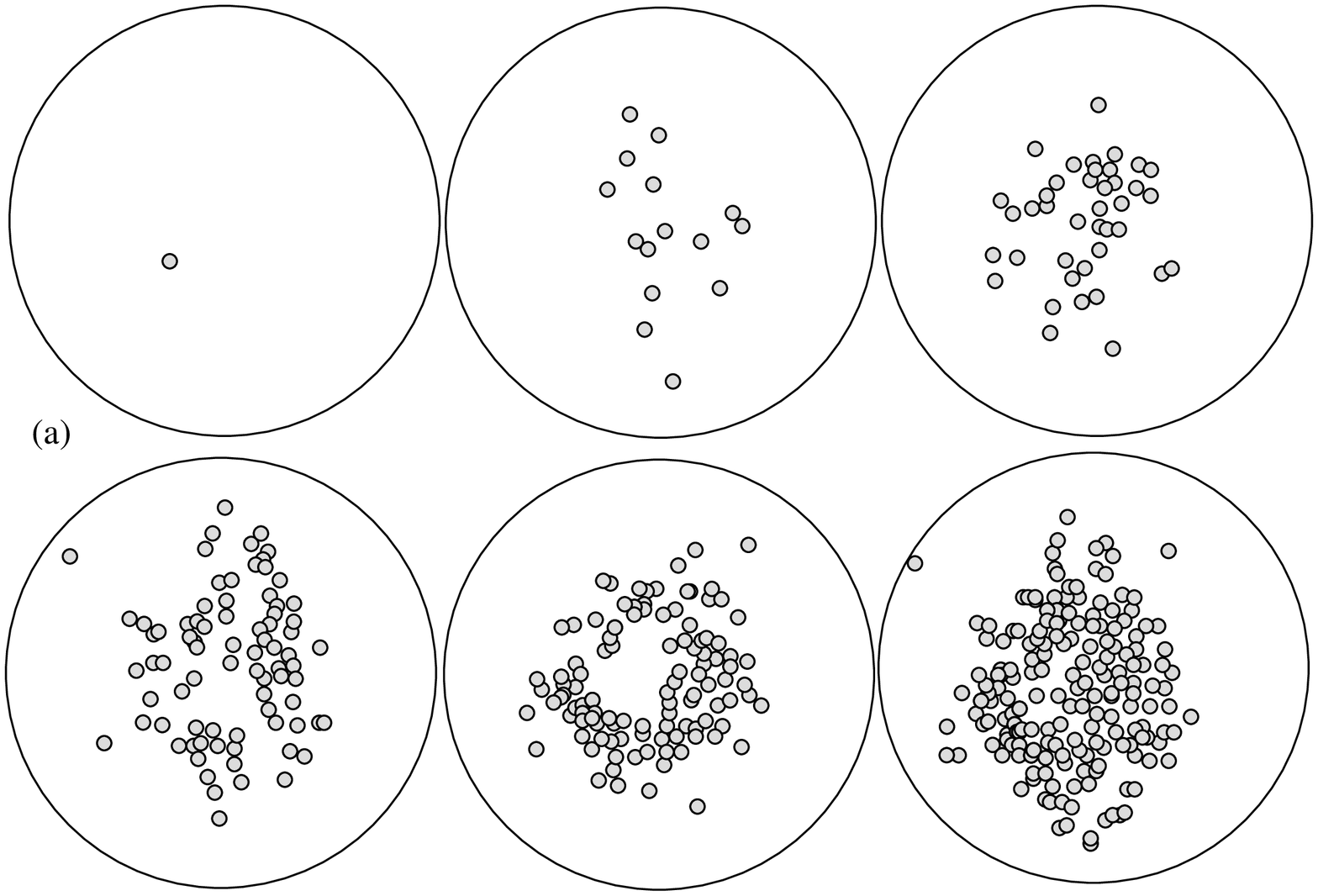}
\end{center}
\vspace{0.5cm}
\begin{center}
\includegraphics[width=8cm]{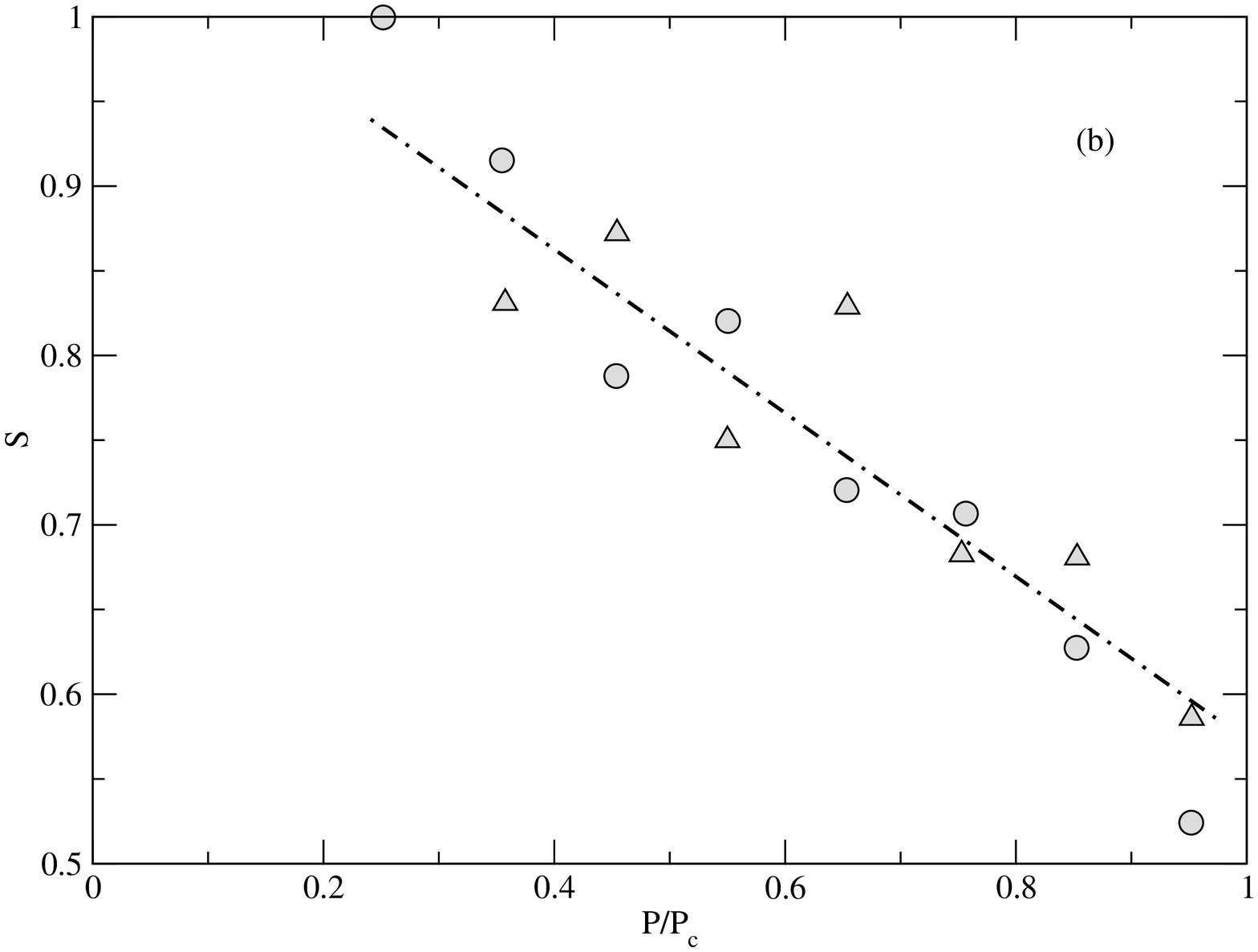}
\end{center}
 \caption{a): the microfractures localize (in 2D) as the control
parameter is increased. The final panel has all the located events
from the five equidistant intervals in five subplots 
(from Ref. \protect\cite{guarino98}). Note that any localization experiment
typically misses some events that can not be placed with certainty.
b): The entropy, which decays roughly linearly from one (corresponding
to random locations) as the final fracture is approached for
the pressure (control parameter) value of $P/P_c=1$.}
\label{fig:guarino}
\end{figure}

Another consideration is
to link the AE activity to final crack formation and
properties,
but unfortunately so far this has not been attempted
(in paper, one can relate the jerky advancement of a crack tip
- which is easy to demonstrate, see Ref.~\cite{balankin01} -
to AE activity, though).
Instead, several authors have considered the AE energy
release rate, or the integrated energy as a simplest 
way to see the approach to final failure. This naturally
tells about the correlations in the damage development,
and the nature of the final crack growth. Conflicting
evidence exists, and Figure \ref{fig:enediv} shows that
in some cases experimental data can be used to justify
the possibility of a ``critical divergence'', in that
the data seems to fit a scenario where $\partial E /\partial t
\sim (t_c - t)^{-a}$, where $t_c$ is a failure mode 
-dependent critical time (e.g. equivalent to 
a critical strain). In some cases, this seems not
to work so well (an exponential increase is rather
indicated) \cite{rosti06}, moreover such data has also
been used to justify the presence of collective dynamics
in the process via ``log-periodic oscillations'',
\cite{anifrani95,johansen00}. Such collective phenomena,
of unknown exact origin, would be of importance since they
would allow to consider AE as a diagnostic for sample life-time
or failure predictability.

\subsection{Time-dependent fracture and plasticity}
\label{sec:timedep}
%skip the Ciliberto time-dependent stuff.
Creep, fatigue and ductile failure extend the 
theoretical questions one meets within brittle or quasi-brittle
fracture. Dislocations, activated dynamics, and 
frictional response (fiber pull-out in fiber composites,
crack surface interactions) can be combined
with the presence of ``disorder'' (the emphasis of this
article) to repeat many of the relevant questions:
how to describe the statistics of strength, how to understand
the properties of crack surfaces, and how to interpret
damage mechanics via acoustic emission. In fact in
many of the experimental references cited above one can
not so easily label the material or test at hand as
purely brittle or ``ductile''. In some cases the physics
is changed ``relatively little'': the existence of ductility
and plastic deformation is interpretable in terms of a FPZ
size $\xi_{FPZ}$ inside of which the divergence of the stress-field
of LEFM is cut off.

The two classical problems of fatigue and creep should
also be affected by disorder. Indeed, the ``Paris' law''
which states that the lifetime $t_f$ of a sample in 
stress-cycled experiment, with a variation $\Delta K$
for the stress intensity factor scales as $t_f \sim (\Delta K)^n$
where $n$ is the Paris' law exponent, can be measured
in various non-crystalline materials which are certainly
inhomogeneous \cite{suresh98}. Likewise, the physics of
creep in disordered materials as paper or composites is
affected in some yet to be understood way by the randomness.
This can be seen in the stick-slip -like advancement of cracks in creep
\cite{santucci04}  and in the statistics of creep times
and the related acoustic emission \cite{nechad05}.
In these directions, much more careful experimental
work is called for.

In the case of pure dislocation plasticity, outside of
``fracture'' there are collective effects (see the review
of Zaiser for an outline \cite{zaiser06}), which often
need to be treated with the tools of statistical mechanics. 
Examples of collective behavior are the Portevin-LeChatelier
effect \cite{hahner02}, and  the temporal and
spatial fluctuations in dislocation dynamics \cite{deshpande01,weiss03}.
One would like to understand the laws of damage
accumulation and how to define a RVE or correlation
length 
\cite{nematnasser99,needleman00,main00,mahnken02,bazant02,clayton04}. 
The plastic deformation
develops in non-crystalline materials patterns, whose
statistical structure is largely not understood
(see e.g.. Ref.~\cite{ostojastarzewski05}) though simple
models have been devised \cite{baret02}. These questions
could also be studied again via statistical mechanics
ideas and models.

\section{Statistical models of failure}
\label{sec:models}
In the following we establish the rules of the game.
We consider models with interacting elements, whose
coarse-grained behavior captures the key 
features of fracture processes in the presence of 
{\it randomness}, that can be either {\it annealed}
(thermal) or  {\it quenched}
(frozen). Moreover one may consider models in which
one incorporates a history effect. A typical case would
be where the  damage accumulated by individual elements
would be a monotonically increasing stochastic function
of its stress-history.  The simple rules used are then
intended to capture the behavior of materials on the
mesoscopic level ranging from brittle to perfectly
plastic to viscoelastic. Another extension concerns
models with inertia, so that sound waves are included.
We also briefly take up the issue of connecting these
abstract models to the finite element simulations 
practiced in engineering fracture mechanics and to atomistic
simulations. Note again that in this review we on purpose leave aside
the well-established field of dielectric breakdown,
in spite of its close connections to scalar fracture
\cite{duxbury88}.

\subsection{Random fuse networks: brittle and plastic}
\label{sec:rfm_plastic}

\begin{figure}[t]
\begin{center}
\includegraphics[width=10cm]{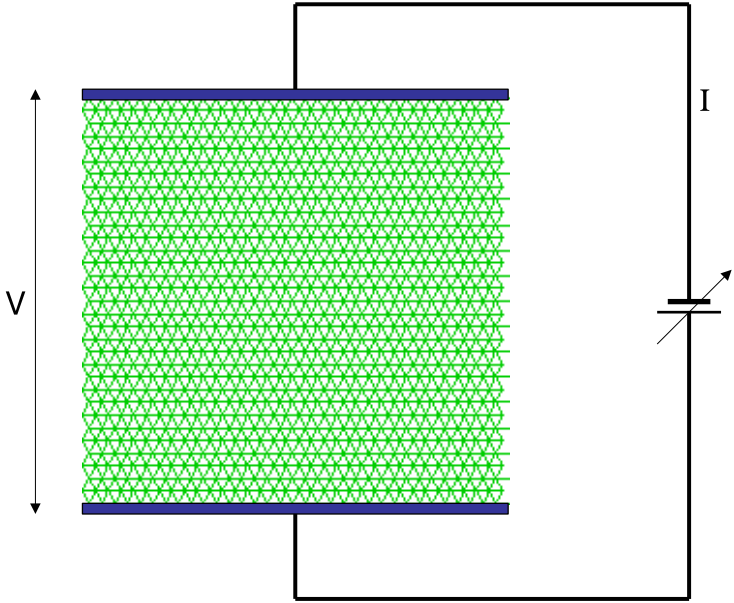}
\end{center}
 \caption {Random thresholds fuse network with triangular lattice topology.  Each of
the bonds in the network is a fuse with unit electrical conductance and a
breaking threshold randomly assigned based on a probability distribution (for example, 
a uniform distribution).
The behavior of the fuse is linear up to the breaking threshold. Periodic
boundary conditions are applied in the horizontal direction and a unit voltage
difference, $V = 1$, is applied between the top and bottom of lattice system
bus bars. As the current $I$ flowing through the lattice network is
increased, the fuses will burn out one by one.}
\label{fig:rfn}
\end{figure}

\begin{figure}[htb]
\begin{center}
\includegraphics[width=10cm]{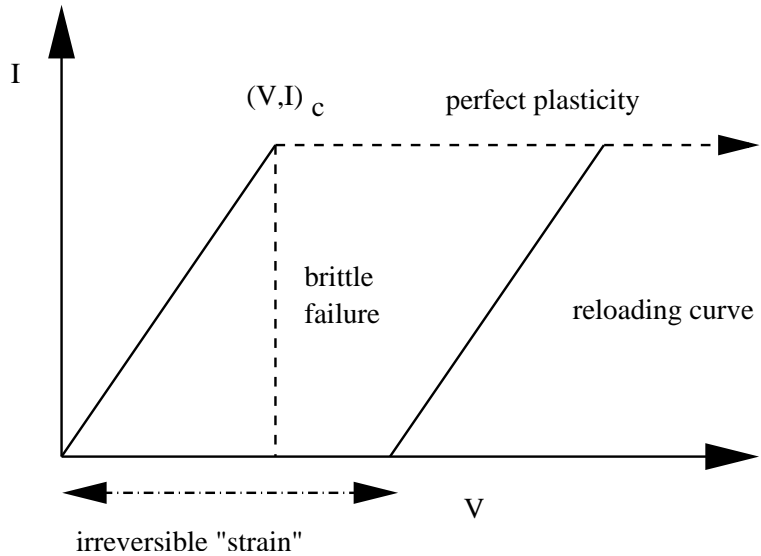}
\end{center}
 \caption {Examples of perfectly brittle and plastic
fuses. A fuse undergoes an irreversible transformation
from $(V,I)_c$ to $(V_c,0)$ in the case of a brittle fuse or keeps its current at the
yield current ($I_c$ or $I_y$) in the case of a plastic fuse. In the brittle fuse case
the response is linear up to $(V,I)_c$. On the other hand, in the plastic fuse case, the 
unloading tangent modulus is same as the original tangent modulus, and in the 
fully unloaded state (zero current) there is a remnant ``strain'' or voltage in the bond.}
\label{fig:fuse}
\end{figure}

The simplest truly interacting fracture models
are variations of the scalar analogy of fracture captured by
the random fuse model (RFM) \cite{deArcangelis85,herrmann90}.
The equations governing the RFM can be considered as a discretization
of the continuum Laplace equation
\begin{equation}
\nabla^2 V = 0
\end{equation}
with appropriate boundary conditions (see Fig.~\ref{fig:rfn}). 
This scalar electrical analogy of elasticity is 
a simplification over the Lam\'e equations (Eq. (\ref{lame})) described in 
Sec.~\ref{sec:frac-mec}, and formally corresponds to an antiplanar
shear deformation scenario. In the discrete version, the
currents and voltages over the nodes $x_{ij}$ 
satisfy the Kirchhoff and Ohm's laws and depend on 
the local conductivities $\sigma_{ij}$ for the links 
(or resistors between the nodes). The fuse is then simply an
element with a property of irreversibility (see Fig.~\ref{fig:fuse}): once a critical current
(or similarly a voltage drop across the fuse, but if $\sigma_{ij} \equiv 1$ the two 
cases are perfectly equivalent) $i_c$ is reached, the conductivity goes
to zero. Clearly, a model with such fuses is appropriate
to describe the failure of very {\em brittle} media. 
The failure becomes more gradual in the 
presence of disorder, since not all the links would fail 
at the same time. Although much of our focus is on the RFM, many of the issues that 
we consider are directly applicable to 
other discrete models as well.

Quenched randomness is usually introduced in the RFM in different ways:
\begin{enumerate}
\item {\it random thresholds}, extracting the thresholds $ i_{c,ij}$ of each
link from a probability distribution $p(x)$;
\item {\it random dilution}, removing a fraction $p$ of the links at
the beginning of the simulation;
\item {\it random conductivities}, extracting the conductivities $\sigma_{ij}$ of each
link from a probability distribution $p_{\sigma}(x)$.
\end{enumerate}
All these three methods allow to tune the amount of disorder present in
the system by changing the distributions or the dilution ratio.

The crossover
from weak to strong disorder is mostly studied in the framework {\it random thresholds} (case 1) using
either a uniform distribution in $[1-R,1+R]$ \cite{kahng88} or a power law distribution
$p(x) \sim x^{-1+1/\Delta}$ in $[0,1]$ \cite{hansen91}. The first case interpolates between
no disorder  ($R=0$) to strong disorder ($R=1$), while the second case allows for 
extremely strong disorder, since for $\Delta\to \infty$ the distribution becomes
non-normalizable. The case $R=1$ is probably the most 
studied in the literature. Alternatively it is also possible in principle to connect 
the disorder in the model to a realistic material microstructure by prescribing 
the fuse properties based on the morphology of the material microstructure 
(see Sec. \ref{sec:latticetofem}). The effect of percolation type disorder 
({\it random dilution}, case 2) on fracture will be explored in Sec.~\ref{sec:perc}. 
The scenario of quenched {\it random conductivities} poses an interesting
question on how different disorder phases would need to
be defined in the case of a RFM \cite{ootani92}. In general, analogous to elastic media with locally
varying elastic constants \cite{hristopulos04} or simply percolative
disorder, quenched conductivity variations create correlations among local currents $i$ even in
the linear IV-regime. The largest current or stress variations are related to 
specific, ``funnel''-like defects (see e.g. \cite{chen02}), which
plays an important role in the weak disorder limit.
Similarly, the use of irregular lattices is analogous to a percolation
type disorder. These type of irregular lattice systems have been carried out using 
spring and beam-type lattice models \cite{schlangen96,bolander05}.

The brittle RFM scenario is usually considered in the
limit of time scale separation. This  implies that the equilibration time
of the currents is much faster than the ramp-up rate of external potential
or current. Moreover, any type of possible overshoots in the local currents 
$i_{t,ij}$ are neglected (see Sec. \ref{sec:dynamics} for discussion on 
inertial effects). The dynamics is then defined by finding the quantity
\begin{equation}
\mathrm{max}_{ij}\,\,\, \frac{i_{t,ij}}{i_{c,ij}}
\label{max}
\end{equation}
at each timestep $t$. This is an example of {\em extremal}
dynamics, similar to many toy-models that result in
``avalanches'' of activity. Here the physics ensues
from the combination of local current enhancements and 
disorder, in the thresholds $i_c$. The question is then,
whether the latter plays any role or not, and, if it does, 
what are the consequences for a statistical and dynamical
properties of fracture. Instead of successive bond failures of 
one at a time, for computational convenience, it is also 
possible to increase the external voltage 
in steps, and then let {\em all} the fuses for which the ratio
of Eq. (\ref{max}) exceeds unity fail at once. This of course neglects
the small adjustments of the static electric/elastic field 
that would take place in the usual one-by-one fuse failure dynamics.

\begin{figure}[t]
\begin{center}
\includegraphics[width=10cm]{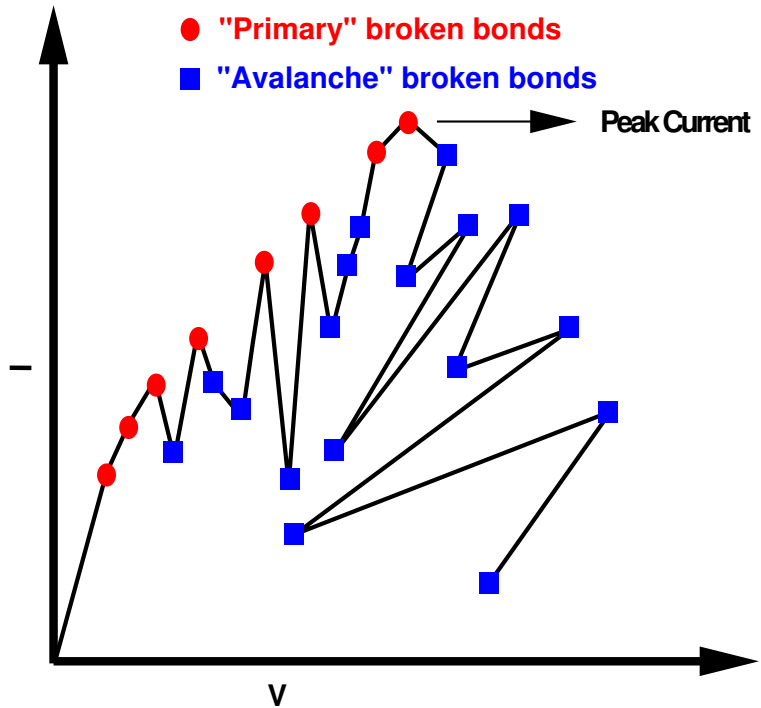}
\end{center}
 \caption {The schematic of a current-voltage curve using RFM. 
Two alternatives, current- and voltage-driven,
are outlined. The schematic corresponds to a failure process in 
which one bond at a time fails.}
\label{sscurve}
\end{figure}

This discussion should at this point be made more precise by defining
the pertinent quantities. Figure \ref{sscurve} shows a schematic 
of the $I-V$ behavior of a RFM. Note that the series of fuse failures 
defined by Eq.~(\ref{max}) is unique given the set of thresholds $i_c$.
The $I-V$ characteristics depends, however, on the way the system is driven:
two different $I-V$ curves ensue depending on whether the system is driven
by constantly increasing current or voltage.
The figure implies that in the voltage driven scenario, there is the 
possibility of a weakening tail (the regime of $I < I_{max}$ and
$V > V(I_{max})$), whereas in the current driven scenario, the curve stops at $(V,I)_{max}$. 
Hence, depending on the loading scenario, there are two failure points. 
Under current control
all the fuses in the post-peak regime (beyond the peak current $I_{max}$) break at once, 
whereas in the voltage case fuses in the post-peak regime fail gradually. The schematic 
presented in Fig. \ref{sscurve} corresponds to a typical simulation in which 
failure of one bond at a time occurs. In order to obtain the average $I-V$ response, 
a naive averaging of the $I-V$ responses at constant number of broken bonds may not be the 
best option since such an averaging will 
smoothen out the maximum, exactly as if many such systems were in parallel.
One way to overcome this problem would be to center first the voltage axis on the breakdown
point $V(I_{max})$ of each realization and then average. Alternatively, an average of the 
envelopes of the $I-V$ curves can be performed at constant voltage values (see Sec. \ref{sec:iv}).

An interesting variant of the RFM 
is given by the perfectly plastic one.  The corresponding
fuse response is pictured in Fig.~\ref{fig:fuse}: 
a blown fuse, instead of failing, ``yields'' so that its current
is set to a constant. In this case, the yield iteration
proceeds until there is a connected yield path crossing
the system, which then blocks any further increase.
These systems can be simulated via a ``tangent-algorithm''
envisioned by Roux and Hansen. The crucial point is that
after a fuse yields any further progress along the 
$I-V$ curve can be considered in a state in which the 
yielded fuses are simply removed. Then, the $I-V$ behavior
will be composed of piece-wise linear parts.
The important thing here is the $i_{t,ij}/i_{y,ij}$ ratio, and during 
each step one subtracts from $i_{y_ij}$
the already existing current.
Meanwhile permanent strain (or actually voltage) $V_{pl}$ 
develops. The tangent modulus or conductivity stays
still at the initial value, which means that if
the system is allowed to unload to zero current then there
will be a remnant voltage, $V_{pl}$, in each of the plastic fuses. Consequently, 
local {\em internal stresses} develop in each of the fuse of the lattice system. 
All this rich behavior arises in a model 
with simple scalar plasticity and without any time-dependent deformation.

Since the final yield current is
connected to the presence of a yield surface, it is
possible to understand its properties {\em without}
resorting to actual RFM simulations. The 
path is given by the solution of the following optimization
problem: minimize the value of the path-cost, over all
possible configurations, when the path energy is given by
\begin{equation}
\langle \sigma_y \rangle L = \sum_{path} \sigma_{ij \in path}.
\end{equation}
This is exactly the energy of the random bond Ising Hamiltonian
at zero temperature (groundstate), so a number of properties
are apparent (see \cite{alava00} for a review).
The yield surfaces in $2D$ and $3D$ are self-affine,
with the scaling exponents $\zeta_{2D} = 2/3$ and $\zeta_{3D} 
\sim 0.41$. These can be affirmed by using a mapping (in $3D$)
to the max-flow/min-cut problem of combinatorial optimization,
and in $2D$ the task maps to the directed polymer in a 
random medium (DPRM) partition function, 
which can be connected to the Kardar-Parisi-Zhang equation (KPZ)
of kinetic interface roughening \cite{kardar86}. The KPZ problem is exactly
solvable, and thus the $\zeta_{2D}$ follows.
Further consequences for yield surfaces are the fact
that the yield strength fluctuates, with the standard
deviation scaling as $L^\theta$, where $2\theta = 2\zeta+d-3$, and $d$ is the 
dimension ($d = 2$ for 2D and $d = 3$ for 3D).
Also, the exact nature of the probability distribution
function (pdf) of fuse yield thresholds is irrelevant, as long
as the thresholds are not too strongly spatially correlated, and 
the pdf does not have too fat a tail (power-law distribution,
scaling as $P(i_y) \sim i_y^{-a}$ with $a<4$). 

Finally, it should be noted that simulations 
of brittle-plastic networks of the kind depicted
in Fig.~\ref{fig:fuse} have not been carried out so far. 
Such networks would present interesting
problems to analyze from the theory viewpoint since
the scaling properties of fracture surfaces,
say $\zeta_{2D}$, are expected to be different. From a 
computational point of view, one problem 
is that unloading due to fuse failure causes yielded
elements to return to elastic/linear behavior. Furthermore,
returning to zero external load nevertheless creates 
{\em internal stresses} along with irreversible deformation \cite{alava95}.

It is interesting to mention here that there is a way to interpolate
between fully brittle and plastic response. Instead of removing
a fuse, it is possible to just reduce its conductivity (akin to softening) 
by  $\sigma_{ij} \to (1-D) \sigma_{ij}$, where $D$ refers to damage. This
leads naturally to crack arrest \cite{li88}, and thus to distributed
damage. If one iterates the procedure (\cite{zapperi97b}) so that
continuous damage (Eq.~(\ref{damage})) is created by several fracture events 
per bond, one can observe an effectively plastic $I-V$ curve. Note that the
tangent modulus is reduced, but there is no irreversible deformation.
Models of this type have been found effective to model biological materials \cite{nukala05b}, 
but could also be useful for modeling softening of concrete.

Finally we notice that there are some 
experimental studies of systems corresponding
to realizations of the RFM. Otomar et al. build a 
network of fuses similar to the numerical models considered in Sec. \ref{sec:simul} \cite{otomar06},
while a thermal fuse model has been
worked out experimentally in Ref. \cite{lamaignere96}.
Such attempts, while being laudatory, also highlight
the difficulty of reaching substantial system
sizes. Otomar et al. have studied the $I-V$ curves
and the dependence of the strength on the effective
disorder, which was varied by putting together samples differently. 
Their experimental results are all comparable to the numerical results
presented in Sec. \ref{sec:simul}.

\subsection{Tensorial models}
\label{sec:tensorial}
Following the general strategy of the RFM, several models with more realistic elastic 
interactions (see Sec.~\ref{sec:lin-elast}) have been proposed in the past. 
The simplest possibility is provided by central
force systems, where nodes are connected by elastic springs
\cite{hansen89C,sahimi931,sahimi932,nukala05}. The random spring model 
(RSM) is defined by the Hamiltonian
\begin{equation}
H = \sum_{ij} \frac{K}{2} (\vec{u}_i - \vec{u}_j)^2,
\label{CF}
\end{equation}
where $\vec{u}_i$ is the displacement of node $i$ and
$K$ the spring constant. The elastic equilibrium is obtained
by minimizing Eq.~(\ref{CF}), and disorder is introduced in
the standard way by either random dilution, random threshold, or random elasticity. 
Despite the fact that the RSM suffers from the caveats associated
with the presence of soft modes, it is still considered quite often in the context 
of fracture. In terms of continuum 
elasticity, the discrete triangular central-force lattice model represents an isotropic elastic medium with 
a fixed Poisson's ratio of $1/3$ in two-dimensions,
and $1/4$ in three-dimensions.

A similar slightly more complicated model is the Born model, 
defined by the Hamiltonian
\begin{equation} 
H = \frac{1}{2} \sum_{(i,j)} \left[ \alpha 
(\vec{u_i} - \vec{u_j})_{\parallel}^2
+ \beta (\vec{u_i} - \vec{u_j})_{\perp}^2 \right].
\end{equation}
where $(i,j)$ represents the nearest neighbors of the node $i$, and 
the summation is carried out over all the nodes in the lattice system. 
The quantities $(\vec{u_i} - \vec{u_j})_{\parallel}$ and 
$(\vec{u_i} - \vec{u_j})_{\perp}$ represent the relative 
displacement of the node $j$ in the directions parallel 
and perpendicular to the bond $(i,j)$ respectively. 
In this case, there is a primitive competition between
stretching and bending the ``bond'' between two nodes
on a lattice. Note that in
the limit $\alpha = \beta$ the Born model reduces to
the random spring network (if $u$ is scalar, then it reduces to 
random fuse network), but otherwise it still
lacks one fundamental property of real elastic systems that 
local rigid rotations cost energy (see Refs. \cite{feng,hassold89}).

The most used (and most physically correct) models
in this respect are the so-called beam and bond-bending models. 
One may view this kind of a system as a lattice of massless
particles interacting (or connected) by a beam to their nearest neighbors 
\cite{roux85}. In the beam models, each beam is capable of sustaining  
longitudinal ($F$) and shear forces
($S$), and a bending moment ($M$). Following 
the notation of Ref. \cite{herrmann89},  one can define three
effective parameters $a=l/EA$, $b=l/GA$, and $c=l^3 /EI$. For a more generic 
treatment of 2D and 3D beam models, refer to Ref. \cite{nukalaijnme}. 
Using standard beam theory, it is possible to relate the beam displacements and rotations  
($x$ and $y$ displacements and rotation) to the three
forces $F$, $S$, and $M$ using the parameters $a$, $b$, and $c$. 
The out-of-plane deformations or "buckling" of $2D$ beam lattice models 
have however not been dealt thoroughly in the literature so far. 
In the literature, the beam models have often been used to describe the fracture of random
fiber networks, as lattice models of composites and paper. In terms of 
continuum elasticity, the beam models correspond to discretization of 
Cosserat continuum elasticity.
The bond-bending model, on the other hand, attempts to be
a generalization of the central force model  
\cite{kantor,sahimi86,sahimi932}, whose Hamiltonian is given by 
\begin{equation} 
H = \sum_{ij} \frac{K}{2} (\vec{u_i}-\vec{u_j})^2 +
\sum_{ijk} \frac{B}{2} (\Delta \theta_{jk})^2 
\end{equation}
where $K$ and $B$ are extensional and rotational rigidities. In the above Hamiltonian, 
the first term corresponds to the usual extensional energy between 
neighbors $i$ and $j$, and the second term corresponds to the changes
in the angle between two ``neighboring bonds'' emanating
from $i$ to $j$ and $j$ to $k$. This has of course the effect of
making such angles rigid and thus resisting local rotations
of bonds. The corresponding reactions of the bond bending model are 
longitudinal ($F$) and shear forces
($S$) that can be sustained by each of the elements of the lattice, and a twisting moment or 
torque ($M$) at each of the nodes of the lattice system.

The inclusion of bond-bending terms into a triangular lattice system 
result in a conventional 2D isotropic elastic medium with 
varying Poisson's ratio values \cite{monette}. In particular, even the square 
lattice system with bond bending terms and 
next to nearest neighbor interactions is equivalent to a 2D isotropic elastic 
continuum \cite{monette}. On the other hand, 
the triangular beam model lattice system represents a Cosserat continuum.

The failure condition of these bond-bending and beam lattice models 
depends on the combination of local torque, longitudinal
and  shear forces in beam models, or on central and angular forces in the bond
bending  model. In the beam case a typical fracture criterion is of the type
\begin{equation}
[\frac{F}{t_F}]^2  + \frac{\mathrm{max} (|M_i|, |M_j|)}{t_M} \ge 1
\label{beamcrit}
\end{equation}
where $t_F$ and $t_M$ are force and moment thresholds respectively, which can be chosen separately
for each bond. Similar to the RFM, one
can then tune the system by scaling the external load
so that only one bond fails at each time. In addition,
it is possible to control the relative roles of the 
two fracture mechanisms by either adjusting the magnitudes of
$t_F$ and $t_M$ or adjusting the beam elasticity
(slenderness or relative extensional and bending rigidities of the beams). 
It should be noted that Eq.~(\ref{beamcrit}) resembles the
von Mises yield criterion. In this respect it is worth pointing 
out once again that there are no studies of plasticity or elastic-plastic
fracture using these vectorial formulations for interactions.

\subsection{Discrete Lattice versus Finite Element Modeling of Fracture}
\label{sec:latticetofem}
Traditionally both discrete lattice models and standard continuum finite 
element models are often used for modeling progressive damage evolution 
in quasi-brittle materials, leading to macroscopic fracture 
or crack propagation.
In the discrete lattice models, the continuum is approximated {\it a priori} 
by a system of discrete elements such as central-force spring and 
beam elements. 

With regard to fracture, the standard continuum based finite elements are 
especially suitable for modeling progressive damage evolution or fracture of 
homogeneous (at least at the length scale of interest) materials. 
Simulation of fracture in heterogeneous materials is however complicated by 
the presence of disorder, whose presence naturally leads 
to statistical distributions of failure stresses, accumulated damage,
acoustic activity, crack shapes and so on. 
Standard continuum constitutive equations are based on 
mean-field or homogenization considerations, and hence do not 
capture the effect of fluctuations as has been briefly discussed 
in Sec. \ref{sec:frac-mec}. A few of the recent continuum based investigations 
however have made progress in explicitly modeling   
the material disorder to account for its influence on dynamic 
crack propagation \cite{xpcu,needleman,vikas1,vikas2}. 

In the discrete lattice modeling of fracture, the disorder can be
explicitly modeled. As discussed in Sections \ref{sec:rfm_plastic} and 
\ref{sec:tensorial}, the medium is described 
by a discrete set of elastic bonds with randomly distributed failure thresholds,
or in the electrical analogy via the fuse model. In the elastic case
one can resort according to the continuum elastic description desired
to bond-bending models, the RSM, or the beam-type models.
These have been used extensively in the mesoscopic modeling of 
progressive damage evolution in  quasi-brittle materials. 
Sections \ref{sec:rfm_plastic} and 
\ref{sec:tensorial} describe how statistical effects of fracture 
and disorder may be modeled using these discrete lattice systems. 
Applications of  central-force spring elements to model fracture of 
homophase and heterophase 
composite materials is presented in Refs. \cite{chung01,chung02,chung02b}. 
Similarly, Refs.  
\cite{vanmierbook,bazantbook,vanmier03,lilliu03} present the application of 
beam lattice models for modeling fracture of concrete structures. 

On the other hand, in the continuum mechanics based finite element models, 
the disorder is quite often implicitly 
modeled using the homogenization concepts over RVE and by degrading the 
overall  constitutive response (stiffness) of the material based on effective 
stress and 
damage variable definitions \cite{Lemaitre,krajcinovic,bazantbook}. 
In the continuum sense, the fracture of quasi-brittle materials is studied 
from two different 
perspectives, namely, the continuum damage mechanics and the continuum
fracture mechanics. 
The continuum damage mechanics deals with the study of crack formation and 
growth 
from an initially flaw or defect free structure. In particular, it
describes the progressive loss of material  
integrity due to the propagation and coalescence of microcracks, microvoids, 
and similar defects.
On the other hand, the continuum fracture mechanics studies how a preexisting 
crack-like flaw or defect would grow. 

Two continuum damage mechanics based approaches are commonly employed at the 
macroscopic level for modeling damage evolution. 
In these approaches, the bulk behavior of the material is 
described by either a phenomenological model \cite{Lemaitre,krajcinovic,bazantbook} 
or a homogenized micro-mechanics based model. 
As described in Sec. \ref{sec:frac-mec}, damage is described by a scalar 
damage state variable $D$ (or a generic tensorial ${\bf D}$). 
The phenomenological models are based on the notion of 
an effective stress and the constitutive response of the bulk material is 
described by a 
phenomenological strain energy function. 
An evolution law for damage is prescribed along the gradient of a 
damage loading surface. Broadly speaking, 
the continuum damage mechanics formulation 
parallels that of standard plasticity theories except that the 
effects of strain 
softening must be considered in modeling damage evolution leading to fracture. 

Homogenized micro-mechanics based damage models 
\cite{kunin,frantziskonis,frantziskonis1,budiansky}
consider an averaged response of the material 
microstructural features. 
In these models, the macroscopic damage description is 
obtained by homogenizing the material response over a {\it 
representative volume element} (RVE) unit cell. 
The homogenization of the material response, however, is 
valid only as long as the material is statistically homogeneous. 
While accounting for short-range 
local stress fluctuations and multiple crack interactions, 
the homogenized micro-mechanics 
models suffer from the same shortcomings as the phenomenological damage 
models. 
In particular, these formulations  do not include long-range spatial 
correlations of cracks and stress fluctuations \cite{kunin}. 
In the crack growth and coalescence stages of 
damage evolution, statistical fluctuations become dominant and homogenization 
based approaches are not applicable. 

It is useful again to compare the RVE size, $a_{RVE}$, to
the correlation length, 
$\xi$. In practice, one needs $\xi \ll a_{RVE}$ for the 
homogenization to be accurate. Beyond the 
correlation length, $\xi$, the material can be regarded as statistically 
local  and homogeneous. With the increasing value of 
correlation length, i.e., as $\frac{\xi}{a_{RVE}} \rightarrow 1$, 
finite size corrections are necessary for homogenization, and the material response  
changes from a {\it local} to {\it nonlocal} behavior \cite{kunin}. 
Within the scale $\xi$, statistical fluctuations are dominant, and 
hence homogenization is not valid. 
Close to macroscopic fracture, the extreme moments of 
statistical distribution of microcracks and the corresponding correlation 
lengths dominate the material response. Under these circumstances, the densities become 
irrelevant and the concept of RVE ceases to make sense. This drawback 
in homogenized methods has enabled the more recent investigations to explicitly model the microstructural 
disorder in finite element simulations \cite{needleman,vikas1,vikas2}. 

An example of continuum damage mechanics based finite element model 
applied to the fracture of 
disordered materials such as concrete is the smeared crack 
approach \cite{bazantbook}, 
wherein the cracks are not explicitly modeled. 
Instead, the constitutive 
properties of the finite elements where a crack is supposed to develop are 
degraded according 
to some continuum law. 
That is, instead of considering the crack explicitly as a discrete 
displacement jump, 
the effect of crack is smeared over all the Gauss points of the finite 
elements that contain the 
crack by degrading the corresponding material properties at the Gauss points. 
Many such 
finite element models based on continuum damage mechanics are in common 
use for modeling 
damage evolution including strain softening behavior of materials.

On the other hand, in continuum fracture mechanics based finite element models 
such as the 
the discrete crack method \cite{bazantbook}, an explicit modeling of the 
crack is considered using appropriate 
interface elements and crack-tip finite elements. 
The zero thickness interface elements are 
in general embedded between the edges of two adjacent finite elements and/or 
along probable 
crack paths that are decided upon {\it a priori}. 
The constitutive behavior of an interface element is governed by a 
traction-displacement cohesive law, which is used to evaluate interface 
traction in terms of 
normal and tangential displacement jumps, 
or a non-dimensional displacement jump parameter
which is zero at the traction free state and equals unity at 
interface failure. The progressive failure of these interface elements 
simulates crack propagation. 
When the remeshing is not performed during crack propagation, the crack 
path is restricted along the 
edges of the finite elements. This results in a finite element 
mesh-orientation dependency of the 
numerical solution. On the other hand, remeshing in each load step may be 
considered to alleviate the 
problem of crack path propagation on mesh layout. However, remeshing 
techniques \cite{ted99}, which model 
a discontinuity or crack by modifying the finite element mesh topology to 
explicitly capture a 
discontinuity, are computationally expensive and not readily applicable for 
non-linear problems. 
Alternative fictitious crack model \cite{hillerborg76} 
approaches based on ideas using Dugdale-Barenblatt plastic crack-tip zones are 
also in common use; however these approaches suffer from similar drawbacks. 

During strain localization and softening, the deformation concentrates in 
regions of finite size. The modeling of this localization process using 
continuum finite elements is fraught with 
serious difficulties, both from mathematical and numerical  
points of view. Mathematically, the onset of strain softening leads 
to ill-posedness of the rate boundary value problem of the continuum equations 
\cite{deborst}. 
In particular, the governing set of partial differential equations in the 
standard 
rate-independent continuum lose their ellipticity for quasi-static 
loading conditions, while hyperbolicity may be lost locally for 
dynamic conditions \cite{deborst}. 
In analytical solutions, strain softening is 
characterized by the appearance of infinite strains of a set of measure zero. 
When the problem is modeled and discretized with finite elements, 
the localization concentrates in the volume of material capable of 
representing the set of measure zero, which is a one element 
in one-dimensional problems or a one element wide line in 
two-dimensional problems. 
From a numerical perspective, this ill-posedness manifests itself in 
severe mesh dependence of the finite element solution in terms of mesh 
size and alignment. Hence, the problem of mesh dependency is not 
intrinsically numerical but stems from the change of the 
character of the underlying differential equations of the 
continuum description.

In order to allow for modeling of localization and strain softening in the 
continuum description, 
approaches based on strain localization limiters are proposed. 
The essential idea 
is to change the character of the equations so that the region of localization 
does not generate to the set of measure zero but to the physical region that 
can be deduced from the experimental procedures. In the literature, several 
regularization techniques (localization limiters) based on spatial nonlocal 
continuum concepts \cite{pcabot}, higher-order gradient theories 
\cite{peerlings}, micro-polar or Cosserat 
continuum descriptions \cite{deborst}, viscoplastic regularization 
methodologies \cite{deborst1}, and the 
regularized discontinuous displacement approximations \cite{oliver} have 
been presented for 
capturing the localization phenomena. 
These approaches ensure that the numerical 
solutions remain meaningful in the sense that a finite width of the 
localization 
zone and finite energy dissipation are obtained even when the localization 
zone is fully developed. The above localization limiters (regularization 
techniques) 
can be classified as integral, differential and rate limiters. 
Integral or non-local 
limiters are based on integral strain measures over a finite domain. 
Differential 
limiters include strain or stress measures with derivatives of order higher 
than one. 
In rate limiters, the time dependence of the strain softening is built 
into the constitutive equations. Although, each of the above methodologies 
has its own advantages and disadvantages, from an implementation point of view, 
differential and rate limiters are the realistic candidates for implementation 
into existing codes. Non-local limiters 
are more complex both theoretically and implementation-wise.

In order to eliminate the effects of mesh size and alignment dependencies in 
the  numerical simulation of fracture, finite element methods 
based on so-called embedded discontinuities \cite{oliver,gnwells} within the 
finite elements are also 
used. These finite element formulations are based on enhanced assumed strain 
(EAS) formulations \cite{simo90}, and the displacement jump across a crack is 
incorporated into a finite element 
as an incompatible mode in the strain field. Recently, extended finite element 
methods (XFEM) \cite{ted00} based 
on the partition of unity approach \cite{babuska} are also employed for 
simulating the fracture of 
quasi-brittle materials. Although the XFEM has similarities in the modeling of 
cracks using discontinuous 
displacement interpolations within the elements, the XFEM is far superior 
to the 
discrete discontinuous displacements based finite element approach. 
To start with, XFEM is capable of representing arbitrary propagation of cracks 
within the elements with or without bisecting the elements completely 
\cite{ted00}. For example, in the 
mesoscopic fracture simulations, this feature allows for the simulation of 
transgranular crack propagation in addition to the usual intergranular crack 
propagation \cite{sukumar}. 
Furthermore, XFEM 
allows for the presence of multiple cracks with secondary branches including 
their mutual 
interaction thereby determining the change in the crack orientation. It is 
possible to enhance 
these discontinuous interpolations with a suitably chosen near crack-tip 
fields to achieve a 
superior coarse mesh accuracy in the simulations. Representing the cracks 
using a signed 
distance function based on level set methodology facilitates the evolution of 
cracks and 
hence alleviates the necessity to use automatic remeshing techniques and 
explicit 
representation of cracks \cite{ted00}. 

Even with these vastly improved finite element methodologies and the recent 
trend in explicitly modeling the material disorder \cite{xpcu,needleman,vikas1,vikas2}, 
fracture simulation of disordered quasi-brittle materials with many interacting microcracks, and 
long-range spatial crack correlations and stress fluctuations poses a daunting task for 
current finite element methodologies. In particular, when the macroscopic 
fracture is preceded by many competing and interacting microcracks, both 
mathematical and numerical issues arise in assessing which crack should propagate and by how 
much should it propagate \cite{budyn}. In this sense, the rigor and accuracy of these homogenized 
finite element models based on effective medium (or mean-field) continuum 
theories is limited to moderate damage levels, whereas those based on continuum fracture mechanics 
is limited to the regime where the behavior is dominated by a single (or few) dominating crack. 
Currently, the most fundamental challenge is to bridge the gap between the 
discrete and continuum descriptions of brittle fracture. Recent finite element 
studies \cite{xpcu,needleman,vikas1,vikas2} that explicitly include the material disorder effects are very promising in this direction.

\subsection{Dynamic effects}

\label{sec:dynamics}
\subsubsection{Annealed disorder and other thermal effects}

In reality, on scales that are supposed to be described
by the quasi-static models introduced above, fracture
processes may also be influenced by time-dependent details. 
A possibility that has been widely explored in the past is to move from a deterministic 
dynamics in a disordered system to a stochastic dynamics in
an homogeneous system \cite{cafiero97,caldarelli}.

Time dependent randomness has been implemented in two
different ways in the RFM: the failure probability can be chosen
to be a function of the instantaneous fuse current, 
or a functional (such as an integral) that depends on the current history.
The first case corresponds to {\em annealed} disorder, where
probability of failure of fuses is given by \cite{niemayer84} 
\begin{equation}
P_{ij} \sim (i_{t,ij})^\eta,
\label{eq:anneal}
\end{equation}
properly normalized over all $ij$, such
that $\eta$ measures the relative role of current
enhancements. In the limit $\eta=0$, the model is 
equivalent to screened percolation. In the opposite case of 
$\eta \rightarrow \infty$, the current enhancements
always dominate over any randomness (or disorder). It is commonly believed that
quenched and annealed disorder lead to similar behavior in
such a scenario. The simplest explanation is that one can
always find a set of quenched thresholds such that it
will exactly reproduce the same sequence of failures
as any stochastically induced history \cite{hansen91}.
The behavior of the RFM and RSM for various $\eta$ values has been
explored both using simulations and mean-field theory
\cite{curtin91,curtin91b,curtin92,curtin93,roux93,curtin93b,curtin93c,curtin97a,curtin97b}. 
A critical value seems to be $\eta_c =2$ that arises due to a competition
of damage accumulation and crack growth.
Many issues still remain open, of those to be discussed in
Sec. \ref{sec:simul}, the crack roughness is an example.

In the second case (history dependent failure probability scenario), 
one may consider the case where the local
current $i_{t,ij}$ is followed as a function of time.
For example, consider the integrated current, or some
power thereof to a fixed threshold, and then use a 
rule similar to Eq.~(\ref{eq:anneal}). Under these circumstances, the 
scenario mimics a creep-like failure due to the external timescale
that is now introduced by the integrated history. Consequently, 
it is most natural to study such models
at constant current or voltage since a ramp-up in loading 
will only complicate the behavior further.

Alternatively, the dynamics may be studied by considering the 
evolution of an extra state variable, $T$, defined in a manner 
similar to temperature \cite{sornette92b}. The dynamics of $T$ is supposed to obey a state 
equation given by  
\begin{equation}
C \partial_t T = R i^b - aT.
\end{equation}
Now, if we suppose that a fuse fails whenever $T>T_c$, where the critical
value $T_c$ may be distrbuted uniformly, one again observes two regimes: percolation 
when $b \rightarrow 0$ and strict brittle fuse networks when  
$b\rightarrow \infty$ \cite{sornette92b}. For $b\rightarrow \infty$, the studied examples
involve quenched disorder via local conductance variations.
In the intermediate range of $b$'s, the
networks develop interesting damage dynamics.

\subsubsection{Sound waves and viscoelasticity}

The dynamics just discussed  do not explicitly involve the two relevant time scales of
{\em loading} and {\em relaxation}. The relaxation time scale in general encompasses 
two aspects; namely, the time scales that arise due to transient dynamics, i.e, as the microscopic damage takes place
(``bonds break'') the stress field deviates from its static or asymptotic stationary value, 
and the relaxation time scales inherently associated with the 
local dissipative mechanisms such as viscoelasticity. In the lattice models, these local 
dissipation mechanisms may be introduced locally on an individual element level.

The transient dynamic effects consider coarse-grained
sound waves together with inertial effects.  
The {\em perturbations} induced by the
dynamics into a quasi-static model can be considered in 
a time-dependent model through a dissipation mechanism. As an example, let us consider 
a simple scalar $2D$ model
\begin{equation}
M \ddot{u}_{i,j}= -K\sum_{(l,m)} (u_{i,j}-u_{l,m})-\Gamma\dot{u}_{i,j},
\label{dyn}
\end{equation}
This could be considered
as a dynamical version of a RFM with the masses $M$ and the damping
$\Gamma$ included, or a bare-bones simplification of a central-force
one. Here the sum of site $(i,j)$ runs over the nearest neighbors $(l,m)$. 
A constant loading rate (velocities) of $V$ and $-V$ is applied at the 
two opposite boundaries so that the boundaries move in opposite directions. 

\begin{figure}[t]
\begin{center}
\includegraphics[width=10cm]{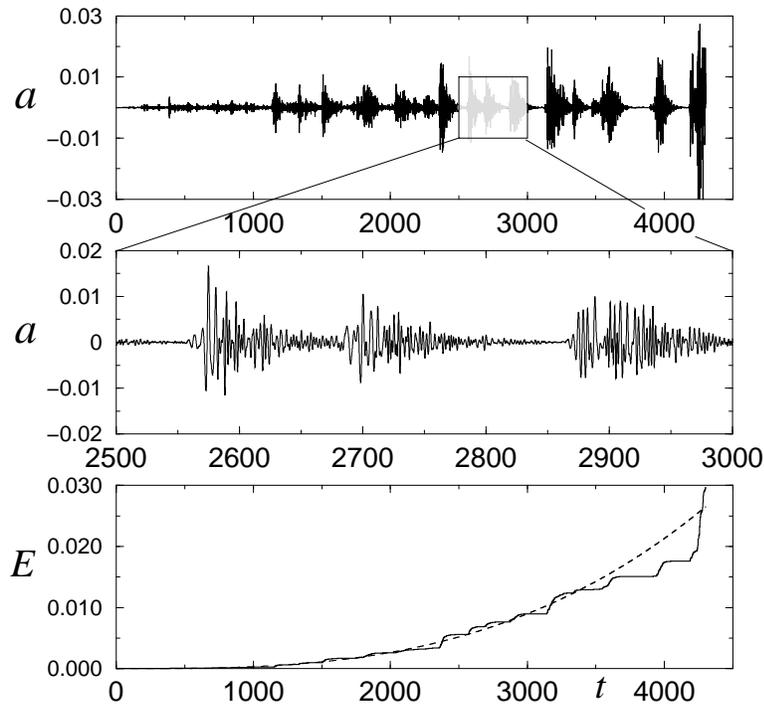}
\end{center}\label{Minozzi}
 \caption {An example of a local amplitude at a site $i$. 
Clearly, individual events can be distinguished with an envelope
that decays as a function of time. The lowest panel presents 
the integrated wave energy as a function of time. 
With linearly increasing applied strain, the
increase in local amplitude as the failure point is approached 
is also visible in the time-series of ``events''
presented in the topmost panels. (see Ref.~\cite{minozzi03})}.
\end{figure}

The transient dynamics of the system are followed by numerically 
integrating the Eq.~(\ref{dyn}), which is a damped wave equation. Each time
a bond is stretched beyond its threshold $u_{c,ij}$, the bond is removed
from the lattice system. In
this scalar model one has only transverse waves, with the
sound speed $c=\sqrt{K/\rho}$. With such dynamics, one may monitor
the local signal (Fig.~\ref{Minozzi}) from an ``event'' which 
does have some qualitative resemblance with real acoustic events
from experiments (Sec. \ref{sec:exp}). One of the complications 
associated with transient dynamics simulations is that 
wave interference and reflections significantly disrupt the 
accuracy of numerical solution. Consequently, the 
damping coefficient $\Gamma$ must be chosen 
such that the waves are damped strongly enough to avoid interference and reflections. This becomes 
particularly important in problems involving dynamic fragmentation. 
Another problem commonly associated with transient dynamics is also that close to surface boundaries incoming and
outgoing reflections may interfere disrupting the solution. 

In the context of elastic waves excited by energy release,
there is a multitude of possible mechanisms that are of 
interest:
\begin{itemize}
\item
local, transient stresses - due to a wave passing over 
a point $\vec{x}$ - may  induce local failures, similar to 
a "beam" failure in a lattice beam model.
\item
interference from several waves and from the descendant
fronts of the same wave as it gets reflected by already
existing microcracks \cite{valstos03}. The main point is that 
how these wave interferences amplify the dynamical, fluctuating
stress $\sigma(\vec{x},t) > \sigma_{static}$ beyond its static value.
\end{itemize}

The consequences of dynamics in controlled (if not
completely quasi-static) fracture would most likely
be felt in the advancement of crack lines in $3d$ systems;
analysis of the related effects indicates that this
may induce crack front roughening beyond the otherwise
at most logarithmic scaling \cite{bouchaud00,ramanathan97b}.
In models with distributed damage, the consequences thereof
would be rather unclear. One must also recall the 
attenuation of waves from single source event with distance.
It seems to us that it would be of interest to investigate
systematically the role of deviations from the quasi-static
limit, in particular as the failure point is approached.

The lattice model discussed above is one of many 
examples used to study dynamical
processes. Some of these we have encountered above
in the context of DEM, the Discrete Element Method.
There is a rather large literature on applying the DEM 
(also under the pseudonym Particle Method) to various
materials problems, in particular to that of fracture behavior of concrete.

A more physics-related application is the study
of fast, dynamic crack propagation in $2d$ (for an overview,
see \cite{cox05}). Originally,
the question arose how such simple models \cite{marder93}
would relate to the crack velocity vs. time, its asymptotic
value in relation to the Rayleigh velocity, and the possible
pattern formation. The models are studied either by taking
the continuum limit, by simulations, or by considering
in detail the effects induced by the discreteness. The phenomenon 
of lattice trapping is an important consequence of such studies.

Various authors have elaborated on the crack branching,
in particular on the role of crack velocity
oscillations, damping, and the exact
microscopic force relation and so forth. There is however
little overlap with statistical mechanics aspects such as 
the disorder in the samples, thermal effects and the
thermodynamic aspects (where does the released energy go?). 
The same holds for various phase-field
models of fracture \cite{karma01,aranson00,eastgate02,karma04}. 
In these models, one combines two fields:
one for the elastic deformations such that a minimal
choice for elasticity (e.g. in $2d$) holds, and another
one that allows for a order parameter $\rho(\vec{x})$
that continuously interpolates between undamaged ($\rho=1$)
and failed ($\rho=0$) states. The models are designed 
in order to simulate fast 
fracture propagation and the concomitant pattern
formation. In the simplest form, phase-field models
do have the problem that a crack is formed due to
the presence of ``sharp interfaces'', with an
intrinsic thickness or length-scale. These interfaces
separate regions of the phase field $\rho=1$ where the
material is intact from completely broken ones ($\rho=0$).
The interface width is a fixed parameter which gives a minimum FPZ scale.

There are no attempts to apply such models to situations
with randomness, and perhaps one may imagine that there
are difficulties in justifying physically the choices
applied for the noise that one would add. For example, it is 
conceivable that one might start 
from a fourth-order Ginsburg-Landau -style free energy
describing the intact and failed phases, and couple that
to a field term that establishes a local preference
(``toughness'' or local strength). However, it is not
clear how to adjust the noise strength in relation to 
the intrinsic FPZ size.

One further possibility is to induce dissipation in lattice models 
such that the time relaxation effects are directly incorporated into the equations
of motion. This can be done simply on the model level by applying rules
from dashpot-style models of viscoelasticity 
\cite{rautiainen95,heino96,heino97}. One
possibility is the Maxwell model in which the 
force $f_{ij}$ acting through bond $ij$ is given by 
\begin{equation}
\frac{\partial f_{ij}}{\partial t}=\frac{\partial f_H}{\partial t} 
- \frac{1}{\tau} f_{ij}
\label{eq:maxwell}
\end{equation}
Here $\tau$ is a phenomenological damping constant, and has the
effect of relaxing the local forces. $f_H$ is the contribution of 
a Hamiltonian, induced by the displacements of nodes. One could use
any particular choice for it, scalar or vector.
The system behavior is controlled by the 
ratio of two timescales: $\tau$ and the strain-rate (or rate of loading). 
In a model that includes inertia effects, there exists 
another time scale related to the dynamics of the system, and hence many kinds of behavior
could be foreseen. Other possible examples instead of
Eq.~(\ref{eq:maxwell}) would be Kelvin and Voigt -types, which include
damping in parallel to a spring.  The corresponding response functions are of the
form
\begin{eqnarray}
\dot{\epsilon} &=& \dot{\sigma}/E + \sigma/\tau, \mathrm{Maxwell}\\
\sigma  &=& E \epsilon + \tau \dot{\epsilon}, \mathrm{Kelvin-Voigt}\\
\sigma + \tau_1 \dot{\sigma} &=& E (\epsilon + \tau_2 \dot{\epsilon}),
\mathrm{3-parameter model}
\end{eqnarray}
with appropriate boundary conditions. All these models result in so-called
creep and relaxation functions after the application of a 
unit stress or strain, as can be seen by the presence of the
appropriate relaxation times $\tau$, $\tau_1$ and $\tau_2$ in
the equations.
Qualitative simulations of Maxwellian models have witnessed
as one could expect increased tolerance to damage and crack
meandering. This scenario arises due to the blunting of microcracks,
and the fact that the local stresses are dissipated in the
course of time. However, these models face certain challenges in 
dealing with boundary conditions, and stress-waves that originate
from microfailures.

\subsection{Atomistic simulations}

In all materials the ultimate process leading to fracture
is of course the breaking of bonds at the level of individual
atoms. This is true regardless of whether one considers metals
with a crystal lattice, amorphous materials or fiber-matrix
composites. In the fiber-matrix composites, it may be that 
the important phenomena take place on a mesoscopic scale,
however for crystalline solids (with dislocations present)
it is clear that the fracture energy and the favored 
mechanisms of crack advancement depend in a fundamental way on 
atomistic effects. Since fracture is also crucially dependent on long-range elastic fields, 
one needs to consider processes on large, coarse-grained
scales. The main question is thus how to bridge the length-scales
from atomistic to continuum elasticity. A current
trend is to consider {\em multi-scale modeling} in which one
uses microscopic, even quantum-mechanical input to set up 
final simulations using the finite element method (see for example, Ref. \cite{moriarty02}). 

\begin{figure}[t]
\begin{center}
\includegraphics[width=10cm]{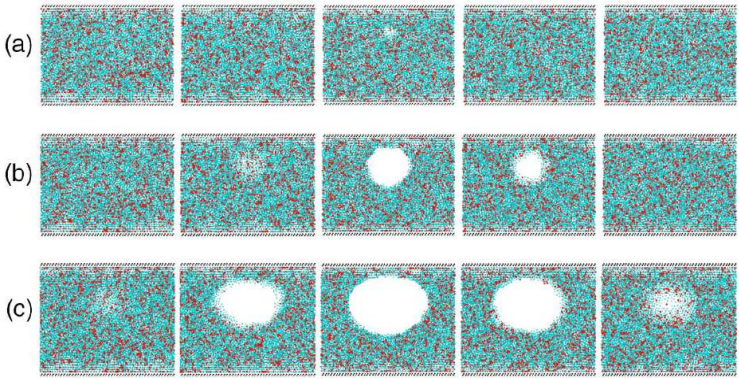}
\end{center}
 \caption {Void initiation and subsequent growth in
a binary LJ mixture. Each of the three rows represents five
cross-cuts across a 3D system, while the rows themselves correspond to 
snapshots taken beyond the maximum stress  
(from Ref.~\cite{lorenz03}).}
\label{fig:LJ}
\end{figure}

A more fundamental problem is the following: since damage and
deformation on the atomistic level typically involve thermally activated
or Arrhenius-processes, it is technically difficult to 
simulate a sample long enough to reach realistic conditions. 
Numerically accessible timescales are in fact in the range of picoseconds, 
much smaller than the normal laboratory scales on which fracture 
operates. One may of course try to circumvent this problem by using 
idealized systems like Lennard-Jones crystals 
\cite{selinger91,selinger91b,wang91}. This
has been somewhat popular for the last 20 years, and
the crux of such simulations is that a prepared LJ
crystal is subjected to shear or tension, leading
finally to failure exactly as in ordinary lattice
models. In this case, however, one may incorporate a
finite temperature and rudimentary dislocations and
their dynamics.  Very large scale simulations by
Abraham et al. \cite{abraham02} have demonstrated many fundamental
phenomena in single crystal failure, namely, dislocation
emission, sound waves and crack acceleration. 
As an example of recent application of this approach
to phenomenological studies \cite{lorenz03,buxton04}, we show in Fig.~\ref{fig:LJ} 
the formation of voids in a three-dimensional study of a binary amorphous LJ system.

More elaborate potentials are then needed to make prediction on
any real material, but the scenario gets even more complicated. 
For the elastic modulus, in many cases the available semi-empirical (to
say nothing about full quantum mechanical ab-initio calculations) 
potentials allow relatively good quantitative accuracy. 
It is not clear, however, to determine the best possible way to 
define the condition for the failure of an atomic ``bond'' when 
a particular N-body potential is used. On a
quantum mechanical level this concept makes no sense,
and on a classical level one has to resort to ad hoc techniques. 
In addition, anharmonicity plays a significant role in such calculations. 

In spite of these difficulties, in recent times, there have been significant developments in using such 
atomistic simulations to study void growth in dynamic
fracture of ductile metals, void growth ahead of the crack tip in silicon oxide and carbide, and crack propagation in single-crystal silicon 
\cite{seppala04,kalia97,moriarty02,rountree02,kalia03}. Another
theme where progress is being made is in the context of failure of nanotubes \cite{lu05}.
However, such atomistic calculations are not yet at
the level where they can be connected with mesoscale
disorder, not even in the case of carbon nanotubes where 
fundamental issues still remain. The exception might be simulations of 
dynamic fracture, where in spite of the simplifications of
the potentials used one can nevertheless demonstrate fundamental
phenomena \cite{kalia03}. We would however like to underline
that even in such cases there have not been many studies of
fracture as a stochastic phenomenon, with all the phenomenology
that it involves.

\section{Statistical theories for fracture models}
\label{sec:theory}
The lattice models for fracture introduced in the previous section are
reminiscent of other models studied in non-equilibrium statistical
mechanics.  It is thus tempting to apply standard theoretical tools to
understand their behavior and explore the analogies with other
systems. This program turns out to be extremely complicated, due to a
complex interplay between long range interactions, irreversibility and
disorder. We discuss here the solution of simplified models (fiber
bundles) and draw analogies between fracture and phase transitions.
The fracture process is conveniently analyzed in the framework of
interface depinning when disorder is weak, while the case of strong
disorder is better viewed in the context of percolation. While these
theoretical approaches provide a useful guidance, a complete understanding
of the problem has not yet been reached.

\subsection{Fiber bundle models}
\label{sec:fiber}
One of the simplest approaches to analyze fracture in disordered media
is represented by the study of fiber bundle models 
\cite{pierce26,daniels45}. These models were
originally introduced to describe fibrous materials, schematizing
the sample as a set of brittle fibers loaded in parallel. As in 
other lattice models, the failure threshold of each fiber is 
randomly chosen from a distribution. Next, one has to impose 
a rule for load redistribution after each failure. The simplest
possibility is the case of an equal load sharing (ELS), in which each 
intact fiber carries the same fraction of the load. This case
represents a sort of mean-field approximation and allows for a 
complete analytic treatment 
\cite{daniels45,hemmer92,hansen94,sornette92,kloster97}.
 At the other extreme lies the local load sharing model (LLS) where
the load of a failed fiber is redistributed to the intact neighboring
fibers
\cite{harlow78,smith80,smith81,beyerlein96,phoenix97,phoenix92,curtin91,curtin93,zhou95,leath94,zhang96}

These simplified models serve as a basis for more realistic damage
models such as the micromechanical models of fiber reinforced
composites which take into account stress localization
\cite{harlow78,smith81,beyerlein96}, 
the effect of matrix material between fibers
\cite{harlow78,smith81,phoenix97,phoenix92,curtin91,curtin93,zhou95},
and possible non-linear behavior of fibers \cite{krajcinovic82}.  
Other generalizations of the fiber bundle model include viscoelastic
couplings \cite{fabeny96,hidalgo02,kun03}, 
continuous damage \cite{kun00}, plasticity \cite{raischel06} 
and thermally activated fracture
\cite{roux00,scorretti01,politi02}. 
Fiber bundle models have been  used in the past to address the macroscopic 
constitutive behavior, the reliability and size scaling of 
material strength, and the avalanches of fiber breaks preceding 
ultimate failure \cite{hemmer92,hansen94,sornette92,kloster97}.

\subsubsection{Equal load sharing fiber bundle models}
\label{sec:fb_els}
We consider first the case of ELS fiber bundles, in which $N$ fibers
of unitary Young modulus $E=1$ are subject to an uniaxial load $F$. Each
fiber $i$ obeys linear elastic equation up to a critical load $x_i$,
which is randomly distributed according to a distribution $p(x)$. When
the load on a fiber exceeds $x_i$, the fiber is removed. Due to the
ELS rule, when $n$ fibers are present each of them carries a load $F_i
= F/n$ and consequently a strain $\epsilon=F/n$.
The constitutive law for ELS fiber bundles can be easily be obtained
from a self-consistent argument. At a given load $F$, the
number of intact fibers is given by
\begin{equation}
n= N\left(1-\int^{F/n}_0 p(x) dx \right).
\label{eq:fbm_sc}
\end{equation}
Rewriting Eq.~(\ref{eq:fbm_sc}) as a function of the strain, we obtain the
constitutive law
\begin{equation}
 \label{eq:constone}
  \frac{F}{N} = \epsilon (1-P(\epsilon)),
\end{equation}
where $P(x)$ is the distribution obtained from $p(x)$.
As a simple illustration, we consider a uniform distribution
in $[0,1]$, so that Eq.~(\ref{eq:constone}) becomes
\begin{equation}
 \label{eq:constone_unif}
  f = \epsilon (1-\epsilon),
\end{equation}
where $f\equiv F/N$. Similarly we can obtain the fraction of intact
fibers $\rho\equiv n/N$ from Eq.~(\ref{eq:fbm_sc})
$\rho= 1-f/\rho$ which can be solved to yield
\begin{equation}
\rho=(1+\sqrt{1-4f})/2.
\label{eq:fbm_rho}
\end{equation}
This equation shows that as the load is increased $\rho$ decreases
up to $f_c = 1/4$ at which $\rho=\rho_c=1/2$. For larger loads  
Eq.~(\ref{eq:fbm_rho}) displays no real solution, indicating the onset
of catastrophic failure. It is interesting to rewrite Eq.~(\ref{eq:fbm_rho}) as 
\begin{equation}
\rho=\rho_c + A(f_c-f)^{1/2},
\label{eq:fbm_rho2}
\end{equation}
with $\rho_c=1/2$, $f_c=1/4$ and $A=1$. In fact, this form is
generically valid for most distributions $p(x)$. To see this
we rewrite Eq.~(\ref{eq:fbm_sc}) as $f=x (1-P(x))$ where $x\equiv f/\rho$ is
the load per fiber. Failure corresponds to the maximum $x_c$ of the 
left-hand side, after that there is no solution for $x(f)$. Expanding
close to the maximum we obtain $f\simeq f_c+B(x-x_c)^2$, which then
leads to Eq.~(\ref{eq:fbm_rho2}) as long as the distribution is 
sufficiently regular (for a discussion of the limits of validity see 
Ref.~\cite{silveira98}). This law  also implies that the average rate of bond
failures increases very rapidly before fracture:
\begin{equation}
{d\rho \over df} \sim (f_c-f)^{-1/2}.
\label{eq:fbm_rho3}
\end{equation}

The preceding discussion focused only on average quantities, but the
presence of a random distribution of failure thresholds necessarily
leads to fluctuations. Due to the ELS rule, however, strength
fluctuations are not particularly interesting and vanish in the limit
of large bundles. In particular, for any threshold distribution such
that $1-p(x)$ goes to zero faster than $1/x$ for $x\to \infty$, the
strength distribution is Gaussian with average $f_c=x_c(1-p(x_c))$ and
standard deviation $\sigma=x_c p(x_c)(1-p(x_c))/\sqrt{N}$
\cite{daniels45}.  More interesting fluctuations are found in the
precursors: the load redistribution after a failure event can lead to
avalanches of subsequent failures. This process can be treated
analytically considering the sequence of external loads $\{F_k\}$ at
which some fiber fails \cite{hemmer92,hansen94,sornette92}. 
After $k-1$ failure event, the next fiber will
break at a load $F_k=(N-k+1)x_k$, where the thresholds $\{ x_k \}$
have been ordered. On average we have
\begin{equation}
\langle F_{k+1}-F_k \rangle = (1-P(x_k)) p(x_k),
\end{equation}
and close to global failure it vanishes as
$\langle F_{k+1}-F_k \rangle \sim x_c-x_k$. In addition, one
can see that $\langle (F_{k+1}-F_k)^2 \rangle$ 
goes to a constant as $x\to x_c$. Hence, the external load $\{F_k\}$
performs a biased random walk, and the bias vanishes as $x\to x_c$.
When the load is fixed, an avalanche is defined by the return of the
variable $F_k$ to its initial value. For instance, given that the first fiber
fails at load $F_1$, the avalanche proceeds until $F_k > F_1$. The process
is then repeated from $F_k$, obtaining a series of avalanches whose
size $s$ is given by the first return time of the $\{F_k\}$ stochastic process.
Thus the  avalanche size distribution is equivalent 
to the first return time distribution of a biased random walk \cite{sornette92}. 
For a unbiased random walk, this distribution is given by $D(s)=s^{-3/2}$
so that the avalanches just before global failure decay as a power
law. When the load is smaller than the critical load, the power
law is  exponentially cut off and the distribution becomes
\begin{equation}
D(s,f)\sim s^{-3/2} \exp(-s/s_0),
\label{eq:psf_fbm}
\end{equation}
where $s_0 \sim (x_c-x)^{2}\sim (f_c-f)^{-1}$, diverging as $f \to
f_c$ is the limit $N\to \infty$. Eq.~(\ref{eq:psf_fbm}) refers to the
avalanches at a fixed value of the load $f$. If instead we consider
the integrated avalanche distribution obtained when the load is increased from
zero to complete failure, we obtain
\begin{equation}
D_{int}(s)\sim s^{-5/2},
\label{eq:psint_fbm}
\end{equation}
as can be seen integrating Eq.~(\ref{eq:psf_fbm}) over $f$.
Recently the difference between the integrated distribution,
decaying as $s^{-5/2}$, and the distribution sampled in a small bin,
decaying as $s^{-3/2}$, has been rediscovered and reinterpreted 
as a crossover indicating imminent failure \cite{pradhan05b}. 
In fact there is no crossover: the distribution {\it at each load} $f<f_c$
has an exponent $3/2$ as stated Eq.~(\ref{eq:psf_fbm}), but the cutoff
prevents to see in practice the power law unless $f \simeq f_c$.

\subsubsection{Local load sharing fiber bundle models}

The case of LLS represents another extreme case in which
the effect of long-range stress field is completely neglected
but stress enhancement around the crack is treated in the
simplest way. Consider for instance a one dimensional series
of fibers loaded in parallel with random breaking threshold
from a distribution $p(x)$. When the load on a fiber exceeds
the threshold its load is redistributed to the neighboring
intact fibers. Thus the load on a fiber is given by
$f_i= f(1+k/2)$, where $k$ is the number of failed 
fibers that are nearest neighbors of the fiber $i$
and $f=F/N$ is the external load \cite{harlow78}. Even for this apparently
 simple one dimensional model a closed form solution is not
available, but several results are known from numerical simulations,
exact enumeration methods or approximate analytical calculations.
The analysis of the LLS model is extremely complicated and we thus
list here the main results obtained, referring the reader 
to the relevant literature for the details \cite{harlow78,smith80,smith81,beyerlein96,phoenix97,phoenix92,curtin91,curtin93,zhou95,leath94,zhang96}. 

\begin{figure}[htb]
\begin{center}
\includegraphics[width=10cm]{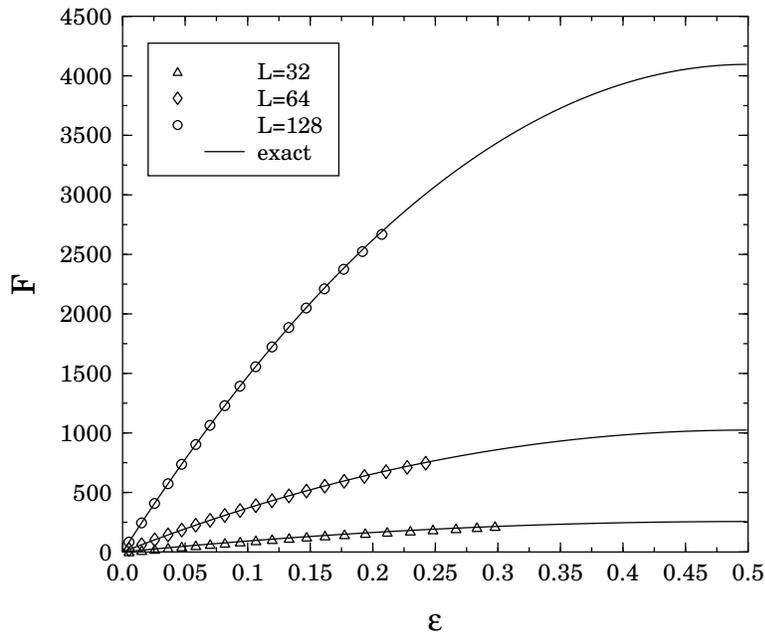}
\end{center}
\caption{The constitutive law for the ELS
model Eq.\ {\protect (\ref{eq:constone_unif})}
(solid line) is compared with the local stress transfer model.
Increasing the system size $L$, the failure stress decreases in the LLS
model (from Ref.~\cite{kun00}).}
\label{fig:size_bundle}
\end{figure}

Contrary to the ELS model, LLS fiber bundles normally exhibit non-trivial
size effects as could be anticipated from general consideration of extreme
value statistics. In particular, the average bundle strength decreases
as the bundle size grows as 
\begin{equation}
f_c \sim 1/\log(N), 
\end{equation}
so that an infinitely large bundle has zero strength.
The difference in the size effects between LLS and ELS is 
apparent in Fig.~\ref{fig:size_bundle} where the constitutive
law of ELS fiber bundles is compared with that of an equivalent
LLS model for different fiber bundle system sizes. The LLS follows the ELS law at small strains but fails before the ELS critical load. In addition, the failure point
in the LLS model decreases as the lattice size is increased.
In the limit of large $N$, it has been shown that
the strength distribution should follow the form
\begin{equation}
W(f)=1-[1-C(f)]^N,
\label{eq:str_LLS}
\end{equation}
where $C(f)$ is a characteristic function, close to the Weibull 
form, but difficult to determine exactly \cite{harlow78}. The existence
of a limit distribution has been recently proved under very generic
conditions for the disorder distribution in Ref.~\cite{mahesh04}. 
This result implies that in LLS bundles
there is no disorder induced transition from brittle failure, ruled
by extreme value statistics, to a more gradual tough regime
where fibers break gradually. The physics and mathematics of this
problem is ruled by the determination of critical defects, and
it is often so that one needs very large samples to meet some of these 
important critical defects (see Ref. \cite{mahesh04} for a discussion about
various LLS models). 

The avalanche behavior has also been studied both numerically  \cite{hansen94} 
and by analytical methods \cite{kloster97}. While initially it was
believed that LLS would also exhibit a power law avalanche distribution similar to 
ELS, it was later shown that the power law is only apparent and restricted
to a small region. In fact, the integrated avalanche distribution is well
approximated by 
\begin{equation}
D_{int}(s) \sim s^{-4.5}\exp(-s/s_0)
\end{equation}
where $s_0$ is independent on $N$.

\subsubsection{Generalizations of fiber bundle models}
\label{sec:gen-fbm}
Fiber bundle models with ELS and LLS represent two extreme
idealizations of fibrous materials that provide some insight on
the general failure properties of disordered systems. These models
have been improved in various ways, to obtain a more realistic
representation of fibrous composites through a more detailed
description of their mechanics and by including the effects of disorder on
more complicated constitutive laws. There is a vast literature on the
effect of the fiber arrangement, for instance two dimensional arrays
rather than a one dimensional chains have been considered. Furthermore, it is possible to
interpolate between LLS and ELS behavior through a long-range load
transfer rule \cite{kun00,hidalgo02b}, or by a mixed mode in which a
fraction of the load is transfered locally and the rest globally
\cite{pradhan05}.  In both cases one observes a crossover 
between local and global behavior as a function of the parameters
of the model.  In particular, in Ref.~\cite{hidalgo02b} the authors use a load 
transfer law decaying $1/r^{\gamma}$, where $r$ is the distance
from the broken fiber, and study the failure process as a function
of $\gamma$. For $\gamma$ close to zero, they recover the ELS result,
while the LLS results is found for large $\gamma$. The two regimes are
separated by the point $\gamma_c \simeq 2$. This model is interesting 
because it makes contact with lattice models for fracture where stress is generally
transfered in a non-local, but geometrically dependent, fashion. Interestingly,
for the two dimensional random fuse model the current is redistributed
with a law decaying as $1/r^{2}$, exactly the crossover point between
local and global behavior. We will elaborate further on this relation in 
the next section.

The variable load transfer rule also allows to study 
the patterns of damage using $\gamma$ as a control parameter, 
as does the version where perfect plasticity
is included \cite{raischel06}. In the latter case, with LLS one can
have a growing crack which turns out to be compact, with a rough
perimeter. The connected cluster of broken fibers can also percolate
forming a particular case of a percolation transition. This is slightly 
different from the damage patterns observed in more realistic - even RFM -
models, which are isotropic. Last, in the partly plastic FBM the avalanche
size distributions of bursts deviate from their LLS/ELS behaviors
\cite{raischel06}: the scale-free nature vanishes with ELS, while
there is a special value of remnant load in a yielded fiber,
that seems to induce a power-law even with LLS.

\begin{figure}[ht]
\begin{center}
\includegraphics[width=10cm]{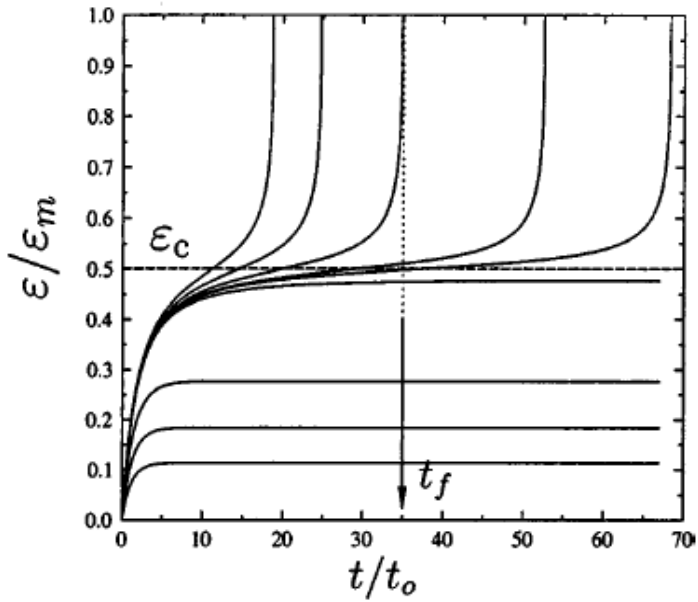}
\end{center}
\caption{The time evolution of the strain 
of a viscoelastic fiber bundle for different values of the load $f$.
For $f<F_c$ the strain accelerates up to a time to failure $t_f$ 
(from Ref.~\cite{kun03}).}
\label{fig:visco_bundle}
\end{figure}

Next, we discuss in more detail some time 
dependent generalizations of fiber bundle models.
In viscoelastic fiber bundles \cite{fabeny96,hidalgo02,kun03} 
each fiber obeys a  time dependent constitutive equation of the type
\begin{equation}
f = \beta \dot \epsilon + \epsilon,
\end{equation}
where $f$ is the external load, $\beta$ is a damping coefficient
and the Young's modulus is unitary.
In a creep test (i.e. constant load test), 
one thus finds that the strain evolves according to 
\begin{equation}
\epsilon(t)= f (1-\exp(-t/\beta))+\epsilon_0 \exp(-t/\beta)
\end{equation}
where $\epsilon_0$ is the initial strain. Fibers fail when the local
strain $\epsilon$ exceeds a random threshold $x_i$, taken from a
distribution $p(x)$, and its load is redistributed. Here we discuss
the ELS case, the generalization to LLS being straightforward.  
The asymptotic behavior of the model is the same as in the time
independent model since the Hooke's law is eventually valid when $t\gg \beta$,
but the order of individual failures is different.
The time dependent constitutive behavior of the model is obtained
by solving the self-consistent equation
\begin{equation}
f = (\beta \dot \epsilon + \epsilon)(1-P(\epsilon)),
\label{eq:visco}
\end{equation}
reducing to Eq.~(\ref{eq:constone}) in the steady state regime.
A typical example of the time evolution for different values
of the loads is reported in Fig.~\ref{fig:visco_bundle}. 
From this we can expect that the bundle will eventually fail for $f>f_c$
and resist otherwise. The time to failure can be evaluated from
Eq.~(\ref{eq:visco}) and is found to diverge as
$t_f \sim (f-f_c)^{-1/2}$ as $f\to f_c$. Notice that in the LLS version
of the model one still observes a transition but  
$t_f$ does not diverge as $f\to f_c^+$ \cite{hidalgo02}.

Another interesting time dependent generalization of the fiber bundle
model considers thermally activated fracture
\cite{roux00,scorretti01,politi02}. The additional ingredient that is
added to the model, typically in the ELS approximation, is the
presence of an uncorrelated Gaussian noise term $\eta(t)$ acting
independently on each fiber. In particular the noise distribution
is given by
\begin{equation}
p(\eta)={1 \over \sqrt{2T} } \exp \left(-{\eta^2 \over 2T}\right),
\end{equation}
where $T$ is the temperature in appropriate units.
Contrary to the viscoelastic case, the
noise has the effect to induce fracture even in subcritical
conditions, when $f<f_c$. It is instructive to consider first a bundle
without disorder, in which $p(x)=\delta(x-1)$. Even in this simple
case, it is not possible to obtain a general closed form for the
failure time, but in the low temperature limit $T << 1-f$ case, one can
obtain an approximate expression
\begin{equation}
\langle t_f \rangle \approx \sqrt{2\pi T \over  f }\exp \left( {(1-f)^2 \over 2T} \right).
\label{eq:fbm_th}
\end{equation}
In the presence of disorder the problem becomes clearly more complicated,
but there is an indication that in the low temperature case, the failure
time follows a law similar to Eq.~(\ref{eq:fbm_th}) with an important
difference: the temperature $T$ is replaced by an effective
temperature $T_{eff}=T+\Theta$, where $\Theta$ depends on
disorder. Hence the net effect of disorder is to enhance thermal
activation. It is remarkable that an analogous result was found in
Ref.~\cite{arndt01} for crack nucleation in a disordered medium.

\subsection{Statistical mechanics of cracks: fracture as a phase transition}

The idea that there is a relation between fracture and phase
transitions has a long history.  For instance, the
Griffith theory of fracture is very similar in spirit to the classical
theory of nucleation in first order phase transitions.  In bubble
nucleation, a critical droplet will form when the loss in free energy
due to the bulk forces exceeds the increase in the interfacial
energy. Similarly fracture occurs if the external stress prevails over
the resistance at surface of the crack. Further analogies come from
the scaling behavior observed in fracture experiments, such as for the
crack roughness and the acoustic emission distributions.
In the following we provide a basic introduction to phase transitions
and discuss its relation to fracture using lattice models
as an illustration.

\subsubsection{Generalities on phase transitions}
\label{sec:gen_pt}
Phase transitions are characterized by changes in the internal
symmetries of a material as external control parameters are varied.
Familiar examples are the melting of a crystal, or the ferromagnetic
transition in a magnet. In the first example, we have an abrupt
first-order phase transition, with latent heat, coexistence and no
precursors, while the latter is a continuous second-order transition. 
Notice that one can also observe a first or a second
order transition in the very same system, depending on the external
conditions, the typical example being the gas-liquid transition that
is first order at low temperatures and pressures and becomes second
order in a particular point of the PT diagram.

To be more quantitative, one typically distinguishes a phase by an
order parameter whose value is related to the internal symmetries of
the system. For instance, a ferromagnet acquires a non-zero
magnetization when the spin rotational symmetry of the paramagnetic
phase is broken. Since the transition is continuous, the magnetization
vanishes as the transition is approached. This is unlike first order
transitions, where the order parameter is discontinuous across the
transition. This case is typically associated also with metastability:
when the transition point is reached, the system needs some time to
cross into the new equilibrium state. It does so by nucleation: small
regions of the new phase are formed, until they become large enough to
transform the system completely in a rapid event.  In second order
transitions, the buildup of the new phase instead occurs 
before the transition point is reached through the formation of larger and
larger correlations that span the entire system exactly at the
critical point. This process is encoded in scale invariance: 
thermodynamic quantities around the critical point are described
by homogeneous scaling functions.

To be more concrete and set up the notation, we consider the 
example of a uniaxial ferromagnetic system. The control parameters, in
this case, are the temperature $T$, the magnetic field $H$ and the
average magnetization $M$, which is the order parameter of the system. At high
temperatures, fluctuations dominate and the magnetization is zero on an 
average. As the temperature is lowered, the local magnetization
becomes more and more correlated until the critical point $T=T_c$ is
reached. The correlation length $\xi$ then diverges as
\begin{equation}
\xi \sim (T-T_c)^{-\nu},
\end{equation}
where $\nu$ is a critical exponent. The average susceptibility 
$\chi\equiv dM/dH$ also diverges, indicating that the system is extremely
prone to changes, and it does so as
\begin{equation}
\chi \sim (T-T_c)^{-\gamma}.
\end{equation}
The order parameter $M$ becomes non-zero in the ferromagnetic phase
and close to the critical point follows
\begin{equation}
M \sim (T-T_c)^{\beta},
\end{equation}
defining another critical exponent. The presence of a magnetic field
destroys the transition, forcing a particular direction for
the magnetization, and thus defines another scaling law
at $T=T_c$
\begin{equation}
\phi \sim H^{1/\delta}.
\end{equation}
We have defined a certain number of critical exponents, which are
however not all independent, as can be shown by a more detailed 
analysis. Here, we do not want to go further into these details but
just highlight the fact that second-order phase transitions are
associated with scaling and a diverging correlation length $\xi$.

A ferromagnet can also be used to illustrate the idea
of nucleation, which is at the core of first order transitions.
Imagine that the temperature is fixed below the critical point
and we instead vary the field $H$. When the field is suddenly 
increased from a negative to a positive value the magnetization
should also change sign, but if the value of the positive field
is not so strong then the state with negative magnetization is still
metastable. Thus small local changes of the magnetization will be
reabsorbed, while large sudden changes will be less favorable 
for entropic reasons. The system will spend a considerable time
in this metastable state before a large enough change occurs.
To be more precise, the free energy cost of a spherical droplet
of size $l$ is the sum of two contributions: a bulk decrease
due to the magnetic field and a surface energy increase: i.e.
\begin{equation}
\Delta F = - 4\pi^2 H l^3+\pi J l^2, 
\end{equation}
where $J$ is the surface energy. If $l > l_c$ the droplet
is unstable and a further increase in droplet size would 
result in a decrease in its free energy. Thus
the nucleation time can be estimated as the time needed to form
the first unstable droplet. The analogy with Griffith theory is
evident and one could thus associate crack growth with a first
order transition from an elastic body to a fractured solid
driven by stress.

\begin{figure}[t]
\begin{center}
\includegraphics[width=10cm]{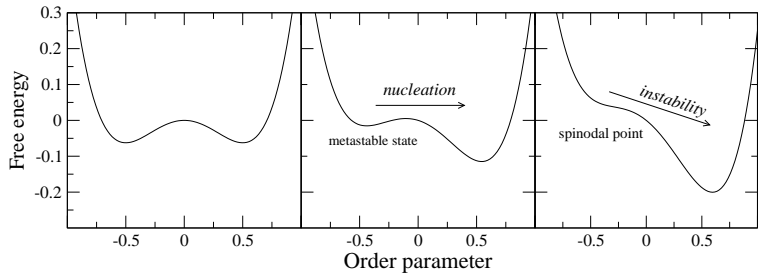}
\end{center}
 \caption {A free energy representation of nucleation. In the absence
of a field the system has two equivalent stable states (left). In the presence 
of a weak positive field one of the two minima becomes metastable and
thermally activated nucleation will eventually occur (middle). When the field
reaches the spinodal value, the negative minimum becomes unstable (right).}
\label{fig:doublewell}
\end{figure}

The difference between first and second order behavior becomes more
nuanced when nucleation takes place close to a spinodal point,
corresponding to the field at which the metastable state becomes
unstable and thus the system is forced to reverse. Strictly speaking,
the spinodal point only exists in mean-field theory, but its
signatures can be observed even when interaction is long-ranged.  The
theoretical description of homogeneous spinodal nucleation is based on
the Landau-Ginzburg free energy of a spin system in the presence of an
external magnetic field.  When the temperature is below the critical
value, the free energy has the typical two-well structure as depicted
in Fig.~\ref{fig:doublewell}.  In the
presence of an external magnetic field, one of the wells is depressed
with respect to the other, which represents therefore the metastable
state. The system must cross a free energy barrier to relax into the
stable phase. When the external field is increased, this nucleation
barrier decreases, eventually vanishing at the spinodal. The approach
to the spinodal is characterized by scaling laws, analogous to
critical phenomena, but the reversal is discontinuous and the
transition is first-order.  The magnetization or the order parameter $M$
scales with the external field $H$ as
\begin{equation}
M-M_s \sim (H_s-H) ^{1/2}
\label{eq:op_spinodal}
\end{equation}
where $M_s$ and $H_s$ are the values of the order parameter
and the field at the spinodal. This law implies a divergence
of the quasi-static susceptibility
\begin{equation}
\chi\equiv\frac{dM}{dH} \sim (H_s-H) ^{-\gamma};~~~~~~\gamma=1/2
\label{eq:susc_spinodal}
\end{equation}
The fluctuations in the order parameter can be related to suitably
defined droplets, whose sizes turn out to be power law distributed
with an exponent $\tau=3/2$.  For finite-dimensional short-range
models, this mean-field picture is expected to fail, since the system
will nucleate before reaching the spinodal point. Thus no scaling
would be  observed  in this case, as expected for first-order 
transitions.

\subsubsection{Disorder induced non-equilibrium phase transitions}

We have discussed above the properties of nucleation and
phase-transitions in thermally activated homogeneous systems.
In the case of fracture, however, we often deal with a mechanically
driven disordered system.  An interesting
analogy can be made with the zero-temperature 
random-field Ising model (RFIM) proposed by Sethna et al.
\cite{sethna93,dahmen96,sethna01} in the context of magnetic hysteresis.  
In this model, a set spins $s_i = \pm 1$ are assigned to the sites $i$ of a
lattice. The spins interact with their
nearest-neighbors by a ferromagnetic coupling $J$
and are subject to an external field $H$.  In addition, to each site of the
lattice is associated a random field $h_i$ taken from a Gaussian
probability distribution with variance $R$, such that 
$p(h)=\exp(-h^2/2R^2)/\sqrt{2\pi}R$.  The Hamiltonian reads
\begin{equation} 
E = -\sum_{\langle i,j \rangle}Js_i s_j
-\sum_i(H+h_i)s_i,
\label{eq:rfim}
\end{equation}
where the first sum is restricted to nearest-neighbors pairs.

\begin{figure}
\begin{center}
\includegraphics[width=10cm]{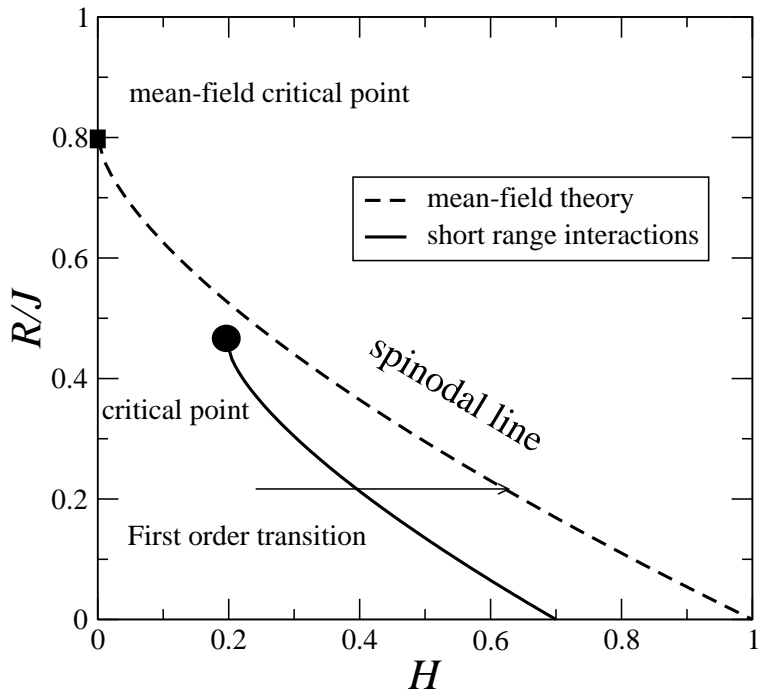}
\end{center}
 \caption{The phase diagram of the non-equilibrium RFIM. The dashed line
represents the location of the spinodal, which is obtained from the
mean-field theory. When the interactions are short-ranged the reversal
occurs on the solid line through a first-order transition. The first-order
line ends at a critical point, above which the reversal is continuous.}
\label{fig:Phasediag}
\end{figure}

The system is started from a saturated negative state and 
the external field is increased from a large negative value.
At each time the spins $s_i$ take the sign of the 
local field
\begin{equation} 
h^{(eff)}_i\equiv-\frac{\delta E}{\delta s_i}=J\sum_j s_j+h_i+H,
\end{equation}
where the sum is over the nearest neighbor sites $j$.
When a spin $i$ flips, the effective field of the neighbors $j$ is
increased by $2J$. This can lead the effective field to change sign
inducing an avalanche. For small values of the order parameter $\Delta
\equiv R/J$, flipping a few spins generates a big avalanche whose size
is comparable to the system size, leading to a discontinuous
magnetization reversal. On the other hand, a large disorder prevents
the formation of large avalanches and the magnetization reversal is
a smooth process. The two regimes are separated by a critical point $\Delta_c$,
where the avalanches are distributed as a power law. The behavior of
the model in the ($\Delta$, $H$) parameter space  
is very similar to that of standard equilibrium 
phase transitions as it is shown in Fig.~\ref{fig:Phasediag}.
In particular, all the usual exponents discussed in
Sec.~\ref{sec:gen_pt} can be defined replacing $T$ with $\Delta$.

In the mean-field theory, the approach to the instability $H_c$, is
characterized by the same scaling laws and exponents of spinodal
nucleation, as reported in 
Eqs.~(\ref{eq:op_spinodal})--(\ref{eq:susc_spinodal}) \cite{dahmen96}.  
In addition, the avalanche size distribution is
described by a scaling form
\begin{equation} 
P(s)\sim s^{-\tau}f[s(H_c-H)^\kappa],
\end{equation}
with $\tau=3/2$ and $\kappa=1$.  It is worth noting that these scaling
exponents coincide with the mean-field exponents for the distribution of
droplets in homogeneous spinodal nucleation.  {}From these studies it
appears that the behavior of thermally activated homogeneous spinodal
nucleation is similar to the approach to the instability in disordered
systems driven at zero temperature.  However, one should bear in mind
that for a given realization of the disorder the dynamics is completely
deterministic in the second case.  Concepts such as metastability and
nucleating droplets are formally not defined in this context.
This picture is only valid in mean-field theory or when interactions
are long-ranged. As in conventional equilibrium phase transitions,
when interactions are short-ranged nucleation takes place before the
spinodal line and scaling laws can be observed only at the critical point.

\subsubsection{Phase transitions in fracture models}

The interpretation of fracture as a phase transition is a highly
controversial question. While at first glance the analogies are
striking, a precise relation is hampered by several difficulties and
no consensus has been reached in the literature on this subject.  A
series of studies consider the case of crack nucleation in
homogeneous media, elaborating on the analogy between the classical
theory of nucleation and Griffith theory. Selinger et
al. \cite{selinger91,selinger91b,wang91} have used mean-field theory
and numerical simulations to show that a solid under stress is in a
metastable state, becoming unstable when the external stress reaches a
spinodal point. This is best illustrated with a simple linear chain
of particles. It breaks down once one of the pairs
gets separated due to a thermal fluctuation that makes the
inter-bond to fail.
Rundle and Klein \cite{rundle89} have proposed an
equation for the growth of a single crack, deriving scaling laws for
spinodal nucleation.  The nature of the nucleation process in a
stressed solid was further analyzed in
Refs.~\cite{golubovic91,golubovic95} using Monte Carlo simulations.
More recently, it was shown more rigorously in the framework of
elastic theory, that the point of zero external stress corresponds to
the condensation point in gas-liquid first order transitions
\cite{buchel97}. It is important to remark 
that these studies deal with the
thermally activated fracture where quenched disorder is not
considered. The presence of disorder, in the form of vacancies or
microcracks, strongly affects the nucleation process
\cite{selinger91b,golubovic91,golubovic95}, providing in some cases 
an effective disorder-induced correction to the temperature \cite{arndt01}
as pointed out above in Sec. \ref{sec:gen-fbm}.

Several authors have tried to relate the scaling laws observed in
fracture morphologies and acoustic emission distributions to an
underlying critical point, basing their analysis on experiments 
\cite{garcimartin97,guarino02}
and models \cite{hansen94,zapperi97,zapperi99,andersen97,barthelemy02,toussaint05}. 
A natural starting point is provided by a lattice model, which could be mapped into
other models, typically drawn from ferromagnetism, where the presence
of a phase transition is well established 
\cite{zapperi97,zapperi99,barthelemy02,toussaint05}.  It is instructive to
consider first the fiber bundle model, since exact results are
available and the nature of the hypothetical phase transition can be
clearly identified \cite{zapperi97,zapperi99}. To make contact with
the RFIM, we can assign to each fiber a spin variable $s_i$, whose
value depends on whether the fiber is intact ($s=-1$) or broken
($s=1$).  The effective field is formally given by
\begin{equation}
h^{(eff)}_i=F_i(\{s_j\},F)-x_i
\label{eq:eff_field}
\end{equation}
where $x_i$ is the fiber threshold, $F_i$ is the load on the fiber $i$, which depends on the
external load $F$ and on the state of the other fibers. The dynamics
is now similar to the one of the RFIM since each ``spin'' follows the
local field (i.e.  it breaks when $F_i > x_i$).  Thus changing the
disorder strength and the nature of the load sharing we could obtain a
phase diagram of the type depicted in Fig.~\ref{fig:Phasediag}.  

In the case of ELS fiber bundles, the load on each fiber is given by $F_i=
2 F/(N-\sum_j s_j)$, while for LLS $F_i$ is a complicated function of
the state of the neighboring fibers: if a fiber is close to a crack of
length $k$ it carries a load of $F_i = F(1+k/2)$.  The crucial
difference between these load sharing rules and ferromagnetic
interactions is that for fibers the increase of the effective field
due to other fiber breaking grows as the fracture process progresses, 
while in the RFIM this quantity is always given by $2J$.  For instance
in the case of ELS, the breaking of a single fiber increases the
effective field of the others by a quantity $\Delta F =F/(n(n-1))$
which increases as the fraction of intact fibers $n$
decreases. Similarly in the case of LLS each time we break a fiber, we
increase the effective field of the neighbor by a quantity $\Delta F
=kF/2$ which becomes larger as more and more fibers break. 

This difference has a profound effect on the phase diagram, because it
implies that interactions always prevail over disorder. Thus, as
the system size increases, the system will necessarily hit the low
disorder first-order transition line. The only exception is provided
by the extreme case of a non-normalizable disorder distribution, where
we can always find some strong enough bond to resist the load
amplification \cite{andersen97,silveira98}. In more general cases, however, it is
not possible to fine tune the ratio between disorder width and
interactions as for the RFIM in order to find a critical point where the two
effects are perfectly balanced.  Thus the ``phase transition''
observed in fiber bundle models is always first-order, with a
catastrophic nucleation event preceded by some small precursors. The
scaling observed in the ELS models is the one associated with a
spinodal instability, similar to what is found in the low disorder
phase of the RFIM in mean-field theory. When interactions are local,
such as for the LLS model, the spinodal line can not be reached and
the transition is simply first order. With this idea in mind we can
reinterpret the results presented in Fig.~\ref{fig:size_bundle}:
as the range of interaction is increased nucleation occurs closer and
closer to the spinodal point, as in conventional first-order phase
transitions.

Having settled the question of fiber bundle models, we can now turn
our attention to more complicated lattice models, such as the random
fuse model. The attempts made in the past to map fracture models into
spin models are typically based on the lattice Green function
formalism: removing a bond $i$ in the lattice implies that the load in
the other bonds has to be changed by a quantity $G_{ij}$ (the Green
function).  Thus we can rewrite the effective field in
Eq.~(\ref{eq:eff_field}) in terms of $G_{ij}$ and obtain a suitable spin
model.  For instance, in the random-fuse model $G_{ij}$ can be
approximated as the dipolar kernel, but only in the limit where only a
few bonds are broken. It is interesting to remark that in these conditions
the Green function decays as $r^{-2}$ in two dimensions, which is
exactly the crossover point between local and global behavior
in the FBM. This could imply that the spinodal point, inaccessible
under LLS, could be reached in the RFM, as is also suggested by 
the simulations discussed in Sec.~\ref{sec:rfm_ava}.

One should take these considerations with due care,
since for a generic damage configuration, finding
$G_{ij}$ amounts to solving the full problem (i.e. the Kirchhoff
equations for the random fuse model) and is thus practically
impossible.  Thus we can rigorously map fracture models into spin
models only in the dilute limit, which is not, however, the limit of
interest in fracture. Extrapolating the results of spin models outside
their regime of validity leads to inconsistent results, because the
stress amplification at the crack tips is not properly captured by the
dilute Green function. For instance, the authors of
Ref.~\cite{barthelemy02} treat in mean-field theory the spin model
obtained from Green function formalism.  Neglecting the load
amplification in their analysis (i.e. the load transfer is chosen to
be independent of the state), they obtain a spurious phase diagram
similar to that obtained for the RFIM. Similar problems affect the analysis
presented in Ref.~\cite{toussaint05}.

\subsection{Crack depinning}
\label{sec:depinning}
We have discussed in Section \ref{sec:exp} 
that experiments in several materials
under different loading conditions, lead to a rough crack surface
described by self-affine scaling.  The simplest theoretical approach
to the problem identifies the crack front with a deformable line
pushed by the external stress through a random toughness landscape and
the trail of the line yields the fracture surface \cite{bouchaud93,daguier97}.  
The roughness of
the line comes from the competition between the deformations induced
on the line by the materials inhomogeneities and the elastic
self-stress field that tries to keep the front straight. This problem
has been studied in full generality and the basic phenomenology applies
to a variety of systems including domain walls in ferromagnetic and
ferroelectric materials, dislocations gliding through solute atoms,
flux lines in superconductors and others. 

\begin{figure}
\begin{center}
\includegraphics[width=10cm]{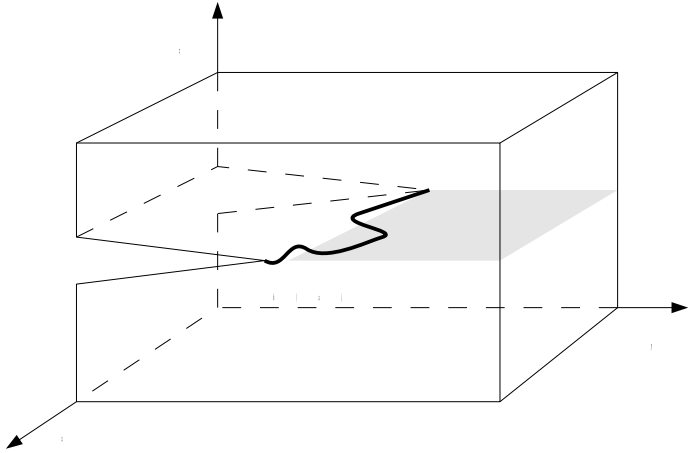}
\end{center}
\caption{The geometry of a deformed planar crack}
\label{fig:planarcr}
\end{figure}

For simplicity, it is instructive to consider first the case of a
planar crack front, corresponding to the experiments reported in
Refs.~\cite{schmittbuhl97,delaplace99,maloy01}.  
In this case we can schematize the crack as a line moving on the
$xy$ plane with coordinates $(x,h(x,t))$ (see Fig.~\ref{fig:planarcr})
\cite{schmittbuhl95,ramanathan97,ramanathan98}. 
An equation of motion for
the deformed line position is obtained by computing, from the theory of
elasticity, the variations to the stress intensity factor induced by
the deformation of the front. In the quasistatic scalar approximation,
this is given by \cite{gao89}
\begin{equation}
K(\{h(x,t)\})=K_0 \int dx' \frac{h(x',t)-h(x,t)}{(x-x')^2},
\end{equation}
where $K_0$ is the stress intensity factor for a straight crack.
The crack deforms because of the inhomogeneities present in
the materials, and these inhomogeneities give rise to fluctuations in the local
toughness $K_c(x,h(x,t))$.  These  ingredients
can be joined together into an equation of motion of the type
\begin{equation}
\Gamma \frac{\partial h}{\partial t}=K_{ext}+K(\{h(x,t)\})+K_c(x,h(x,t)),
\label{eq:depinning}
\end{equation}
where $\Gamma$ is a damping term and $K_{ext}$ is the stress intensity
factor corresponding to the externally applied stress 
\cite{schmittbuhl95,ramanathan97,ramanathan98}.
Eq.~(\ref{eq:depinning}) belongs to a general class of interface models
which have been extensively studied numerically and analytically. For
low stress the crack is pinned, and there is a critical threshold $K_c$
above which the crack advances at constant velocity
\begin{equation}
v \sim (K_{ext}-K_c)^\beta, 
\end{equation}
where $\beta$ is a scaling exponent, whose value is estimated
numerically to be $\beta \simeq 0.68$, while a renormalization
group analysis yields $\beta=2/3$ \cite{ertas94}. 
Close to the depinning transition
crack motion becomes correlated, with avalanches of activity.
The distribution of avalanche lengths $l$ follows a power law
distribution with a cutoff
\begin{equation}
P(l) \sim l^{-\kappa} f(l/\xi),
\end{equation}
where $\kappa \simeq 1$ and the correlation length diverges at 
the transition as
\begin{equation}
\xi \sim (K_{ext}-K_c)^{-\nu}, 
\end{equation}
with an exponent estimated as $\nu\simeq 1.52$ \cite{ramanathan98}. 
In addition, the
crack front is self-affine, with a roughness exponent given by
\begin{equation}
\zeta=1-1/\nu.
\end{equation}
Numerical estimates of the roughness exponents range between 
$\zeta=0.34$ \cite{schmittbuhl95,ramanathan97} to $\zeta=0.39$
\cite{rosso02}.

The results of the planar crack propagation model can be compared with
the experiments reported in
Refs.~\cite{schmittbuhl97,delaplace99,maloy01}.  It is apparent that
there is significant discrepancy between theory and experiments, since the
roughness exponent is considerably larger in the latter case.  Several
mechanisms have been proposed in the literature to account for this
discrepancy, but none of them so far is entirely convincing or
commonly accepted. The arguments involve correlated disorder
\cite{schmittbuhl99}, elastodynamic effects \cite{ramanathan98}, crack front waves
\cite{bouchaud00} and microcrack nucleation ahead of the main crack
\cite{zapperi00,astrom00,schmittbuhl03,bouchbinder04}.  

The more general problem of three dimensional crack surface roughening
can also be recast as a depinning problem, with similar difficulties
\cite{ramanathan97b}. It is possible to derive an equation of motion,
based on linear elasticity, for the evolution of the crack
surface. With respect to the planar crack case, we have an additional
equation for the out of plane component $h_\perp (x,y)$ of the crack, while
the in-plane component $h_\parallel (x)$ follows Eq.~(\ref{eq:depinning}). The
equation for $h(x,y)$ and the resulting roughness exponent depend on
the particular fracture mode imposed in the experiment. Without
entering into the details of the derivation, which can be found in
Ref.~\cite{ramanathan97b}, we report here the result. In mode I the
out of plane roughness is only logarithmic, while an exponent
$\zeta_\perp=1/2$ is found in mode III.  Both results are quite far
from the experimental measurement $\zeta \simeq 0.78$, although the
mode III result could be in line with the value $\zeta =0.5$,
sometimes measured at small scale (consider also Ref. \cite{bonamy06}).  
Unfortunately, experimental
results are normally obtained under mode I loading conditions, at least
effectively,  while mode III cracks tend to be unstable.

Given the results discussed above, we would conclude that crack 
roughening is probably more complicated than a simple depinning
problem of a linear elastic line moving in a disordered landscape. 
Nonetheless the depinning picture is quite appealing, since it relates the 
observed scaling behavior to a non-equilibrium critical point. To obtain
a quantitative explanation of the experiments, it
is necessary to analyze in detail further modifications 
of the basic depinning problem, including for instance the combined effects of the anelastic 
fracture process zone, the formation of microcrack, plastic flow, 
or elastodynamics. Among these proposals the most promising pathway 
for future investigation is probably represented by the effect of non-linear elasticity.

The depinning of a non-linear long-range elastic line has been recently 
studied by renormalization group for contact line depinning, a problem
that is formally equivalent to a planar crack \cite{ledoussal04}. It
was shown that adding a non-local KPZ-type  non-linearity \footnote{The standard KPZ 
term is $\lambda (\nabla h)^2$, a non-local KPZ term  is written in
Fourier space as $\lambda q^\alpha |h(q)|^2$}
to Eq.~\ref{eq:depinning} raises the roughness exponent to $\zeta\simeq 0.45$.
This value is a little closer to the experimental result, but needs further
confirmation, since it is based on a one-loop $\epsilon$ expansion. 
A more striking similarity is obtained if we consider a non-linear,
but local, elastic kernel in Eq.~\ref{eq:depinning}. In this case,
numerical simulations indicate that the exponent is $\zeta=0.63$ \cite{rosso03b}
which is in perfect agreement with the experiments. This universality
class corresponds to the pinning of the front by a directed percolation
path \cite{tang92,buldyrev92}. In order to accept this explanation, we have
to understand the mechanism responsible for the screening of
long-range forces, resulting in a
short-range, but non-linear, kernel. In this respect, recent
experiments on amorphous Silica show that the roughness
correlation length $\xi$ is of order of the size of the 
FPZ $\xi_{FPZ}$ \cite{celarie03,marliere03,prades04}. By definition, linear elasticity is not applicable inside the FPZ, where the crack roughening process would takes place. Hence, this
observation could be used to justify why the effective kernel is 
local and non-linear, accounting for the observed value of $\zeta$. 
While this appears to be a plausible explanation, 
further work is needed to substantiate this claim.

\subsection{Percolation and fracture}
\label{sec:perc}

\subsubsection{Percolation scaling}

\begin{figure}[t]
\begin{center}
\includegraphics[width=10cm]{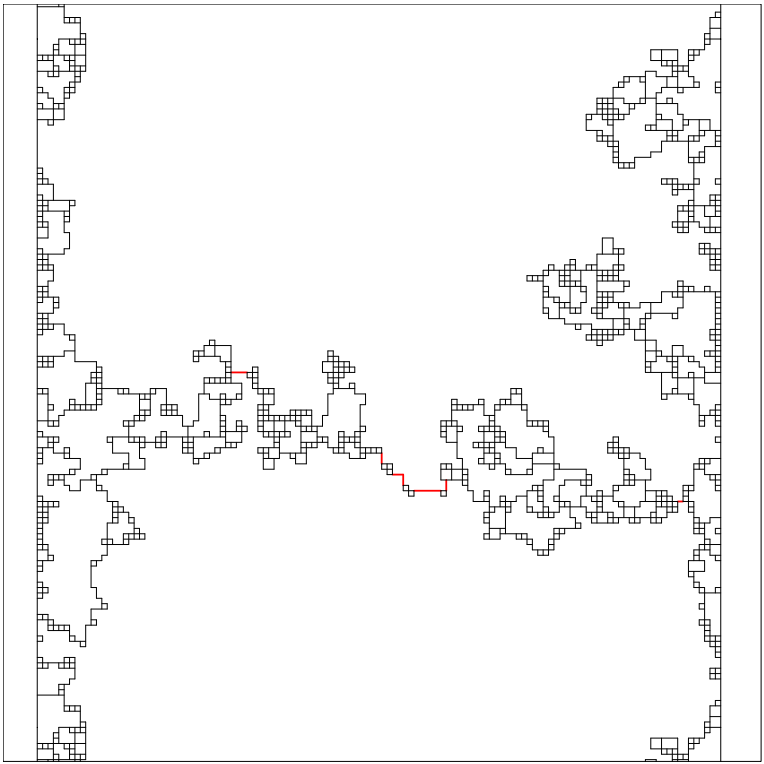}
\end{center} 
\vspace{1cm} 
\caption {The backbone of a bond percolation problem in 
$2d$: only the bonds belonging to it are depicted. The 
``red bonds'' are marked, naturally, with red.
So-called busbar boundary conditions are used, which
makes the backbone more dense close to the boundaries.}
\label{fig:backbone}
\end{figure}

Adding randomness to brittle lattice models of fracture can be done in
many ways, the simplest being the random removal of a fraction of the
elements. This process is best understood using  the concept of percolation, a
particular second-order phase transition that exhibits the full
spectrum of usual characteristics, such as scaling and critical
exponents as discussed in Sect. \ref{sec:gen_pt} \cite{stauffer}.
A percolation transition signifies that a structure becomes
{\em connected} for certain values of the control parameter (i.e. the fraction of
intact bonds). The usual example, demonstrated in Fig. \ref{fig:backbone},
is that of bond percolation in a two-dimensional square lattice,
where a fraction $p$ of the bonds have been removed. The question
is now whether the remaining ones form a connected (or spanning)
cluster that extends through the lattice, connecting for instance the
two horizontal edges. In this particular example, the spanning cluster
is found, in the limit of large lattices, for $p<p_c=1/2$. 
As in other second-order phase-transitions,  the location of the critical 
point 
(i.e. the value of $p_c$) depends on the type of lattice,
but the critical exponents are universal.

An important quantity for fracture in the percolation 
context is the order parameter, usually defined as the probability 
$\Pi$ that a bond belongs to the spanning cluster.
Below the percolation threshold $p_c$, the order parameter
scales as
\begin{equation}
\Pi \sim (p-p_c)^\beta,
\end{equation}
with $\beta=5/36$ in $d=2$, while $\Pi=0$ for $p>p_c$.  
The essential part of describing percolation  as a second-order phase transition is that
there is a divergent correlation length $\xi$ scaling as
\begin{equation}
\xi \sim (p-p_c)^{-\nu},\label{eq:nuperc}
\end{equation}
with $\nu=4/3$ in $d=2$. This correlation length can be measured analyzing the
statistics of the clusters of connected bonds for $p<p_c$. The cluster radius 
distribution is a power law up to a cutoff lengthscale $\xi$. For lengthscales larger  
than $\xi$ (for the system size $L\gg \xi$ or for distances between 
two points $r\gg \xi$) the system is out of the critical regime and is either 
connected or disconnected. For scales 
such that $r<\xi$, the physics of percolation is governed in general simply by the
{\em singly connected} or ``red'' bonds. They are - in the bond
percolation language - those that make the spanning cluster connected,
so that removing any of them also destroys connectedness, or cuts the
continuous, spanning path across the sample into two. The sensitivity 
to the lattice to small perturbations, like the removal of a red bond, is 
a signature of criticality.

Red bonds also play a crucial role on various important properties. For instance,
if the bonds are taken to be fuses in a RFM, then red bonds will all carry the same current.
Thus the conductivity (or equivalently the elastic stiffness in tensorial models) and
strength are going to relate to the properties of such bonds. This
allows to write down in the critical regions the scaling Ansatz for
the conductivity $\Sigma$ 
\begin{equation}
\Sigma \sim (p-p_c)^{t_\Sigma} f_{perc}(\xi/L^{1/\nu})
\label{eq:percansatz}
\end{equation}
such that for $L \gg \xi$ or $p$ large enough but still close to the
critical regime, the scaling function $f$ becomes a constant. We have
formulated Eq.~(\ref{eq:percansatz}) in terms of a (general) transport
exponent $t_\Sigma$, which can also describe any particular physical
property from conductivity to maximum current to strength, if the
bonds are taken to be beams with certain elasticity properties. The ratio of 
the conductivity exponent $t_\Sigma$ to $\nu$ is for instance known to be 
$t_\Sigma/\nu = 0.9826 \dots$ \cite{grassberger99} in $2d$, whereas in $3d$ the
ratio is about $2.3$ \cite{clerc00}. 
These questions are often analyzed either numerically via
transfer matrix calculations on strips, by renormalization group
arguments, or by real-space renormalization, and it appears that the
numerical value of most any exponent $t_G$ is non-trivial in the sense
that it is not simply related to $\nu$ and the physical dimension.

\subsubsection{Variations of the percolation problem}
\label{sec:rig}

Interesting physics is encountered in several separate limits, which
concern the nature of the interaction Hamiltonian imposed on the
bonds. The difference between elastic and scalar/electric percolation should also 
be visible in the strength case. At the minimal level, this follows from the scaling of
the stiffness and conductivity with $p-p_c$. A fundamental special case arises if the angular or 
bond-bending part is taken to be zero, exactly, so  that only central forces remain. Then the main effect
is that single-connectedness is no longer sufficient
since simple Maxwell-counting demonstrates that the
degrees of freedom are not balanced by the constraints
at hand. Several often-used lattices become unstable (e.g.
simply cubic ones, like the square one) against deformation. 
The critical behavior is now called {\em rigidity percolation}, and its
nature is still a not completely resolved question, e.g. with regards
to its order in two dimensions - whether it is simply second-order similar to 
percolation or not \cite{moukarzel97}. For studies of central-force
system fracture in diluted systems the implications
are not quite clear either. With increasing damage, at some point, 
the system will have a vanishing (linear) elastic modulus, while still
being geometrically connected.

Another example of percolation-related processes
in the fracture context is given by continuum percolation,
where one considers the transport properties of a medium
with a set of voids or defects. The difference to the
lattice case is now due to the fact that the ``bottlenecks''
of the spanning cluster can have a (naturally) varying
distribution of properties. A typical example is the
so-called Swiss Cheese -model, where one introduces
spherical or circular defects \cite{feng87,stenull01}. 
From geometrical considerations,
one can now compute the distribution of minimum widths
for the narrowest ligaments that one finds on the
spanning cluster. The result is that transport
properties are changed. In this example, the conductivity
 is expected to follow $\Sigma \sim (p-p_c)^{T_{sc}}$
with $T_{sc} = (d-2)\nu + \mathrm{max} [\theta, (1-a)^{-1}]$.
Here $\theta$ is the resistance exponent of normal percolation,
describing what happens between two terminals on the same cluster.
The exponent $a$ depends on  model  and dimension,  and
describes the power-law distribution of the conductivities of
the bottle-necks. One can derive the failure properties easily
by noting that all the red bonds carry the same current, but
the result depends on what one assumes about the failure thresholds
(see also Ref.~\cite{alava03}).

Finally, an opposite limit to the diluted case is
given by the {\em infinite disorder} case, in which
the medium fails following two rules: one picks always
the bond with the smallest threshold, and excludes in
the process those that do not carry current. Evidently
in the beginning this is just dilution, as in geometric
percolation, but at some point the {\em screening} will
start to play an increasing role \cite{roux88}. In this context and
in studying fracture close to $p_c$ it might be helpful
to use recent algorithmic developments in pruning 
percolation clusters \cite{moukarzel98}. These allow to consider only the
backbone, and to identify the red, critical bonds, 
with polynomial time scaling in $L$ and could thus be
used to obtain starting configurations for simulations.

\subsubsection{Strength of diluted lattices}

\begin{figure}[t]
\begin{center}
\includegraphics[width=10cm]{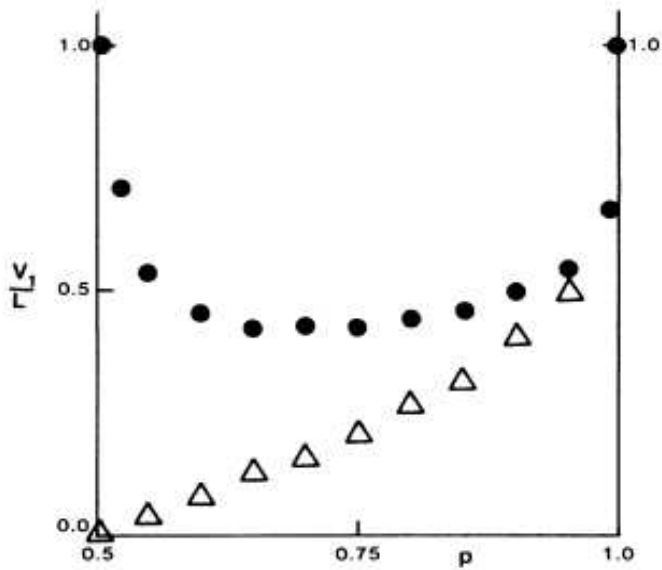}
\end{center} 
 \caption {The scaling of both conductivity and
strength in $2d$ RFM as a function of the 
bond probability $p$ (from Ref. \cite{duxbury87}).}
\label{fig:cond}
\end{figure}

Above the critical region, when the lattice is connected, one can 
consider the scaling of strength as a function of $L$ and $p$. For 
simplicity, we restrict our discussion to the scalar case.
As noted above, the random removal of bonds naturally leads
to a macroscopic reduction of the conductivity from the 
homogeneous system value. Figure \ref{fig:cond} shows an example of the behavior in two
dimensions, with fixed $L$ \cite{duxbury87}.  
The crucial argument that relates  the strength $\sigma_c (p,L)$ 
to the extremal statistics arguments underlying 
the scaling of brittle fracture is due again to Duxbury,
Beale, and Leath \cite{duxbury86,duxbury87}. 
The most critical defect - this holds also for the elastic
case - is a linear one, a row of $n$ missing bonds in two dimensions. 
In higher dimensions, the corresponding situation is given
by a penny-shaped crack and the fuse current will scale
as $i_{tip} \sim i_0 (1+k_d n^{1/(2d-1)})$. This results
from noticing that the total current increase will scale
as $n^{1/2}$, but it is shared by a number of fuses that
depends on dimension (e.g. $d = 2$ in $2d$). 

\begin{figure}[t]
\begin{center}
\includegraphics[width=10cm]{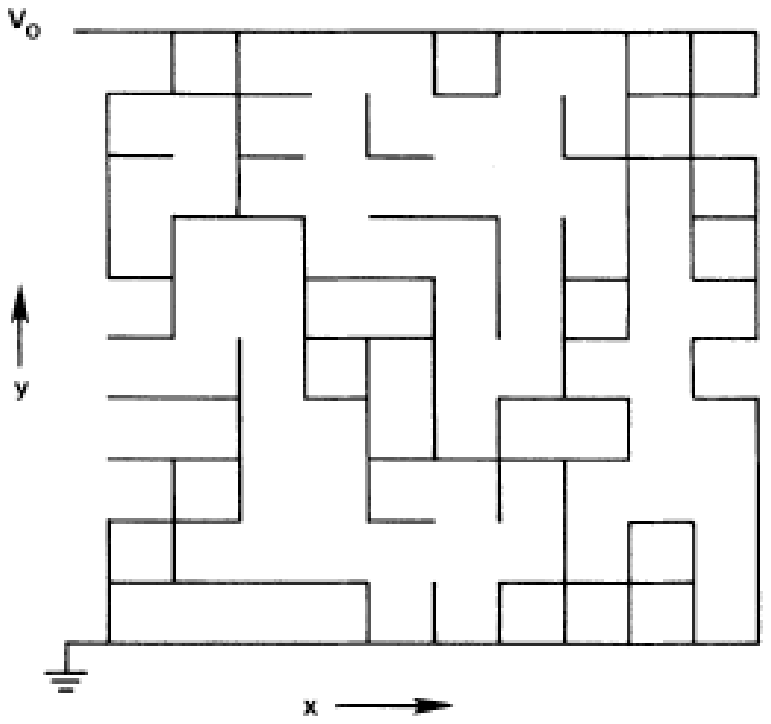}
\end{center}\label{fig:p075}
 \caption {Example of a $2d$ fuse network for $p=0.75$.
What is the largest linear defect in such a system?}
\end{figure}

Forgetting about rarer configurations where two such cracks are found
next to each other, the probability of having a defect of
given length is clearly proportional to $P_n \sim p^2(1-p)^n$,
times a prefactor which is $L$-dependent from the row length.
Thus a connection can be established non-rigorously to the
basics of extremal statistics distributions outlined earlier: 
the critical defect is the biggest found in a sample of transverse size
($L$, usually) and consisting of (again) $L$ independent
subvolumes. 

From the damage mechanics viewpoint it is useful
to take note that in this variant of the RFM the fracture 
process zone is practically non-existent; thus there is
no corresponding lengthscale $\xi_{FPZ} \gg 1$ that might depend on
system size. Furthermore, for any $p$ the damage or number
of fuses that fails before the maximum current is very small.
Thus in these systems there is no ``nucleation'', the largest
defect in the original system dictates the strength.

In particular, the defect population with
dilution or percolation disorder is
{\em exponential}, and thus the limiting distribution will
be of the double-exponential or Gumbel-form, of extremal
statistics \cite{gumbel}. One can write
for the integrated distribution
\begin{equation}
P(i) = 1 - \exp{(-cL^d \exp{[-k i_b^{-1/\alpha}]})},
\end{equation}
where $\alpha$ is a constant that
depends on $p$ (recall again this is for percolative disorder),
as are $c$ and $k$. For the average strength it
follows that there is a logarithmic size-effect, 
$\langle \sigma_c (L) \rangle \sim 1/\log{L}^a$ where
$a \sim \mathcal O (1)$ and is difficult to establish rigorously.

\subsubsection{Crack fronts and gradient percolation}
\label{sec:gradper}
An interesting variation of the percolation problem is {\em gradient
percolation} in which the symmetry is broken by letting the occupation
probability of bonds to be a function of the vertical coordinate $y$
so that $p \equiv p(y)$ \cite{sapoval85}. 
In the simplest case, $p(y)$ is a linearly
decreasing function interpolating between $p=1$ and $p=0$, but other
choices do not change the basic physics. This model has applications
in several mean-field -like fracture models in which the gradient is
imposed into damage by an external displacement or potential field
\cite{zapperi00,astrom00,schmittbuhl03}.

\begin{figure}[t]
\begin{center}
\includegraphics[width=10cm]{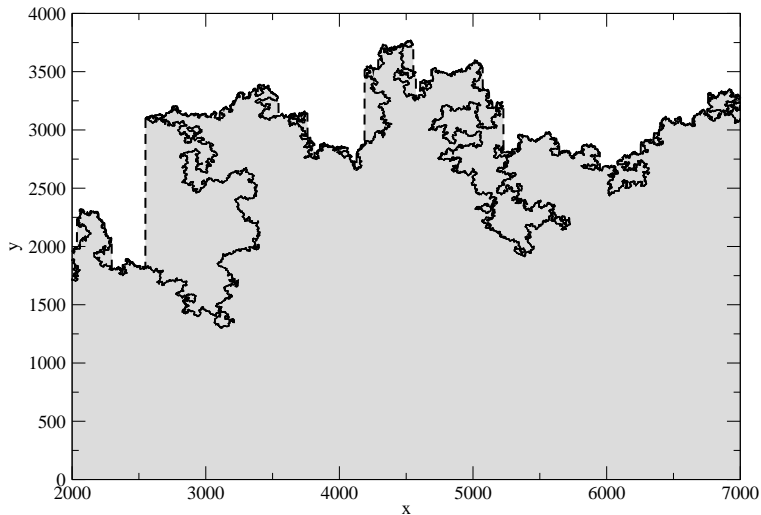}
\end{center}
\vspace{0.5cm}
 \caption {A fractal front in gradient percolation. We also display 
the solid-on-solid projection of the front which is 
single valued but not self-affine.
Courtesy of A. Baldassarri, see also \cite{asikainen02}.}
\label{fig:cluster-gradper}
\end{figure}

The essential feature of gradient percolation is that an occupied 
region is separated from an empty region by a corrugated front.
The front occurs around the region where $p(x) \sim p_c$ and displays 
a self-similar geometry at short scales, crossing over to a flat
profile at larger scale. The crossover scale $\xi_c$ is set by the 
value of concentration gradient $\nabla p$ around $p_c$. Using
scaling arguments it is possible to show that 
\begin{equation}
\xi_c \sim (\nabla p)^{-\alpha} ~~~~~ \alpha=\frac{\nu}{\nu+1},
\label{eq:gradper}
\end{equation}
where $\nu$ is the correlation-length critical exponent of percolation
defined in Eq.~(\ref{eq:nuperc}). The width of the separation front
scales linearly with $\xi_c$ and thus increases as the gradient
becomes weaker. For lengthscales smaller than $\xi_c$ the front is
essentially equivalent to the perimeter of a percolation cluster and
has thus a fractal dimension $D=4/3$ (or $D=7/4$ depending on the 
definition of the front) \cite{baldassarri02}. 
An example of a gradient percolation
cluster is reported in Fig.~\ref{fig:cluster-gradper}.

The relation between gradient percolation and fracture is based
on the study of models of planar cracks. An approach alternative
to the crack line depinning scenario discussed in Sec. \ref{sec:depinning} considers
damage accumulation in a stress gradient. Thus the crack is not
considered as line, but as the separation front between damaged and  
intact areas, allowing for microcrack formation 
and coalescence even ahead of the main crack. The question is
whether this effect can account for the discrepancy between 
the experiments and the crack depinning predictions.

Different types of lattice models have thus been simulated 
with boundary conditions inducing a stress gradient. Simulations
of a two dimensional RFM in the planar crack geometry (i.e. considering
two conducting plates connected by fuses) yield a gradient
percolation front \cite{zapperi00}. A beam model studied in a similar geometry,
recovers instead the crack line depinning exponents, indicating
that damage nucleation is irrelevant for crack roughness in that model \cite{astrom00}.
On the other hand, a long-range scalar model, simulating planar cracks in a three
dimensional geometry, yields a larger exponent $\zeta\simeq 0.6$
which could explain the experiments \cite{schmittbuhl03,schmittbuhl04}.  
The validity of this result is questionable based on the reasoning given  
below \cite{alava04}.

\begin{figure}[t]
\begin{center}
\includegraphics[width=10cm]{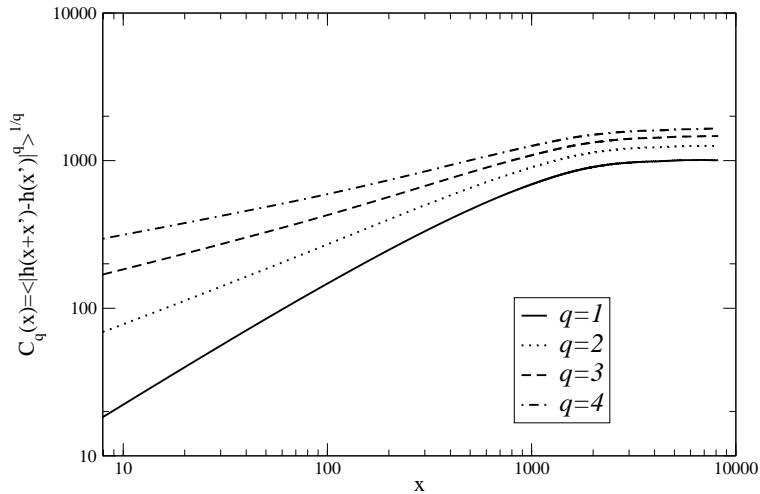}
\end{center}
 \caption {The high order correlation function for the solid-on-solid
projection of a gradient percolation front. Since the original front
is self-similar, the projection displays an artificial multi-affinity
with effective exponents $\zeta_q=1/q$ for $q>1$.
Courtesy of A. Baldassarri, see also \cite{asikainen02}.}
\label{fig:multiscaling}
\end{figure}

In the model of Refs.~ \cite{schmittbuhl03,schmittbuhl04}, 
the width of the front $W$ increases with the external load 
approximately as a power law, and eventually saturates. As in gradient percolation,
the saturated width $W^*$ scales with the gradient of the damage
profile (see Eq.~(\ref{eq:gradper})).  Since in Ref.~\cite{schmittbuhl03} $\xi
\sim L$, where $L_x$ is the lattice size parallel to the front, one can combine
the initial dynamic scaling with the saturated width into
a ``Family-Vicsek''-like scaling form $W(L_x,t)= L^\alpha f(t/L^z)$,
and conclude that the fronts are self-affine interfaces with $\zeta=\alpha$.  
In gradient percolation, however, $\alpha$ can {\it not} be interpreted as a
roughness exponent, since as we discussed above the front is
{\it self-similar} \cite{alava04}.  
If we remove the overhangs from the front through a
solid-on-solid projection, we obtain a single valued interface $h(x)$
that can be analyzed using conventional methods. The result, shown in
Fig.~\ref{fig:multiscaling} is an artificial multiscaling for the
correlation functions
\begin{equation}
\langle |h(x+x')-h(x')|^q \rangle^{1/q} \sim x^{\zeta_q},
\end{equation}
where $\zeta_q=1$ for $q \le 1$ and $\zeta_q=1/q$ for $q > 1$ The
multiscaling behavior is easily understandable: for $q<1$ the scaling
is sensitive to the fractal, hence $\zeta_q=1$ as expected for
isotropic scaling. For $q>1$ the scaling is dominated by the big
jumps, which are also apparent in Fig.~\ref{fig:cluster-gradper} 
\cite{asikainen02}.

In summary, if the front is due to a gradient percolation mechanism then
its geometry is not self-affine, even when the gradient scales with
$L$ as in some fracture models \cite{schmittbuhl04}.
The same reasoning applies to the theoretical arguments put forward in
Ref.~\cite{hansen03} to explain the roughness of the cracks in the
random fuse models. The scaling theory of Ref.~\cite{hansen03} rests 
on a number of assumptions: 
\begin{enumerate}
\item in the limit of strong disorder fracture occurs at a percolation
(possibly correlated) critical point;
\item for weak disorder crack localization occurs, producing a quadratic 
damage profile;
\item the width of the damage profile is a linear function of the lattice size; and
\item weak and strong disorder are characterized by the same $\nu$ exponent.
\end{enumerate}
Given these assumptions, the authors of Ref.~\cite{hansen03} argue
that in the localization regime the crack is formed as a front in
gradient percolation, leading to a scaling relation $\alpha=
2\nu/(1+2\nu)$, where the factor $2$ (not present in
Eq.~(\ref{eq:gradper})) comes from assumptions 2 and 3. As discussed
above, the identification of $\alpha$ with $\zeta$ is incorrect.
In addition, the analysis of large scale numerical simulations yields
a series of additional problems: (i) the existence of a reliable
percolation scaling is dubious; (ii) the damage profile is not
quadratic; (iii) the width of the damage profile is not a linear
function of $L$ \cite{nukala04}.

\section{Numerical simulations}
\label{sec:simul}

Here we review the main results, obtained from numerical
simulations of statistical models for fracture for the
kind of models discussed in Section \ref{sec:models}.
We present a detailed discussion of the algorithms employed for
such simulations in the appendix. The main
work horse for these studies has traditionally been the RFM,
which therefore will garner the most attention in what follows.
The main motif of this section is to summarize the phenomenology
of the results, from the RFM and similar models, and to emphasize
the relationship with the fracture mechanics framework outlined in
Sec.~\ref{sec:frac-mec} and the theoretical approaches reported in 
Sec.~\ref{sec:theory}. For simplicity we defer a general
comparison between simulations and experiments
discussed in Sec.~\ref{sec:exp} to the conclusions.

We first discuss the scaling of the stress-strain characteristics (Sec.~\ref{sec:iv}) ,
and its interpretation in terms of damage mechanics (Sec.~\ref{sec:damage_dist}). 
One issue is whether the numerical data can be interpreted in terms of a scaling
picture or framework from statistical physics. Damage as a thermodynamic
quantity is then further explored analyzing the spatial distributions
of local damage, the development of localization and internal correlations.
The two central questions here concern the nature of the approach
to peak load, what damage can tell about it, and the role of disorder
in the process.

Next, we discuss the topics that are more relevant for statistical
experiments: fracture
strength, crack morphology and avalanche precursors. In particular 
\begin{itemize}
\item[(i)] We analyze the properties of the strength distribution and  
the associated size effects, including a discussion of the effect of disorder in notched specimens. Such results are of importance due to the relation to weakest-link
theory and due to the engineering applications (Sec.~\ref{sec:fract_str}). 
\item[(ii)] The surface properties of cracks are considered from the viewpoint
of establishing the true scaling indicated by the data: what is the
minimal set of exponents that suffices, and what is the role of disorder
and dimensionality (Sec.~\ref{sec:rough}). 
\item[(iii)] The avalanches in fracture models are discussed
based on general scaling relations for systems with crackling noise.
Again, it is important to understand how the specific scenario (i.e., 
type of interactions, disorder, dimensionality) influence the 
avalanche behavior (Sec.~\ref{sec:rfm_ava}).
\end{itemize}

\subsection{The I-V characteristics and the damage variable}
\label{sec:iv}
The first basic information that can be obtained from numerical simulation
is the constitutive response of the model, in the form of a stress-strain curve
for tensorial models or, more simply, of a current-voltage ($I-V$) curve for the RFM.
This quantity is particularly important in connection with damage mechanics 
(see Sec.~\ref{sec:damage}), since lattice models allow to test possible relations 
between the microscopic damage state and the macroscopic constitutive behavior. 
Here for the sake of clarity  we focus on the scalar case (i.e. the RFM) 
where the damage variable $D$ is simply defined. The same discussion
can be translated to generic (tensorial) fracture models considering the
stress strain characteristics. These models have in general the important
feature, that they allow in contrast to mean-field -like approaches the
existence of damage correlations, which can thus be explored.

\begin{figure}[hbtp]
\includegraphics[width=10cm]{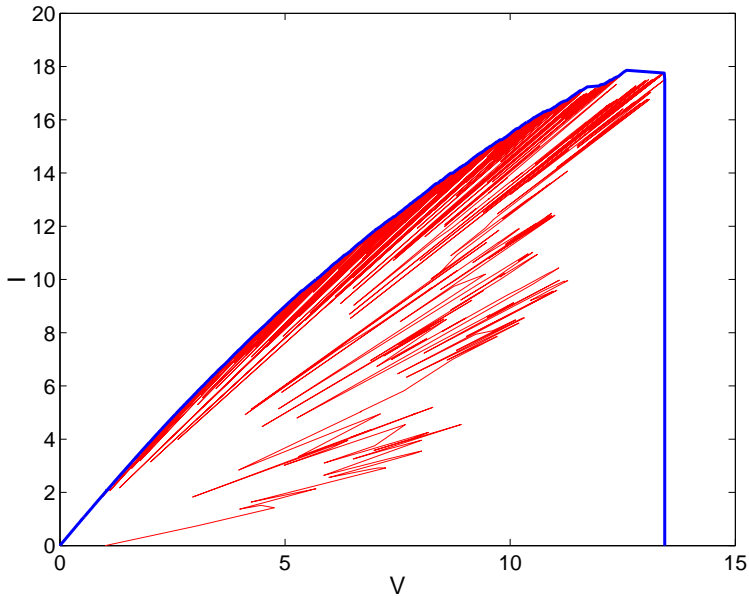}
\caption{Typical $I-V$ response of a 2D RFM triangular lattice system of size $L = 64$. The thin
line is the evolution that follows a quasistatic dynamics in  which bonds are broken one at
a time. The envelope of the $I-V$ response (thick line) represents the voltage controlled response.}
\label{fig:fdenv}
\end{figure}

Figure \ref{fig:fdenv} presents a typical $I-V$ response, from a 2D RFM (triangular lattice), with a system size $L = 64$. 
One can see damage accumulation, and then signatures of crack growth which becomes 
unstable after the peak current. While initially the stress concentrations are masked by
the disorder, in the last stage one of the microcracks wins over the others and dominates.

To obtain a faithful representation of the macroscopic response one needs to perform
an ensemble average over many sample configurations. This can naturally 
be done fixing three different control parameters: current, voltage,
or accumulated damage (i.e. number of broken bonds). Appropriate rescaling
of the curves is then necessary to obtain intensive quantities
(i.e. independent of the system size). For this reason the voltage
$V$ is the most natural parameter, since the current excludes the post-peak load
part of the $I-V$ response, and the damage is - as is discussed below -
not an intensive variable, at least in the post peak regime. 

An old idea, inspired from self-affine interface growth, was 
to attempt a Family-Vicsek (FV) \cite{deArcangelis89} scaling
for the $I-V$ curve (see also Ref.~\cite{krajcinovic05b}). 
This entails the scaling form
\begin{eqnarray}
I & = & L^{\Omega_1} \phi(\frac{V}{L^{\Omega_2}}) \label{IV1}.
\end{eqnarray}
The scaling function $\phi(x)$ has to fulfill the condition
that for small $V$ the response is independent of system
size. There is an important detail, not fully appreciated in the 
literature: usually in the context of surface growth the use of Eq. (\ref{IV1}) implies that 
$\phi \to const >0$ for $x \gg 1$. In fracture, however, the current becomes
zero for large enough $V$: in that limit $\phi$ vanishes and
is thus not monotonic. For small $x$ the function has to
scale as power-law of its argument, which directly implies
that the FV collapse {\em is not applicable} unless the 
peak load has a power-law scaling form ($I_{max} \sim L^y$, for some $y$), 
and the same holds
also for the corresponding voltage. In addition, the scaling of $\phi$
couples the exponents $\Omega_1$ and $\Omega_2$:
since for small $V$, $I \propto V$, it follows that $\phi \sim x$
and $\Omega_1 = \Omega_2$. While this discussion is valid in general,
for arbitrary models and experiments, 
we consider as an example the numerical data for the RFM.

\begin{figure}[hbtp]
\includegraphics[width=8cm]{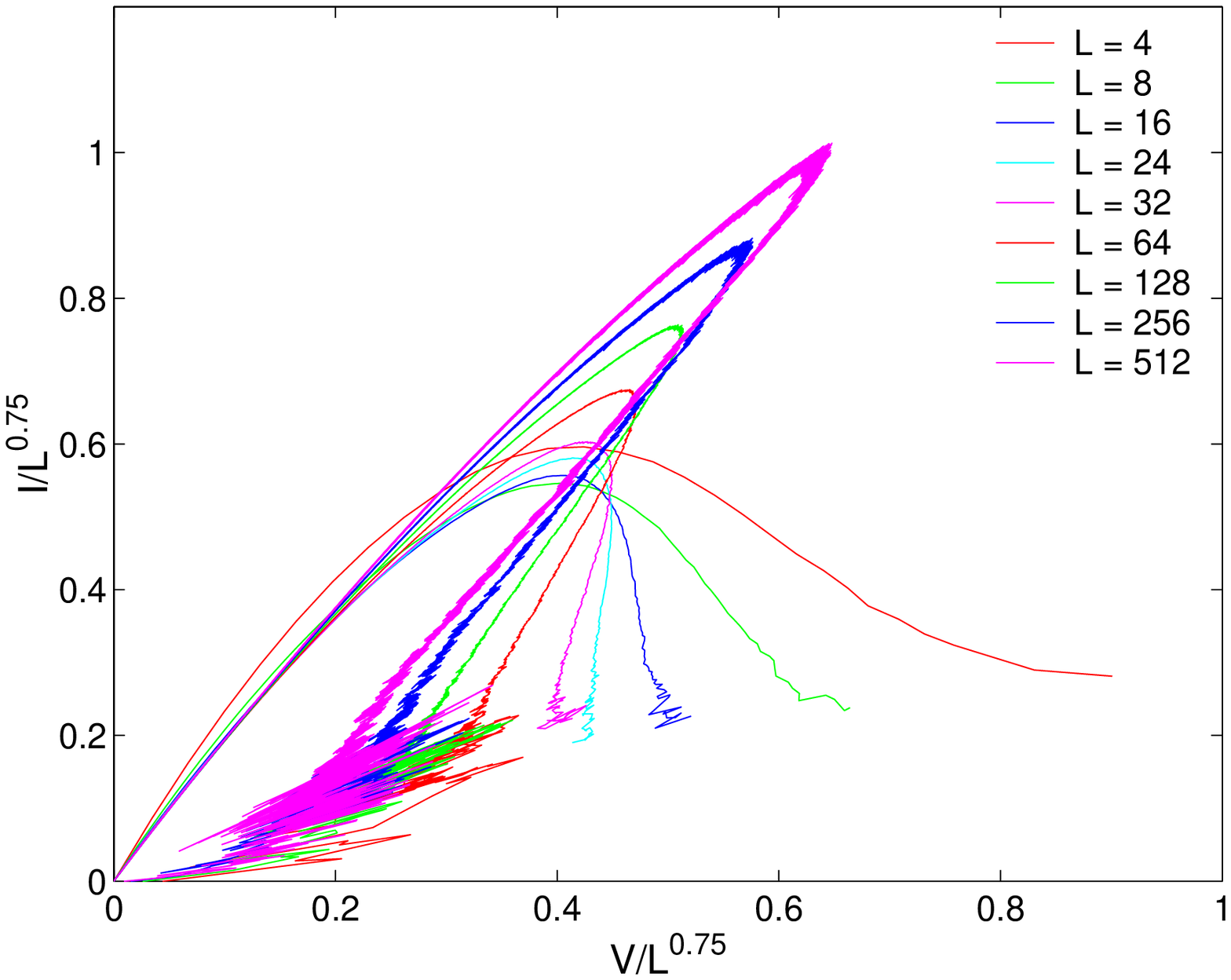}
\includegraphics[width=8cm]{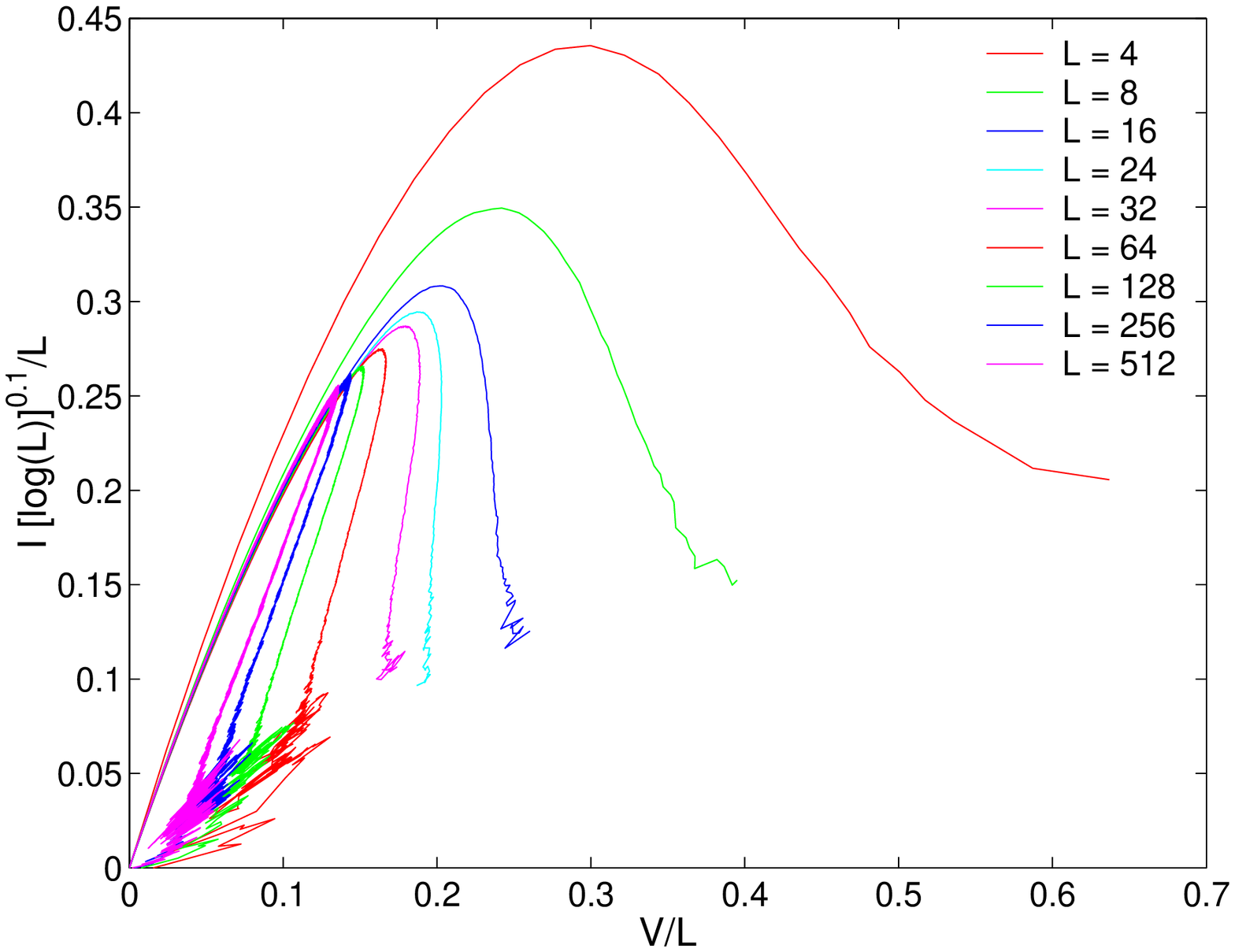}
\caption{Scaling of $I-V$ response for different triangular lattice system sizes of 2D RFM. 
(a) Family-Vicsek scaling as proposed in 
Ref. \cite{deArcangelis89} (see Eq. (\ref{IV1})) with $\Omega_1 = \Omega_2 = 0.75$. 
(b) Scaling form proposed in Refs. \cite{sahimi933,sahimi98} (see Eq.
(\ref{IV2}) with $\Omega_1 = 1.0$ and $\psi = 0.1$).}
\label{fig:IV1}
\end{figure}

Figure \ref{fig:IV1}(a) presents the scaling of $I-V$ 
response based on the FV scaling. The plots in Fig. \ref{fig:IV1}(a) correspond to 
$\Omega_1 = \Omega_2 = 0.75$ as proposed in Ref. \cite{deArcangelis89}. 
On the other hand, an alternative ad-hoc scaling form 
\begin{eqnarray}
I & = & \frac{L^{\Omega_1}}{(log L)^\psi} \phi_2(\frac{V}{L^{\Omega_1}}) \label{IV2}
\end{eqnarray}
has been proposed by Sahimi and collaborators, in Refs. 
\cite{sahimi933,sahimi98}. The values indicated are
$\Omega_1 \simeq 1 \pm 0.1$ 
and $\psi \simeq 0.1$. 
Figure \ref{fig:IV1}(b) presents the corresponding scaling of $I-V$ curves
based on Eq. (\ref{IV2}). 
This suggests that scaling form based on Eq. (\ref{IV2}) is de facto
more appropriate than the FV one, but of course one notes that
the scaling is far from perfect. The logarithmic factor in 
Eq. (\ref{IV2}) arises from an attempt to correctly take into account the
strength size effect. For such a scaling to hold the scaling function $\phi_2$ would also need to
contain logarithmic terms so that the small-voltage limit 
is independent of $L$. Their origin would be unknown.
Notice that the analysis presented here is also valid
for experiments, and for instance questions the application
of FV scaling to the experiment of Otomar et al. \cite{otomar06}.

Notice that for the data in Figs.~\ref{fig:IV1} ensemble averaging
is performed using the accumulated damage as a control parameter as originally
done in Refs.~\cite{deArcangelis89,sahimi90,sahimi933,sahimi98}. This produces an
unphysical hysteresis in the curves, which corresponds to unstable crack growth in
voltage or current controlled cases. This effect was not apparent for the small
lattices sizes at hand in Refs.~\cite{deArcangelis89,sahimi90,sahimi933,sahimi98}, but
becomes clear for larger lattices employed here. To avoid this problem, it is more
natural to average the envelopes of the $I-V$ curves, using the voltage as a control parameter.
The total current $I$ flowing through the
lattice  in the RFM is given by $I = \Sigma~V$
where $\Sigma$ is the total conductance of the lattice system expressed as
\begin{equation}
\Sigma = \frac{\left[\sum_{ij} \sigma_{ij}~V_{ij}^2\right]}{V^2},
\end{equation}
and $\sigma_{ij}$ is the local conductivity of the fuse between nodes $i$ and $j$.
A simple normalized form of the $I-V$ curve may then be obtained as
\begin{equation}
\frac{IL}{N_{el}} = \frac{\left[\sum_{ij} \sigma_{ij}~s_{ij}^2\right]}{N_{el}}~\frac{V}{L}, \label{eqI1}
\end{equation}
where $s_{ij}$ denotes the enhancement of current (stress concentration factor) due to fuse
burnouts and $\frac{V}{L}$ is voltage drop per unit length.
It should be noted that $s_{ij} = 0$ for all the {\it dangling} or {\it dead ends}
(using percolation theory language) that do not carry any current (stress) even though they are not broken.

\begin{figure}[hbtp]
\begin{center}
\includegraphics[width=10cm]{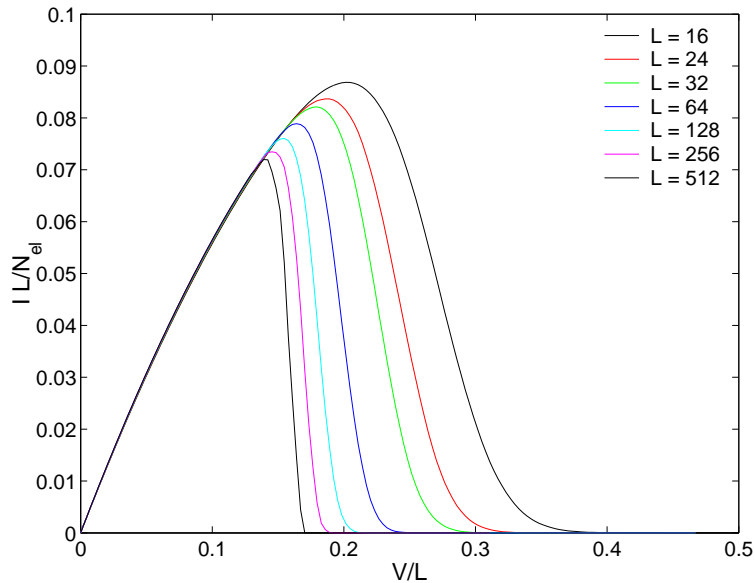}\end{center}
\caption{Normalized ensemble averaged $I-V$ response based on Eq. (\protect\ref{eqI1}) for different 
2D RFM triangular  lattice systems of sizes $L = \{4, 8, 16, 24, 32, 64, 128, 256, 512\}$.}
\label{fig:IV2}
\end{figure}

\begin{figure}[hbtp]
\begin{center}
\includegraphics[width=10cm]{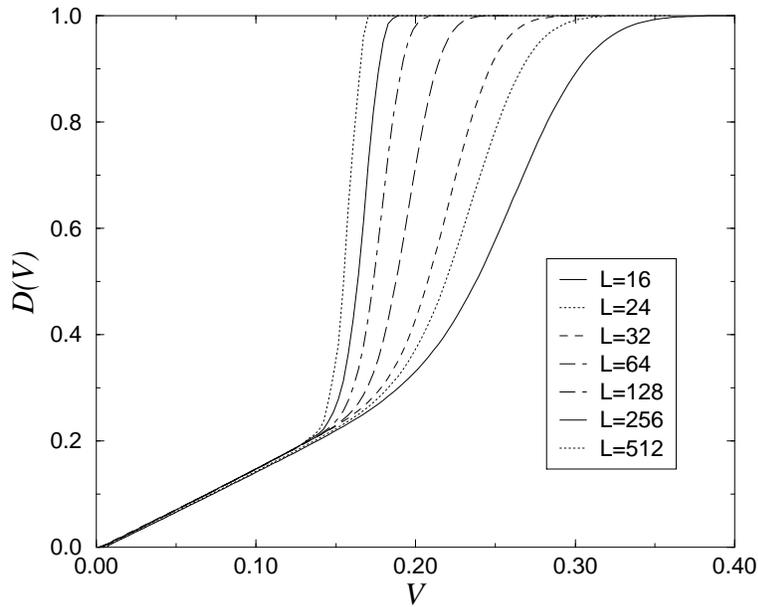}\end{center}
\caption{Ensemble averaged  damage variable $D(V)$ for different 
2D RFM triangular  lattice systems of sizes $L = \{4, 8, 16, 24, 32, 64, 128, 256, 512\}$.
After peak load the damage variable is not intensive. Also, the nonlinearity 
in the plots of $D(V)$ close to $1.0$ is an artifact of $I-V$ averaging close to fracture.}
\label{fig:DV}
\end{figure}

Figure ~\ref{fig:IV2} presents the ensemble averaged $I-V$ response normalized as in Eq. (\ref{eqI1}),
showing a perfect collapse up to peak load, with strong deviations thereafter. One can easily extract from
this curve the behavior of the damage variable $D(V) \equiv 1-\Sigma/\Sigma_0$, where $\Sigma_0$
is the conductivity of the intact lattice.  Figure~\ref{fig:DV} shows that damage variable $D(V)$ grows linearly 
and remains an intensive variable up to the peak load, beyond which size dependent effects become apparent.
Similar damage variable definitions are often used in the engineering literature \cite{krajcinovic}. It is clear that they are
suitable to represent the extent of damage only within the 
early stages of damage evolution, where the damage levels are moderate and the damage is not localized, but
become clearly inadequate around the peak load and especially in the softening regime of the stress-strain curve.  
In these regimes, $D$ is strongly size-dependent and hence 
cannot be considered as a thermodynamics based {\it intensive damage state variable}. Of course, one should keep in mind that due to the ``size-effect''
asymptotically $I_{max}(L)$ may go to zero anyway, which implies that
$D(V)$ would not exist in a well-defined sense for any particular
value of $V$. 

\begin{figure}[hbtp]
\includegraphics[width=8cm]{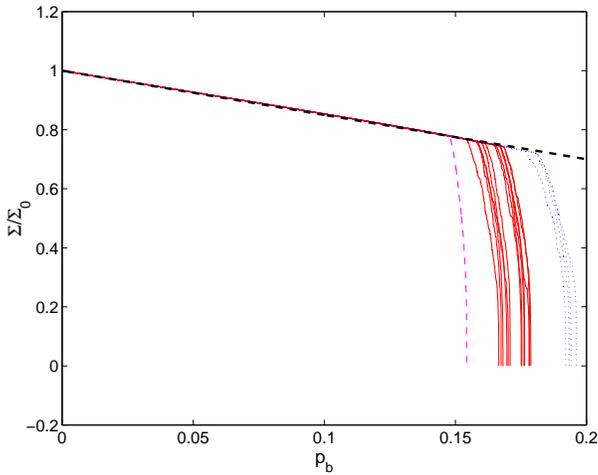}
\caption{Typical conductivity (stiffness) degradation in various 2D RFM samples: 
triangular lattices of sizes $L = 256$ (dotted), $512$ (solid), $1024$ (thin dashed).  The thick 
dashed line corresponds to the 
degradation of the conductivity predicted by effective medium theory. }
\label{fig:stif_samples}
\end{figure}

In order to better understand the behavior of the damage variable, it is instructive to 
observe the degradation of the conductivity of a single sample as a function of the
fraction of broken bonds $p_b$. As is apparent looking at Fig.~\ref{fig:stif_samples}, the 
conductivity before the peak load follows nicely the prediction of effective-medium theory,
considering the contribution to $p_b$ coming both from broken and dangling bonds. 
This means that there are no strong sample-to-sample variations (nor
as we discuss below a strong $L$-dependence).
The agreement with the conjecture implies that damage is spread on large scales
homogeneously in the sample 
with no localization or strong correlations. Hence the definition of a RVE is possible and a continuum damage description is appropriate. The lack of
a ``critical scenario'' with a divergent correlation length $\xi$ in any
case implies that the scalar damage variable $D$ could be simply
written as a power series in the control parameter $V$,
$D = d_0 V+d_1V^2 \dots$, with $I = (1-D(V)) \Sigma_0 V$. Then, 
the study of damage becomes that of the coefficients $d_i$ and 
their dependence on model, dimensionality, and disorder.

\begin{figure}[hbtp]
\includegraphics[width=8cm]{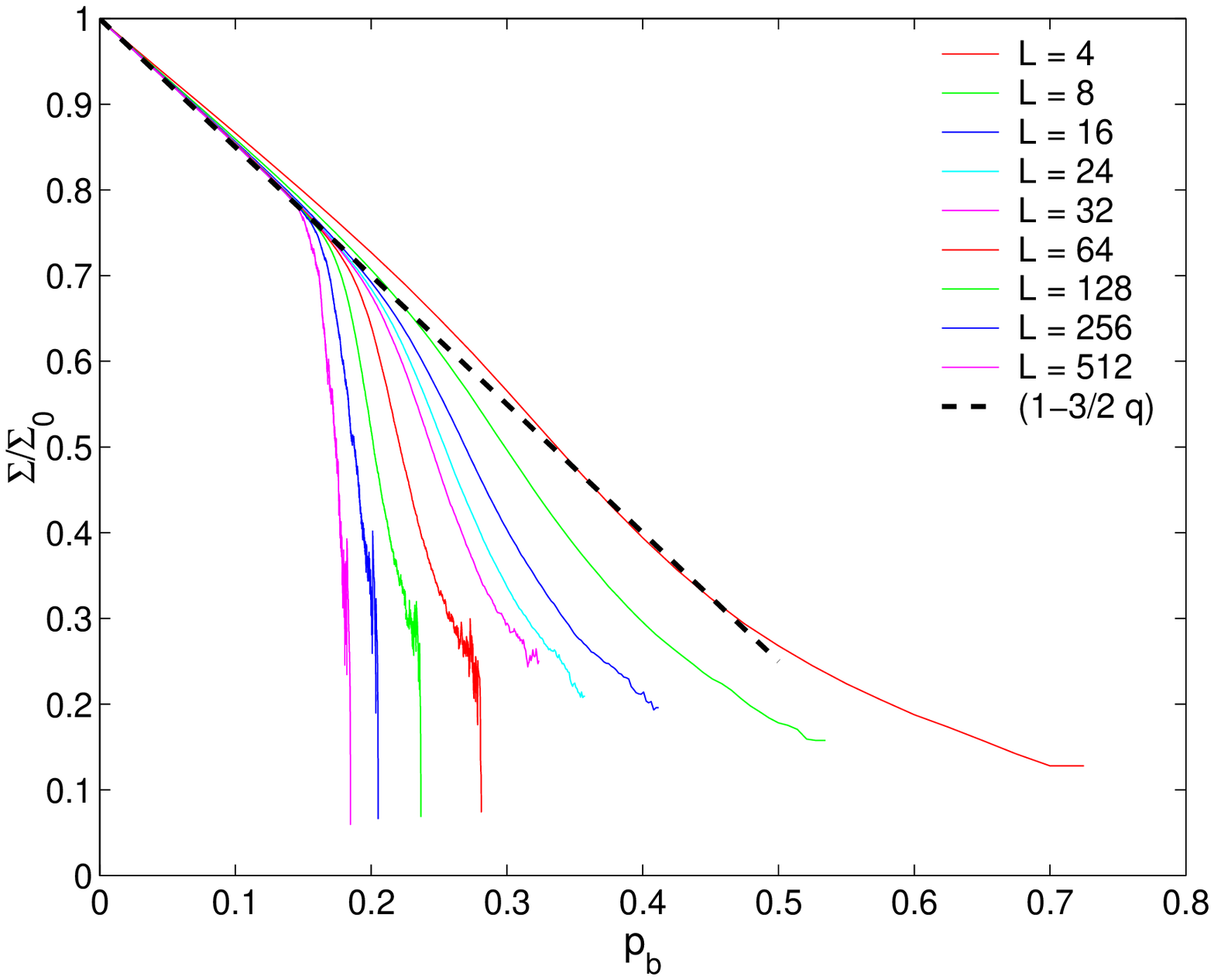}
\includegraphics[width=8cm]{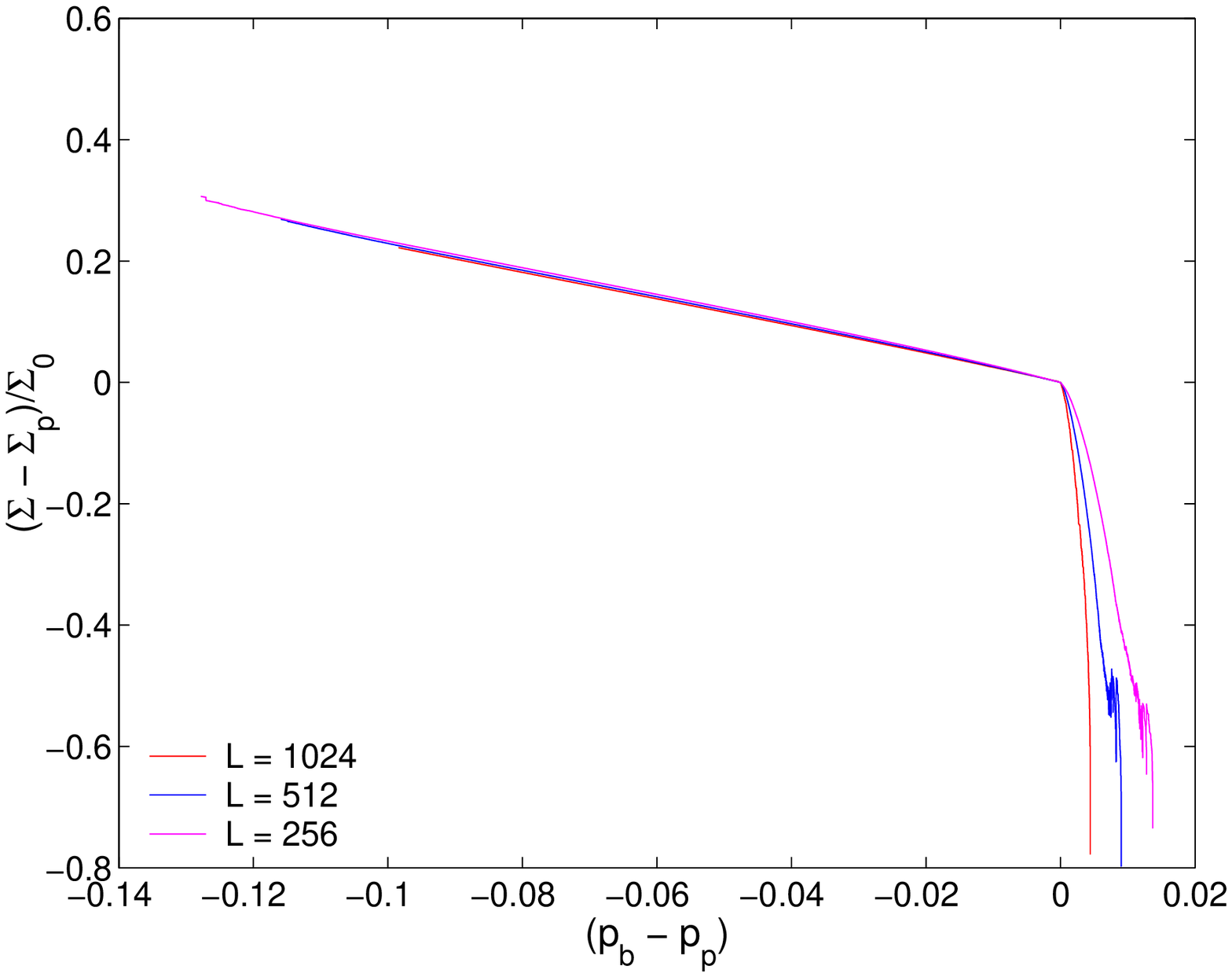}
\caption{Average conductivity scaling in a 2D RFM triangular lattice system. 
(a) Averaging performed at constant $n_b$ number of broken bonds. 
(b) Averaging by first shifting each of the sample responses by their
respective peak load positions.}
\label{fig:kstif}
\end{figure}

In the post peak regime, however, the conductivity deviates abruptly from
the effective medium prediction, with strong sample-to-sample and lattice size dependent
fluctuations. This brings back the question of ensemble averaging: by averaging together
several curves as the one reported in Fig.~\ref{fig:stif_samples}, the abrupt character of the
discontinuity is partly lost and one obtains smooth curves as the ones reported in Fig.~\ref{fig:kstif}(a).
This effect becomes less and less important as the sample size is increased, but has lead
to some confusion in the early literature, when only small samples were available.
Alternatively, the averaging of the conductivity degradation responses may be 
performed by first shifting the data by the respective peak load locations 
$(p_p,\Sigma(p_p))$. In this way, one preserves the main character of the curves
as shown in Fig. \ref{fig:kstif}(b).

\subsection{Damage Distribution}
\label{sec:damage_dist}

Fuse networks can be used to study the microscopic dynamics
of damage. In the Sec.~\ref{sec:iv} we discussed the macroscopic
characteristics of the RFM in terms of the I-V curves.  
The increase in the damage variable $D$
(in $I = (1-D(V)) \Sigma_0 V)$) is concomitant to the growth of correlations
due to microcracking controlled by the distribution of quenched disorder. 
When the disorder is  narrowly distributed
(weak disorder),  it is known that damage is localized and the network
breaks down without significant precursors \cite{kahng88}.   
For infinitely strong disorder, the damage accumulation process can exactly be
mapped onto an uncorrelated percolation problem \cite{roux88}. The interesting
situations are to be found for strong but finite disorder, where
a substantial amount of damage is accumulated prior to failure. 
In these conditions, one can try to study several quantities such 
as the scaling of damage at maximum current, or the geometric characteristics
of the ensembles of broken fuses along $I-V$ curves. 
One interesting question is in addition
related to the fluctuations of damage at the $I-V$ curve maximum.

\begin{figure}[hbtp]
\includegraphics[width=10cm]{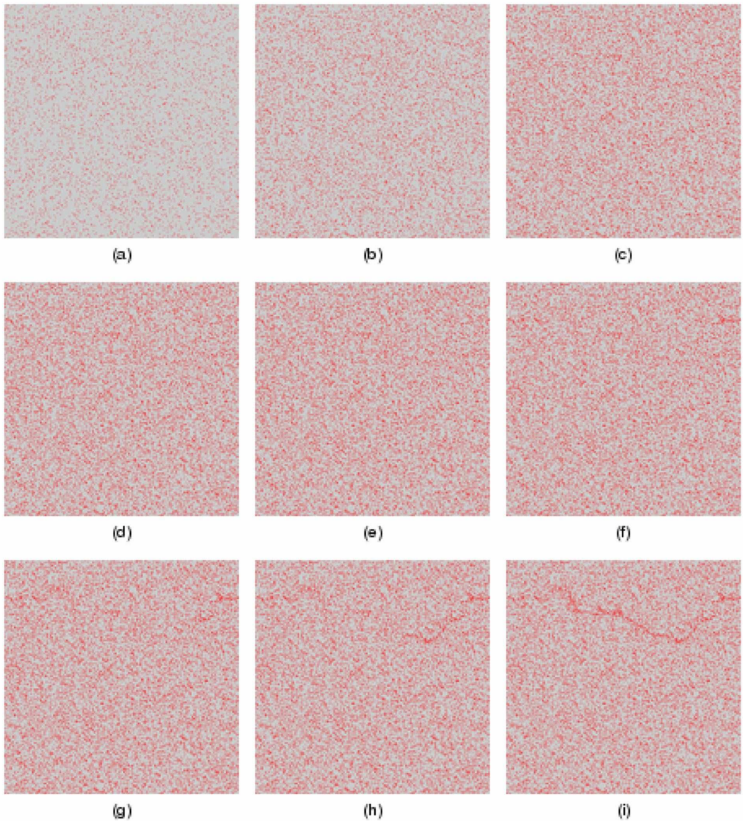}
\caption{Snapshots of damage in a typical 2D triangular lattice system of size $L = 512$. 
Number of broken bonds at the peak load and at failure are 83995 and
89100, respectively.  (a)-(i) represent the snapshots of damage after
breaking $n_b$ number of bonds.  (a) $n_b = 25000$ (b) $n_b = 50000$
(c) $n_b = 75000$ (d) $n_b = 80000$ (e) $n_b = 83995$ (peak load) (f)
$n_b = 86000$ (g) $n_b = 87000$ (h) $n_b = 88000$ (i) $n_b = 89100$
(failure)}
\label{fig:fig1a}
\end{figure}

Figure \ref{fig:fig1a} presents the snapshots of damage evolution in a
typical 2D RFM simulation of size $L = 512$ and uniform disorder 
distribution in $[0,1]$.  From this figure, at
least visually, it is apparent that damage evolves in a fashion
that does not foretell the final crack up to the peak load 
(see Fig. \ref{fig:fig1a}e), beyond which the localization of damage occurs. 
Next we discuss various results concerning the distributions and
fluctuations of damage, both at peak load and at final failure.

\subsubsection{Scaling of Damage Density}

We first consider the evolution of the number of broken bonds at peak
load, $n_p$, which excludes the bonds broken in the last catastrophic
event, and the number of broken bonds at failure, $n_f$. The quantity
$n_p$ is of interest since it tells how the damage is distributed before
failure occurs. If the damage were to be localized along
a single row, the number of broken bonds at peak load and failure would be
$O(1)$ and $O(L)$, respectively, whereas screened percolation would result in $n_f \sim N_{el}$,
where $N_{el}$ is the total number of bonds in the intact lattice.

\begin{figure}[hbtp]
\begin{center}\includegraphics[width=10cm]{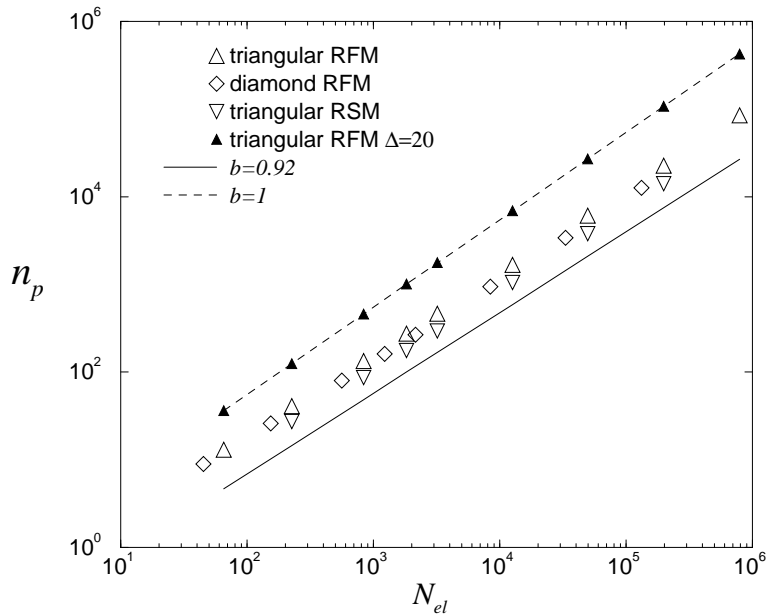}\end{center}
\caption{Scaling of number of broken bonds at peak load for two-dimensional 
triangular and diamond random fuse lattices, and 
triangular spring lattices. In all these models disorder is uniformly distributed in $[0,1]$  and
the scaling exponents are very close to each other. The last curve represents data obtained using a power low disorder distribution: the exponent changes to $b=1$.}
\label{fig:np}
\end{figure}

Figure \ref{fig:np} presents $n_p$ as a function of the lattice size
$N_{el}$ for 2D RFM and for 2D RSM with uniform threshold
disorder. The data displays a reasonable power law behavior $n_p \sim
N_{el}^b$, with $b=0.91-0.93$ 
\cite{nukala04,nukala05,delaplace96,deArcangelis88,deArcangelis89}. 
In other words, the damage is sub-extensive reflecting a size-effect
at the peak load that decreases with $N_{el}$.
The difference between the RFM and RSM models is marginal.
Some systematic deviations from the scaling form $n_p \sim
N_{el}^b$ can perhaps be seen by plotting $n_p/N_{el}^b$ vs $N_{el}$ (see
Ref.~\cite{nukala04}). Such attempts would indicate that the damage
data be equally well fit by a linear law times a logarithmic
correction $n_p \simeq N_{el}/\log(N_{el})$ as suggested in
Ref.~\cite{mikko04}. This also implies that in the thermodynamic limit
the peak load damage vanishes. 
For strong disorder (power law distributed with $\Delta=20$, see Sec.~\ref{sec:rfm_plastic})  
the data depicted in Fig. \ref{fig:np} implies a volume-like $b\simeq 1$, indicating 
behavior similar to the percolation limit in the range of $L$ used. 
The three dimensional results for the RFM
are similar \cite{zapperi_ijf,raisanen98}, with
$n_p \sim N_{el}^b$ and $b=0.72$ for weak disorder
(a uniform distribution in $[1/2,3/2]$ \cite{raisanen98}) and
$b=0.99$ for stronger disorder (a uniform distribution in $[0,1]$ 
\cite{zapperi_ijf}). The weak disorder scaling can also be recast as 
$n_p \simeq L^2$ corresponding to a localization along a plane.

The other interesting quantity is the 
average number of broken bonds in the post-peak regime,
$(n_f-n_p)$. This corresponds to the size of the last catastrophic
avalanche, which is trivially so in particular in the current-driven case. 
It also exhibits 
a power law scaling $N_{el}^{\theta}$ with $\theta=0.72$ 
in 2D RFM triangular lattices  and $\theta=0.69$ in 2D RFM diamond lattices \cite{nukala04},
while $\theta=0.68$ in 2D RSM triangular
lattices \cite{nukala05}, and $\theta=0.81$ in 3D RFM cubic lattices
\cite{zapperi_ijf}. If instead one uses again the $\Delta=20$ case depicted
above, the scaling  with $N_{el}$ seems to be linear ($\theta=1$), 
i.e., extensive similar to $n_p$ \cite{nukala04} whereas for weaker 
disorder one has $\theta <b$.

If $n_f$ scales extensively, one may then ask what is the finite-size
correction to an asymptotic density defined as 
$p_c = \lim_{L\rightarrow \infty}
n_f/N_{el}$, in analogy to percolation as argued by Hansen
and co-workers  \cite{hansen03,ramstad}.  
A percolation-like scaling for the finite-$L$  fraction of broken bonds
would be given by
\begin{eqnarray}
p_{f} - p_c & \sim & L^{-\frac{1}{\nu}} \sim N_{el}^{-\frac{1}{2\nu}}
\label{pf}
\end{eqnarray}
Here, $p_f = \frac{n_f}{N_{el}}$, and $\nu$ would define an exponent
similar to that of the correlation length exponent in percolation.
While small system size simulations could be consistent with 
Eq.~(\ref{pf}), Ref.~\cite{nukala04} shows that percolation
scaling can be excluded and there is no hidden divergent
lengthscale defining an exponent $\nu$. This is in
agreement with the first-order, nucleation picture of fracture
that one gains from mean-field -like models (Sec.~\ref{sec:fiber}). 
A similar conclusion is corroborated by considering other measures related to
a characteristic lengthscale (which may be called a correlation length or a ``RVE
size'', as discussed in Sec.~\ref{sec:frac-mec}) \cite{delaplace96,mikko04}. 

%\subsubsection{Scaling of Damage Density Probability Distributions}
%remove this subsubsection

\begin{figure}[hbtp]
\begin{center}
\includegraphics[width=10cm]{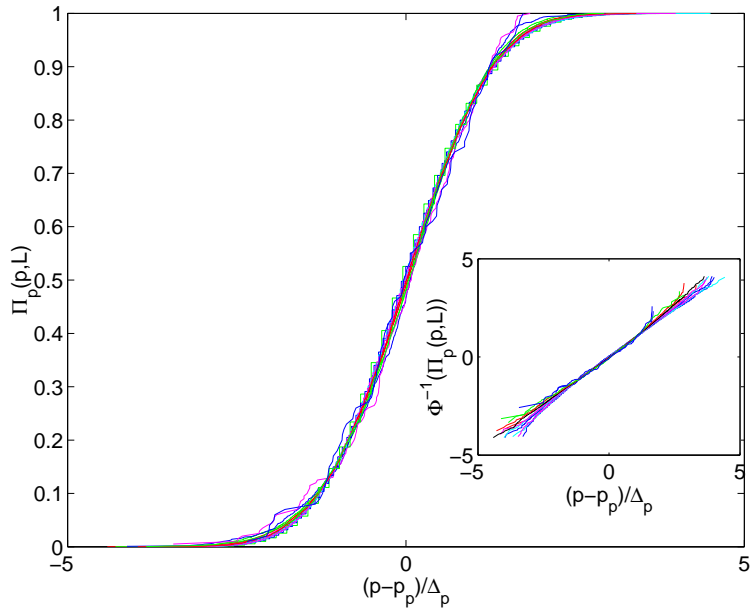}
\end{center}
\caption{Comparison 
between the cumulative probability distributions of the fraction of
broken bonds at the peak load is presented for the RSM and RFM.  For
the RSM, triangular lattices of sizes ($L = 8,16,24,32,64,128,256$),
and for the RFM, triangular lattices of sizes ($L =
16,24,32,64,128,256,512$) are plotted. The normal distribution fit 
in the re-parameterized form is presented in the inset. In the RFM case, collapse of
cumulative probability distributions at the peak load for different
lattice topologies (triangular and diamond) and different disorder
distributions (uniform and power law) is presented in
Ref. \cite{nukala04}.}
\label{fig:RSM6}
\end{figure}

To study further the question of damage fluctuations at peak load and failure
we next consider the distributions ensemble averaged over samples.
The cumulative
probability distribution for the damage density at the peak load is
defined as the probability $\Pi_{p} (p_b,L)$ that a system of size $L$
reaches peak load when the fraction of broken bonds equals $p_b =
\frac{n_b}{N_{el}}$, where $n_b$ is the number of broken bonds.
Scaling such distributions using the average and standard deviation
one obtains a good collapse of the distributions (Fig. \ref{fig:RSM6})
for both random fuse and random spring models in $d=2$ and it seems that
this behavior is independent of lattice structure and disorder
\cite{nukala04}. Similar results are found in three dimensions.

The cumulative probability distribution for the damage
density at failure is defined similarly,  $\Pi_{f} (p_b,L)$
\cite{nukala04}, and there seems to be no essential difference between
the scaled forms of $\Pi_{f}$ and $\Pi_{p}$. This is in contrast
with some claims in the literature wherein a percolation scaling $\Pi_{f} (p_b,L) =
\Pi((p_b-p_c)L^{1/\nu})$ was suggested for $\Pi_{f} (p_b,L)$, and the width of the distribution is
assumed to originate from a correlation length-like scaling \cite{hansen03}.
In the random fuse model, the scaling
forms of cumulative distributions of damage density at peak load and
failure seem to follow the same scaling function, $\Pi (\bar{p}) = \Pi_{f}
(\bar{p}_f) = \Pi_{p} (\bar{p}_p)$ \cite{nukala04}. This does not seem to hold
for spring lattice systems \cite{nukala05}, possibly due to the non-local
character of the loss of rigidity in central-force models (Sec. \ref{sec:tensorial}).
We note finally, that the scaling functions display Gaussian
forms in the case of the RFM (see the inset of Fig. \ref{fig:RSM6}
and Ref.~\cite{nukala05} for the peak load of 2D
RSM). The Gaussian distribution is consistent with the absence of a divergent correlation 
length at failure \cite{delaplace96,mikko04}, which can be understood by 
observing that the damage density $p_f$ is the sum of local random variables (the local
damage, $s_i$), each originating from the contribution of a single RVE.
The central limit theorem yields then a Gaussian distribution, only if
correlations in the field $s_i$ are short-ranged as is the case if a 
RVE can be defined.

%This is an interesting
%theoretical question since in 2D, say, the $s_i$ form a 2D matrix $s_{xy}$,
%whose elements have anisotropic
%correlations (due to the dominant crack growth direction). Given
%a $s_{xy}$, one needs a formal criterion to establish what kind
%of correlations are necessary for the central limit theorem to hold.

\subsubsection{Damage Localization}
\label{sec:localize}

The question of how damage localizes in lattice models is a problem
with many facets. An important question is how the damage variable $D$
is connected to $n_b/N_{el}$ and to the effective RVE size. 
One has to separate here
two issues, the first of which is the development of a macroscopic damage
profile in a sample, and the second being how local microfailures become
correlated. The first issue can be resolved by evaluating the disorder averaged
damage profiles from simulations. However, a posteriori analysis of 
damage profiles is tricky since the final crack position fluctuates, i.e, it 
can occur anywhere along the loading direction of the lattice system, and the 
fact that the final crack position is in general surrounded by a damage ``cloud''.
To overcome this problem, one can use different techniques to determine
the location of the ``crack'' along the macroscopic 
current flow direction. Some of these techniques are based on shifting
individual damage profiles by the center of mass of the damage, or averaging
the power spectrum (square of magnitude of the Fourier
transforms) of individual damage profiles 
\cite{nukala04}.  

\begin{figure}[hbtp]
\centerline{\includegraphics[width=10cm]{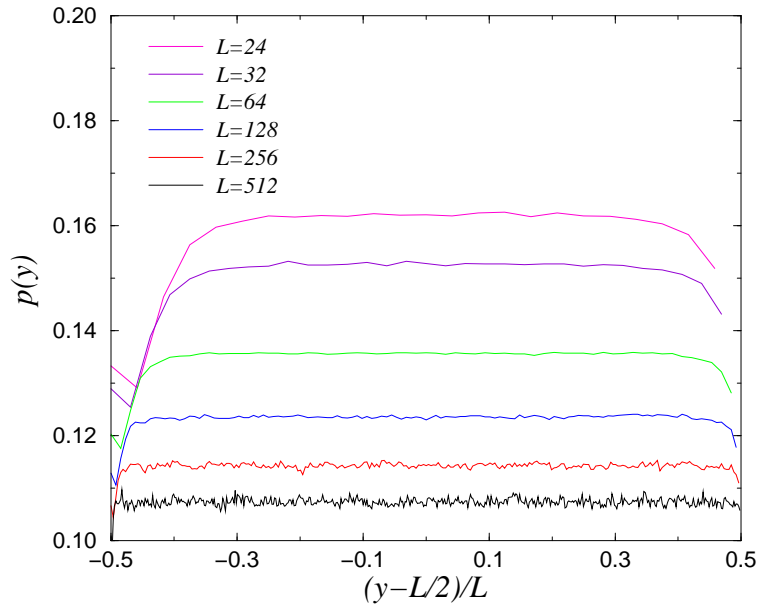}}
\caption{Average damage profiles at peak load obtained by first centering the data
around the center of mass of the damage and then averaging over different samples.
Data are for RFM simulations with uniform thresholds disorder.}
\label{fig:loc2}
\end{figure}

\begin{figure}[hbtp]
\begin{center}
\centerline{\includegraphics[width=10cm]{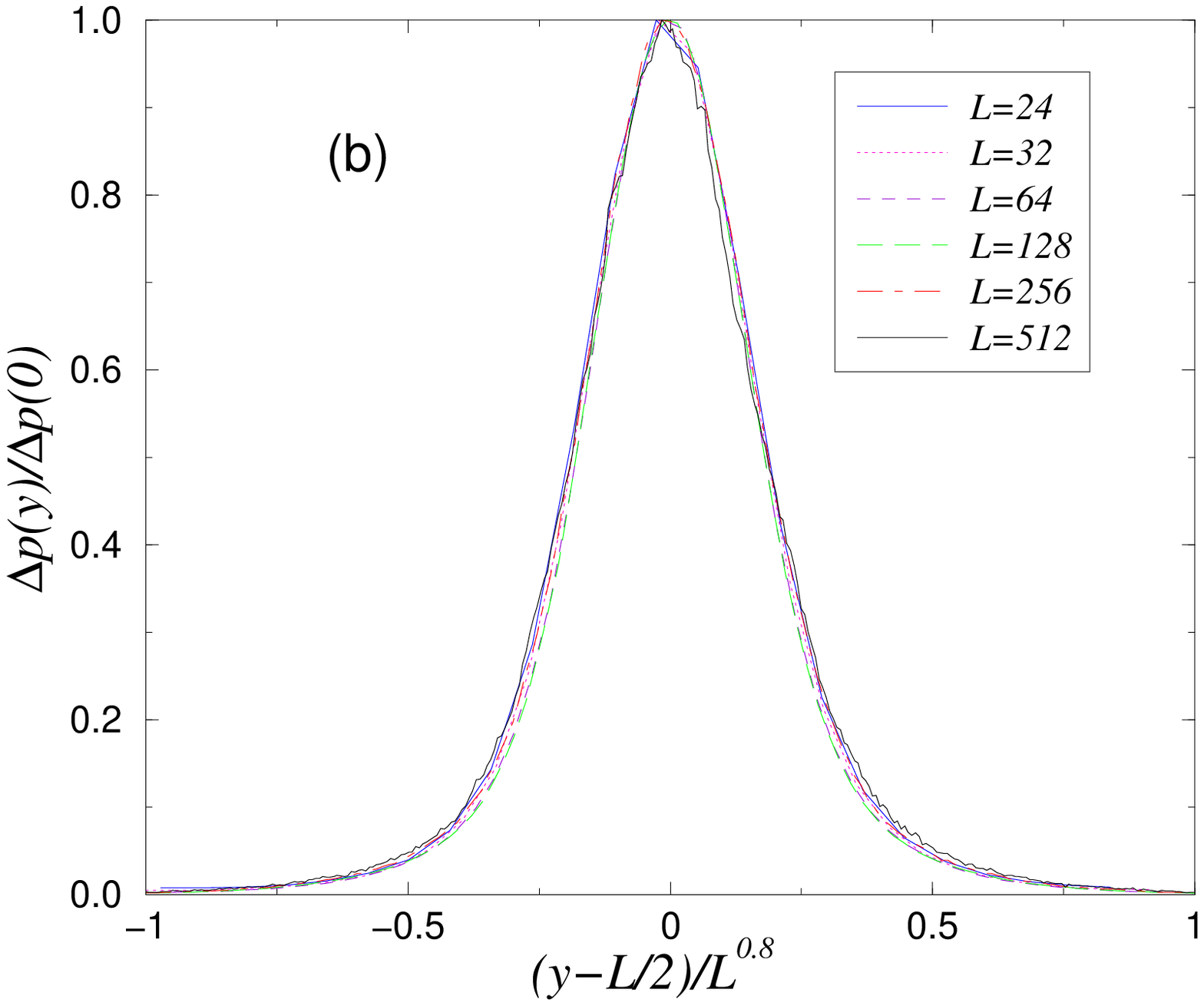}}
\end{center}
\caption{Data collapse of the localization profiles 
for the damage accumulated between peak load and failure. The profiles
show exponential tails. Data are for RFM simulations with uniform thresholds disorder.}
\label{fig:profCM_coll}
\end{figure}

Figure \ref{fig:loc2} presents the average damage profiles for the
damage accumulated up to the peak load by first shifting the damage
profiles by the center of mass of the damage and then averaging over
different samples. This result, displayed in Fig. \ref{fig:loc2} for 2D RFM model 
with uniform disorder distribution, clearly shows that there is no
localization at peak load.  Similar behavior has been observed in
various other cases; namely, the 2D RFM simulations with power law
threshold distribution ($\Delta = 20$) \cite{nukala04}, 
the 2D RSM simulations with uniform thresholds
distribution \cite{nukala05}, and 3D RFM
simulations with uniform thresholds distribution. 
For small system sizes, $L$, there are residual effects due to 
the non-periodicity of the lattice system in the loading direction.
All in all, such profiles suffice to show that the
localization profiles measured at failure in the literature
\cite{hansen03,ramstad,bakke03,mikko04} are due to the damage
accumulated between the peak load and failure. Thus, in the following, we consider
the damage profiles $\Delta p(y)$ for the damage 
accumulated between the peak load and failure.
Figure \ref{fig:profCM_coll} presents a data collapse of the average 
profiles, using the form
\begin{equation}
\langle \Delta p(y,L) \rangle/\langle \Delta p(0) \rangle= f(|y-L/2|/\xi),
\label{eq:prof2}
\end{equation}
where the damage peak scales as $\langle \Delta p(0) \rangle =
L^{-\chi}$, with $\chi=0.3$  and the profile width scales as $\xi \sim L^\alpha$,
with $\alpha=0.8$ (see Fig. \ref{fig:profCM_coll}).  
Similar results are obtained for the 2D RSM \cite{nukala05} and 3D
RFM \cite{nukala_fuse3d} models: the profiles display exponential tails in both cases,
with the exponents for the peak being $\chi=0.37$ and $\chi=0.15$ and
for the profile width $\alpha=0.65$ and $\alpha=0.5$, respectively. The $\alpha$-exponents clearly have to do with the final crack roughness, which is discussed in 
Sec.~\ref{sec:rough}. It is an open question as to how the dynamics in
this catastrophic phase should be characterized, in order to explain
the value of $\alpha$ (or possibly that of $\zeta$, conversely),
and more work is needed in this respect. This would also call for
studies with simplified models (Sec. \ref{sec:fiber}).

One may also consider as a measure the ``width of the damage
cloud'', which is defined as
$W \equiv (\langle (y_b-\bar{y_b})^2 \rangle)^{1/2}$. 
Here $y_b$ is the $y$ coordinate of a broken bond
and a disorder average is performed. At peak load,
$W \sim L/\sqrt(12)$ for both 2D RFM and RSM lattice systems 
\cite{nukala04,nukala_fuse3d,nukala05}, consistent with  
uniformly distributed damage profiles at the peak load.
In the post-peak regime, if the average 
is restricted only to the bonds broken in the last catastrophic failure,
one has $W \sim L^{0.81}$ for 2D RFM \cite{nukala04},
and $W \sim L^{0.7}$ for the 3D RFM. 
These non-trivial scaling exponents in the post-peak regime indicate 
that the damage profiles exhibit a localized behavior in the post-peak regime
in agreement with the results presented above.

\begin{figure}[hbtp]
\includegraphics[width=8cm]{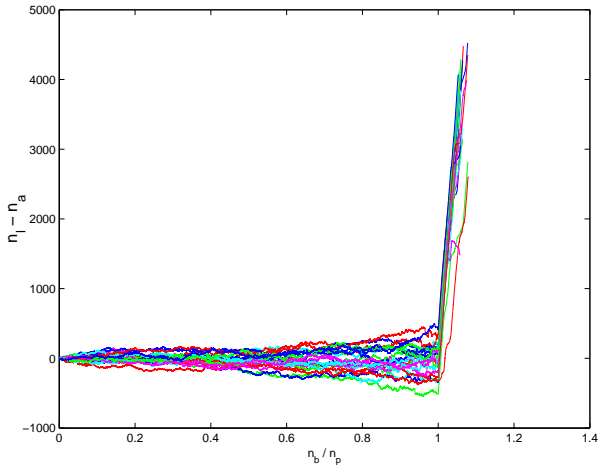}
\caption{Localization of damage $(n_l-n_a)$ in a 2D RFM simulation of 
lattice system size $L = 512$ 
  averaged over 200 samples. Localization of 
damage occurs abruptly at the peak load ($n_b/n_p = 1$).}
\label{fig:ent_local}
\end{figure}

The presence of an abrupt localization can also be seen 
by considering the difference in the accumulated damage 
between the window containing the final crack (and damage)
and any other arbitrary window.
Let $n_l$ and $n_a$ denote the associated numbers of broken bonds.
Figure \ref{fig:ent_local} presents the evolution of $(n_l-n_a)$ 
with damage accumulation for 2D RFM lattice system of size $L = 512$, averaged over 
$200$ samples.  This result shows that $(n_l-n_a) \approx 0$ 
up to the peak load, and a steep increase in $(n_l-n_a)$ in the post-peak regime
indicating abrupt localization. 

\subsubsection{Crack clusters and damage correlations}
\label{sec:corrindamage}

An analysis of the clusters of connected broken bonds is the easiest
way to consider the correlation properties of damage.
This topic is also of interest due to the relation between the largest defect cluster and the 
fracture strength distributions discussed in Sec.~\ref{sec:extr_stat}
\cite{duxbury94,duxbury94b,chakrabarti,sahimi98}. 
In the RFM with strong disorder damage accumulation is non-trivial.
One could postulate that the size-scaling and strength distribution
correspond to the statistics of the largest crack present in
the system at peak load, while crack represents the 
result of damage accumulation prior to the peak load. 

The relevant questions are: (i) what kind of distribution is 
followed by the crack clusters in the strong disorder case 
just before the appearance of an unstable crack,
and (ii) whether there exists a relationship between the distribution of 
maximum cluster size at peak load and the fracture strength distribution?

\begin{figure}[hbtp]
\includegraphics[width=8cm]{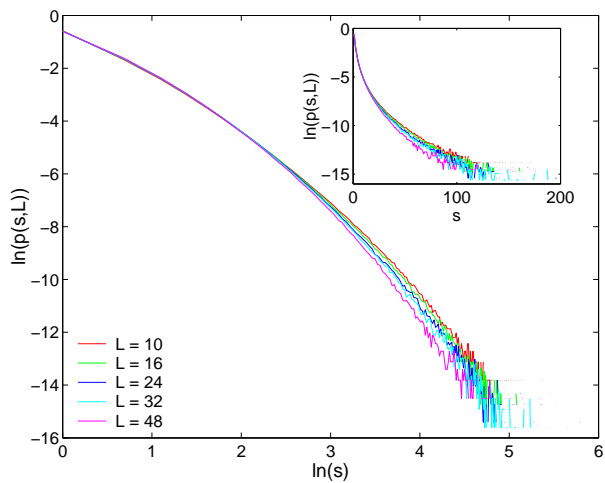}
\caption{Damage cluster distribution 
at the peak load in cubic lattices of system sizes $L$ = \{10, 16, 24, 32, 48\}
in log-log and log-linear scales (inset).  
The cluster distribution data for different lattice system sizes is neither a power-law or
an exponential distribution. }
\label{fig:cluster}
\end{figure}

\begin{figure}[hbtp]
\begin{center}
\includegraphics[width=8cm]{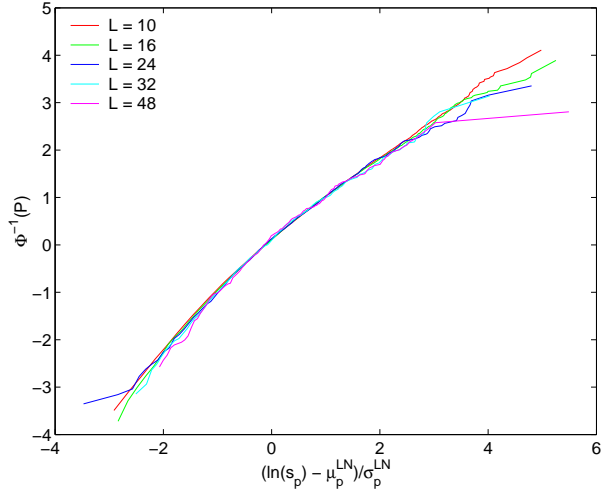}
\end{center}
\caption{Re-parameterized form of 
Lognormal fit for the cumulative probability distribution of largest crack cluster size 
at the peak load in cubic lattices of system sizes $L$ = \{10, 16, 24, 32, 48\}. 
$\mu_p^{LN}$ and $\sigma_p^{LN}$ refer to
the mean and the standard deviation of the logarithm of largest cluster sizes, $s_p$, at peak load. }
%(b) Re-parameterized form of 3 parameter Weibull fit. $\gamma_p$ refers to shift parameter (right).}
\label{fig:maxcluster}
\end{figure}

Figure \ref{fig:cluster} presents the connected cluster size 
distribution at the peak load in a 
3D RFM for various $L$. The plots 
indicate that simple power-law and exponential representations are not 
adequate, and a similar result is obtained in 2D as well 
\cite{nukala_cluster}. For completeness, one should note that the distribution 
of the {\em largest}
crack cluster at the peak load seems to follow a Lognormal
distribution as demonstrated in Fig. \ref{fig:maxcluster} for 
various system sizes. It is interesting to note that the fracture strength data 
presented in the following subsection indicate that a Lognormal distribution 
represents an adequate fit for the fracture strength 
distribution in broadly disordered materials. At first glance, this result appears to 
be trivial: assuming that the stress concentration around the largest crack 
cluster is given by a power law, the fracture strength follows a Lognormal 
distribution since the power of a Lognormal distribution is also a 
Lognormal distribution. However, as noted in Ref. \cite{nukala_cluster}, 
a close examination of the largest crack cluster size data at peak load and the fracture strength 
data reveals that the largest crack cluster size and the fracture strength 
are uncorrelated \cite{nukala_cluster}. Indeed, quite often, the final spanning crack 
is formed not due to the propagation of the largest crack at the peak load, but instead 
due to coalescence of smaller cracks (see Figs. 6 and 7 of Ref. \cite{nukala_cluster}). 

As a further method to assess the presence of correlations, we 
consider the ensemble averaged response of entropy, which has
also been used in experiments (see Fig. \ref{fig:guarino} in Sec. \ref{sec:exp}).
The entropy is defined as 
\begin{equation}
S = -\langle\sum_i q_i ~log(q_i)\rangle,
\end{equation}
where $q_i$ is the fraction of broken bonds within the $i^{th}$ 
window (or section). The lattice system is divided into $i=1,\ldots,m_{box}$ independent measurement windows. 

A useful choice is to take the windows to be of width 
$\Delta y$, along the current flow direction \cite{mikko04}. 
We now present data for the load-displacement response of a typical RFM 
simulation, divided 
into 12 segments, with six equal segments each before and after the
peak load. Let $D_I = 1,2,\ldots,12$ refer to each of these segments. 
The entropy $S$ of the damage cloud is calculated at the end of each of these $D_I$ stages 
using different section widths. 
For each system size $L$, we considered $N_{box}$ number of different section widths, where 
$N_{box} = \frac{log(L/8)}{log(2)} + 1$. Each of the different window sizes 
is denoted by a box size index $B_I = (0,1,\cdots,N_{box}-1)$, 
with the corresponding section widths $B_{size} = 2^{B_I}$. Thus, 
there are $m_{box} = \frac{L}{\Delta y}$ number of sections (boxes) over which 
$\sum_i$ is performed (i.e., $i = 1,\ldots,m_{box}$). 

\begin{figure}[hbtp]
\includegraphics[width=8cm]{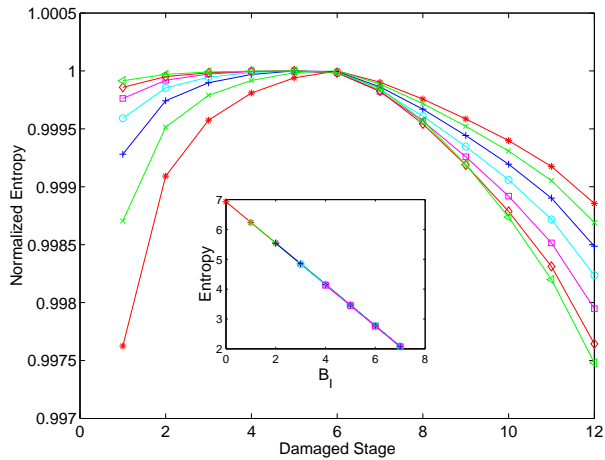}
\caption{Entropy $S$ versus the damaged stage index $D_I$ for a triangular lattice of size $L = 512$ with uniform threshold distribution ($\Delta = 1$) threshold distribution. Symbols
are $B_I = (0,1,\cdots,N_{box}-1) = (*,x,+,\circ,\square,\diamond,\triangleleft)$. 
The inset presents the entropy at the peak load versus the box size index for uniform 
and power law threshold distributions for 
$L = (64,128,256,512,1024)$. The data for smaller systems is shifted such that 
the points correspond to the same relative box size. The slope of the graph in the inset is equal to 
$-log(2) = -0.693$ corresponding to randomly distributed damage.}
\label{fig:entropy}
\end{figure}

Fig.~\ref{fig:entropy} presents the entropy $S$ versus $D_I$,
calculated using different section widths for a uniform threshold distribution 
($\Delta = 1$).  For damage accumulation up to the peak load, 
the entropy increases and attains a maximum 
($S_{max} \approx log(m_{box})$) corresponding to a completely random spatial 
damage distribution, followed at the peak load by a localization phase in 
which the entropy decreases.  In the inset of Fig.~\ref{fig:entropy} 
the entropy at the peak load versus the box size index $B_I$ is plotted 
for uniform and power law threshold distributions,
for system sizes $L = (64,128,256,512,1024)$. 
The slope of the graph in the inset equals $-log(2) = -0.693$, 
which matches exactly with the slope of the maximum entropy $S_{max}$ versus $B_I$ since 
$S_{max} =  log(m_{box}) = log(L/B_{size}) = log(L/2^{B_I}) = 
log(L) - B_I log(2)$.

\subsection{Fracture strength}
\label{sec:fract_str}
\subsubsection{The fracture strength distribution}

Before discussing the results of numerical simulations let us
recall the main theoretical points of the strength of disordered
materials. As discussed in Sec.~\ref{sec:extr_stat}, 
the widely used Weibull and modified Gumbel 
distributions  naturally arise from the extreme-value statistics
if one assumes a particular form of the defect cluster size
distribution in a randomly diluted network: 
(1) defect clusters are independent of each other, i.e., they do not interact with one another;
(2) system failure is governed by the "weakest-link" hypothesis, and 
(3) there exists a critical defect cluster size below which the system does not fail, and 
it is possible to relate the critical size of a defect cluster to the material strength.
In particular, if the defect cluster size distribution
is described by a power-law, then the fracture strength obeys Weibull
distribution, whereas an exponential defect cluster size distribution
leads to the Gumbel distribution for fracture strengths
\cite{duxbury94,duxbury94b,chakrabarti,sahimi98}.
Theoretical arguments along these lines have been put forward in
the context of lattice models, as discussed in Sec.~\ref{sec:perc}.
A different theoretical approach is based fiber bundle models which yields
in the  case of ELS \cite{daniels45}, a Gaussian strength distribution
with variance proportional to $1/N$, where $N$ is the number
of fibers \cite{daniels45}. The more interesting case of LLS
\cite{harlow78}, yields a mean fracture strength $\mu_f \equiv \langle \sigma_f \rangle$
decreasing as $1/\log(N)$, such that the strength of an infinite bundle
system is zero. The fracture strength distribution follows
Eq. (\ref{eq:str_LLS}), and is close to a Weibull form although it is
difficult to determine its exact form
\cite{harlow78}.  Using renormalization techniques, an asymptotic size
effect of the form $\mu_f^{-\delta} \sim \log(\log(L))$ was proposed in
Refs. \cite{smalley,newman91,newman94,duxbury94} for LLS models.

\begin{figure}[t]
\begin{center}
\includegraphics[width=8cm]{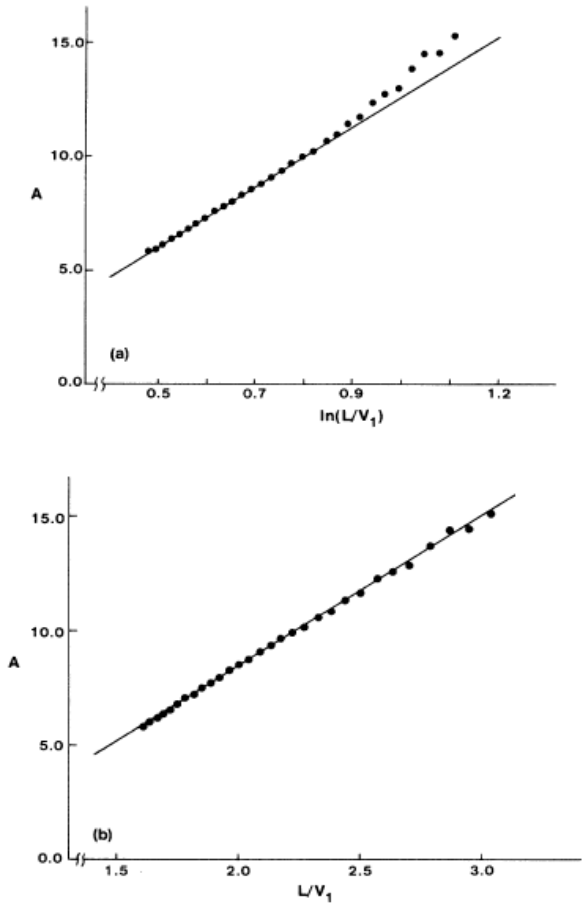}
\end{center}
\caption {Examples of how the Gumbel and Weibull forms
fit into the percolation disorder RFM data (from Ref. \cite{duxbury87}).}
\label{fig:weibull}
\end{figure}

Duxbury et al. \cite{duxbury86,duxbury87,duxbury88} studied the
distribution of fracture thresholds in the randomly diluted fuse
networks with bond percolation disorder (see Sec. \ref{sec:perc}).  
Fig. \ref{fig:weibull} shows simulation results \cite{duxbury87}
compared with the Gumbel-Ansatz. Subsequently, the fracture strength
distribution was studied in a variety of lattice models with different
local behavior including central-force spring models
\cite{beale88,sahimi933}, Born models \cite{hassold89}, and bond-bending models \cite{sahimi933}.  These studies along with the analytical investigations based on largest stable defect size 
\cite{kahng88} concluded that the fracture threshold decreased with the system 
size $L$ as a power of $\log(L)$ and the fracture threshold
distribution is best described by a double exponential (modified
Gumbel) distribution.  In these randomly diluted disorder problems,
the defect cluster size distribution is exponential far away from the
percolation threshold and follows a power law close to the percolation
threshold. Consequently, in general, a Gumbel distribution better fits
the fracture strengths distribution far away from the percolation
threshold and a Weibull distribution provides a better fit close to
the percolation threshold
\cite{duxbury94,duxbury94b,chakrabarti,sahimi98}.

In the case of RFM models with weak disorder, failure is dominated by
the stress concentration effects around the crack tip, and hence the
"weakest-link" hypothesis should be valid. Thus the fracture strength follows
extreme-value theory  and is distributed according to 
the Weibull or modified Gumbel distributions.  The interesting
scenario corresponds to the case of strong disorder, where neither
Weibull nor Gumbel distribution applies well
(see Ref. \cite{nukalaepjb} for 2D RFM, Ref. \cite{nukala05} for 
2D RSM, and Ref. \cite{zapperi_ijf} for 3D RFM). 
There are two main reasons behind this behavior:
\begin{itemize}
\item[(i)] in the "weakest-link" hypothesis, the fracture strength is
determined by the presence of few critical defect clusters, and is
defined as the stress required for breaking the very first
bond in the system. In materials with broad disorder, the
breaking of the very first bond ("weakest-link") does not usually lead
to the entire system failure (as is manifested with the scaling
 $n_p \sim N_{el}^b$ with $b=0.93$ for 2D RFM with strong disorder), 
and hence a fracture strength distribution
based on the importance of very first bond failure may not be applicable.

\item[(ii)] As discussed above in Sec. \ref{sec:corrindamage}, the actual geometric
crack cluster size distribution in the RFM at the peak load develops its
own form, quite different from the initial defect cluster
size distribution in randomly diluted networks. In general at the peak load of RFM simulations, the crack cluster distribution  is represented neither by a power law nor by an 
exponential distribution (see Fig. ~\ref{fig:cluster} and Ref.~\cite{nukala_fuse3d}).
Hence, neither the Weibull nor the Gumbel distributions fit the
fracture strength distribution accurately.
\end{itemize}

\begin{figure}[hbtp]

\includegraphics[width=8cm]{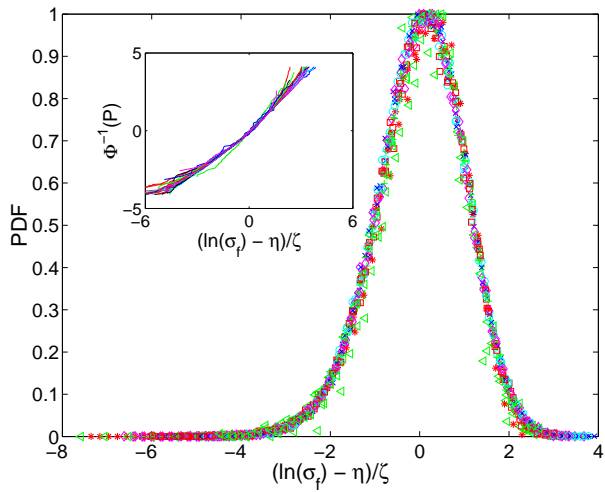}
\caption{Universality of fracture strength distribution in the random thresholds fuse and spring models.  Data are for different lattice system
sizes, $L$, corresponding to triangular fuse lattice with uniform
disorder ($L = \{4,8,16,24,32,64,128\}$), diamond fuse lattice with
uniform disorder ($L = \{4,8,16,24,32,64,128\}$), triangular fuse
lattice with power law disorder ($L = \{4,8,16,24,32,64,128\}$), and
 triangular spring lattice with uniform disorder ($L =
\{8,16,24,32,64,128\}$). The inset reports a Lognormal fit for the 
corresponding cumulative distributions.}
\label{fig:univ_pdf}
\end{figure}

Figure~\ref{fig:univ_pdf} shows the fracture strength density
distributions for random thresholds fuse and spring models using the
standard Lognormal variable, $\bar{\xi}$, defined as $\bar{\xi} =
\frac{Ln(\sigma_f) - \eta}{\zeta}$, where $\sigma_f$ refers to the
fracture strength defined as the peak load divided by the system size
$L$, and $\eta$ and $\zeta$ refer to the mean and the standard
deviation of the logarithm of $\sigma_f$. The excellent collapse of the data for
various fuse and spring lattices clearly indicates the universality of
the fracture strength density distribution. 

In addition, the collapse of the data in Fig. \ref{fig:univ_pdf}
 suggests that $P(\sigma \le \sigma_f) =
\Psi(\bar{\xi})$, where $P(\sigma \le \sigma_f)$ refers to the cumulative
probability of fracture strength $\sigma \le \sigma_f$, $\Psi$ is a
universal function such that $0 \le \Psi \le 1$, and $\bar{\xi} =
\frac{Ln(\sigma_f) - \eta}{\zeta}$ is the standard Lognormal variable.
The inset of Fig. \ref{fig:univ_pdf} presents a Lognormal fit for
fracture strengths, tested by plotting the inverse of the cumulative
probability, $\Phi^{-1}(P(\sigma_f))$, against the standard Lognormal
variable, $\bar{\xi}$. In the above description, $\Phi( \cdot )$ denotes the
standard normal probability function.  As discussed
in Ref. \cite{zapperi_ijf}, a Lognormal distribution is 
an adequate fit for 3D random fuse models as well.
This further confirms the notion of universality of fracture strength
distribution in broadly disordered materials. 

The origin of a Lognormal strength distribution is not entirely
clear from the theoretical point of view 
(for a heuristic argument see Ref. \cite{nukalaepjb}).
It is worth remembering that what determines the survival
of a system is a quantity with strong sample to sample
variations: the microcrack with the largest stress enhancement.
In contrast to percolation disorder, this critical crack grows from
the no-damage state in these simulations.
In Sec.~\ref{sec:corrindamage}, we have discussed also the scaling of the connected
crack clusters. It is worth remembering that the stability
of a given configuration results from two contradictory
effects: stress enhancements and configurations (of fuses
in the RFM) that are able to arrest the crack and make it
stable (as is the case in LLS fiber bundles \cite{mahesh04}).
In experiments on paper, there are results that point out in a similar vein 
the combined role of local strength and local stresses,
which can vary not only because of microcracking but also because
of structural inhomogeneity \cite{hristopulos04}.

\subsubsection{Size effects}
\label{sec:size-effect}

Two forms have been proposed in the literature to
express the size effects on mean fracture strength in
randomly diluted lattice systems (percolation
disorder): a simple power law scaling \cite{chakrabarti},
 consistent with the Weibull statistics,
\begin{equation}
\mu_f \sim L^{-\frac{2}{\bar{m}}},
\label{eq:sigmaf1}
\end{equation}
or a more complicated form \cite{duxbury86,duxbury87,duxbury88,beale88,sahimi933,duxbury94,sahimi98,seppala00},
consistent with a modified Gumbel description, 
\begin{equation}
\mu_f = (\frac{1}{A_1 ~+~ B_1 ~ln ~L})^{1/\delta} = 
(\frac{1}{ln(A_2 L^{B_1})})^{1/\delta},
\label{eq:sigmaf2}
\end{equation}
where $\bar{m}$ is the Weibull exponent, and $\delta$, $A_1=\log A_2$ and $B_1$ are
constants.

\begin{figure}[hbtp]
\includegraphics[width=8cm]{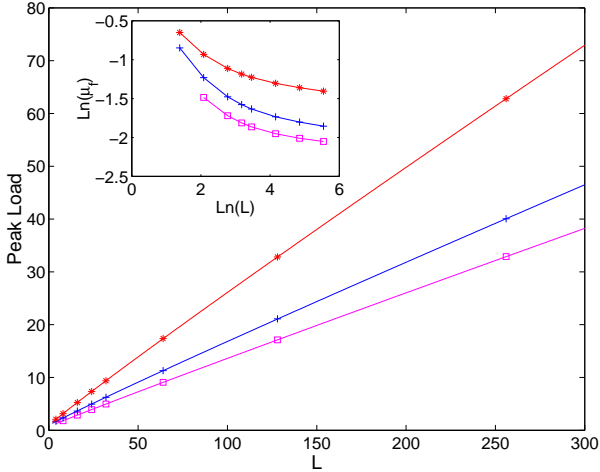}
\caption{Proposed scaling law for the mean fracture strength ($F_{peak} = C_0 L^{\alpha} + C_1$). 
(1) Triangular spring network (symbol: -$\square$-): $\alpha = 0.97$. 
(2) Triangular fuse network (symbol: -*-): $\alpha = 0.956$; 
(3) Diamond fuse network: (symbol: -+-): $\alpha = 0.959$; 
The corresponding Weibull fit for the mean fracture strength is shown in the inset.} 
\label{fig:wei_mean}
\end{figure}

The inset of Fig. \ref{fig:wei_mean} presents the
scaling of mean fracture strength, $\mu_f$, based on power law scaling
$\mu_f \sim L^{-\frac{2}{\bar{m}}}$.  From the nonlinearity of the plot, 
it is clear that the mean fracture strength
does not follow a simple power law scaling. On the other hand, the
scaling form $\mu_f^\delta = \frac{1}{A_1 ~+~ B_1 ~ln ~L}$ may
represent the mean fracture strengths reasonably well even though the
Gumbel distribution is inadequate to represent the cumulative fracture
strength distribution. However, the fit results in coefficients that are extremely 
sensitive; namely, $\delta =-2.45$, $A_1 \approx -18.3$ ($A_2 \approx 10^{-9}$) 
and $B_1 \approx 20$ (see for
instance, Fig. 5b of Ref. \cite{nukalaepjb} for 2D RFM data)
consistent with the divergent behavior reported in
Refs. \cite{duxbury88,beale88} for randomly diluted systems. 

Alternatively, Ref. \cite{nukalaepjb} suggests a scaling form for the
peak load (peak current), $\bar{I}_{peak}$, given by ${\bar{I}}_{peak}
= C_0 L^{\alpha} + C_1$, where $C_0$ and $C_1$ are constants.
Correspondingly, the mean fracture strength defined as $\mu_f =
\frac{{\bar{I}}_{peak}}{L^{d-1}}$, is given by $\mu_f = C_0~L^{\alpha
- d + 1} + \frac{C_1}{L^{d-1}}$, where $d = 2$ in 2D and $d = 3$ in
3D. Using this scaling law, the two-dimensional RFM and RSM fracture
strength data  are plotted
as shown in Fig. \ref{fig:wei_mean}. The results presented in
Fig. \ref{fig:wei_mean} indicate that the exponent $\alpha = 0.97$ for
RSM and $\alpha = 0.96$ for RFM using both triangular and diamond
lattice topologies. 

The results for three-dimensional random fuse models are similar.  A
power law scaling form $\mu_f \sim L^{-\frac{2}{\bar{m}}}$ consistent with a
Weibull distribution does not represent an adequate fit 
(see Ref. \cite{zapperi_ijf}). On the other hand, a scaling of the
form $\mu_f^\delta = \frac{1}{A_1 ~+~ B_1 ~ln ~L}$
results in a very high value of
$\delta = -7.5$ for the exponent and the high values for the
coefficients $A_1 = -156380$ and $B_1 = 85050$  
indicating the divergence behavior
similar to that observed in 2D RFM and in randomly diluted systems
\cite{duxbury88,beale88}.  Alternatively, using the scaling form
${\bar{I}}_{peak} = C_0 L^{\alpha} + C_1$, the mean fracture strength
data for 3D RFM  results in a value of
$\alpha=1.95$ \cite{zapperi_ijf}.

Since a very small negative exponent $(\alpha - d + 1)$ is equivalent
to a logarithmic correction, i.e., for $(d-1-\alpha) << 1$, $L^{\alpha
- d + 1} \sim (ln ~L)^{-\psi}$, an alternative expression for the
mean fracture strength may be obtained as $\mu_f =
\frac{\mu^\star}{(ln ~L)^\psi} + \frac{c}{L}$, where $\mu^\star$
and $c$ are constants that are related to the constants $C_0$ and
$C_1$.  This suggests that the mean fracture strength of the lattice
system decreases very slowly with increasing lattice system size, and
scales as $\mu_f \approx \frac{1}{(ln ~L)^\psi}$, with $\psi \approx
0.15$, for very large lattice systems. The same asymptotic scaling 
relation (i.e., $\mu_f \approx \frac{1}{(ln ~L)^\psi}$) 
was also conjectured in Ref. \cite{sahimi933,sahimi98} for broadly disordered 
materials. 

\subsubsection{Strength of notched specimens}
\label{sec:notch}
Above, we have investigated the fracture of unnotched specimens, but
from an engineering point of view, scaling of fracture 
strength in pre-notched specimens is of significant interest 
\cite{bazant99,bazant02,bazant04,bazant04b,bazant05}. The size effect 
in this case is often given by a scaling of the form 
$\mu_f \propto a^{-1/2}$, where $a = a_0 + c_f$ is the effective crack size, 
$a_0$ is the initial crack (notch) size, and $c_f$ is the FPZ size surrounding the crack tips
\cite{bazant99,bazant02,bazant04,bazant04b,bazant05}. As we have noted in the 
previous section, a logarithmic size effect is observed in 2D and 3D simulations using the random 
thresholds models (RFM and RSM) with both uniform ($\Delta = 1$) and power law ($\Delta = 20$) 
thresholds disorder. The difference between $a^{-1/2}$ size effect in the engineering literature 
\cite{bazant99,bazant02,bazant04,bazant04b,bazant05} and the typical logarithmic 
size effect in the statistical physics literature is due to two aspects: i) initial relative 
crack size $a_0/L$, and ii) disorder.

\begin{figure}[hbtp]
\includegraphics[width=10cm]{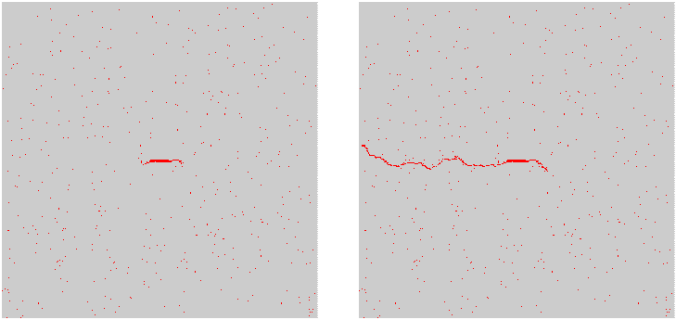}
\caption{Evolution of damage in a notched sample. Although diffuse damage is present,
a wandering fracture line starts from the notch breaking the entire lattice.}
\label{fig:notch}
\end{figure}

A simulation of an initially cracked (or notched) lattice system is reported
in Fig. \ref{fig:notch} for a 2D RFM. It illustrates the possible
role of disorder in determining the ultimate sample strength and
its scaling with $L$ and/or $a_0/L$. The scenario corresponds to some initial distributed
damage followed by the crack roughening, which at the trivial level indicates  
the effect of crack arrest due to strong bonds, followed finally by the 
point of catastrophic failure.

\begin{figure}[hbtp]
\includegraphics[width=10cm]{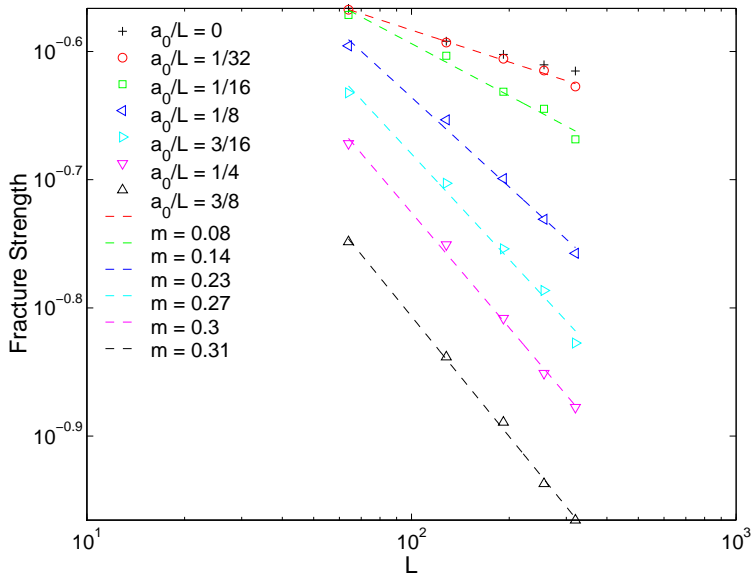}
\caption{Size effects in notched specimens with uniform thresholds disorder. 
The effective scaling exponent of mean fracture strength depends on the ratio $a_0/L$.}
\label{fig:FvsL}
\end{figure}

The size effects obtained in such notched specimens using the 2D RFM 
with uniform thresholds disorder are presented in Fig. \ref{fig:FvsL}
showing the scaling of fracture strength with system size for a fixed 
initial crack size, $a_0/L$. The results can be fitted by a power law of
the type 
\begin{equation}
\mu_f\propto L^{-m}, 
\end{equation}
where the scaling exponent $m$ is significantly influenced by $a_0/L$: for small $a_0/L$ values, 
$m$ is very small, and is equivalent to a logarithmic correction as in Sec.~\ref{sec:size-effect} while 
for large $a_0/L$ values, $m$ approaches $1/2$ as predicted by LEFM \cite{bazant99}. The reason for
this behavior is that in the small $a_0/L$ regime, fracture is dominated by disorder, whereas in the large $a_0/L$ regime, fracture  is controlled by the initial crack. 

A weak disorder leads to the standard $m=1/2$ scaling, while for
strong disorder we have a logarithmic size effect.
One way to understand the data - qualitatively - is to take note of the
fact that for sure a typical sample can accumulate damage upto a number
of broken bonds $n_p$ for which the notch-induced crack is still stable.
E.g. for a crack of size ``$a_0=1$'' this implies, that $n_p$ can be 
interpreted roughly as a density of broken fuses, for which the integral
of the failure threshold distribution corresponds to the typical strength
of the system if the bulk damage is excluded. This, in turn, is given
by $\pi/4$ times the mean of the threshold distributions times a
numerical $O(1)$ factor, coming from the number
of fuses most affected by the current enhancement (at most two).
The damage will then produce a number of microcracks competing with
the flaw, which is of size $a_0$. Increasing the control parameter $a_0/L$ 
implies, 
that the damage $n_p$  will decrease since the current enhancement increases. 
Eventually, the notch will win over the other flaws, but this is a slow
process.

\subsection{Crack roughness}
\label{sec:rough}
Lattice models for fracture have been used in the past to analyze the roughness of the crack surfaces in various 
geometries. The RFM has been numerically simulated in two \cite{hansen91b,seppala00,raisanen98,zapperi05} 
and three dimensions \cite{batrouni98,raisanen98b} using various types of disorder. Other lattice models 
have also been simulated for this purpose, including the beam model \cite{skjetne01}, 
the RSM and the Born model \cite{caldarelli,parisi00}. Interfacial (planar) cracks have been 
studied with the RFM \cite{zapperi00}, the beam model \cite{astrom00}, and a long-range 
scalar model \cite{schmittbuhl04}. Due to numerical limitations, earlier numerical simulations 
mostly considered global measurements for the crack roughness, while more recent results 
explored other quantities and the possibility of anomalous scaling.

Let us first discuss the widely studied RFM \cite{hansen91b,seppala00,raisanen98,zapperi05,batrouni98,raisanen98b}. As discussed in Sec.~\ref{sec:rfm_plastic} the plastic version of the RFM yields on minimum energy surfaces, which in two dimensions corresponds to the directed polymer, characterized by $\zeta=2/3$. The results of numerical simulations
of the brittle RFM yield values close to $\zeta \simeq 0.7$ for a variety of disorder distributions \cite{hansen91b} suggesting that indeed the fracture surface would result from minimum energy considerations \cite{raisanen98}. High quality data for the surface roughness in $d=2$ have been recently published in Ref.~\cite{zapperi05} ruling out this interpretation. The only caveat would be if the effective disorder would become correlated due the microcracking, 
which can change the $\zeta$ \cite{seppala00}.

\begin{figure}[hbtp]
\centerline{\includegraphics[width=10cm]{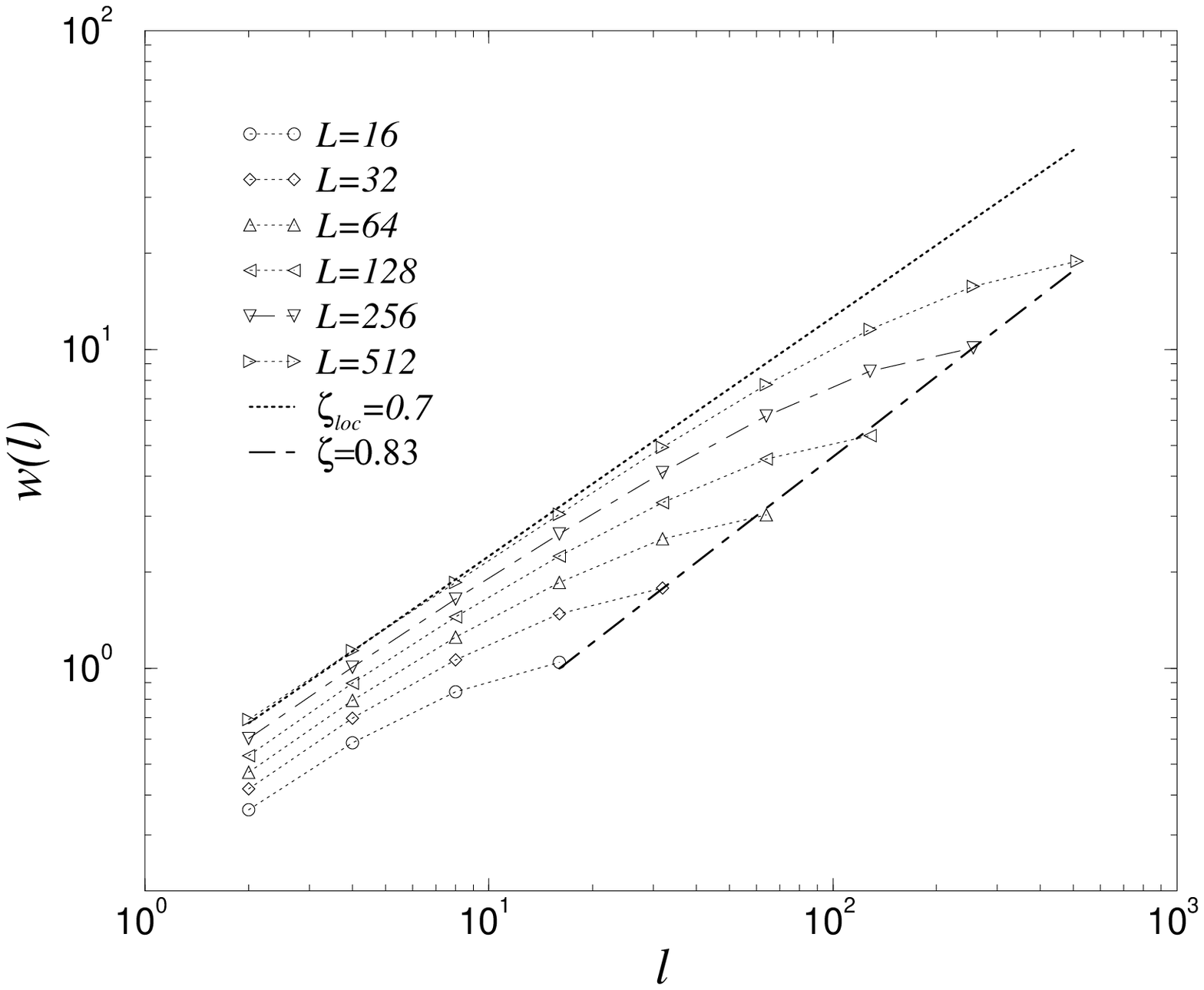}}
\centerline{\includegraphics[width=10cm]{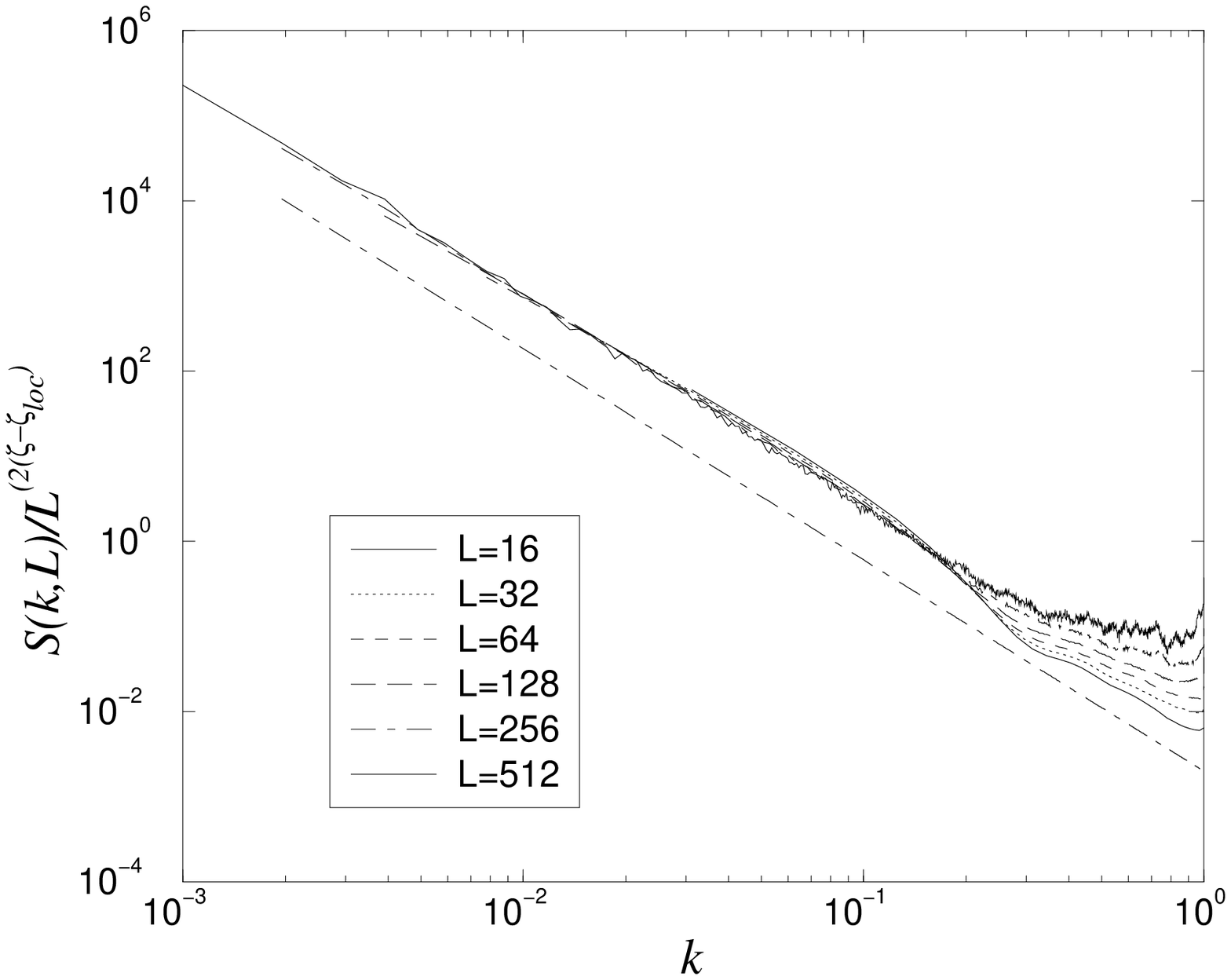}}
\caption{(a) The local width $w(l)$ of the crack surface for different lattice sizes of 2D RFM 
triangular lattice systems plotted in log-log scale. A line with the local exponent $\zeta_{loc}=0.7$ is plotted for reference.
The global width displays an exponent $\zeta>\zeta_{loc}$.(b) The corresponding power spectrum in log-log scale. 
The spectra for all of the different lattice sizes can be collapsed indicating anomalous scaling.}
\label{fig:RFM_2d_rough}
\end{figure}

Several methods have been devised to characterize the roughness of an interface and their reliability has been tested against synthetic
data \cite{schmittbuhl95}. If the interface is self-affine all the methods should yield the same result in the limit of large samples. In Fig.~\ref{fig:RFM_2d_rough}a we report the local width for triangular random fuse lattices for different sizes $L$. The curves for different system sizes do not overlap even for $l \ll L$ 
which would be the scenario 
when anomalous scaling is present (Sec. \ref{sec:selfaff}). 
The global width scales with an exponent $\zeta=0.83 \pm 0.02$ ($\zeta =0.80 \pm 0.02$, 
obtained for diamond lattices \cite{zapperi05}). On the other hand the local width increases 
with a smaller exponent, that can be estimated for the larger system sizes as $\zeta_{loc} \simeq 0.7$. 
A more precise value of the exponents is obtained from the power spectrum, 
which is expected to yield more reliable estimates.  The data reported in Fig.~\ref{fig:RFM_2d_rough}b are 
collapsed using  $\zeta-\zeta_{loc}=0.13$ ($\zeta-\zeta_{loc}=0.1$ for diamond lattices). 
A fit of the power law decay of the spectrum yields instead $\zeta_{loc}=0.74$ ($\zeta=0.7$ for diamond lattice), 
implying $\zeta=0.87$ ($\zeta=0.8$ for diamond lattice). The results are close to the real 
space estimates and we can attribute the differences to the bias associated to the 
methods employed \cite{schmittbuhl95}.
Although the value of $\zeta-\zeta_{loc}$ is small, it is significantly larger 
than zero so that we would conclude that anomalous scaling is present. This implies
in analogy to other theoretical scenarios that there is an external transverse
lengthscale, increasing as a function of $L$. The question now becomes,
what about the origins of such a scale?

While the local exponent is close to the directed polymer value $\zeta=2/3$, the global value is much higher. It is
also interesting to remark that for directed polymers anomalous scaling is not expected to be present, even with noise which has a more complex character \cite{barabasi95}. The result obtained in other lattice models are partly similar, posing the
question whether there is a broad universality class for fracture 
surfaces of disordered lattices. In particular, some measurements reported in the literature are $\zeta=0.7$ for a RFM with dilution disorder \cite{seppala00}, 
$\zeta=0.87$ for the beam model in the limit of strong disorder \cite{skjetne01}, $\zeta_{loc}=0.66$ in the Born model for a variety of parameters \cite{caldarelli}. 

\begin{figure}[hbtp]
\centerline{\includegraphics[width=10cm]{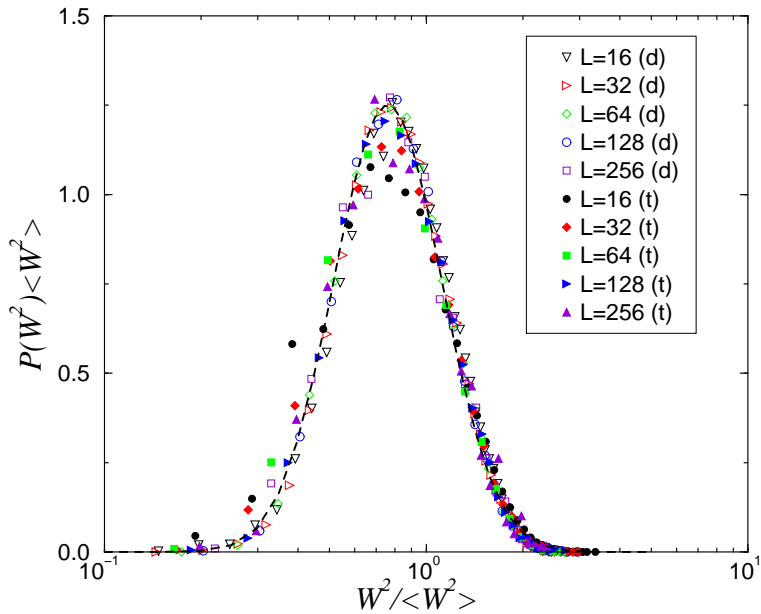}}
\caption{The distribution of the crack width is universal for diamond and triangular
random fuse lattices. A fit with a log normal distribution is shown by a dashed line.}
\label{fig:histW}
\end{figure}

An additional tool to characterize the roughness is provided by the distribution 
$p(W)$ of the crack global width.
This distribution has been measured for various interfaces in models and experiments 
and typically rescales as \cite{foltin94,seppala00,rosso03}
\begin{equation}
p(W)=p(W/\langle W\rangle)/\langle W\rangle,
\label{eq:widthdist}
\end{equation}
where $\langle W \rangle \sim L^{\zeta}$ is the average global width. 
As shown in Fig.~\ref{fig:histW} the
distributions can be collapsed well using Eq.~(\ref{eq:widthdist}) for diamond and triangular lattices \cite{zapperi05},
providing additional evidence for universality. Moreover, the scaling function follows a Lognormal distribution.

\begin{figure}[hbtp]
\centerline{\includegraphics[width=10cm]{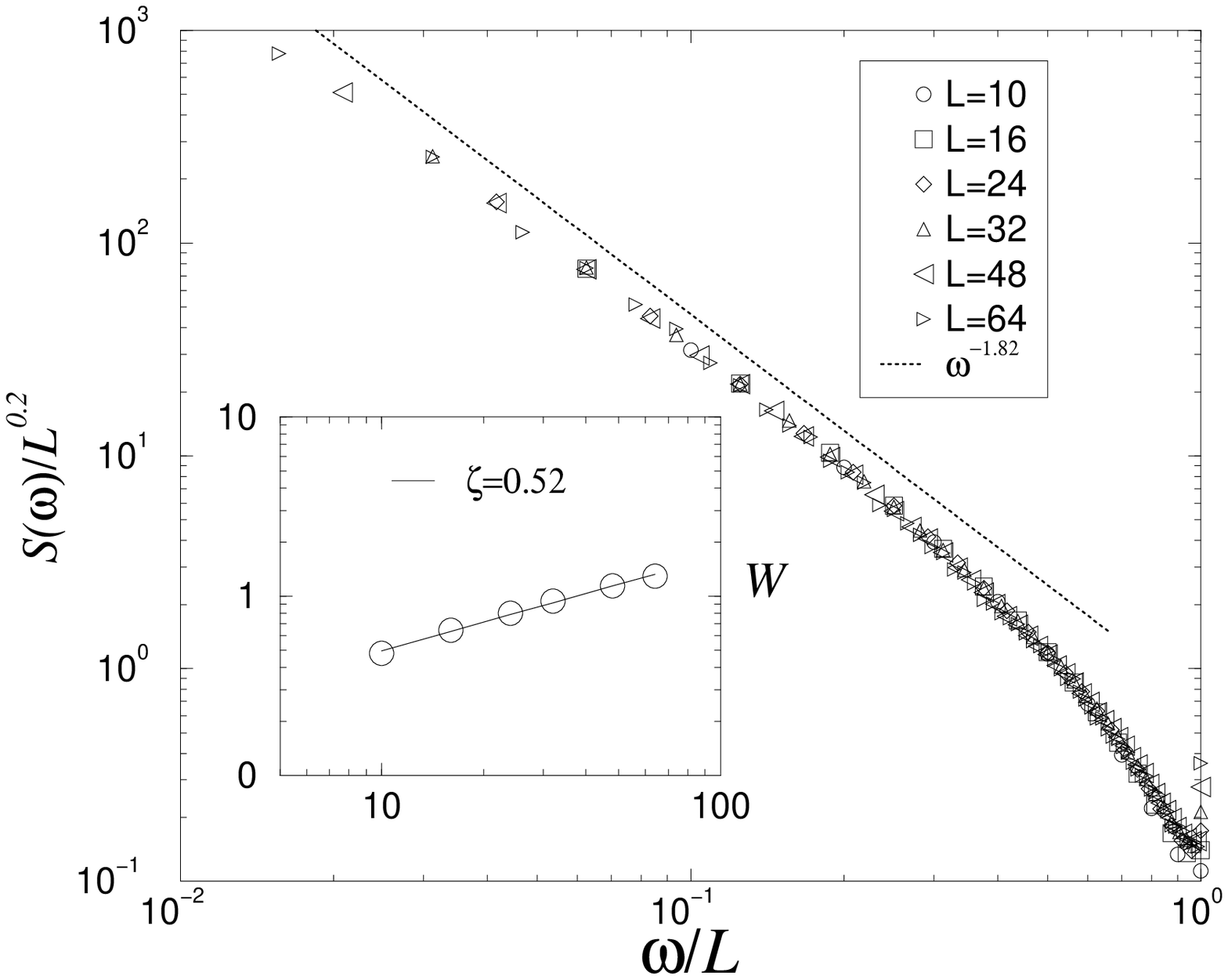}}
\caption{The power spectrum of the crack surface of the three dimensional RFM. The spectrum exponent $2\zeta+1=1.82$, corresponds to $\zeta_{loc}=0.41$. The global width scales with $\zeta=0.52$ (see inset), so that the power spectrum
should be normalized by $2(\zeta-\zeta_{loc})\simeq 0.2$}
\label{fig:rough3d}
\end{figure}

The results of simulations in $d=3$ are more difficult to interpret coherently since the numerical data is limited to 
small system sizes and fewer statistical samples. The higher computational cost 
of 3D simulations compared to 2D simulations restricts the consideration of 3D simulations 
to relatively small system sizes and smaller statistical
sampling. It was debated whether the fracture roughness of the RFM could be described by a minimum energy surface, corresponding to $\zeta\simeq 0.41$. Simulations of the model with weak disorder lead to
a similar value, but the exponent increases for stronger disorder \cite{raisanen98,raisanen98b}. Ref.~\cite{batrouni98}
proposes an extrapolation leading to $\zeta=0.62$, but the numerical estimates fluctuate considerably, depending on the method employed and on the considered disorder distribution. 
Recently performed extensive simulations of 3D RFM have obtained similar results. In particular, an analysis of the 
width and the power spectrum data results in a value of $\zeta\simeq 0.52$, with possible anomalous scaling corresponding 
to a local value of $\zeta_{loc}\simeq 0.42$ (see Fig. \ref{fig:rough3d}).
Finally, results of three dimensional simulations have also been reported for the Born model (on FCC lattices), yielding $\zeta_{loc} \simeq 0.5$ from power spectrum and local width calculations.

An interesting point to note is the consistency between the scaling of 
final crack width and the scaling of localization length $\xi$ discussed in Sec. \ref{sec:localize}. 
For the 2D RFM triangular lattice system, we have $\zeta=0.83 \pm 0.02$, which is 
consistent with the scaling of the localization length $\xi \sim L^{0.8}$ 
used in the data collapse of damage profiles in the post-peak regime. 
The center of mass shifting of the damage profiles for the purpose of averaging 
the damage profiles makes this consistency 
between the final crack width and the localization length $\xi$ of the damage profiles 
self-evident. In addition, this agreement supports the idea that 
the post-peak damage profile is predominantly due to the localization 
produced by the catastrophic failure, which at the same time results in the 
formation of the final crack. Similar behavior is observed in three-dimensional RFM; namely, 
the scaling of final crack width 
$\zeta \simeq 0.52$ consistent with 
the scaling of localization length $\xi \sim L^{0.5}$ \cite{nukala_fuse3d}  
of the post-peak damage profiles. Interestingly, the same consistency in scaling between the  
final crack width and the localization length of the post-peak damage profiles is observed 
in the 2D random spring models as well. The final crack width scales as 
$W \sim L^{0.64}$ (see Ref. \cite{caldarelli}), and the localization 
length $\xi$ of the post-peak damage profiles scales as $\xi \sim L^{0.65}$ 
(see Fig. 6 of Ref. \cite{nukala05}).

\subsection{Avalanches}
\label{sec:rfm_ava}
The loading curve of fracture models, such as the RFM, is characterized by avalanches of failure events. As discussed in Sec.~\ref{sec:fiber}, fiber bundle models represent a useful idealization providing a qualitative description of the avalanche phenomenology and their statistics. On this basis one would expect to find a power law distribution with
a cutoff changing with the applied load. An important theoretical question is whether the scaling exponents measured in lattice models are universal, i.e. they are the same in different models. On the experimental side, 
the main challenge is to relate the avalanche exponents to the acoustic emission statistics recorded in experiments. In both cases, one can ask what is the relation
of the avalanches, and the underlying dynamics, to crack formation and damage
mechanics.
In lattice model simulations one typically defines the avalanche size $s$ as in the FBM (see Sec-~\ref{sec:fb_els}) by 
counting the number of broken bonds that fail without any further increase in the applied load. 
Although this avalanche size is not directly equivalent to the acoustic 
emission energy, under some assumption, it is possible to relate the two distributions as discussed in the following.

The avalanche statistics of the RFM has been reported  for  two dimensional \cite{hansen94,zapperi97,zapperi99,zapperi05} and three dimensional simulations \cite{raisanen98b,zapperi05b}.
While initially it was believed that FBM and RFM were in the same universality class \cite{hansen94,zapperi97,zapperi99} with exponent $\tau=5/2$, recent larger scale simulations
displayed deviations from the expected behavior. The puzzling fact is, however, that the estimated
value of the exponent $\tau$ was found to be dependent on the lattice topology. This is contrary
to any expectation, since in critical phenomena the exponents depend only on symmetries and
conservation law, but not on the type of lattice or similar small scale details. An opposite
situation is found in $d=3$ where early simulations reported a strong deviation from mean-field, $\tau \sim 2 < \tau_{FBM}=5/2$ \cite{raisanen98b}
while more recent results seem in agreement with this value \cite{zapperi05b}. Let us discuss all
these results in detail.

First, one can define avalanches increasing the voltage (deformation) or the
current (force), but we do not expect strong differences for these two cases, so in the following
we always refer to the latter. If one simply records the distribution considering all the avalanches, one 
observes a power law decay culminating with a peak at large avalanche sizes. As in the FBM, the peak is due to the last catastrophic event which can thus be considered as an outlier and analyzed separately, avoiding possible
bias in the exponent estimate  (this problem is present in Ref.~\cite{meacci04} reporting simulations
of a central force model).  When the last avalanche is removed from the distribution the peak disappears
and one is left in these models with a power law (see Ref.~\cite{zapperi05}). On the other hand, the size distribution of the last avalanche is Gaussian \cite{zapperi05,nukala05,caldarelli99}.

\begin{figure}[hbtp]
\begin{center}
\includegraphics[width=10cm]{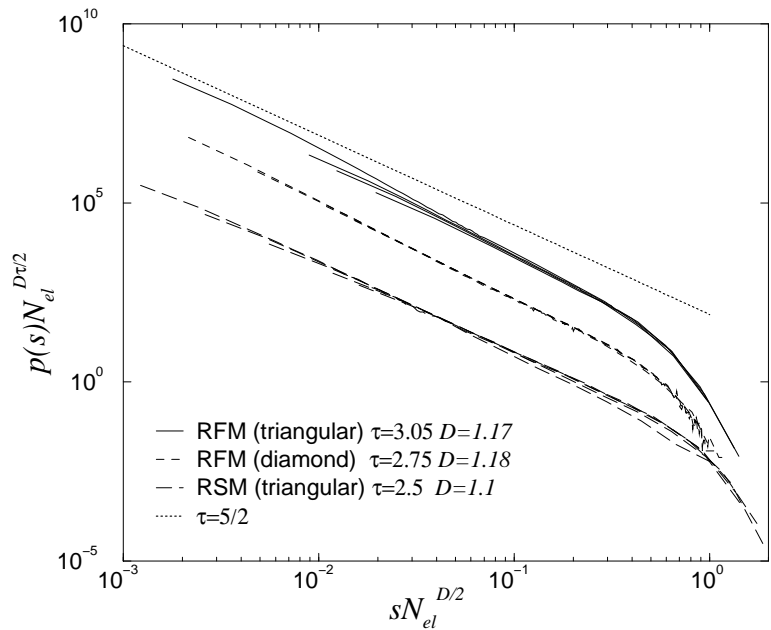}
\end{center} 
 \caption {A summary of two-dimensional results for avalanche distributions integrated
along the entire loading curve. The results are for the RFM on triangular and diamond
lattices \cite{zapperi05} and for the RSM \cite{nukala05}. Data for different 
lattice sizes have been collapsed according to 
Eq.~\protect(\ref{eq:ps_N}). A line with exponent $\tau=5/2$ is reported for reference.}
\label{fig:avalanche2d}
\end{figure}

\begin{figure}[hbtp]
\begin{center}
\includegraphics[width=10cm]{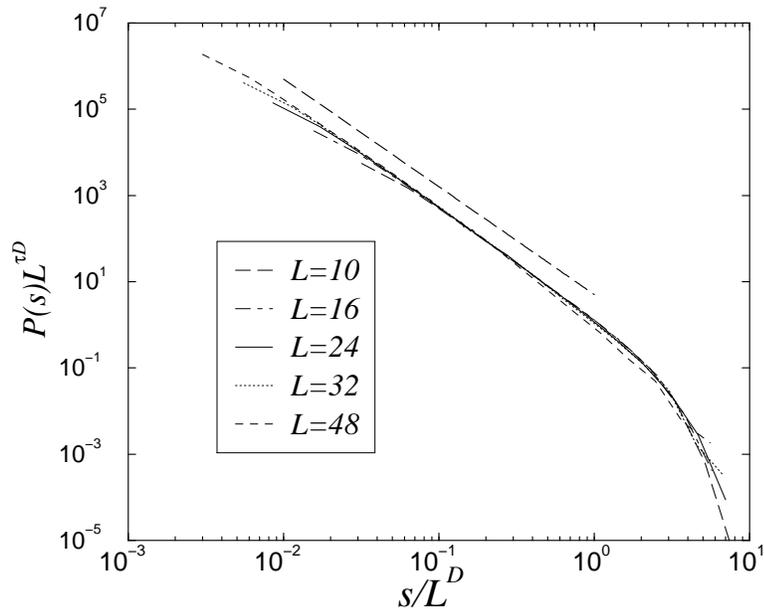}
\end{center}
\caption{Data collapse of the integrated avalanche size distributions for the three
dimensional RFM. The exponents used for
the collapse are $\tau=2.5$ and $D=1.5$. A line with a slope $\tau=5/2$ 
is reported in Ref.~\cite{zapperi05}.}
\label{fig:avalanche3d}
\end{figure}

Next, as discussed in Sec.~\ref{sec:fb_els},
one should distinguish between distributions sampled over  the entire load history, and those recorded
at a given load. In the first case, the integrated distribution has been described by a scaling form \cite{zapperi05}
\begin{equation}
P(s,L)=s^{-\tau} g(s/L^D),
\end{equation}
where $D$ represents the fractal dimension of the avalanches. To take into
account the different lattice geometries, it is convenient to express
2D scaling plots in terms of $N_{el}$ rather than $L$
\begin{equation}
P(s,N_{el})=s^{-\tau} g(s/N_{el}^{D/2}).
\label{eq:ps_N}
\end{equation}
A summary of the results for two-dimensional lattices are reported in Fig.~\ref{fig:avalanche2d}, 
showing that in the RFM  $\tau > 5/2$, with variations depending on the lattice topology, while
the result reported for the RSM is instead very close to $\tau = 5/2$. As for the avalanche
fractal dimension, $D$ varies between $D=1.10$ and $D=1.18$, indicating almost linear crack
(which would correspond to $D=1$). Note that the last avalanche 
has been studied earlier in the context of damage, and its geometrical
properties in the context of crack roughness. The variation of $D$
should be compared to that of $\zeta$ in the same systems.
Three dimensional results are plotted in Fig.~\ref{fig:avalanche3d}
for the RFM \cite{zapperi05b}. Again the agreement with FBM results is quite remarkable.

\begin{figure}[hbtp]
\begin{center}
\includegraphics[width=10cm]{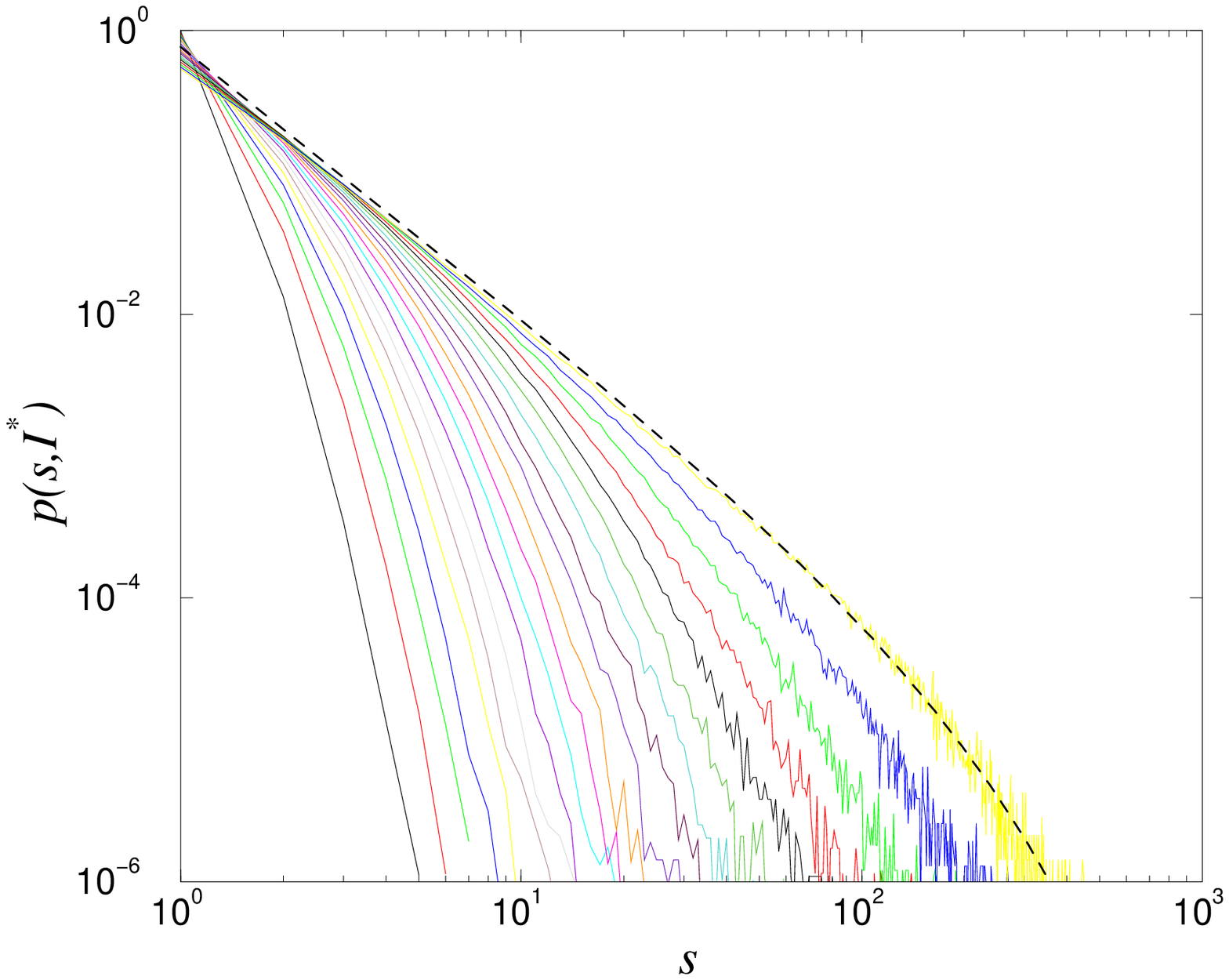}
\end{center}
\caption{The avalanche size distributions sampled over a small bin of the reduced current
$I^*$ for a diamond lattice of size $L=128$.
The dashed line represents a fit according to Eq.~\protect(\ref{eq:binsize}) with $\gamma=1.9$ 
(from Ref.~\cite{zapperi05}).}
\label{fig:binned_dist}
\end{figure}

The distribution of avalanche sizes sampled at different
values of the current $I$  is obtained by normalizing the current by its peak value $I_c$
and dividing the $I^*=I/I_c$ axis into several bins. 
The distribution follows then a law of the type
\begin{equation}
p(s,I^*)= s^{-\gamma}\exp(-s/s^*),\label{eq:binsize},
\end{equation}
where the cutoff $s^*$ depends on $I^*$ and $L$, and can be conjectured to scale as
\begin{equation}
s^* \sim \frac{L^D}{(1-I^*)^{1/\sigma}L^D+C},
\label{eq:fss}
\end{equation}
where $C$ is a constant. Integrating Eq.~(\ref{eq:binsize})
we obtain
\begin{equation}
P(s,L) \sim s^{-(\gamma+\sigma)} \exp[-sC/L^D].
\end{equation}
which implies $\tau=\gamma+\sigma$. We recall here that in the FBM $\gamma=3/2$ and $\sigma=1$
(see Sec.~\ref{sec:fb_els}). Fig.~\ref{fig:binned_dist} reports a series of distributions obtained
for the RFM for different values of the currents \cite{zapperi05}. The best fit for the scaling
of the distribution moments as a function of $I$ and $L$ (Eq.~(\ref{eq:fss})), and a direct fit on the curves
suggest $\gamma \approx 1.9-2$, $\sigma \approx 1.3-1.4$ in $d=2$ and $\gamma \approx 1.5$, $\sigma\approx 1$
in $d=3$.

In summary, the avalanche distribution for RFM is qualitatively close to the prediction
of the FBM (i.e. mean-field theory). Apparently non-universal (lattice type dependent) corrections
are present in $d=2$, but the problem disappears in $d=3$. This could be expected building on
the analogy with critical phenomena, where exponents are expected to approach the mean-field
values as the dimension is increased. Simulations of other models have been reported in the literature:
the distribution obtained in Ref.~\cite{nukala05} for the RSM is very close to the FBM result (see Fig.~\ref{fig:avalanche2d}). The avalanche distribution has been also reported for the Born model 
($\tau\approx 2$ \cite{caldarelli96}), the bond-bending model 
($\tau\approx 2.5$ \cite{zapperi99}), the continuous damage model ($\tau\approx 1.3$ \cite{zapperi97b}).

To conclude, we discuss the issue of the relation between the AE distribution and
and the avalanche size distribution. These two are, contrary to usual
indications in the literature, not the same. The same holds even more
clearly for the acoustic amplitude distribution.
Acoustic waves were incorporated in a scalar model in Ref.~\cite{minozzi03},
and the AE energy distribution was measured, with an exponent $\beta \simeq 1.7$. This exponent can
be related to $\tau$ if we assume that the typical released AE energy of an avalanche of size $s$
scales as $\epsilon \sim s^\delta$, with $\delta=2$ (i.e. the scaling relation is $\tau=1+\delta(\beta-1)$). 
This was verified in the model of Ref.~\cite{minozzi03}, but also in
the RFM defining $\epsilon$ as the dissipated electric energy during one event
seems to agree \cite{salminen02}.

\section{Discussion and outlook}
\label{sec:concl}
We would like to conclude the present review by trying to answer two essential questions: 
(i) what can be learned from statistical models in order to
reach a better understanding of fracture in  
quasi-brittle materials and interpret relevant experiments and (ii) 
what kind of developments 
can be foreseen to improve the current theoretical/modeling framework. 
The first question involves as well an assessment of the possible advantages
of the statistical mechanics strategy with respect to traditional damage mechanics approaches. On the other hand, answering the second question 
will also help to clarify the limitations of statistical lattice models
and their role in the future.

The macroscopic behavior of lattice models is well described
using the terminology of phase transitions and the ensuing
picture is conceptually rather simple: in the stress-controlled
case crossing the peak-load implies a first-order transition
to a ``broken state''. This means that regardless of system
size and disorder strength, the correlations in damage are local albeit with 
a growing characteristic length. This lengthscale does not
diverge however as the peak-load is approached in contrast to that of a 
second-order transition. 
Such a picture has automatic consequences for the scaling of 
thermodynamic quantities as the conductivity or stiffness, damage 
and strength distributions. In particular, scaling laws inspired from critical
phenomena typically fail to describe the size dependence of these quantities.
The statistical properties are instead well captured by relatively simple
laws, such as Gaussian or Lognormal distributions, whose precise origin, 
however, is not always clear and remains a topic for future research.  
There are still interesting issues left, such as the most 
appropriate mathematical framework for measuring the damage anisotropy and 
correlations, or understanding the dynamics of microcrack populations. 
It would be important to  carry out an extensive analysis for vectorial fracture models,
 even though such quasistatic brittle fracture models are expected to have the same kind 
of ``phase diagram'' in the statistical mechanics language. Presumably
allowing for several modes of element failure will also influence
the right description of damage, summarized by the classical
damage mechanics question whether it makes sense to use 
scalar damage parameter $D(\vec{x})$. 

Next we would like to summarize with a few points the
picture obtained from the theoretical
approaches discussed in Sec.~\ref{sec:theory} and the numerical
simulations reported in Sec.~\ref{sec:simul} and compare 
the results with the experimental data analyzed in Sec.~\ref{sec:exp}.
An additional question that naturally follows from statistical models 
is the role of internal disorder for macroscopic damage laws used in 
engineering applications.

\subsection{Strength distribution and size effects}

One of the successes of statistical models of fracture has
been to shed light into the scaling of strength with sample
size and the associated distributions. There are indications that the
asymptotics of extremal statistics are not the only alternative to 
explain the strength distribution, which results instead from 
cooperative phenomena or fluctuations in the damage accumulation.
While for weak disorder the Weibull distribution provides an adequate
fit to the data, for strong disorder we observe a Lognormal 
strength distribution. Whether this is a general result valid at
all scales or the signature of a crossover behavior to Gumbel
or Weibull scalings, still remains to be explored. 

A possible pathway to understand these results could
stem from fiber bundle models (see Sec.~\ref{sec:fiber}). Under 
LLS conditions the model displays an extremal type distribution
and a Logarithmic size effect. Conversely under ELS the distribution
becomes Gaussian and size effects are absent. A clever way to interpolate
between LLS and ELS was suggested by Hidalgo et al. \cite{hidalgo02b} 
through a long-range rule in which the stress is transfered as 
$1/r^\gamma$. The crossover between LLS and ELS is identified
with $\gamma \simeq d$, which corresponds with the exponent of the
Green function obtained from lattice models, such
as the RFM. Hence these models lie exactly at the boundary 
between LLS and ELS and the Lognormal distribution could then 
arise from the competition between local and global effects in
a strongly disordered environment. Here, we would like to call
for further basic science -oriented experiments. In the materials
science -community it is customary to use the Weibull scaling to
describe strength. However, it is clear that it is very rare to
have a situation in which one would a priori expect that the conditions
underlying the Weibull statistics are valid.

Lattice models are also useful to analyze the effect on strength coming 
from the interplay between elastic interactions and disorder in notched samples. 
For instance, the data displayed in Sec. \ref{sec:notch} show
the various mechanisms by which disorder can mask the classical LEFM prediction 
involving a $1/\sqrt{a_0}$-type scaling. The RFM has typically a rather 
small fracture process zone which decreases in size, 
for a fixed notch-to-sample size -ratio $a_0/L$. The simple lesson we can draw
from the model is that in practice one may, for a limited range of sample sizes,
expect to get almost any strength exponent (in the law $\sigma_f \propto L^{-m}$ )
between $m=0$ (or logarithmic behavior) and the $m=1/2$ prediction indicated by LEFM. 

It is interesting to close the present discussion by analyzing the predictions for
strength coming from two other theoretical frameworks commonly 
used to interpret statistical properties of fracture, namely percolation
(Sec.~\ref{sec:perc}) and depinning (Sec.~\ref{sec:depinning}).
Percolation describes rigorously the infinite disorder
limit of lattice models, where stress enhancement is irrelevant,
the bonds fail one by one, and no real size effect is expected
besides the small corrections associated with finite size scaling. 
The infinite disorder limit is useful as a theoretical
idealization but is not relevant in practical cases, where the
disorder distribution might be broad but remains normalizable. 
In that case, only the initial stage of damage accumulation resembles
percolation, but eventually a catastrophic process is initiated
and size effects, both in damage and strength are observed in agreement
with experiments. 

The depinning approach is applicable instead to the weak disorder limit, single crack limit. Strong inhomogeneities would break up the single valued crack interface that is the starting assumption of the theory. According to the depinning scenario, the crack interface moves, 
eventually leading to the failure of the sample, when the stress intensity
factor on the crack overcomes a critical threshold $K_c$. Since we deal
here with a non-equilibrium phase transition, as in percolation, 
we do not expect to see size effects in $K_c$, besides the  
usual finite size corrections (i.e. $K_c>0$ for $L\to\infty$). 
Other possible size effects are hidden in the definition of the stress intensity factor. In other words,
a depinning transition alone can not account for the observed size effects
but can at best be used to describe the crack dynamics and morphology 
once crack growth has been initiated.

\subsection{Morphology of the fracture surface: roughness exponents}

We have discussed both the experimental (in Sec.~\ref{sec:rough-exp}), theoretical
(in Sec.~\ref{sec:depinning}) and simulation-based (in Sec.~\ref{sec:rough}) 
crack roughness analysis. As we discussed extensively there are several 
possible definitions of the roughness exponent, depending on the experimental
conditions (dimensionality of the sample, loading mode) 
and on the measured surface properties 
(in plane or out of plane, parallel or perpendicular, local or global roughness). 
This variety of exponents makes a direct comparison between theory and experiments
quite involved. In order to simplify this task we report in Table~\ref{table:exp_rough}
the values of the roughness exponents measured in experiments, as well as the
predictions of the models.

\begin{table}
\begin{tabular}{|l|c|c|c|c|}
\hline
 exponent & exp. & RFM & DT-LR & DT-SR \\ \hline
$\zeta$ 2D local (i),& 0.6-0.7 & 0.7 $\pm$ 0.1 &  - &  - \\ \hline
$\zeta$ 2D global (i) & - & 0.8 $\pm$ 0.1 & - & - \\ \hline
$\zeta$ 3D planar crack (ii)& 0.6 & fractal & 0.39 &  0.63 \\ \hline
$\zeta_\perp$ 3D local (iii) & 0.4-0.5 (ss) 0.8 (ls) & 0.4 & log & - \\ \hline 
$\zeta_{\parallel}$ 3D local (iii) &  0.6 & 0.4 & - & - \\ \hline
$\zeta_{\perp}$ 3D global  (iii) & 1.2-1.3 & 0.5 & - & - \\ \hline
\end{tabular}

\caption{A summary of the values of the roughness exponent measured in experiments, in the RFM and in the depinning transition (DT) scenario,
both with long-range (LR) and (SR) interactions. Refer
to Sec.~\protect\ref{sec:selfaff} for the definition of the exponents, to Sec.~\protect\ref{sec:rough-exp} for details about experiments,
to Sec.~\protect\ref{sec:rough} for the RFM results and to
Sec.~\protect\ref{sec:depinning} for the depinning transition.
}
\label{table:exp_rough}
\end{table}

A direct comparison of the exponent values leads to two main conclusions: 
first, the exponents agree ``roughly'' in two dimensions 
(i.e. $\zeta\simeq 0.7$) - depending on the case considered - but in
three dimensions the numerical value is significantly smaller than the
experimental one, $\zeta_{\rm loc} \simeq 0.4$ and $\zeta \simeq 0.5$ in  
simulations while  $\zeta_{\rm loc} \simeq 0.8$  and $\zeta \simeq 1.2$
in experiments. This discrepancy is quite striking given the large range 
over which $\zeta_{\rm loc}$ can be observed in experiments.
Second, The {\em physical mechanism} to crack roughening is still not
understood even in the case of the RFM, not to mention other more realistic 
models and of course experiments. 

From the experimental point of view,
presently it is not completely clear if the 3D fracture surface 
exponents are universal, and what kind of regimes can be found. 
The crucial lengthscales seem to be that of FPZ \cite{ponson06,bonamy06}, 
$\xi_{FPZ}$, and in some cases one related to
voids formed ahead of the crack tip \cite{celarie03}: their density, and typical
size. From the model point of view, it is crucial to underline here the main implication 
of the observed catastrophic failure occurring after peak-load and the lack of localization 
before this point: the roughness of the crack arises in the last abrupt ``avalanche''.
This has important consequences for the modeling strategies, since quasi-static
models such as the ones we are discussing here may become inappropriate when
the crack grows unstable. It would be necessary to incorporate a more realistic
dynamics, possibly including sound waves or other dynamical effects, or some
process leading to residual strength.

A further relevant theoretical point is the question of the universality of the
roughness exponent, an hypothesis supported by ample experimental evidence.
Early RFM simulations suggest that $\zeta$ does not depend on the disorder 
strength \cite{hansen91b}, while recent results indicate as well an independence
of the lattice type \cite{zapperi05}. It would be very interesting to consider
in more detail comparisons to/within truly vectorial (beam or bond-bending)
models to see their roughness characteristics. This also holds
for crack growth experiments from notches, in which the interaction
of dominating crack with diffuse damage could be investigated.
An important aspect stems  from the connection
between fracture energy (or toughness) and the crack shape. 
It should be mentioned that the amplitude $A$ of the 
roughness (as in $w \propto A L^\zeta$) usually increases as
the prefactor of strength decreases if the disorder is varied.
Thus we may conclude that ``rough'' does not necessarily imply ``tough'', to answer 
the original question posed 20 years ago in Ref.~\cite{mandelbrot84}.

\subsection{Crack dynamics: avalanches and acoustic emission}

The statistical properties of acoustic emission also remain to be explained,
 completely.
Quasistatic lattice models qualitatively predict the correct behavior:
a power law distribution of avalanche events, with an activity increasing as
the failure point is approached. The scaling behavior is well captured by mean-field
theory, as exemplified by fiber bundle models, with some deviations from the
analytical predictions in the two-dimensional RFM (where $\tau>5/2$), but almost
perfect agreement in $d=3$ and for the RSM. The apparent success of mean-field theory is
probably related to the occurrence of a first-order transition in presence of
long-range elastic interactions. In these conditions, the spinodal instability with
the related mean-field scaling is accessible even in low dimensionality.
From the point of view of the experimental comparison, however, lattice models typically
predict a too large energy exponent $\beta$. It is not known as to whether this problem results
from any of the known possible caveats, such the brittle nature of the models,
the dispersion of the acoustic waves and data in experiments, or the 
possibility that in reality one can not separate the time-scales
of stress relaxation and loading, as assumed in the quasi-static models.
For this reason we would like to call for more experiments
in various loading conditions (tensile, creep, fatigue etc.), for
varying materials so that the influence of plasticity, 
time-dependent response, and localization can be studied. 

Brittle fracture models are due to the first-order nature
of the final failure such that their predictability is 
inherently weak, though there have been (mainly experimental) attempts
to show the opposite \cite{johansen00,garcimartin97,andersen05}.
For the practically minded scientist, the main lessons are that
brittle materials have a tendency to fail abruptly,  exhibit critical-looking noise, 
and their fracture statistics depend crucially on the presence of large flaws and 
on the actual disorder. 

\subsection{From discrete models to damage mechanics}

Lattice models have been used to describe qualitatively the stress-strain
curves of materials as noted earlier. Here, we have overviewed, mainly in
the context of the RFM,  the basic tenants of damage 
accumulation. The crucial questions are: how to average over samples,
what kind of correlations arise in the damage, including localization,
and in general what one learns from a representative volume element
-idea.

Large-scale simulations provide evidence for strong self-averaging
for the $D$, damage variable, in the RFM. This is connected with the
notion that while the damage accumulated becomes anisotropic, it still
has only a finite correlation length $\xi$ (both parallel
and perpendicular to the load) up to the peak load. Meanwhile, as one takes the
limit $L\rightarrow \infty$ the strength decreases and actually the
self-averaging  gets stronger since $L/\xi$ remains
large even up to the peak load. Again, it would be useful to test
these conclusions in vectorial models. One question not addressed
here in detail is the probability distribution of $P(D,V)$, which is related
to $\xi(V)$ via a simple application of the central limit
theorem. Again, we would like to call for experiments, including
looking at damage patterning and localization to also stimulate the
further development of models.

\subsection{Concluding remarks and perspectives}

We think that there are substantial advances to be made
within the confines of lattice fracture models with inelastic,
or time-dependent effects incorporated. Even simulations of
elastic-plastic fracture in this respect are still awaiting,
though these have no rate-dependence.
As was very briefly mentioned in Sec. \ref{sec:timedep}, there
are statistical laws in fracture and deformation processes that
are not simply ``brittle''. For many of these cases we expect
that models akin to the RFM and its generalizations will
prove to be useful. 

There are some simple ingredients that ``improved'' models
might contain and some simple developments they might lead to.
For one, we would like to see the invention of lattice models
that allow for true damage localization before the peak load.
This would for instance produce damage dynamics comparable to the 
experimental ``b-value'' analysis of AE, to name but one issue. Similarly,
one would need models that allow for stable crack growth 
beyond the peak load ($V$-controlled ensemble for equivalent
RFM's). 

In all, theoretically it is also an interesting 
question how the ``first-order'' picture of fracture changes
in the presence of other screening mechanisms for the long-range
interactions. The implications of statistical fracture on other
lengthscales, as in multi-scale modeling, are then dependent on
what is the appropriate coarse-grained description. The resulting
picture of the emerging physics should be connected to what we
have presented in this work.

{\bf Acknowledgments:}\\
MJA would like to thank the Centre of Excellence program of the
Academy of Finland  and TEKES (The 
National Technology Agency of Finland) for financial support. 
A number of students, colleagues, and partners in research
(K. Niskanen, V. R\"ais\"anen, L. Salminen, E. Sepp\"al\"a,  J. Rosti, J. Lohi)
have contributed to the related research.
PKVVN is sponsored by the
Mathematical, Information and Computational Sciences Division, Office
of Advanced Scientific Computing Research, U.S. Department of Energy
under contract number DE-AC05-00OR22725 with UT-Battelle, LLC. In addition, 
PKVVN acknowledges the discussions with and support received from Dr. Srdjan 
Simunovic. SZ whishes to thank A. Baldassarri, D. Bonamy, A. Hansen, H. J. Herrmann, 
F. Kun, M. Minozzi, S. Roux, A. Vespignani and M. Zaiser for  
collaborations, discussions and remarks on the topics discussed in this review.

\appendix
\section{Algorithms}
As mentioned earlier, fracture of disordered
quasi-brittle materials is often studied using discrete lattice networks, 
wherein the damage is accumulated progressively by breaking 
one bond at a time until the lattice system falls apart. 
Each time a bond is broken, it is necessary to redistribute the 
stresses to determine the subsequent bond failure. 
Algebraically, this process 
of simulating fracture by breaking one bond at a time is 
equivalent to solving a new set of linear equations (Kirchhoff equations in the 
case of fuse models)
\begin{equation}
{\bf A}_n {\bf x}_n = {\bf b}_n, ~~~~~ n = 0,1,2,\ldots, \label{intro1}
\end{equation}
every time a new lattice bond is broken. In Eq. (\ref{intro1}), 
each matrix ${\bf A}_n$   
is an $N \times N$ symmetric and positive definite matrix (the lattice conductance matrix in the case of 
fuse models and the lattice stiffness matrix in the case of spring and beam models), 
${\bf b}_n$ is the $N \times 1$ (given) applied nodal current or force vector, ${\bf x}_n$ is the 
$N \times 1$ (unknown) nodal potential or displacement vector, and $N$ is the number of degrees of freedom 
(unknowns) in the lattice. The subscript $n$ in Eq. (\ref{intro1}) indicates that 
${\bf A}_n$ and ${\bf b}_n$ are evaluated after the $n^{th}$ bond is broken. The solution ${\bf x}_n$, obtained 
after the $n^{th}$ bond is broken, is used in determining the subsequent ($(n+1)^{th}$) bond 
to be broken. 

Numerical simulations of fracture using large discrete lattice networks are often
hampered due to the high computational cost associated with solving a
new large set of linear equations every time a new lattice bond is
broken. This becomes especially severe with increasing lattice system size, $L$, as 
the number of broken bonds at failure, $n_f$, follows a power-law distribution 
given by $n_f \sim O(L^{1.8})$ in 2D and $n_f \sim O(L^{2.7})$ in 3D. In addition, since the response of the lattice
system corresponds to a specific realization of the random breaking
thresholds, an ensemble averaging of numerical results over
$N_{config}$ configurations is necessary to obtain a realistic
representation of the lattice system response. This further increases
the computational time required to perform simulations on large
lattice systems.

\subsection{Fourier Acceleration Method}
Traditionally, iterative techniques based on preconditioned conjugate gradient (PCG) 
method have been used to simulate fracture using fuse networks (see Ref. \cite{templates} 
for an excellent review of iterative methods; see Ref. \cite{multigrid} for a 
review of multigrid method). However, 
large-scale numerical simulations using iterative solvers have often been hindered due 
to the {\it critical slowing down} associated with the iterative solvers 
as the lattice system approaches macroscopic fracture. That is, 
as the lattice system gets closer to macroscopic fracture, the condition
number of the system of linear equations increases, thereby increasing
the number of iterations required to attain a fixed accuracy.  This
becomes particularly significant for large lattices. As a remedy, 
Fourier accelerated PCG iterative
solvers \cite{batrouni86,batrouni88,batrouni98} have been suggested 
to alleviate the critical slowing down. The Fourier acceleration algorithm 
proposed in Refs. \cite{batrouni86,batrouni88} for solving the 
system of equations ${\bf A} {\bf x} = {\bf b}$ can be expressed as shown 
in Algorithm 1. In this subsection, we do not explicitly write the 
subscript $n$ to refer to the system of equations ${\bf A}_n {\bf x}_n = {\bf b}_n$ obtained after 
the failure of $n$ bonds. Instead, the subscript $n$ is implicitly understood. 

\begin{algorithm}
%\caption{{\it Algorithm 1} Fourier Acceleration Method \cite{batrouni86,batrouni88}}
\begin{algorithmic}[1]
\STATE Compute ${\bf r}^{(0)} = {\bf b} - {\bf A} {\bf x}^{(0)}$ for some initial guess ${\bf x}^{(0)}$
\FOR{$k = 1,2,\ldots$}
	\STATE Solve: ${\bf M} {\bf z}^{(k-1)} = {\bf r}^{(k-1)}$ 
	\STATE ${\bf x}^{(k)} = {\bf x}^{(k-1)} + {\bf z}^{(k-1)}$
	\STATE ${\bf r}^{(k)} = {\bf r}^{(k-1)} - {\bf A} {\bf z}^{(k-1)}$
	\STATE Check convergence; continue if necessary
\ENDFOR

\end{algorithmic}

\end{algorithm}

In Algorithm 1, the matrix ${\bf M}$ is referred to as the preconditioner, and the 
superscript in the parenthesis refers to the iteration number. The steps in the Algorithm 1 can 
be combined into an iterative scheme as shown below
\begin{eqnarray}
{\bf x}^{(k+1)} & = & {\bf x}^{(k)} + {\bf M}^{-1} {\bf r}^{(k)} \label{FFT}
\end{eqnarray}
where ${\bf r}^{(k)} = {\bf b} - {\bf A} {\bf x}^{(k)}$. Denoting the error ${\bf e}$ (or {\it algebraic error})
as ${\bf e} = {\bf x} - \bar{\bf x}$, where $\bar{\bf x}$ is the exact solution such that 
${\bf A} \bar{\bf x} = {\bf b}$, the errors in the $k^{th}$ and $(k+1)^{th}$ iterations 
may be related as
\begin{eqnarray}
{\bf e}^{(k+1)} & = & ({\bf I} - {\bf M}^{-1} {\bf A}) {\bf e}^{(k)} \label{error}
\end{eqnarray}
Consequently, after $k$ number of relaxation sweeps, the error in the $k^{th}$ approximation 
is given by ${\bf e}^{(k)} = {\bf Q}^k {\bf e}^{(0)}$, where ${\bf Q} = ({\bf I} - {\bf M}^{-1} {\bf A})$. 
Choosing a particular vector norm and its associated matrix norm, it is possible to bound the error 
after $k$ iterations by $\|{\bf e}^{(k)}\| \le \|{\bf Q}\|^{k} \|{\bf e}^{(0)}\|$. From this relation, it can be 
shown that the iteration associated with the matrix ${\bf Q}$ converges for all initial guesses if and only if 
$\rho({\bf Q}) < 1$, where $\rho({\bf Q})$ is the spectral radius defined as 
\begin{eqnarray}
\rho({\bf Q}) & = & \mbox{max}_{j} |\lambda_j({\bf Q})|
\end{eqnarray}
and $\lambda_j({\bf Q})$ denotes the $j^{th}$ eigenvalue of ${\bf Q}$ given by 
$\lambda_j({\bf Q}) = 1 - \lambda_j({\bf M}^{-1} {\bf A})$. Hence, convergence of Algorithm 1 
is fast whenever the eigenvalues of ${\bf M}^{-1} {\bf A}$ are clustered around one, i.e., 
when ${\bf M}^{-1}$ is an approximation of the inverse of ${\bf A}$. 

It should be noted that Eq. (\ref{FFT}) 
is similar to the iterative algorithm based on the steepest descent (or alternatively, the 
modified Newton's method or the method of deflected gradients \cite{luenberger}), which is given by 
\begin{eqnarray}
{\bf x}^{(k+1)} & = & {\bf x}^{(k)} + \alpha^{(k)} {\bf M}^{-1} {\bf r}^{(k)} \label{steep}
\end{eqnarray} 
where $\alpha^{(k)}$ is a non-negative scalar chosen to minimize the standard quadratic function 
$f({\bf x}^{(k+1)})$ defined as
\begin{eqnarray}
f({\bf x}) & = & \frac{1}{2} {\bf x}^T {\bf A} {\bf x} - {\bf b}^T {\bf x} \label{quadf}
\end{eqnarray}
The minimization of $f({\bf x})$ is equivalent to the solution of the system of equations 
${\bf A} {\bf x} = {\bf b}$. 
In this sense, a variety of iterative algorithms such as steepest descent and quasi-Newton methods 
\cite{luenberger} may be employed to solve ${\bf A} {\bf x} = {\bf b}$ using the iterative scheme 
based on Eq. (\ref{steep}). For instance, ${\bf M}^{-1} = {\bf I}$ results in a steepest descent 
algorithm, whereas if ${\bf M}^{-1}$ is the inverse of the Hessian of $f$ then we obtain the Newton's 
method. In general, ${\bf M}^{-1}$ is chosen to be an approximation to the inverse of the 
Hessian. In addition, in order to guarantee the iterative process defined by Eq. (\ref{steep}) to be a 
descent method for small values of $\alpha^{(k)}$, 
it is in general required that ${\bf M}$ be positive definite \cite{luenberger}.  
For computational purposes, ${\bf M}$ is chosen such that its inverse can be computed cheaply 
so that the solution of ${\bf M} {\bf z} = {\bf r}$ can be computed with least computational effort. 

The Fourier acceleration algorithm 
proposed in Refs. \cite{batrouni86,batrouni88} chooses an 
ensemble averaged matrix $\bar{\bf A}$ \cite{procaccia1,procaccia2} as the preconditioner 
in Algorithm 1, where $\bar{\bf A}(i,j) = r^{(d_w - d_f)}$,  
$r = |i-j|$, the distance between the nodes $i$ and $j$, and $d_f$ and $d_w$ refer to the fractal dimension of the 
current-carrying backbone and the random-walk dimension respectively. 
This ensemble averaged matrix $\bar{\bf A}$ is 
a symmetric Toeplitz matrix and hence the preconditioned system ${\bf M} {\bf z} = {\bf r}$ 
in Algorithm 1 can be solved in $O(N~log~N)$ operations by using FFTs of size $N$. 
Compared with the unconditioned 
CG methods, the Fourier accelerated technique based on ensemble averaged 
circulant preconditioner significantly reduced the computational time 
required to simulate fracture using random fuse networks \cite{batrouni86,batrouni88}. 
Although the Fourier accelerating 
technique significantly improves the condition number of the unconditioned system 
of equations, these methods still exhibit 
{\it critical slowing down} close to macroscopic fracture. Consequently, 
the numerical simulations in the past were often limited to systems of sizes $L = 128$. 

Mathematically, the ensemble averaged matrix $\bar{\bf A}$ is not the optimal 
circulant preconditioner in the sense that it does not minimize the norm 
$\|{\bf I} - {\bf C}^{-1} {\bf A}\|_F$ over all non-singular circulant matrices ${\bf C}$, 
and hence is expected to take more CG iterations than the {\it optimal} \cite{tchan88,chan89,chan921,chan96} and
{\it superoptimal} \cite{tyrty,chan96} circulant preconditioners. 
In the above description, $\| \cdot \|_F$ denotes the Frobenius norm \cite{golub96}. 
It should be noted that $\|{\bf I} - {\bf C}^{-1} {\bf A}\|_F$ determines the 
convergence rate of Algorithm 1 as evident in Eq. (\ref{error}). The discussion of 
{\it optimal} \cite{tchan88,chan89,chan921,chan96} and
{\it superoptimal} \cite{tyrty,chan96} circulant preconditioners 
is presented in detail in Section A.4. 
In addition, the Fourier acceleration technique is not effective when
fracture simulation is performed using central-force and bond-bending
lattice models \cite{batrouni88}. In the case of central-force and bond-bending lattice models, 
each of the lattice sites has more than one degree of freedom. This results in a 
matrix ${\bf A}_0$ that is block-circulant. This explains why a naive circulant preconditioner 
is not effective in the case of central-force and bond-bending lattice models. A discussion on 
block-circulant preconditioners is presented in Section A.4.

\subsection{Naive Application of Sparse Direct Solvers}
Alternatively, since each of the matrices ${\bf A}_n$ for $n = 0, 1, 2, \dots,$ is sparse,  
it is possible to use any of the existing sparse direct solvers \cite{wsmp,mumps} 
to solve each of the system of equations formed by ${\bf A}_n$. 
It should be noted that the Cholesky factorization of ${\bf A}_n$ for any particular $n$ can be done 
efficiently. However, factorization of each of the ${\bf A}_n$ matrices for $n_f \sim O(L^{1.8})$ 
(or $n_f \sim O(L^{2.7})$ in 3D)
number of times is still a computationally daunting task. For example, in the case 
of a 2D triangular lattice system of size $L = 128$, the algorithm based on 
factorizing each of the ${\bf A}_n$ matrices using supernodal sparse direct solver \cite{taucs} 
took on an average 8300 seconds compared to an average of 7473 seconds taken 
by the {\it optimal} circulant preconditioner based CG algorithm. It should however be noted that, in general, 
the {\it optimal} circulant preconditioner based CG algorithm used in the above comparison is much faster 
than the Fourier accelerated CG algorithm described in Ref. \cite{batrouni88}. 
However, as the following description elaborates, 
the computational advantages of sparse direct solvers are not fully realized when the 
algorithm is based on factorizing each of the ${\bf A}_n$ matrices. 

\subsection{Multiple-rank sparse Cholesky downdating algorithm}
An important feature of fracture simulations using discrete lattice models is that, 
for each $n = 0, 1, 2, \dots$, 
the new matrix ${\bf A}_{n+1}$ of the lattice system after the $(n+1)^{th}$ broken bond 
is equivalent to a rank-$p$ downdate of the matrix ${\bf A}_n$ \cite{nukalajpamg1,nukalaijnme}. 
Mathematically, in the 
case of the fuse and spring models, the breaking of a bond is equivalent to a rank-one 
downdate of the matrix ${\bf A}_n$, whereas in the case of beam models, it is 
equivalent to multiple-rank (rank-3 for 2D, and rank-6 for 3D) downdate 
of the matrix ${\bf A}_n$. Thus, 
an updating scheme of some kind is therefore likely 
to be more efficient than solving the new set of equations formed by 
Eq. (\ref{intro1}) for each $n$. One possibility is to use the 
Shermon-Morrison-Woodbury formula \cite{golub96} 
to update the solution vector ${\bf x}_{n+1}$ 
directly knowing the sparse Cholesky factor ${\bf L}_m$ of ${\bf A}_m$, 
for any $m \le n$. That is, 
given the Cholesky factorization ${\bf L}_m$ of ${\bf A}_m$ for $m = 0,1,2,\cdots$, 
it is possible to obtain a direct updating of the solution ${\bf x}_{n+1}$ for 
$n = m, m+1, m+2, \cdots$ based on $p = (n-m)$ {\it saxpy} vector updates 
\cite{nukalajpamg1,nukalaijnme}. The resulting algorithm is equivalent to a 
dense matrix update algorithm proposed in the Appendix C of Ref. \cite{nukalaijnme}. 
In Ref. \cite{nukalaijnme}, it has been shown that for two-dimensional fracture simulations, 
any of these sparse solver updating algorithms are significantly superior to the traditional 
Fourier accelerated CG iterative schemes. The best performing sparse solver for two-dimensional 
fracture simulations using the discrete lattice models (fuse, spring and beam models) is 
based on the multiple-rank sparse Cholesky downdating algorithm, which is briefly described 
in the following.

Consider the Cholesky factorizations
\begin{equation}
{\bf P} {\bf A}_n {\bf P}^t = {\bf L}_n {\bf L}_n^t \label{pivot}
\end{equation}
for each $n = 0,1,2,\cdots$, where ${\bf P}$ is a permutation matrix chosen to preserve the sparsity 
of ${\bf L}_n$. Since the breaking of bonds is equivalent to 
removing the edges in the underlying graph structure of the matrix ${\bf A}_n$, for each $n$, the 
sparsity pattern of the Cholesky factorization ${\bf L}_{n+1}$ of the matrix 
${\bf A}_{n+1}$ must be a subset of the sparsity pattern 
of the Cholesky factorization ${\bf L}_n$ of the matrix ${\bf A}_n$. 
Hence, for all $n$, the sparsity pattern of ${\bf L}_n$ is contained in that of ${\bf L}_0$. That is, 
denoting the sparsity pattern of ${\bf L}$ by ${\mathcal L}$, we have
\begin{eqnarray}
{\mathcal L}_{m} \supseteq {\mathcal L}_{n} ~~~\forall ~m < n
\end{eqnarray}
Therefore, we can use the sparse Cholesky factorization downdate algorithm of Davis and Hager 
\cite{tdavis1,tdavis2} to successively downdate the Cholesky factorizations ${\bf L}_n$ of ${\bf A}_n$ 
to ${\bf L}_{n+1}$ of ${\bf A}_{n+1}$, i.e., ${\bf L}_n \rightarrow {\bf L}_{n+1}$ for $n = 0,1,2,\cdots$. 
Since ${\mathcal L}_{n} \supseteq {\mathcal L}_{n+1}$, it is necessary to modify only a part of 
the non-zero entries of ${\bf L}_n$ in order to obtain ${\bf L}_n \rightarrow {\bf L}_{n+1}$. This 
results in a significant reduction in the computational time. 
In addition, the rank-$p$ downdating vectors are very sparse 
with only a few non-zero entries (at most two for fuse models; six for spring models; 
and twelve for beam models). Consequently, the computational effort involved in 
downdating the Cholesky factors using the sparse downdating vectors is once again 
significantly reduced \cite{nukalajpamg1,nukalaijnme}. 

\subsection{{\it Optimal} and {\it superoptimal} circulant preconditioners}
Although the algorithms based on sparse direct solvers with multiple-rank 
Cholesky downdating schemes achieve superior performance 
over iterative solvers in 2D lattice simulations (see Section A.7), 
the memory demands brought about by the amount of fill-in during sparse Cholesky factorization 
poses a severe constraint over the usage of sparse direct solvers for 3D lattice simulations. 
Hence, iterative solvers are in common use for large-scale 3D lattice simulations. The 
performance of iterative solvers depends crucially on the condition number and clustering of 
the eigenvalues of the preconditioned system. In general, the more 
clustered the eigenvalues are, the faster the convergence rate is. 
Hence, the main focus of any iterative solution technique is choosing an optimum preconditioner.

The main observation behind developing preconditioners for the iterative schemes 
is that the operators on discrete lattice 
network result in a circulant block structure. For example, the Laplacian operator 
(Kirchhoff equations in the case of random fuse model) on the initial uniform grid 
results in a Toeplitz matrix ${\bf A}_0$. Hence, a fast Poisson type 
solver with a circulant preconditioner can be used to obtain the 
solution in $O(N ~log N)$ operations using FFTs of size $N$, where $N$ denotes the number of 
degrees of freedom. However, as the lattice bonds are broken 
successively, the initial uniform lattice grid becomes a diluted network. Consequently, although the matrix 
${\bf A}_0$ is Toeplitz (also block Toeplitz with Toeplitz blocks) initially, the subsequent matrices 
${\bf A}_n$, for each $n$, are not Toeplitz matrices. However, 
depending on the pattern of broken bonds, ${\bf A}_n$ may still possess block 
structure with many of the blocks being Toeplitz blocks.

The {\it optimal} circulant 
preconditioner $c({\bf A})$ \cite{tchan88,chan89,chan921,chan96} of a matrix ${\bf A}$ is defined as the minimizer of 
$\|{\bf C} - {\bf A}\|_F$ over all $N \times N$ circulant matrices ${\bf C}$. Given a 
matrix ${\bf A}$, the {\it optimal} circulant preconditioner $c({\bf A})$ is 
uniquely determined by 
\begin{eqnarray}
c({\bf A}) & = & {\bf F}^{\ast} \delta\left({\bf F} {\bf A} {\bf F}^{\ast}\right) {\bf F}
\end{eqnarray}
where ${\bf F}$ denotes the discrete Fourier matrix, $\delta\left({\bf A}\right)$ denotes the 
diagonal matrix whose diagonal is equal to the diagonal of the matrix ${\bf A}$, and 
$\ast$ denotes the adjoint (i.e. conjugate transpose). It should be noted that the diagonals  
of ${\bf F} {\bf A} {\bf F}^{\ast}$ represent the 
eigenvalues of the matrix $c({\bf A})$ and can be obtained in $O(N ~log ~N)$ 
operations by taking the FFT of the first column of $c({\bf A})$. The first 
column vector of T. Chan's {\it optimal} 
circulant preconditioner matrix that minimizes the norm $\|{\bf C} - {\bf A}\|_F$ is given by
\begin{eqnarray}
c_i & = & \frac{1}{N} \sum_{j=1}^{N} a_{j,(j-i+1)~\mbox{mod}~N} \label{tchaneq}
\end{eqnarray}
If the matrix ${\bf A}$ is 
Hermitian, the eigenvalues of $c({\bf A})$ are bounded below and above by 
\begin{equation}
\lambda_{min}({\bf A}) \leq \lambda_{min}(c({\bf A})) \leq \lambda_{max}(c({\bf A})) \leq \lambda_{max}({\bf A}) \label{pspect}
\end{equation}
where $\lambda_{min}(\cdot)$ and $\lambda_{max}(\cdot)$ denote the minimum and maximum eigenvalues, 
respectively. Based on the above result, if ${\bf A}$ is positive definite, then 
the circulant preconditioner $c({\bf A})$ is also positive definite. The computational cost associated with 
the solution of the preconditioned system $c({\bf A}) {\bf z} = {\bf r}$ is the initialization cost of 
$nnz({\bf A})$ for setting the first column of $c({\bf A})$ using Eq. (\ref{tchaneq}) 
during the first iteration, and $O(N ~log ~N)$ during every iteration step.

The {\it superoptimal} circulant preconditioner $t({\bf A})$ \cite{tyrty} is based on the idea of 
minimizing the norm $\|{\bf I} - {\bf C}^{-1} {\bf A}\|_F$ over all nonsingular 
circulant matrices ${\bf C}$. In the above description, $t({\bf A})$ is {\it superoptimal} 
in the sense that it minimizes $\|{\bf I} - {\bf C}^{-1} {\bf A}\|_F$, and is equal to
\begin{equation}
t({\bf A}) = c({\bf A} {\bf A}^{\ast}) c({\bf A})^{-1} \label{tyrtyeq}
\end{equation}
The preconditioner obtained by Eq. (\ref{tyrtyeq}) is also positive definite if  
${\bf A}$ itself is positive definite. Although the preconditioner $t({\bf A})$ is 
obtained by minimizing the norm $\|{\bf I} - {\bf C}^{-1} {\bf A}\|_F$, the asymptotic 
convergence of the preconditioned system is same as $c({\bf A})$ for large $N$ system.

\subsection{Block-circulant preconditioner}
Alternatively, block-circulant preconditioners may also be used as preconditioners 
since many of the block matrices in ${\bf A}_n$ for $n > 0$ may still retain the circulant or Toeplitz 
property. Let the matrix ${\bf A}_n$ be partitioned into $r$-by-$r$ blocks such that each block is 
an $s$-by-$s$ matrix. That is, $N = rs$. Since each of the blocks of ${\bf A}_n$ are Toeplitz 
(or even circulant), the block-circulant preconditioner is 
obtained by using circulant approximations for each of the blocks of ${\bf A}_n$. It is the minimizer of 
$\|{\bf C} - {\bf A}_n\|_F$ over all matrices ${\bf C}$ that are $r$-by-$r$ block matrices 
with $s$-by-$s$ circulant blocks. In addition, we have 
\begin{equation}
\lambda_{min}({\bf A}_n) \leq \lambda_{min}(c_B({\bf A}_n)) \leq \lambda_{max}(c_B({\bf A}_n)) \leq \lambda_{max}({\bf A}_n) \label{bspect}
\end{equation}
In particular, if ${\bf A}_n$ is positive definite, then the block-preconditioner 
$c_B({\bf A}_n)$ is also positive definite.
In general, the average computational cost of using the block-circulant preconditioner per iteration is 
$O(rs~log ~s) + delops$, where $delops$ represents the operational cost associated 
with solving a block-diagonal matrix with $r \times r$ dense blocks. For 2D and 3D discrete lattice network 
with periodic boundary conditions in the horizontal direction, this operational cost reduces significantly. 
The reader is referred to \cite{nukalajpamg2,nukalaijnme} for further details on {\it optimal} and 
block-circulant preconditioners for discrete lattice networks.

\subsection{Summary of algorithms}
To summarize, we use the following two algorithms based on sparse direct solvers for modeling 
fracture using discrete lattice systems. 

\begin{itemize}
\item {\it Solver Type A}: Given the factorization ${\bf L}_m$ of 
${\bf A}_m$, the rank-1 sparse Cholesky modification is 
used to update the factorization ${\bf L}_{n+1}$ for all subsequent values of $n = m,m+1, \cdots$.
Once the factorization ${\bf L}_{n+1}$ of ${\bf A}_{n+1}$ is obtained, the solution vector 
${\bf x}_{n+1}$ is obtained from 
${\bf L}_{n+1} {\bf L}_{n+1}^t {\bf x}_{n+1} = {\bf b}_{n+1}$ by 
two triangular solves \cite{nukalajpamg1,nukalaijnme}.

\item {\it Solver Type B}: Assuming that the factor ${\bf L}_m$ of 
matrix ${\bf A}_m$ is available after the $m^{th}$ fuse is burnt, the solution 
${\bf x}_{n+1}$ after the $(n+1)^{th}$ fuse is burnt, for $n = m,m+1,\cdots,m+maxupd$,
is obtained through $p = (n-m)$ {\it saxpy} vector updates using an algorithm 
based on Shermon-Morrison-Woodbury \cite{golub96} formula.
Either a completely new factorization of the matrix ${\bf A}$ is performed or 
a multiple-rank downdate of the factorization ${\bf L}_m \rightarrow {\bf L}_{m+maxupd+1}$ 
is performed every $\mathit{maxupd}$ steps \cite{nukalajpamg1,nukalaijnme}.
\end{itemize}

In addition to the above two algorithms, the simulations are carried out using the 
iterative solvers based on {\it optimal} 
circulant and block-circulant preconditioners \cite{nukalajpamg2,nukalaijnme}. 

\subsection{Performance comparison}
For two-dimensional lattice systems, we choose a 
triangular lattice topology, whereas a cubic lattice topology is used in 
three-dimensional cases. The random thresholds fuse model (rank-one downdate with $p = 1$) 
is used as a basis of performance comparison to evaluate the numerical efficiency of various 
algorithms. The algorithmic framework for central-force (spring) and beam models is 
presented in Refs. \cite{nukalajpamg1,nukalaijnme}. 

In the numerical simulations using solver types A and B based on sparse direct solvers, 
the maximum number of vector updates, $\mathit{maxupd}$, is chosen to be a constant 
for a given lattice size $L$. We choose $\mathit{maxupd} = 25$ for $L = \{4, 8, 16, 24, 32\}$, 
$\mathit{maxupd} = 50$ for $L = 64$, and $\mathit{maxupd} = 100$ for $L = \{128, 256, 512\}$. For 
$L = 512$, $\mathit{maxupd}$ is limited to 100 due to memory constraints. 
By keeping the $\mathit{maxupd}$ value constant, it is possible to compare realistically 
the computational cost associated with different solver types. Moreover, the 
relative CPU times taken by these algorithms remain the same even when the simulations 
are performed on different platforms.

The convergence rate of an iterative solver depends on 
the quality of starting vector. When PCG iterative solvers are 
used in updating the solution ${\bf x}_n \rightarrow {\bf x}_{n+1}$, 
we use either the solution from the previous converged state, i.e., 
${\bf x}_n$, or the zero vector, ${\bf 0}$, as the starting vectors for 
the CG iteration of ${\bf x}_{n+1}$. The choice of using either ${\bf x}_n$ or ${\bf 0}$ 
as the starting vectors depends on which vector is closer to ${\bf b}_{n+1}$ in 2-norm. 
That is, if $\|{\bf b}_{n+1} - {\bf A}_n {\bf x}_n\|_{2} \le \|{\bf b}_{n+1}\|_{2}$, then 
${\bf x}_n$ is used, else ${\bf 0}$ is used as the starting vector for the ${\bf x}_{n+1}$ 
CG iteration. A relative residual tolerance $\epsilon = 10^{-12}$ is used in CG iteration. 

The numerical simulations are 
carried out on a single 1.3 GHz IBM Power4 processor. 
Tables A1 and A2 present the CPU times taken for one configuration (simulation) using the 
sparse direct solver algorithms A and B, respectively.  
These tables also indicate the 
number of configurations, $N_{config}$, over which ensemble averaging of the 
numerical results is performed for performance evaluation. The CPU times taken by the 
iterative solvers are presented in Tables A3-A4. 
For iterative solvers, the number of iterations presented in Tables A3-A4 denote 
the average number of total iterations taken to break one intact lattice configuration 
until it falls apart. In the case of iterative solvers,
some of the simulations for larger lattice systems were not performed either 
because they were expected to take larger CPU times or the numerical results         
do not influence the conclusions drawn in this study. The results presented in 
Tables A1-A4 indicate that for 2D lattice systems, the multiple-rank Cholesky downdate 
based sparse direct solver  
algorithms are clearly superior to the optimal and block-circulant 
preconditioned CG iterative solvers. In particular, for the lattice size $L = 128$, 
the solver type A took 212.2 seconds, which should be compared with the 7473 and 8300 seconds 
taken respectively by the optimal circulant preconditioner and the naive sparse direct solution technique 
based on factorizing each of the ${\bf A}_n$ matrices.

\begin{table}
%Solver Type A
\begin{center}
\caption{Solver Type A (2D Triangular Lattice)}
  \begin{tabular}{|c|c|c|}\hline
  Size  & CPU(sec) & Simulations  \\
  \hline
 32 & 0.592 & 20000 \\
 64 & 10.72 & 4000 \\
128 & 212.2 & 800 \\
256 & 5647 & 96 \\
512 & 93779 & 16 \\
  \hline
  \end{tabular}
\end{center}
\end{table}\hfill
\begin{table}
%Solver Type B
\begin{center}
\caption{Solver Type B (2D Triangular Lattice)}
  \begin{tabular}{|c|c|c|}\hline
  Size  & CPU(sec) &  Simulations \\
  \hline
 32 & 0.543 & 20000 \\
 64 & 11.15 & 4000 \\
128 & 211.5 & 800 \\
256 & 6413 & 96 \\
  \hline
  \end{tabular}
\end{center}
\end{table}

\begin{table}
\begin{center}
\caption{Block Circulant PCG (2D Triangular Lattice)}
  \begin{tabular}{|c|c|c|c|}\hline
  Size  & CPU(sec) & Iter & Simulations  \\
  \hline
32 & 10.00 & 11597 & 20000 \\
 64 & 135.9 & 41207 & 1600 \\
128 & 2818 & 147510 & 192 \\
256 & 94717 & & 32 \\ 
  \hline
  \end{tabular}
\end{center}
\end{table}\hfill

\begin{table}
%Solver Type B
\begin{center}
\caption{Optimal Circulant PCG (2D Triangular Lattice)}
  \begin{tabular}{|c|c|c|c|}\hline
  Size  & CPU(sec) & Iter & Simulations  \\
  \hline
32 & 11.66 & 25469 & 20000 \\
 64 & 173.6 & 120570 & 1600 \\
128 & 7473 & 622140 & 128 \\ 
  \hline
  \end{tabular}
\end{center}
\end{table}

For large 3D lattice systems, the advantages exhibited by the multiple-rank Cholesky downdating 
algorithms in 2D simulations vanish due to the amount of {\it fill-in} during 
Cholesky factorization. Tables A5-A8 present the CPU times taken 
for simulating one-configuration using the block circulant, {\it optimal} circulant, 
un-preconditioned, and the incomplete Cholesky iterative solvers, respectively. 
It should be noted that for large 3D lattice systems (e.g., $L = 32$), 
the performance of incomplete Cholesky preconditioner (see Table A8) is similar to that of 
block-circulant preconditioner (see Table A5), even though the performance of  
incomplete Cholesky preconditioner is far more superior in the case of 2D lattice simulations 
\cite{nukalaijnme}. Indeed, for 3D lattice systems of sizes $L = 48$ and $L = 64$, the CPU time 
taken by the sparse direct solver types A and B is much higher than that of block-circulant 
PCG solver.

\begin{table}
\begin{center}
\caption{Block Circulant PCG (3D Cubic Lattice)}
  \begin{tabular}{|c|c|c|c|}\hline
  Size  & CPU(sec) & Iter & Simulations  \\
  \hline
10 & 16.54 & 16168 & 40000 \\
16 & 304.6 & 58756 & 1920 \\
24 & 2154 & 180204 & 256 \\
32 & 12716 & 403459 & 128 \\
48 & 130522 & 1253331 & 32 \\
64 & 1180230 & & 11 \\
  \hline
  \end{tabular}
\end{center}
\end{table}\hfill
\begin{table}
%Solver Type B
\begin{center}
\caption{Optimal Circulant PCG (3D Cubic Lattice)}
  \begin{tabular}{|c|c|c|c|}\hline
  Size  & CPU(sec) & Iter & Simulations  \\
  \hline
10 & 15.71 & 27799 & 40000 \\
16 & 386.6 & 121431 & 1920 \\
24 & 2488 & 446831 & 256 \\
32 & 20127 & 1142861 & 32 \\
48 & 233887 & 4335720 & 32 \\
  \hline
  \end{tabular}
\end{center}
\end{table}

\begin{table}
\begin{center}
\caption{Un-preconditioned CG (3D Cubic Lattice)}
  \begin{tabular}{|c|c|c|c|}\hline
  Size  & CPU(sec) & Iter & Simulations  \\
  \hline
10 & 5.962 & 48417 & 40000 \\
16 & 119.4 & 246072 & 3840 \\
24 & 1923 & 1030158 & 256 \\
32 & 16008 & 2868193 & 64 \\
  \hline
  \end{tabular}
\end{center}
\end{table}\hfill

\begin{table}
%Solver Type B
\begin{center}
\caption{Incomplete Cholesky PCG (3D Cubic Lattice)}
  \begin{tabular}{|c|c|c|c|}\hline
  Size  & CPU(sec) & Iter & Simulations  \\
  \hline
10 & 5.027 & 8236 & 40000 \\
16 & 118.1 & 42517 & 3840 \\
24 & 1659 & 152800 & 512 \\
32 & 12091 & 422113 & 64 \\
  \hline
  \end{tabular}
\end{center}
\end{table}

To summarize, the numerical simulation results presented in Tables A1-A4 for 2D triangular random fuse 
networks indicate that sparse direct solvers based on multiple-rank sparse Cholesky downdating 
are superior to the competing iterative schemes using block-circulant and optimal circulant 
preconditioners. On the other hand, for 3D lattice systems, the block-circulant preconditioner 
based CG solver exhibits superior performance (for system sizes $L > 32$) over the sparse direct solvers and 
the related incomplete Cholesky preconditioned CG solvers. In addition, for both 2D and 3D 
lattice systems, the block circulant 
preconditioned CG is superior to the {\it optimal} circulant preconditioned PCG solver, 
which in turn is superior to the Fourier accelerated PCG solvers used in Refs. \cite{batrouni86,batrouni88}.


\begin{thebibliography}{100}

\bibitem{leonardo40}
L. da~Vinci,  in {\em I libri di Meccanica}, edited by A. Uccelli ( 
    Hoepli Milano 1940).

\bibitem{galilei58}
G. Galilei, {\em Discorsi e dimostrazioni matematiche intorno a due nuove
  scienze} (    Boringhieri, Torino, 1958).

\bibitem{parsons39}
W.~B. Parsons, {\em Engineers and Engineering in the Renaissance} ( 
    MIT Press, Cambridge, 1939).

\bibitem{lund01}
J.~R. Lund and J.~P. Byrne, Civil. Eng. and Env. Syst {\bf 18},  243  (2001).

\bibitem{griffith20}
A.~A. Griffith, Trans. Roy. Soc. (london) A {\bf 221},  163  (1920).

\bibitem{weibull39}
W. Weibull,  in {\em A statistical theory of the strength of materials}, edited
  by G. litografiska~anstalts f\"orlag (  Stockholm, 1939).

\bibitem{timoshenko}
S.~P. Timoshenko,  in {\em History of strength of materials}, edited by M.~G.
  Hill (  Mc Graw Hill New York, 1953).

\bibitem{herrmann90}
H.~J. Herrmann and S. Roux, {\em Statistical Models for the Fracture of
  Disordered Media} (North-Holland, Amsterdam,   1990).

\bibitem{chakrabarti}
B.~K. Chakrabarti and L.~G. Benguigui, {\em Statistical Physics of Fracture and
  Breakdown in Disordered Systems} (Oxford Science Publications, Oxford,
    1997).

\bibitem{hansen00}
A. Hansen and S. Roux,  in {\em Statistical toolbox for damage and fracture},
  edited by D. Krajcinovic and J. van Mier (Springer-Verlag, New York,  
  2000), pp.\ 17--101.

\bibitem{mishna97}
J. L.~L.~Mishnaevsky, Eng. Fract. Mech. {\bf 56},  47  (1997).

\bibitem{deArcangelis85}
L. de~Arcangelis, S. Redner, and H.~J. Herrmann, Journal of Physics (Paris)
  Letters {\bf 46},  585  (1985).

\bibitem{pierce26}
F.~T. Pierce, J. Textile Inst. {\bf 17},  355  (1926).

\bibitem{daniels45}
H.~E. Daniels, Proc.\ R.\ Soc.\ London A {\bf 183},  405  (1945).

\bibitem{krajcinovic00}
D. Krajcinovic and J.~G.~M. van Mier, {\em Damage and fracture of disordered
  materials} (    Springer-Verlag, New York, 2000).

\bibitem{vanmierbook}
J.~G.~M. van Mier, {\em Fracture Processes of Concrete} (CRC Press, Boca Raton,
  USA,   1996).

\bibitem{bazantbook}
Z.~P. Bazant and J. Planas, {\em Fracture and Size Effect in Concrete and Other
  Quasibrittle Materials} (CRC Press, Boca Raton, USA,   1997).

\bibitem{vanmier03}
J.~G.~M. van Mier and M.~R.~A. van Vliet, Eng. Fract. Mech. {\bf 70},  2281
  (2003).

\bibitem{lilliu03}
G. Lilliu and J.~G.~M. van Mier, Eng. Fract. Mech. {\bf 70},  927  (2003).

\bibitem{schlangen96}
E. Schlangen and E.~J. Garboczi, Int. J. Eng. Sci. {\bf 34},  1131  (1996).

\bibitem{niemayer84}
L. Niemeyer, L. Pietronero, and H.~J. Wiesmann, Phys. Rev. Lett. {\bf 52},
  1033  (1984).

\bibitem{duxbury88}
P.~D. Beale and P.~M. Duxbury, Phys. Rev. B {\bf 37},  2785  (1988).

\bibitem{arndt01}
P.~F. Arndt and T. Nattermann, Phys. Rev. B {\bf 63},  134204  (2001).

\bibitem{Lemaitre}
J. Lemaitre and J.~L. Chaboche, {\em Mechanics of Solid Materials} (Cambridge
  University Press, Cambridge,   1990).

\bibitem{krajcinovic}
D. Krajcinovic, {\em Damage Mechanics} (Elsevier, Amsterdam,   1996).

\bibitem{gluzman01}
S. Gluzman and D. Sornette, Phys. Rev. E {\bf 63},  066129  (2001).

\bibitem{shcherbakov03}
R. Shcherbakov and D.~L. Turcotte, Theoretical and Applied Fracture Mechanics
  {\bf 39},  245  (2003).

\bibitem{atkinson87}
B.~K. Atkinson and P.~G. Meredith,  in {\em Fracture Mechanics of Rock}, edited
  by B. Atkinson (    Academic Press, New York, 1987), p.\ 110.

\bibitem{straumsnes97}
S. Straumsnes, P. Dommersnes, E.~G. Flekkoy, and S. Roux, European Journal of
  Mechanics A-Solids {\bf 16},  993  (1997).

\bibitem{barenblatt96}
G.~I. Barenblatt, {\em Scaling, Self-Similarity, and Intermediate Asymptotics}
  (Cambridge Univ. Press, Cambridge, 1996).

\bibitem{family91}
F. Family and T. Viscek, {\em Dynamics of Fractal Surfaces} ( 
    World Scientific, Singapore, 1991).

\bibitem{barabasi95}
A.-L. Barab{\'a}si and H.~E. Stanley, {\em Fractal concepts in surface growth}
  (Cambridge University Press, Cambrdige,   1995).

\bibitem{lopez97}
J.~M. Lopez, M.~A. Rodriguez, and R. Cuerno, Phys. Rev. E {\bf 56},  3993
  (1997).

\bibitem{lopez99}
J.~M. Lopez, Phys. Rev. Lett. {\bf 83},  4594  (1999).

\bibitem{nattermann92}
T. Nattermann, S. Stepanow, L.~H. Tang, and H. Leschhorn, J. Phys. (France) II
  {\bf 2},  1483  (1992).

\bibitem{leschhorn97}
H. Leschhorn, T. Nattermann, S. Stepanow, and L.~H. Tang, Ann. Physik {\bf 7},
  1  (1997).

\bibitem{narayan93}
O. Narayan and D. Fisher, Phys. Rev. B {\bf 48},  7030  (1993).

\bibitem{chauve01}
P. Chauve, P.~L. Doussal, and K. Wiese, Phys. Rev. Lett. {\bf 86},  1785
  (2001).

\bibitem{chauve02}
P.~L. Doussal, K. Wiese, and P. Chauve, Phys. Rev. B {\bf 66},  174201  (2002).

\bibitem{borodich97}
F.~M. Borodich, J. Mech. Phys. Solids {\bf 45},  239  (1997).

\bibitem{vandembroucq97a}
D. Vandembroucq and S. Roux, J. Mech. Phys. Solids {\bf 45},  853  (1997).

\bibitem{vandembroucq97b}
D. Vandembroucq and S. Roux, Phys. Rev. E {\bf 55},  6186  (1997).

\bibitem{barra02}
F. Barra, H.~G.~E. Hentschel, A. Levermann, and I. Procaccia, Phys. Rev. E {\bf
  65},  045101  (2002).

\bibitem{barra02b}
F. Barra, A. Levermann, and I. Procaccia, Phys. Rev. E {\bf 66},  066122
  (2002).

\bibitem{leverman02}
A. Levermann and I. Procaccia, Phys. Rev. Lett. {\bf 89},  234501  (2002).

\bibitem{morel00}
S. Morel, J. Schmittbuhl, E. Bouchaud, and G. Valentin, Phys. Rev. Lett. {\bf
  85},  1678  (2000).

\bibitem{morel02}
S. Morel, E. Bouchaud, and G. Valentin, Phys. Rev. B {\bf 65},  104101  (2002).

\bibitem{bouchaud94}
E. Bouchaud and J.-P. Bouchaud, Phys. Rev. B {\bf 50},  17752  (1994).

\bibitem{weiss01}
J. Weiss, Int. J. Fracture {\bf 109},  365  (2001).

\bibitem{chiaia98}
B. Chiaia, J.~G.~M. van Mier, and A. Vervuurt, Cem. Concr. Res. {\bf 28},  103
  (1998).

\bibitem{vandenborn91}
I.~C. van~den Born, A. Santen, H.~D. Hoekstra, and J.~T. M.~D. Hosson, Phys.
  Rev. B {\bf 43},  R3794  (1991).

\bibitem{bazant99}
Z.~P. Bazant, Arch. Appl. Mech. {\bf 69},  703  (1999).

\bibitem{sutherland99}
L. Sutherland, R. Shenoi, and S. Lewis, Composite Sci. Techn. {\bf 209},  1999
  (1999).

\bibitem{korteoja98}
M. Korteoja, L.~I. Salminen, K.~J. Niskanen, and M.~J. Alava, J. Pulp Paper
  Sci. {\bf 24},  1  (1998).

\bibitem{korteoja99}
M. Korteoja, L.~I. Salminen, K.~J. Niskanen, and M.~J. Alava, Materials Science
  and Engineering A {\bf 240},  173  (1999).

\bibitem{vanvliet00}
M.~R.~A. van Vliet and J.~G.~M. van Mier, Eng. Fract. Mech. {\bf 65},  165
  (2000).

\bibitem{lu}
C. Lu, R. Danzer, and F.~D. Fischer, Phys. Rev. E {\bf 65},  067102  (2002).

\bibitem{doremus}
R.~H. Doremus, J. Appl. Phys. {\bf 54},  193  (1983).

\bibitem{lavoie00}
J.~A. Lavoie, C. Soutis, and J. Morton, Composite Sci. Techn. {\bf 60},  283
  (2000).

\bibitem{foote86}
R.~M.~L. Foote, Y.~W. Mai, and B. Cotterell, J. Mech. Phys. Solids {\bf 34},
  593  (1986).

\bibitem{cotterell87}
B. Cotterell and Y.~W. Mai, J. Mat. Sci. {\bf 22},  2734  (1987).

\bibitem{mai87}
Y.~W. Mai and B.~R. Lawn, J. Am. Ceram. Soc. {\bf 70},  289  (1987).

\bibitem{wisnom91}
M.~R. Wisnom, Composites {\bf 22},  47  (1991).

\bibitem{hu92}
X.~Z. Hu and F. Wittmann, Mat. Struct. {\bf 25},  319  (1992).

\bibitem{li98}
V.~C. Li, Z. Lin, and T. Matsumoto, Int. J. Solids Structures {\bf 1998},  35
  (1998).

\bibitem{tabiei99}
A. Tabiei and J.~M. Sun, Compos. Struct. {\bf 46},  209  (1999).

\bibitem{wisnom99}
M.~R. Wisnom, Composite Sci. Techn. {\bf 59},  1937  (1999).

\bibitem{hu00}
X.~Z. Hu and F. Wittmann, Eng. Fract. Mech. {\bf 65},  209  (2000).

\bibitem{mai02}
Y.~W. Mai, Eng. Fract. Mech. {\bf 69},  219  (2002).

\bibitem{hu02}
X.~Z. Hu, Eng. Fract. Mech. {\bf 69},  555  (2002).

\bibitem{stevanovic03}
M.~M. Stevanovic, A.~A. Elhisnavi, and D.~R. Pesikan, Mat. Sci. Forum {\bf
  413},  207  (2003).

\bibitem{duan03}
K. Duan, X.~Z. Hu, and F.~H. Wittmann, Mat. Struct. {\bf 36},  74  (2003).

\bibitem{bazant04}
Z.~P. Bazant, Probabilistic Engineering Mechanics {\bf 19},  307  (2004).

\bibitem{bazant04b}
Z.~P. Bazant, PNAS {\bf 101},  13400  (2004).

\bibitem{klock05}
W. Klock and S. Aicher, Wood Fib. Sci. {\bf 37},  403  (2005).

\bibitem{rosti01}
J. Rosti {\it et~al.}, Eur. Phys. J. B {\bf 19},  259  (2001).

\bibitem{bouchaud97}
E. Bouchaud, Journal of Physics: Condensed Matter {\bf 9},  4319  (1997).

\bibitem{salminen03}
L.~I. Salminen, M.~J. Alava, and K.~J. Niskanen, Eur. Phys. J. B {\bf 32},  369
   (2003).

\bibitem{kertesz93}
J. Kert\'esz, V.~K. Horvath, and F. Weber, Fractals {\bf 1},  67  (1993).

\bibitem{engoy94}
T. Engoy, K.~J. Maloy, A. Hansen, and S. Roux, Phys. Rev. Lett. {\bf 73},  834
  (1994).

\bibitem{menezessobrinho05}
I.~L. Menezes-Sobrinho, M.~S. Couto, and I.~R.~B. Ribeiro, Phys. Rev. E {\bf
  71},  066121  (2005).

\bibitem{bouchbinder05a}
E. Bouchbinder, I. Procaccia, and S. Sela, cond-mat/0508549  (2005).

\bibitem{schmittbuhl97}
J. Schmittbuhl and K.~J. Maloy, Phys. Rev. Lett. {\bf 78},  3888  (1997).

\bibitem{delaplace99}
A. Delaplace, J. Schmittbuhl, and K.~J. Maloy, Phys. Rev. E {\bf 60},  1337
  (1999).

\bibitem{maloy01}
K.~J. Maloy and J. Schmittbuhl, Phys. Rev. Lett. {\bf 87},  105502  (2001).

\bibitem{maloy03}
K.~J. Maloy, J. Schmittbuhl, A. Hansen, and G.~G. Batrouni, Int. J. Fracture
  {\bf 121},  9  (2003).

\bibitem{maloy06}
K.~J. Maloy, S. Santucci, J. Schmittbuhl, and R. Toussaint, Phys. Rev. Lett.
  {\bf 96},  045501  (2006).

\bibitem{mandelbrot84}
B.~B. Mandelbrot, D.~E. Passoja, and A.~J. Paullay, Nature {\bf 308},  721
  (1984).

\bibitem{ponson06}
L. Ponson, D. Bonamy, and E. Bouchaud, Phys. Rev. Lett. {\bf 96},  035506
  (2006).

\bibitem{boffa98}
J.~M. Boffa, C. Allain, and J. Hulin, Eur. Phys. J. AP {\bf 2},  281  (1998).

\bibitem{daguier97}
P. Daguier, B. Nghiem, E. Bouchaud, and F. Creuzet, Phys. Rev. Lett. {\bf 78},
  1062  (1997).

\bibitem{bonamy06}
D. Bonamy {\it et~al.}, preprint  (2006).

\bibitem{morel04}
S. Morel, T. Lubet, J.-L. Pouchou, and J.-M. Olive, Phys. Rev. Lett. {\bf 93},
  065504  (2004).

\bibitem{plouraboue93}
F. Plouraboue {\it et~al.}, Phys. Rev. E {\bf 53},  277  (1996).

\bibitem{celarie03}
F. C\'elari\'e {\it et~al.}, Phys. Rev. Lett. {\bf 90},  075504  (2003).

\bibitem{marliere03}
C. Marli\'ere {\it et~al.}, Journal of Physics: Condensed Matter {\bf 15},
  S2377  (2003).

\bibitem{bouchaud93}
J.~P. Bouchaud, E. Bouchaud, G. Lapasset, and J. Plan{\`e}s, Phys. Rev. Lett.
  {\bf 71},  2240  (1993).

\bibitem{lopez98}
J.~M. Lopez and J. Schmittubhl, Phys. Rev. E {\bf 57},  6405  (1998).

\bibitem{morel98}
S. Morel, J. Schmittbuhl, J.~M. Lopez, and G. Valentin, Phys. Rev. E {\bf 58},
  6999  (1998).

\bibitem{mourot05}
G. Mourot, S. Morel, E. Bouchaud, and G. Valentin, Phys. Rev. E {\bf 71},
  016136  (2005).

\bibitem{bouchbinder05b}
E. Bouchbinder, I. Procaccia, S. Santucci, and L. Vanel, cond-mat/0508183
  (2005).

\bibitem{hull99}
D. Hull, {\em Fractography: Observing, Measuring and Interpreting Fracture
  Surface Topography} (Cambridge University Press, Cambrdige,   1999).

\bibitem{kouzeli01}
M. Kouzeli, L. Weber, S.~C. March, and A. Mortensen, Acta Materialia {\bf 49},
  497  (2001).

\bibitem{kawakata99}
H. Kawakata {\it et~al.}, Tectonophysics {\bf 313},  293  (1999).

\bibitem{lockner93}
D.~A. Lockner, Int. J. Rock Mech. Min. Sci. {\bf 30},  883  (1993).

\bibitem{berkowitz97}
B. Berkowitz and A. Hadad, Journal of Geophysical Research B: Solid Earth {\bf
  102},  12205  (1997).

\bibitem{bonnet01}
E. Bonnet {\it et~al.}, Rev. Geophys. {\bf 39},  347  (2001).

\bibitem{rundle99}
J. Rundle, W. Klein, and S. Gross, Pure Appl. Geophys. {\bf 155},  575  (1999).

\bibitem{lyakhovsky01}
V. Lyakhovsky, Geophys. J. Int. {\bf 144},  114  (2001).

\bibitem{goto04}
K. Goto and K. Otsuki, Geophys. Res. Lett. {\bf 31},  L05601  (2004).

\bibitem{weiss97}
J. Weiss, Bull. Seism. Soc. Am. {\bf 87},  1362  (1997).

\bibitem{lysak96}
M.~V. Lysak, Eng. Fract. Mech. {\bf 55},  443  (1996).

\bibitem{ouyang91}
C.~S. Ouyang, E. Landis, and S.~P. Shah, J. Engng. Mech. - ASCE {\bf 117},
  2681  (1991).

\bibitem{harris92}
B. Harris, F. Habib, and R. Cooke, Proc. R. Soc. London A {\bf 437},  1899
  (1992).

\bibitem{mummery93}
P.~M. Mummery, B. Derby, and C.~B. Scruby, Acta Metall. Mater. {\bf 41},  1431
  (1993).

\bibitem{barre94}
S. Barre and M.~L. Benzeggagh, Composite Sci. Techn. {\bf 52},  369  (1994).

\bibitem{degroot95}
P.~J. de~Groot, P.~A.~M. Wijnen, and R.~B.~F. Janssen, Composite Sci. Techn.
  {\bf 55},  405  (1995).

\bibitem{bohse00}
J. Bohse, Composite Sci. Techn. {\bf 60},  1213  (2000).

\bibitem{dzenis02}
Y.~A. Dzenis and I. Saunders, Int. J. Fracture {\bf 117},  L23  (2002).

\bibitem{petri94}
A. Petri {\it et~al.}, Phys. Rev. Lett. {\bf 73},  3423  (1994).

\bibitem{garcimartin97}
A. Garcimartin, A. Guarino, L. Bellon, and S. Ciliberto, Phys. Rev. Lett. {\bf
  79},  3202  (1997).

\bibitem{guarino98}
A. Guarino, A. Garcimartin, and S. Ciliberto, Eur. Phys. J. B {\bf 6},  13
  (1998).

\bibitem{maes98}
C. Maes, A. van Moffaert, H. Frederix, and H. Strauven, Phys. Rev. B {\bf 57},
  4987  (1998).

\bibitem{guarino02}
A. Guarino {\it et~al.}, Eur. Phys. J. B {\bf 26},  141  (2002).

\bibitem{salminen02}
L.~I. Salminen, A.~I. Tolvanen, and M.~J. Alava, Phys. Rev. Lett. {\bf 89},
  185503  (2002).

\bibitem{salminen05}
L.~I. Salminen {\it et~al.}, Europhys. Lett. {\bf 73},  55  (2006).

\bibitem{amitrano03}
D. Amitrano, Journal of Geophysical Research B: Solid Earth {\bf 108},  2044
  (2003).

\bibitem{main91}
I.~G. Main, Geophys. J. Int. {\bf 107},  353  (1991).

\bibitem{main92}
I.~G. Main, Geophys. J. Int. {\bf 111},  531  (1992).

\bibitem{hatton93}
C.~G. Hatton, I.~G. Main, and P.~G. Meredith, J. Struct. Geol. {\bf 15},  1485
  (1993).

\bibitem{main93}
I.~G. Main, P.~R. Sammonds, and P.~G. Meredith, Geophys. J. Int. {\bf 115},
  367  (1993).

\bibitem{reches94}
Z. Reches and D.~A. Lockner, Journal of Geophysical Research B: Solid Earth
  {\bf 99},  18159  (1994).

\bibitem{lyakhovsky97}
V. Lyakhovsky, Y. Ben-Zion, and A. Agnon, Journal of Geophysical Research B:
  Solid Earth {\bf 102},  27635  (1997).

\bibitem{main00}
I.~G. Main, Geophys. J. Int. {\bf 142},  151  (2000).

\bibitem{lei00}
X.~L. Lei {\it et~al.}, Geophys. Res. Lett. {\bf 27},  1997  (2000).

\bibitem{colombo03}
I.~S. Colombo, I.~G. Main, and M.~C. Forde, J. Mat. Civ. Engrg. {\bf 15},  280
  (2003).

\bibitem{aue98}
J. Aue and J.~T. M.~D. Hosson, J. Mat. Sci. {\bf 33},  5455  (1998).

\bibitem{weiss03}
J. Weiss and D. Marsan, Science {\bf 299},  89  (2003).

\bibitem{krysac98}
L.~C. Krysac and J.~D. Maynard, Phys. Rev. Lett. {\bf 81},  4428  (1998).

\bibitem{lockner91}
D.~A. Lockner {\it et~al.}, Nature {\bf 350},  39  (1991).

\bibitem{shah95}
K.~R. Shah and J.~F. Labuz, Journal of Geophysical Research B: Solid Earth {\bf
  100},  15527  (1995).

\bibitem{labuz98}
J.~F. Labuz and L. Biolzi, Int. J. Solids Structures {\bf 45},  4191  (1998).

\bibitem{zang98}
A.~R. Zang {\it et~al.}, Geophys. J. Int. {\bf 135},  1113  (1998).

\bibitem{feng99}
X.~T. Feng and M. Seto, Geophys. J. Int. {\bf 136},  275  (1999).

\bibitem{hayakawa04}
K. Hayakawa, T. Nakamura, H. Yonezawa, and S. Tanaka, Mater. Transac. {\bf 45},
   3136  (2004).

\bibitem{zietlow98}
W.~K. Zietlow and J.~F. Labuz, Int. J. Rock Mech. Min. Sci. {\bf 35},  291
  (1998).

\bibitem{kanit03}
T. Kanit {\it et~al.}, Int. J. Solids Structures {\bf 40},  3647  (2003).

\bibitem{balankin01}
A.~S. Balankin, O. Susarrey, and A. Bravo, Phys. Rev. E {\bf 64},  066131
  (2001).

\bibitem{rosti06}
J. Rosti and M.~J. Alava, preprint  (2006).

\bibitem{anifrani95}
J.~C. Anifrani, C.~L. Floch, D. Sornette, and B. Souillard, J. Phys. I France
  {\bf 5},  631  (1995).

\bibitem{johansen00}
A. Johansen and D. Sornette, Eur. Phys. J. B {\bf 28},  163  (2000).

\bibitem{suresh98}
S. Suresh, {\em Fatigue of Materials} (Cambridge Univ. Press, Cambridge, 1998).

\bibitem{santucci04}
S. Santucci, L. Vanel, and S. Ciliberto, Phys. Rev. Lett. {\bf 93},  095505
  (2004).

\bibitem{nechad05}
H. Nechad, A. Helmstetter, R.~E. Guerouma, and D. Sornette, Phys. Rev. Lett.
  {\bf 94},  045501  (2005).

\bibitem{zaiser06}
M. Zaiser, Adv. Phys. {\bf to appear},    (2006).

\bibitem{hahner02}
P. H\"ahner, A. Ziegenbein, E. Rizzi, and H. Neuh\"auser, Phys. Rev. B {\bf
  65},  134109  (2002).

\bibitem{deshpande01}
V.~S. Deshpande, A. Needleman, and E. van~der Giessen, Scripta Mat. {\bf 45},
  1047  (2001).

\bibitem{nematnasser99}
S. Nemat-Nasser, Mech. Materials {\bf 31},  493  (1999).

\bibitem{needleman00}
A. Needleman and V. Tvergaard, Int. J. Fracture {\bf 101},  73  (2000).

\bibitem{mahnken02}
R. Mahnken, Int. J. Plasticity {\bf 18},  801  (2002).

\bibitem{bazant02}
Z.~P. Bazant {\it et~al.}, Int. J. Fracture {\bf 113},  345  (2002).

\bibitem{clayton04}
J.~D. Clayton and D.~L. McDowell, Mech. Materials {\bf 36},  799  (2004).

\bibitem{ostojastarzewski05}
M. Ostoja-Starzewski, Int. J. Plasticity {\bf 21},  1119  (2005).

\bibitem{baret02}
J.~C. Baret, D. Vandembroucq, and S. Roux, Phys. Rev. Lett. {\bf 89},  195506
  (2002).

\bibitem{kahng88}
B. Kahng {\it et~al.}, Phys. Rev. B {\bf 37},  7625  (1988).

\bibitem{hansen91}
A. Hansen, E.~L. Hinrichsen, and S. Roux, Phys. Rev. B {\bf 43},  665  (1991).

\bibitem{ootani92}
I. Ootani, Y. Ohashi, K. Ohashi, and M. Fukuchi, J. Phys. Soc. Jpn. {\bf 61},
  1399  (1992).

\bibitem{hristopulos04}
D. Hristopulos and T. Uesaka, Phys. Rev. E {\bf 70},  064108  (2004).

\bibitem{chen02}
S.~H. Chen and T.~C. Wang, Acta Mech. {\bf 157},  113  (2002).

\bibitem{bolander05}
J.~E. Bolander and N. Sukumar, Phys. Rev. B {\bf 71},  094106  (2005).

\bibitem{alava00}
M. Alava, P.~M. Duxbury, C. Moukarzel, and H. Rieger,  in {\em Phase
  Transitions and Critical Phenomena} (   Academic Press, New
  York, 2000), Vol.~18.

\bibitem{kardar86}
M. Kardar, G. Parisi, and Y.~C. Zhang, Phys.Rev.Lett. {\bf 56},  889  (1986).

\bibitem{alava95}
M.~J. Alava, M.~E.~J. Karttunen, and K.~J. Niskanen, Europhysics Letters {\bf
  32},  142  (1995).

\bibitem{li88}
Y.~S. Li and P.~M. Duxbury, Phys. Rev. B {\bf 38},  9257  (1988).

\bibitem{zapperi97b}
S. Zapperi, A. Vespignani, and H.~E. Stanley, Nature {\bf 388},  658  (1997).

\bibitem{nukala05b}
P.~K. V.~V. Nukala and S. Simunovic, Phys. Rev. E {\bf 72},  041919  (2005).

\bibitem{otomar06}
D.~R. Otomar, I.~L. Menezes-Sobrinho, and M.~S. Couto, Phys. Rev. Lett. {\bf
  96},  095501  (2006).

\bibitem{lamaignere96}
L. Lamaignere, F. Carmona, and D. Sornette, Phys. Rev. Lett. {\bf 77},  2738
  (1996).

\bibitem{hansen89C}
A. Hansen, S. Roux, and H.~J. Herrmann, J. Physique {\bf 50},  733  (1989).

\bibitem{sahimi931}
S. Arbabi and M. Sahimi, Phys. Rev. B {\bf 47},  695  (1993).

\bibitem{sahimi932}
M. Sahimi and S. Arbabi, Phys. Rev. B {\bf 47},  703  (1993).

\bibitem{nukala05}
P.~K. V.~V. Nukala, S. Zapperi, and S. Simunovic, Phys. Rev. E {\bf 71},
  066106  (2005).

\bibitem{feng}
S. Feng and P.~N. Sen, Phys. Rev. Lett. {\bf 52},  216  (1984).

\bibitem{hassold89}
G.~N. Hassold and D.~J. Srolovitz, Phys. Rev. B {\bf 39},  9273  (1989).

\bibitem{roux85}
S. Roux and E. Guyon, J. Physique Lett. {\bf 46},  L999  (1985).

\bibitem{herrmann89}
H.~J. Herrmann, A. Hansen, and S. Roux, Phys. Rev. B {\bf 39},  637  (1989).

\bibitem{nukalaijnme}
P.~K. V.~V. Nukala, S. Simunovic, and M.~N. Guddati, Int. J. Numer. Meth.
  Engng. {\bf 62},  1982  (2005).

\bibitem{kantor}
Y. Kantor and I. Webman, Phys. Rev. Lett. {\bf 52},  1891  (1984).

\bibitem{sahimi86}
M. Sahimi and J.~D. Goddard, Phys. Rev. B {\bf 33},  7848  (1986).

\bibitem{monette}
L. Monette and M.~P. Anderson, Modelling Simul. Mater. Sci. Eng. {\bf 2},  53
  (1994).

\bibitem{xpcu}
X.~P. Xu, A. Needleman, and F.~F. Abraham, Modelling Simul. Mater. Sci. Eng.
  {\bf 5},  489  (1997).

\bibitem{needleman}
W. Xuan, W.~A. Curtin, and A. Needleman, Engineering Fracture Mechanics {\bf
  70},  1869  (2003).

\bibitem{vikas1}
V. Tomar, J. Zhai, and M. Zhou, International Journal for Numerical Methods
  inEngineering {\bf 61},  1894  (2004).

\bibitem{vikas2}
V. Tomar and M. Zhou, Engineering Fracture Mechanics {\bf 72},  1920  (2005).

\bibitem{chung01}
J.~W. Chung, J.~T. M.~D. Hosson, and E. van~der Giessen, Phys. Rev. B {\bf 64},
   064202  (2001).

\bibitem{chung02}
J.~W. Chung and J.~T. M.~D. Hosson, Phys. Rev. B {\bf 66},  064206  (2002).

\bibitem{chung02b}
J.~W. Chung, J.~T. M.~D. Hosson, and E. van~der Giessen, Phys. Rev. B {\bf 65},
   094104  (2002).

\bibitem{kunin}
I.~A. Kunin, {\em Elastic Media with Microstructure I} (Springer-Verlag,
  Berlin, Germany,   1982).

\bibitem{frantziskonis}
H. Dai and G. Frantziskonis, Mechanics of Materials {\bf 18},  103  (1994).

\bibitem{frantziskonis1}
G.~N. Frantziskonis, A.~A. Konstantinidis, and E.~C. Aifantis, European Journal
  of Mechanics A/Solids {\bf 20},  925  (2001).

\bibitem{budiansky}
B. Budiansky and R.~J. O'Connell, Int. J. Solids Structures {\bf 12},  81
  (1976).

\bibitem{ted99}
T. Belytschko and T. Black, International Journal for Numerical Methods in
  Engineering {\bf 45},  601  (1999).

\bibitem{hillerborg76}
A. Hillerborg, M. Modeer, and P.~E. Petersson, Cement Concrete Res. {\bf 6},
  773  (1976).

\bibitem{deborst}
R. de~Borst, L.~J. Sluys, H.-B. Muhlhaus, and J. Pamin, Engng. Comput. {\bf
  10},  99  (1993).

\bibitem{pcabot}
G. Pijaudier-Cabot and Z.~P. Bazant, ASCE J. Eng. Mech. {\bf 113},  1512
  (1988).

\bibitem{peerlings}
R. Peerlings, R. de~Borst, W.~A.~M. Brekelmans, and M.~G.~D. Geers,
  International Journal for Numerical Methods in Engineering {\bf 39},  3391
  (1996).

\bibitem{deborst1}
M.~A. Gutierrez and R. de~Borst, International Journal for Numerical Methods in
  Engineering {\bf 44},  1823  (1999).

\bibitem{oliver}
J. Oliver, International Journal for Numerical Methods in Engineering {\bf 39},
   3601  (1996).

\bibitem{gnwells}
G.~N. Wells and L.~J. Sluys, International Journal for Numerical Methods in
  Engineering {\bf 50},  2667  (2001).

\bibitem{simo90}
J.~C. Simo and M.~S. Rifai, International Journal for Numerical Methods in
  Engineering {\bf 29},  1595  (1990).

\bibitem{ted00}
C. Daux {\it et~al.}, International Journal for Numerical Methods in
  Engineering {\bf 45},  1741  (2000).

\bibitem{babuska}
J.~M. Malenk and I. Babuska, Computer Methods in Applied Mechanics and
  Engineering {\bf 39},  289  (1996).

\bibitem{sukumar}
N. Sukumar and J.~H. Prevost, International Journal for Numerical Methods in
  Engineering {\bf 40},  7513  (2003).

\bibitem{budyn}
E. Budyn, G. Zi, N. Moes, and T. Belytschko, International Journal for
  Numerical Methods in Engineering {\bf 61},  1741  (2004).

\bibitem{cafiero97}
R. Cafiero {\it et~al.}, Phys. Rev. Lett. {\bf 79},  1503  (1997).

\bibitem{caldarelli}
G. Caldarelli, R. Cafiero, and A. Gabrielli, Phys. Rev. E {\bf 57},  3878
  (1998).

\bibitem{curtin91}
W.~A. Curtin, J. Am. Ceram. Soc. {\bf 74},  2837  (1991).

\bibitem{curtin91b}
W.~A. Curtin and H. Scher, Phys. Rev. Lett. {\bf 67},  2457  (1991).

\bibitem{curtin92}
W.~A. Curtin and H. Scher, Phys. Rev. B {\bf 45},  2620  (1992).

\bibitem{curtin93}
W.~A. Curtin, J. Mech. Phys. Solids {\bf 41},  217  (1993).

\bibitem{roux93}
S. Roux, A. Hansen, and E.~L. Hinrichsen, Phys. Rev. Lett. {\bf 70},  100
  (1993).

\bibitem{curtin93b}
W.~A. Curtin and H. Scher, Phys. Rev. Lett. {\bf 70},  101  (1993).

\bibitem{curtin93c}
W.~A. Curtin and H. Scher, Phys. Rev. Lett. {\bf 70},  519  (1993).

\bibitem{curtin97a}
W.~A. Curtin and H. Scher, Phys. Rev. B {\bf 55},  12038  (1997).

\bibitem{curtin97b}
W.~A. Curtin, M. Pamel, and H. Scher, Phys. Rev. B {\bf 55},  12051  (1997).

\bibitem{sornette92b}
D. Sornette and C. Vanneste, Phys. Rev. Lett. {\bf 68},  612  (1992).

\bibitem{minozzi03}
M. Minozzi, G. Caldarelli, L. Pietronero, and S. Zapperi, Eur. Phys. J. B {\bf
  36},  203  (2003).

\bibitem{valstos03}
S. Vlastos, E. Liu, I.~G. Main, and X.-Y. Li, Geophys. J. Int. {\bf 152},  649
  (2003).

\bibitem{bouchaud00}
E. Bouchaud {\it et~al.}, J. Mech. Phys. Solids {\bf 50},  1703  (2000).

\bibitem{ramanathan97b}
S. Ramanathan, D. Ertas, and D.~S. Fisher, Phys. Rev. Lett. {\bf 79},  873
  (1997).

\bibitem{cox05}
B.~N. Cox, H. Gao, D. Gross, and D. Rittel, J. Mech. Phys. Solids {\bf 53},
  565  (2005).

\bibitem{marder93}
M. Marder and X. Liu, Phys. Rev. Lett. {\bf 71},  2417  (1993).

\bibitem{karma01}
A. Karma, D. Kessler, and H. Levine, Phys. Rev. Lett. {\bf 87},  045501
  (2001).

\bibitem{aranson00}
I.~S. Aranson, V.~A. Kalatsky, and V.~M. Vinokur, Phys. Rev. Lett. {\bf 85},
  118  (2000).

\bibitem{eastgate02}
L.~O. Eastgate {\it et~al.}, Phys. Rev. E {\bf 65},  036117  (2002).

\bibitem{karma04}
A. Karma and A.~E. Lobkovsky, Phys. Rev. Lett. {\bf 92},  245510  (2004).

\bibitem{rautiainen95}
T.~T. Rautiainen, M.~J. Alava, and K. Kaski, Phys. Rev. E {\bf 51},  R2727
  (1995).

\bibitem{heino96}
P. Heino and K. Kaski, Phys. Rev. B {\bf 54},  6150  (1996).

\bibitem{heino97}
P. Heino and K. Kaski, Phys. Rev. E {\bf 56},  4364  (1997).

\bibitem{moriarty02}
J.~A. Moriarty {\it et~al.}, Journal of Physics: Condensed Matter {\bf 14},
  2825  (2002).

\bibitem{selinger91}
R.~L.~B. Selinger, Z.-G. Wang, W.~M. Gelbart, and A. Ben-Saul, Phys. Rev. A
  {\bf 43},  4396  (1991).

\bibitem{selinger91b}
R.~L.~B. Selinger, Z.-G. Wang, and W.~M. Gelbart, J. Chem. Phys {\bf 95},  9128
   (1991).

\bibitem{wang91}
Z.-G. Wang, U. Landman, R.~L.~B. Selinger, and W.~M. Gelbart, Phys. Rev. B {\bf
  44},  378  (1991).

\bibitem{abraham02}
F.~F. Abraham {\it et~al.}, PNAS {\bf 99},  5777  (2002).

\bibitem{lorenz03}
C.~D. Lorentz and M.~J. Stevens, Phys. Rev. E {\bf 68},  021802  (2003).

\bibitem{buxton04}
G.~A. Buxton and A.~C. Balazs, Phys. Rev. B {\bf 69},  054101  (2004).

\bibitem{seppala04}
E. Seppala, J. Belak, and R. Rudd, Phys. Rev. Lett. {\bf 93},  245503  (2004).

\bibitem{kalia97}
R.~K. Kalia {\it et~al.}, Phys. Rev. Lett. {\bf 78},  2144  (1997).

\bibitem{rountree02}
C.~L. Rountree {\it et~al.}, Annual Review of Materials Research {\bf 32},  377
   (2002).

\bibitem{kalia03}
R.~K. Kalia {\it et~al.}, Int. J. Fracture {\bf 121},  71  (2003).

\bibitem{lu05}
Q. Lu and B. Bhattacharya, Eng. Fract. Mech. {\bf 72},  2037  (2005).

\bibitem{hemmer92}
P.~C. Hemmer and A. Hansen, J. Appl. Mech. {\bf 59},  909  (1992).

\bibitem{hansen94}
A. Hansen and P.~C. Hemmer, Phys. Lett. A {\bf 184},  394  (1994).

\bibitem{sornette92}
D. Sornette, J. Phys. I France {\bf 2},  2089  (1992).

\bibitem{kloster97}
P.~M. Kloster, C. Hemmer, and A. Hansen, Phys. Rev. E {\bf 56},  2615  (1997).

\bibitem{harlow78}
D.~G. Harlow and S.~L. Phoenix, J.\ Composite Mater. {\bf 12},  195  (1978).

\bibitem{smith80}
R.~L. Smith, Proc. R. Soc. London A {\bf 372},  179  (1980).

\bibitem{smith81}
R.~L. Smith and S.~L. Phoenix, J. Appl. Mech. {\bf 48},  75  (1981).

\bibitem{beyerlein96}
I.~J. Beyerlein and S.~L. Phoenix, J. Mech. Phys. Solids {\bf 44},  1997
  (1996).

\bibitem{phoenix97}
S.~L. Phoenix, M. Ibnabdeljalil, and C.-Y. Hui, Int. J. Solids Structures {\bf
  34},  545  (1997).

\bibitem{phoenix92}
S.~L. Phoenix and R. Ra, Acta Metall. Mater. {\bf 40},  2813  (1992).

\bibitem{zhou95}
S.~J. Zhou and W.~A. Curtin, Acta. Metall. Mater. {\bf 43},  3093  (1995).

\bibitem{leath94}
P.~L. Leath and P.~M. Duxbury, Phys. Rev. B {\bf 49},  14905  (1994).

\bibitem{zhang96}
S. Zhang and E. Ding, Phys. Rev. B {\bf 53},  646  (1996).

\bibitem{krajcinovic82}
D. Krajcinovic and M.~A.~G. Silva, Int. J. Solids Structures {\bf 18},  551
  (1982).

\bibitem{fabeny96}
B. Fabeny and W.~A. Curtin, Acta Materialia {\bf 44},  3439  (1996).

\bibitem{hidalgo02}
R.~C. Hidalgo, F. Kun, and H.~J. Herrmann, Phys. Rev. E {\bf 65},  032502
  (2002).

\bibitem{kun03}
F. Kun, R.~C. Hidalgo, H.~J. Herrmann, and K.~F. P{\'a}l, Phys. Rev. E {\bf
  67},  061802  (2003).

\bibitem{kun00}
F. Kun, S. Zapperi, and H.~J. Herrmann, Eur. Phys. J. B {\bf 17},  269  (2000).

\bibitem{raischel06}
F. Raischel, F. Kun, and H.~J. Herrmann, cond-mat/0601290  (2006).

\bibitem{roux00}
S. Roux, Phys. Rev. E {\bf 62},  6164  (2000).

\bibitem{scorretti01}
R. Scorretti, S. Ciliberto, and A. Guarino, Europhys. Lett. {\bf 55},  626
  (2001).

\bibitem{politi02}
A. Politi, S. Ciliberto, and R. Scorretti, Phys. Rev. E {\bf 66},  026107
  (2002).

\bibitem{silveira98}
R. da~Silveira, Phys. Rev. Lett. {\bf 80},  3157  (1998).

\bibitem{pradhan05b}
S. Pradhan, A. Hansen, and P.~C. Hemmer, Phys.Rev.Lett. {\bf 95},  125501
  (2005).

\bibitem{mahesh04}
S. Mahesh and S.~L. Phoenix, Phys. Rev. E {\bf 69},  026102  (2004).

\bibitem{hidalgo02b}
R.~C. Hidalgo, Y. Moreno, F. Kun, and H.~J. Herrmann, Phys. Rev. E {\bf 65},
  046148  (2002).

\bibitem{pradhan05}
S. Pradhan, B.~K. Chakrabarti, and A. Hansen, Phys. Rev. E {\bf 71},  036149
  (2005).

\bibitem{sethna93}
J.~P. Sethna {\it et~al.}, Phys. Rev. Lett. {\bf 70},  3347  (1993).

\bibitem{dahmen96}
K. Dahmen and J.~P. Sethna, Phys. Rev. B {\bf 53},  14872  (1996).

\bibitem{sethna01}
J. Sethna, K.~A. Dahmen, and C.~R. Myers, Nature {\bf 410},  242  (2001).

\bibitem{rundle89}
J.~B. Rundle and W. Klein, Phys. Rev. Lett. {\bf 63},  171  (1989).

\bibitem{golubovic91}
L. Golubovic and S. Feng, Phys. Rev. A {\bf 43},  5223  (1991).

\bibitem{golubovic95}
L. Golubovic and A. Pedrera, Phys. Rev. E {\bf 51},  2799  (1995).

\bibitem{buchel97}
A. Buchel and J.~P. Sethna, Phys. Rev. E {\bf 55},  7669  (1997).

\bibitem{zapperi97}
S. Zapperi, P. Ray, H.~E. Stanley, and A. Vespignani, Phys. Rev. Lett. {\bf
  78},  1408  (1997).

\bibitem{zapperi99}
S. Zapperi, P. Ray, H.~E. Stanley, and A. Vespignani, Phys. Rev. E {\bf 59},
  5049  (1999).

\bibitem{andersen97}
J.~V. Andersen, D. Sornette, and K.~W. Leung, Phys. Rev. Lett. {\bf 78},  2140
  (1997).

\bibitem{barthelemy02}
M. Barthelemy, R. da~Silveira, and H. Orland, Europhys. Lett. {\bf 57},  831
  (2002).

\bibitem{toussaint05}
R. Toussaint and S.~R. Pride, Phys. Rev. E {\bf 71},  046127  (2005).

\bibitem{schmittbuhl95}
J. Schmittbuhl, S. Roux, J.~P. Villotte, and K.~J. Maloy, Phys. Rev. Lett. {\bf
  74},  1787  (1995).

\bibitem{ramanathan97}
S. Ramanathan and D.~S. Fisher, Phys. Rev. Lett. {\bf 79},  877  (1997).

\bibitem{ramanathan98}
S. Ramanathan and D.~S. Fisher, Phys. Rev. B {\bf 58},  6026  (1998).

\bibitem{gao89}
H. Gao and J.~R. Rice, J. Appl. Mech. {\bf 56},  828  (1989).

\bibitem{ertas94}
D. Ertas and M. Kardar, Phys. Rev. E {\bf 49},  R2532  (1994).

\bibitem{rosso02}
A. Rosso and W. Krauth, Phys. Rev. E {\bf 65},  025101  (2002).

\bibitem{schmittbuhl99}
J. Schmittbuhl and J.~P. Villotte, Physica A {\bf 270},  42  (1999).

\bibitem{zapperi00}
S. Zapperi, H.~J. Herrmann, and S. Roux, Eur. Phys. J. B {\bf 17},  131
  (2000).

\bibitem{astrom00}
J.~A. Astr{\"o}m, M.~J. Alava, and J. Timonen, Phys. Rev. E {\bf 62},  2878
  (2000).

\bibitem{schmittbuhl03}
J. Schmittbuhl, A. Hansen, and G.~G. Batrouni, Phys. Rev. Lett. {\bf 90},
  045505  (2003).

\bibitem{bouchbinder04}
E. Bouchbinder, J. Mathiesen, and I. Procaccia, Phys. Rev. Lett. {\bf 92},
  245505  (2004).

\bibitem{ledoussal04}
P.~L. Doussal, K.~J. Wiese, E. Raphael, and R. Golestanian, cond-mat/0411652
  (2004).

\bibitem{rosso03b}
A. Rosso, A.~K. Hartmann, and W. Krauth, Phys. Rev. E {\bf 67},  021602
  (2003).

\bibitem{tang92}
L.~H. Tang and H. Leshhorn, Phys. Rev. A {\bf 45},  R8309  (1992).

\bibitem{buldyrev92}
S.~V. Buldyrev {\it et~al.}, 45 {\bf R8313},    (1992).

\bibitem{prades04}
S. Prades {\it et~al.}, Int. J. Sol. Struct. {\bf 42},  637  (2004).

\bibitem{stauffer}
D. Stauffer and A. Aharony, {\em Introduction to Percolation Theory} (Revised
  second edition, Taylor and Francis,   1994).

\bibitem{grassberger99}
P. Grassberger, Physica A {\bf 262},  251  (1999).

\bibitem{clerc00}
J.~P. Clerc, V.~A. Podolskiy, and A.~K. Sarychev, Eur. Phys. J. B {\bf 15},
  507  (2000).

\bibitem{moukarzel97}
C. Moukarzel, P.~M. Duxbury, and P.~L. Leath, Phys. Rev. Lett. {\bf 78},  1480
  (1997).

\bibitem{feng87}
S. Feng, B.~I. Halperin, and P.~N. Sen, Phys. Rev. B {\bf 35},  1897  (1987).

\bibitem{stenull01}
O. Stenull and H.~K. Janssen, Phys. Rev. E {\bf 64},  056105  (2001).

\bibitem{alava03}
M.~J. Alava and C.~F. Moukarzel, Phys. Rev. E {\bf 67},  056106  (2003).

\bibitem{roux88}
S. Roux, A. Hansen, H.~J. Herrmann, and E. Guyon, J. Stat. Phys. {\bf 52},  237
   (1988).

\bibitem{moukarzel98}
C.~F. Moukarzel, Int. J. Mod. Phys. C {\bf 9},  887  (1998).

\bibitem{duxbury87}
P.~M. Duxbury, P.~L. Leath, and P.~D. Beale, Phys. Rev. B {\bf 36},  367
  (1987).

\bibitem{duxbury86}
P.~M. Duxbury, P.~D. Beale, and P.~L. Leath, Phys. Rev. Lett. {\bf 57},  1052
  (1986).

\bibitem{gumbel}
E.~J. Gumbel, {\em Statistics of Extremes} (Columbia University Press, New
  York,   2004).

\bibitem{sapoval85}
B. Sapoval, M. Rosso, and J.~F. Gouyet, J. Physique Lett. (Paris) {\bf 46},
  L149  (19985).

\bibitem{baldassarri02}
A. Baldassarri, G. Gabrielli, and B. Sapoval, Europhys. Lett. {\bf 59},  232
  (2002).

\bibitem{schmittbuhl04}
J. Schmittbuhl, A. Hansen, and G.~G. Batrouni, Phys. Rev. Lett. {\bf 92},
  049602  (2004).

\bibitem{alava04}
M.~J. Alava and S. Zapperi, Phys. Rev. Lett. {\bf 92},  049601  (2004).

\bibitem{asikainen02}
J. Asikainen, S. Majaniemi, M. Dub{\'e}, and T.~A. Nissila, Phys. Rev. Lett.
  {\bf 65},  052104  (2002).

\bibitem{hansen03}
A. Hansen and J. Schmittbuhl, Phys. Rev. Lett. {\bf 90},  045504  (2003).

\bibitem{nukala04}
P.~K. V.~V. Nukala, S. Simunovic, and S. Zapperi, J. Stat. Mech.  P08001
  (2004).

\bibitem{deArcangelis89}
L. de~Arcangelis, A. Hansen, H.~J. Herrmann, and S. Roux, Phys. Rev. B {\bf
  40},  877  (1989).

\bibitem{krajcinovic05b}
D. Krajcinovic and A. Rinaldi, J. Appl. Mech. {\bf 72},  76  (2005).

\bibitem{sahimi933}
M. Sahimi and S. Arbabi, Phys. Rev. B {\bf 47},  713  (1993).

\bibitem{sahimi98}
M. Sahimi, Physics Reports {\bf 306},  213  (1998).

\bibitem{sahimi90}
S. Arbabi and M. Sahimi, Phys. Rev. B {\bf 41},  772  (1990).

\bibitem{delaplace96}
A. Delaplace, G. Pijaudier-Cabot, and S. Roux, J. Mech. Phys. Solids {\bf 44},
  99  (1996).

\bibitem{deArcangelis88}
L. de~Arcangelis and H.~J. Herrmann, Phys. Rev. B {\bf 39},  2678  (1989).

\bibitem{mikko04}
F. Reurings and M.~J. Alava, Eur. Phys. J. B {\bf 47},  85  (2005).

\bibitem{zapperi_ijf}
S. Zapperi and P.~K. V.~V. Nukala, Int. J. Fracture  (in press, 2006).

\bibitem{raisanen98}
V.~I. R\"ais\"anen, E.~T. Seppala, M.~J. Alava, and P.~M. Duxbury, Phys. Rev.
  Lett. {\bf 80},  329  (1998).

\bibitem{ramstad}
T. Ramstad {\it et~al.}, Phys. Rev. E {\bf 70},  036123  (2004).

\bibitem{bakke03}
J.~O.~H. Bakke {\it et~al.}, Physica Scripta {\bf T106},  65  (2003).

\bibitem{nukala_fuse3d}
P.~K. V.~V. Nukala, S. Zapperi, and S. Simunovic, unpublished  (2006).

\bibitem{duxbury94}
P.~M. Duxbury and P.~L. Leath, Phys. Rev. Lett. {\bf 72},  2805  (1994).

\bibitem{duxbury94b}
P.~M. Duxbury, S.~G. Kim, and P.~L. Leath, Materials Science and Engineering A
  {\bf 176},  25  (1994).

\bibitem{nukala_cluster}
S. Nukala, P.~K. V.~V. Nukala, S. Simunovic, and F. Guess, Phys. Rev. E  in
  press  (2006).

\bibitem{smalley}
R.~F. Smalley, D.~L. Turcotte, and S.~A. Solla, Journal of Geophysical Research
  {\bf 90},  1894  (1985).

\bibitem{newman91}
W.~I. Newman and A.~M. Gabrielov, Int. J. Fracture {\bf 50},  1  (1991).

\bibitem{newman94}
W.~I. Newman {\it et~al.}, Physica D {\bf 77},  200  (1994).

\bibitem{beale88}
P.~D. Beale and D.~J. Srolovitz, Phys. Rev. B {\bf 37},  5500  (1988).

\bibitem{nukalaepjb}
P.~K. V.~V. Nukala and S. Simunovic, Eur. Phys. J. B {\bf 37},  91  (2004).

\bibitem{seppala00}
E.~T. Sepp\"al\"a, V.~I. R\"ais\"anen, and M.~J. Alava, Phys. Rev. E {\bf 61},
  6312  (2000).

\bibitem{bazant05}
Z.~P. Bazant and A. Yavari, Eng. Fract. Mech. {\bf 72},  1  (2005).

\bibitem{hansen91b}
A. Hansen, E.~L. Hinrichsen, and S. Roux, Phys. Rev. Lett. {\bf 66},  2476
  (1991).

\bibitem{zapperi05}
S. Zapperi, P.~K. V.~V. Nukala, and S. Simunovic, Phys. Rev. E {\bf 71},
  026106  (2005).

\bibitem{batrouni98}
G.~G. Batrouni and A. Hansen, Phys. Rev. Lett. {\bf 80},  325  (1998).

\bibitem{raisanen98b}
V.~I. R\"ais\"anen, M.~J. Alava, and R.~M. Nieminen, Phys. Rev. B {\bf 58},
  14288  (1998).

\bibitem{skjetne01}
B. Skjetne, T. Helle, and A. Hansen, Phys. Rev. Lett. {\bf 87},  125503
  (2001).

\bibitem{parisi00}
A. Parisi, G. Caldarelli, and L. Pietronero, Europhys. Lett. {\bf 52},  304
  (2000).

\bibitem{foltin94}
G. Foltin {\it et~al.}, Phys. Rev. E {\bf 50},  R639  (1994).

\bibitem{rosso03}
A. Rosso {\it et~al.}, Phys. Rev. E {\bf 68},  036128  (2003).

\bibitem{zapperi05b}
S. Zapperi, P.~K. V.~V. Nukala, and S. Simunovic, Physica A {\bf 357},  129
  (2005).

\bibitem{meacci04}
G. Meacci, A. Politi, and M. Zei, Europhys. Lett. {\bf 66},  55  (2004).

\bibitem{caldarelli99}
G. Caldarelli, C. Castellano, and A. Petri, Phil. Mag. B {\bf 79},  1939
  (1999).

\bibitem{caldarelli96}
G. Caldarelli, F. di~Tolla, and A. Petri, Phys. Rev. Lett. {\bf 77},  2503
  (1996).

\bibitem{andersen05}
J. Andersen and D. Sornette, cond-mat/0508549  (2005).

\bibitem{templates}
R. Barrett {\it et~al.}, {\em Templates for the Solution of Linear Systems:
  Building Blocks for Iterative Methods, 2nd Edition} (SIAM, Philadelphia,
  1994).

\bibitem{multigrid}
W.~L. Briggs, V.~E. Henson, and S.~F. McCormick, {\em A Multigrid Tutorial, 2nd
  Edition} (SIAM, Philadelphia, 2000).

\bibitem{batrouni86}
G.~G. Batrouni, A. Hansen, and M. Nelkin, Phys. Rev. Lett. {\bf 57},  1336
  (1986).

\bibitem{batrouni88}
G.~G. Batrouni and A. Hansen, Journal of Statistical Physics {\bf 52},  747
  (1988).

\bibitem{luenberger}
D.~G. Luenberger, {\em Linear and Nonlinear Programming, 2nd Edition}
  (Addison-Wesley Publishing Company, Reading, 1984).

\bibitem{procaccia1}
O'Shaughnessy and I. Procaccia, Phys. Rev. Lett. {\bf 54},  455  (1985).

\bibitem{procaccia2}
O'Shaughnessy and I. Procaccia, Phys. Rev. A {\bf 32},  3073  (1985).

\bibitem{tchan88}
T. Chan, SIAM J. Sci. Stat. Comput. {\bf 9},  766  (1988).

\bibitem{chan89}
R.~H. Chan, SIAM J. Matrix Anal. Appl. {\bf 10},  542  (1989).

\bibitem{chan921}
R. Chan and T. Chan, Numerical Linear Algebra Applications {\bf 1},  77
  (1992).

\bibitem{chan96}
R.~H. Chan and M.~K. Ng, SIAM Review {\bf 38},  427  (1996).

\bibitem{tyrty}
E. Tyrtyshnikov, SIAM J. Matrix Anal. Appl. {\bf 13},  459  (1992).

\bibitem{golub96}
G.~H. Golub and C.~F. van Loan, {\em Matrix Computations} (The Johns Hopkins
  University Press,   1996).

\bibitem{wsmp}
A. Gupta, M. Joshi, and V. Kumar, Technical Report RC 22038 (98932), IBM T. J.
  Watson Research Center, Yorktown Heights, NY  (2001).

\bibitem{mumps}
P.~R. Amestoy, I.~S. Duff, J.-Y. L'Excellent, and J. Koster, SIAM Journal on
  Matrix Analysis and Applications {\bf 23},  15  (2001).

\bibitem{taucs}
S. Toledo, D. Chen, and V. Rotkin, http://www.tau.ac.il/~stoledo/taucs/
  (2001).

\bibitem{nukalajpamg1}
P.~K. V.~V. Nukala and S. Simunovic, J. Phys. A: Math. Gen. {\bf 36},  11403
  (2003).

\bibitem{tdavis1}
T.~A. Davis and W.~W. Hager, SIAM J. Matrix Anal. Appl. {\bf 20},  606  (1999).

\bibitem{tdavis2}
T.~A. Davis and W.~W. Hager, SIAM J. Matrix Anal. Appl. {\bf 22},  997  (2001).

\bibitem{nukalajpamg2}
P.~K. V.~V. Nukala and S. Simunovic, J. Phys. A: Math. Gen. {\bf 37},  2093
  (2004).

\end{thebibliography}
\end{document}